\documentclass[a4paper,12pt,twoside]{report}

\usepackage[ngerman,english]{babel}
\usepackage[latin1]{inputenc}

\usepackage[intlimits]{amsmath}
\usepackage{amssymb}

\usepackage{fancyhdr}
\usepackage{booktabs}
\usepackage[bf,small]{caption}
\usepackage{graphicx}
\usepackage{color}

\usepackage{axodraw4j}
\usepackage{pstricks}

\usepackage{mathrsfs}
\usepackage{slashed}

\usepackage{cite}

\usepackage[dvips,unicode]{hyperref}

\title{
\vspace*{-4.55cm}
{\normalsize\normalfont
DESY-THESIS-2011-039\hfill\mbox{}\\
November 2011\hfill\mbox{}\\}
\vspace{2.5cm}
\begin{huge}\textbf{Unstable Gravitino Dark Matter}\end{huge}\\[4mm] \begin{LARGE}\textbf{Prospects for Indirect and Direct Detection}\end{LARGE} \vspace{10mm}}

\author{\begin{LARGE}\textbf{Dissertation}\end{LARGE}\\[8mm] \begin{LARGE}\textbf{zur Erlangung des Doktorgrades}\end{LARGE}\\[8mm] \begin{LARGE}\textbf{des Departments Physik}\end{LARGE}\\[8mm] \begin{LARGE}\textbf{der Universität Hamburg}\end{LARGE}\\[16mm] \begin{Large}\textbf{vorgelegt von}\end{Large}\\[6mm] \begin{Large}\textbf{Michael Grefe}\end{Large}\\[6mm] \begin{Large}\textbf{aus Lüneburg}\end{Large}\\[16mm] \begin{Large}\textbf{Hamburg}\end{Large}\\[6mm] \begin{Large}\textbf{2011}\end{Large}\vspace*{-1cm}}

\date{}

\definecolor{bordeauxred}{rgb}{0.4,0,0}

\hypersetup{
 linktocpage=true,
 bookmarksnumbered=true,
 pdftitle = {Unstable Gravitino Dark Matter - Prospects for Indirect and Direct Detection},
 pdfauthor = {Michael Grefe},
 pdfsubject = {Gravitino Dark Matter},
 pdfkeywords = {Gravitino, Dark Matter, R-Parity Violation, Indirect Detection, Direct Detection},
 colorlinks=true,
 linkcolor=bordeauxred,
 citecolor=bordeauxred,
 urlcolor=bordeauxred
}

\numberwithin{equation}{chapter}
\numberwithin{figure}{chapter}
\numberwithin{table}{chapter}

\providecommand{\unit}[1]{\ensuremath{\mathrm{#1}}}
\providecommand{\usk}{\ensuremath{\,}}
\providecommand{\abs}[1]{\ensuremath{\left\lvert #1\right\rvert }}
\providecommand{\MP}{\ensuremath{M_{\text{Pl}}}}
\DeclareMathOperator{\BR}{BR}
\DeclareMathOperator{\Tr}{Tr}
\DeclareMathOperator{\RE}{Re}
\DeclareMathOperator{\IM}{Im}
\DeclareMathOperator{\e}{e}

\allowdisplaybreaks[1]

\begin{document}

\maketitle

\newpage
\thispagestyle{empty}
\section*{}
\vfill
\begin{tabbing}
\hspace{8.5cm} \= \kill
Gutachter der Dissertation: \> Prof.~Dr.~Laura Covi \\[1mm]
\> Prof.~Dr.~Jan Louis \\[1mm]
\> Prof.~Dr.~Piero Ullio \\[3mm]
Gutachter der Disputation: \> Prof.~Dr.~Laura Covi \\[1mm]
\> Prof.~Dr.~Wilfried Buchm\"uller \\[3mm]
Datum der Disputation: \> 6.~Juli 2011 \\[3mm]
Vorsitzender des Pr\"ufungsausschusses: \> Prof.~Dr.~G\"unter H.~W.~Sigl\\[3mm]
Vorsitzender des Promotionsausschusses: \> Prof.~Dr.~Peter H.~Hauschildt\\[3mm]
Dekan der Fakult\"at f\"ur Mathematik, \> \\
Informatik und Naturwissenschaften: \> Prof.~Dr.~Heinrich Graener
\end{tabbing}

\newpage

\begin{abstract}

We confront the signals expected from unstable gravitino dark matter with observations of indirect dark matter detection experiments in all possible cosmic-ray channels. For this purpose we calculate in detail the gravitino decay widths in theories with bilinear violation of $R$ parity, particularly focusing on decay channels with three particles in the final state. Based on these calculations we predict the fluxes of gamma rays, charged cosmic rays and neutrinos expected from decays of gravitino dark matter. Although the predicted spectra could in principal explain the anomalies observed in the cosmic-ray positron and electron fluxes as measured by PAMELA and Fermi LAT, we find that this possibility is ruled out by strong constraints from gamma-ray and antiproton observations.

Therefore, we employ current data of indirect detection experiments to place strong constraints on the gravitino lifetime and the strength of $R$-parity violation. In addition, we discuss the prospects of forthcoming searches for a gravitino signal in the spectrum of cosmic-ray antideuterons, finding that they are in particular sensitive to rather low gravitino masses. Finally, we discuss in detail the prospects for detecting a neutrino signal from gravitino dark matter decays, finding that the sensitivity of neutrino telescopes like IceCube is competitive to observations in other cosmic ray channels, especially for rather heavy gravitinos.

Moreover, we discuss the prospects for a direct detection of gravitino dark matter via $R$-parity violating inelastic scatterings off nucleons. We find that, although the scattering cross section is considerably enhanced compared to the case of elastic gravitino scattering, the expected signal is many orders of magnitude too small in order to hope for a detection in underground detectors.

\end{abstract}

{\thispagestyle{empty}
\cleardoublepage}

{\selectlanguage{ngerman}
\begin{abstract}

Wir konfrontieren die erwarteten Signale von instabiler Gravitino-Dunkler-Materie mit Beobachtungen von Experimenten zur indirekten Suche nach dunkler Materie in allen m\"oglichen Kan\"alen kosmischer Strahlung. Zu diesem Zweck berechnen wir in allen Einzelheiten die Gravitinozerfallsbreiten in Theorien mit bilinearer Verletzung der $R$-Parit\"at, wobei wir uns insbesondere auf Zerfallskanäle mit drei Teilchen im Endzustand konzentrieren. Auf der Basis dieser Berechnungen sagen wir die Fl\"usse von Gammastrahlen, geladenen kosmischen Strahlen und Neutrinos vorher, die aus Zerf\"allen Gravitino-Dunkler-Materie erwartet werden. Obwohl die vorhergesagten Spektren prinzipiell die beobachteten Anomalien in den von PAMELA und Fermi LAT gemessenen Fl\"ussen von Positronen und Elektronen in der kosmischen Strahlung erkl\"aren k\"onnten, beobachten wir, dass diese M\"oglichkeit auf Grund starker Einschr\"ankungen aus den Beobachtungen von Gammastrahlen und Antiprotonen ausgeschlossen ist.

Daher verwenden wir aktuelle Daten von Experimenten zur indirekten Suche nach dunkler Materie um starke Schranken an die Gravitinolebensdauer und die St\"arke der $R$-Parit\"ats-Verletzung zu setzen. Des Weiteren diskutieren wir die Erfolgsaussichten kommender Suchen nach einem Gravitinosignal im Spektrum von Antideuteronen in der kosmischen Strahlung, wobei wir beobachten, dass sie insbesondere f\"ur niedrige Gravitinomassen empfindlich sind. Schlie{\ss}lich er\"ortern wir ausf\"uhrlich die Erfolgsaussichten ein Neutrinosignal aus Zerf\"allen Gravitino-Dunkler-Materie zu messen, wobei wir feststellen, dass die Sensitivit\"at von Neutrinoteleskopen wie IceCube insbesondere f\"ur den Fall schwerer Gravitinos zu Beobachtungen in anderen Kan\"alen kosmischer Strahlung konkurrenzf\"ahig ist.

Dar\"uber hinaus besprechen wir die Erfolgsaussichten f\"ur eine direkte Entdeckung Gravitino-Dunkler-Materie \"uber $R$-Parit\"ats-verletzende inelastische Streuungen an Nukleonen. Wir stellen fest, dass, obwohl der Streuwirkungsquerschnitt gegen\"uber dem Fall elastischer Gravitino-Streuungen deutlich erh\"oht ist, das erwartete Signal um viele Gr\"o{\ss}enordnungen zu klein ist als dass man eine Entdeckung in unterirdischen Detektoren erhoffen k\"onnte.

\end{abstract}
}

{\thispagestyle{empty}
\cleardoublepage}

\thispagestyle{empty}
\vspace*{4cm}
\hfill
\textit{F\"ur Melanie}

\newpage

{\thispagestyle{empty}
\cleardoublepage}

\pagenumbering{roman} 

\phantomsection 
\addcontentsline{toc}{chapter}{Contents}
\tableofcontents

\newpage

\phantomsection 
\addcontentsline{toc}{chapter}{List of Figures}
\listoffigures

\newpage

\phantomsection 
\addcontentsline{toc}{chapter}{List of Tables}
\listoftables

\newpage 

{\thispagestyle{empty}
\cleardoublepage}

\newpage

\pagenumbering{arabic}

\chapter{Introduction}

Almost eighty years after the first evidence for dark matter in the universe\footnote{Although Fritz Zwicky is generally accepted to be the first who discovered dark matter (in a study of radial velocities of galaxies in the Coma cluster in 1933~\cite{Zwicky:1933gu}) the term 'dark matter' was first used as early as 1922 by James Jeans to describe the missing mass found in a study of vertical motions of stars close to the galactic plane~\cite{Jeans:1922}.} its existence is firmly established on the basis of astrophysical and cosmological observations~\cite{Bergstrom:2000pn,Bertone:2004pz,Einasto:2009zd}. However, the question of the nature of dark matter is still one of the biggest unresolved problems in modern cosmology. While it has been proposed that the observed gravitational effects might be explained by a modification of the theory of gravity~\cite{Milgrom:1983ca} or that the dark matter could be composed of non-luminous astrophysical objects in the halo of galaxies~\cite{Paczynski:1985jf,Griest:1990vu}, both explanations are strongly disfavored by current experimental data.

The best candidates for the dark matter are new elementary particles that obey all observational constraints.\footnote{For recent reviews of the particle explanation for dark matter see for instance~\cite{Bertone:2004pz,Feng:2010gw}.} However, no direct evidence for dark matter particles has been found so far on microscopic scales and thus little is known about their properties like their mass or their interaction strength. It is even unclear if dark matter particles are stable or simply very long-lived. Well-motivated particle dark matter candidates can arise from extensions of the standard model of particle physics, the most thoroughly studied candidates being weakly interacting massive particles. These are particles with weak-scale interactions that are stabilized by a symmetry in the particle physics model. The prototype dark matter candidate of this class is the lightest neutralino in supersymmetric theories that is stabilized in models with conserved $R$ parity~\cite{Jungman:1995df}. Another prominent candidate of this type is the lightest Kaluza--Klein particle in theories with universal extra dimensions that is stabilized by the KK parity~\cite{Servant:2002aq}.

An immense experimental effort is undertaken to search for signatures of particle dark matter candidates: Depending on the details of the particle physics model distinct signatures are expected in high-energetic proton-proton collisions at the Large Hadron Collider at CERN. In addition, weakly interacting massive particles of the dark halo of the Milky Way are expected to elastically scatter off nuclei while traversing the Earth, leading to nuclear recoils that might be observable in low-background underground detectors. Another strategy is to search for exotic contributions from the annihilation or the decay of dark matter particles in the galactic halo to the spectra of cosmic rays.

Only a combination of evidence for particle dark matter from signals at colliders and from astrophysical searches in direct and indirect detection experiments will allow to connect the cosmological observation of dark matter with a particle physics explanation, finally leading to an unambiguous identification of the particle nature of the dark matter in the universe. A large portion of the neutralino parameter space is already excluded by direct detection experiments and collider searches and it is expected that within the next decade either neutralino dark matter will be detected or completely excluded.

This is a strong motivation to study more elusive dark matter candidates. In this thesis we will thus concentrate on the well-motivated case of the gravitino. It arises naturally in locally supersymmetric extensions of the standard model as the gauge fermion of supergravity~\cite{Freedman:1976xh,Deser:1976eh}. Depending on the mechanism of supersymmetry breaking, the gravitino can be the lightest supersymmetric particle and thus represent the dark matter of the universe~\cite{Pagels:1981ke}. Due to its extremely weak interactions that are suppressed by the Planck scale it appears to be one of the most elusive dark matter candidates with respect to the prospects for its experimental detection.

The existence of the gravitino in the particle spectrum leads to several cosmological problems, the most severe being that late gravitino decays are in conflict with the successful predictions of primordial nucleosynthesis for the abundances of light elements~\cite{Weinberg:1982zq}. Since gravitinos are produced in thermal scatterings after the end of the inflationary phase in the early universe~\cite{Bolz:2000fu}, compatibility with big bang nucleosynthesis puts strong upper limits on the reheating temperature of the universe~\cite{Kawasaki:1994af}. However, the observation of small, nonvanishing neutrino masses strongly supports the mechanism of thermal leptogenesis as the origin of the baryon asymmetry in the universe~\cite{Fukugita:1986hr}, thus requiring a high value for the reheating temperature~\cite{Davidson:2002qv}. For this reason, there is an apparent conflict between supersymmetry, predicting the existence of the gravitino, and the successful predictions of big bang nucleosynthesis, which also require a mechanism for baryogenesis.

It has been proposed that this problem could be solved if the gravitino is the lightest supersymmetric particle and thus a stable dark matter candidate~\cite{Bolz:1998ek}. However, strong constraints arise also from possible late decays of the next-to-lightest supersymmetric particle which in general still lead to conflicts with big bang nucleosynthesis~\cite{Kawasaki:2008qe}. By contrast, theories with a slight violation of $R$ parity naturally lead to a cosmological scenario that is consistent with thermal leptogenesis and all bounds from big bang nucleosynthesis~\cite{Buchmuller:2007ui}. In this case the next-to-lightest supersymmetric particle decays mainly via $R$-parity breaking interactions into standard model particles well before the time of nucleosynthesis. The decays of the gravitino are then suppressed by the Planck scale and additionally by the small amount of $R$-parity breaking, predicting a gravitino lifetime that exceeds the age of the universe by many orders of magnitude~\cite{Takayama:2000uz}. Therefore, the gravitino remains a perfectly viable candidate for the dark matter in the universe.

An intriguing feature of this scenario is that the gravitino is not that elusive anymore but in contrast exhibits a rich phenomenology. Since the gravitino is unstable it might be observed via its decays in the late universe~\cite{Takayama:2000uz}. Due to the large density of dark matter particles in the universe, which is five times higher than that of ordinary matter, the decay signal might be strong enough to be observed in the isotropic diffuse flux of gamma rays~\cite{Buchmuller:2007ui,Takayama:2000uz,Bertone:2007aw,Ibarra:2007wg,Ishiwata:2008cu,Buchmuller:2009xv,Bomark:2009zm}, in the fluxes of cosmic-ray antimatter~\cite{Ibarra:2008qg,Ishiwata:2008cu,Buchmuller:2009xv,Bomark:2009zm} or in the flux of neutrinos~\cite{Bomark:2009zm,Covi:2008jy}, irrespective of the extremely long lifetime of the gravitino. In addition, this model predicts spectacular observational consequences for collider experiments: Since the decay width of the next-to-lightest supersymmetric particle is suppressed by $R$-parity violating interactions, it might be observable as a long-lived particle leading to displaced vertices or, in the specific case of a charged next-to-lightest supersymmetric particle, to a charged particle track leaving the detector~\cite{Buchmuller:2004rq,Ellis:2006vu,Bobrovskyi:2010ps,Lola:2008bk}.

In particular the field of indirect dark matter searches via cosmic-ray signals has been very active in the past years. Deviations from the expected cosmic-ray spectra from astrophysical processes have been reported already several years ago for the extragalactic diffuse gamma-ray signal as derived from EGRET data~\cite{Strong:2004ry} and for the cosmic-ray positron fraction as measured by the HEAT instrument~\cite{Barwick:1997ig}. These observations led to several studies interpreting the anomalous spectra as a hint for dark matter annihilations or decays~\cite{Bertone:2007aw,Ibarra:2007wg,Ibarra:2008qg,Ishiwata:2008cu,Baltz:1998xv,Kane:2001fz,Baltz:2001ir,deBoer:2005tm}.

However, in particular the observation of a steep rise in the positron fraction at energies above 10\,GeV by PAMELA~\cite{Adriani:2008zr}, confirming the result from HEAT, and the precise measurement of the absolute cosmic-ray electron plus positron flux by Fermi LAT~\cite{Abdo:2009zk} stimulated a lot of activity in the field. A multitude of studies for annihilating~\cite{ArkaniHamed:2008qn,Cholis:2008qq,Nomura:2008ru,Feldman:2008xs,Fox:2008kb,Ishiwata:2008cv,Nezri:2009jd} and decaying~\cite{Buchmuller:2009xv,Ishiwata:2008cv,Yin:2008bs,Chen:2008md,Hamaguchi:2008rv,Ibarra:2008jk,Nardi:2008ix,Hamaguchi:2008ta,Chen:2009ew,Ibarra:2009bm,Shirai:2009fq} dark matter candidates was published, in many cases trying to simultaneously explain all deviations from astrophysical background expectations. One should keep in mind, though, that also astrophysical sources like supernova remnants or pulsars could explain the observations~\cite{Hooper:2008kg,Profumo:2008ms,Shaviv:2009bu,Blasi:2009bd,Barger:2009yt,Grasso:2009ma}.

On the other hand, recent data from Fermi LAT do not confirm an excess in the extragalactic diffuse gamma-ray spectrum as claimed based on EGRET data~\cite{Abdo:2010nz}. In addition, searches for photon lines in the Fermi LAT data have been negative so far, thus providing strong limits on dark matter annihilations or decays predicting monoenergetic photons in the final state~\cite{Abdo:2010nc}. Another interesting channel for the indirect detection of dark matter are neutrinos as they provide directional information and possibly an independent confirmation of an exotic contribution to cosmic rays. Several studies of neutrino signals from annihilating and decaying dark matter candidates have been presented in the literature~\cite{Covi:2008jy,Hisano:2008ah,Liu:2008ci,Hisano:2009fb,Spolyar:2009kx,Buckley:2009kw,Liu:2009ac,Mandal:2009yk}.\smallskip

In the present work we will therefore employ a multi-messenger approach of indirect searches and confront the expected signals from unstable gravitino dark matter with indirect detection data in all possible cosmic ray channels. The intention of this strategy is to place strong constraints on the lifetime of the gravitino and also on the amount of bilinear $R$-parity breaking. For this purpose we will revisit and extend the calculation of gravitino decay channels in scenarios with bilinear $R$-parity violation. As a first step we will re-evaluate the mixing of leptons with gauginos and higgsinos induced by $R$-parity violating operators and find new analytical approximate formulae for the mixing parameters governing the decay of gravitino dark matter particles. Equipped with these results we will present a detailed calculation of gravitino decay widths, upgrading previously obtained two-body decay results by the addition of a set of Feynman diagrams that were neglected in earlier calculations. One of the main results of this work is the extension of the gravitino decay width calculation to include three-body gravitino decay channels for gravitino masses below the threshold for the on-shell production of massive gauge bosons. These contributions are important for the phenomenology of low mass gravitinos as first pointed out in~\cite{Choi:2010xn}. Using the spectra of stable final state particles produced in gravitino decays we will then predict the spectra of gamma rays, positrons, electrons, antiprotons and neutrinos, and compare them to current observations of cosmic rays.

Beyond that, we will study in detail future prospects for the detection of a neutrino signal from gravitino decays. Neutrinos are not directly observed in neutrino telescopes but only via their weak interactions with the material inside or close to the detector. We will discuss the different event topologies that can arise and predict the corresponding signal spectra at detector level. Taking into account the energy resolution of different detection channels we will argue which of the channels provides the best sensitivity to signals from gravitino decays.

In addition, we want to discuss for the first time the antideuteron signal expected from gravitino decays. In~\cite{Donato:1999gy} it was first noted that antideuterons are a convenient cosmic-ray channel for the detection of dark matter signals as the expected astrophysical background can be very low compared to the signal expectation, in particular at the low-energetic end of the spectrum. In the derivation of the antideuteron spectrum from gravitino decays we will employ a Monte Carlo treatment as it was found that the conventional spherical approximation of the coalescence model for antideuteron formation leads to erroneous results~\cite{Kadastik:2009ts}. As no antideuterons have been observed in cosmic rays so far we will present future prospects for the sensitivity of the antideuteron channel based on the projected sensitivity regions of forthcoming antideuteron experiments.

Moreover, we will shortly discuss the prospects for a direct detection of gravitino dark matter. In contrast to the case of weakly interacting massive particles, no observable signal is expected from elastic gravitino--nucleon scatterings due to the Planck-scale suppression of gravitino couplings to matter. This situation could change in the case of broken $R$ parity. In this case the gravitino is expected to scatter inelastically off nucleons with a cross section that is considerably enhanced compared to that of elastic scatterings as it is suppressed by a lower order of the Planck scale. In order to study this effect quantitatively, we will calculate the gravitino--nucleon cross sections via Higgs and $Z$ boson exchange. In addition, we observe the intriguing feature that the gravitino can scatter off nucleons via photon exchange, a channel that in general is not available for weakly interacting massive particles. We expect that the scattering via photon exchange leads to a strong enhancement of the cross section due to the massless photon propagator.\smallskip

Let us summarize the structure of this thesis: In the next chapter we will shortly review the basics of big bang cosmology as well as evidence and constraints for dark matter from cosmological observations. In Chapter~\ref{susysugra} we will introduce supersymmetry and supergravity and discuss the effects of bilinear $R$-parity violation. The gravitino will be discussed in Chapter~\ref{gravitino}: After a short review of the field-theoretical description of the gravitino, we will summarize its cosmological implications. In the main part of the chapter we will study in detail the decay channels of the unstable gravitino, discuss the branching ratios of the different decay channels and determine the spectra of stable final state particles produced in gravitino decays. Chapter~\ref{indirectdetection} contains an extensive discussion of indirect searches for gravitino dark matter, covering the expected signals of gamma rays, positrons, antiprotons, antideuterons and neutrinos. In Chapter~\ref{gravitinoDD} we discuss the cross sections for inelastic gravitino--nucleon scattering and the prospects of detecting gravitino dark matter in direct detection experiments. Finally, we will present our conclusions and an outlook for future directions of investigating gravitino dark matter.

The appendices contain supporting material on the calculations in this work: Appendix~\ref{constants} summarizes the physical constants used in calculations throughout the present thesis. Appendix~\ref{notation} fixes our notation and contains useful formulae for the calculation of matrix elements. In Appendices~\ref{feynmanrules} and~\ref{kinematics} we present, respectively, the Feynman rules and the kinematics needed for the calculation of gravitino decays and scattering processes. Appendix~\ref{gravitinodecay} contains the complete calculation of the gravitino decay widths that are discussed in Chapter~\ref{indirectdetection}, while Appendix~\ref{gravitinoscattering} contains the calculation of the gravitino--nucleon scattering cross sections that are discussed in Chapter~\ref{gravitinoDD}.\smallskip

The results for the detection of neutrinos from gravitino dark matter decays presented in Section~\ref{neutrinos} are based on our study of neutrino signals for generic decay channels of scalar and fermionic dark matter particles that is published in~\cite{Covi:2009xn}. Instead of directly presenting the results of this publication we decided to redo the analysis for the specific case of unstable gravitino dark matter exactly along the lines of the published work, only updating the analysis to the most recent status of neutrino telescopes and their observations.

The discussion of gravitino three-body decays in Section~\ref{gravdecay} and the discussion of signals for the indirect detection of gravitino dark matter in Chapter~\ref{indirectdetection}, in particular the prospects for antideuteron searches, are part of an ongoing project in collaboration with Laura Covi and Gilles Vertongen~\cite{Covi:2011a}.

Similarly, the discussion of the prospects for a direct detection of gravitino dark matter in scenarios with bilinear $R$-parity violation in Chapter~\ref{gravitinoDD} is part of an ongoing project in collaboration with Laura Covi~\cite{Covi:2011b}.

\chapter{Cosmology and Dark Matter}
\label{cosmology}
In this introductory chapter we want to present the basic picture of big bang cosmology, shortly discuss the astrophysical and cosmological evidence for dark matter and summarize the constraints on the dark matter properties coming from cosmological observations. More comprehensive reviews on these topics can be found for instance in~\cite{Bertone:2004pz,Nakamura:2010zzi,Trodden:2004st}.

\section{Big Bang Cosmology}

In this section we will first discuss the dynamics of our expanding universe and then highlight several important stages of its thermal history.

\subsubsection*{Dynamics of the Universe}

The framework for the study of the dynamics of our universe is the theory of general relativity~\cite{Einstein:1916vd}. Einstein's equations,
\begin{equation}
 R_{\mu\nu}-\frac{1}{2}\,g_{\mu\nu}\,R=8\,\pi\,G_NT_{\mu\nu}+\Lambda\,g_{\mu\nu}\,,
\end{equation}
express the connection of the geometry of space-time and the energy content of the universe. In this expression, $R_{\mu\nu}$ and $R$ are the Ricci tensor and Ricci scalar, respectively, while $g_{\mu\nu}$ is the space-time metric, $T_{\mu\nu}$ is the energy-momentum tensor describing the energy content of the universe and $\Lambda$ is a cosmological constant. In order to solve this set of coupled equations it is reasonable to employ symmetries of the universe: Measurements of the Cosmic Microwave Background (CMB) show that the universe is highly isotropic. In addition, galaxy surveys indicate that the universe is homogeneous on large scales ($\mathcal{O}(100)\,$Mpc).

The most general space-time metric compatible with the isotropy and homogeneity of the universe is the Friedmann--Lema\^\i tre--Robertson--Walker metric~\cite{Friedmann:1924bb,Lemaitre:1927,Robertson:1935,Walker:1937}. The line element can be written as
\begin{equation}
 g_{\mu\nu}\,dx^{\mu}\,dx^{\nu}=ds^2=dt^2-a^2(t)\left[ \frac{dr^2}{1-k\,r^2}+r^2\left( d\theta^2+\sin^2\theta\,d\phi^2\right) \right] ,
\end{equation}
where $a(t)$ is the scale factor, $r$, $\theta$ and $\phi$ are the comoving spatial coordinates and the constant $k$ characterizes the spatial curvature of the universe: $k=-1$ corresponds to an open, $k=0$ to a flat and $k=+1$ to a closed universe.

The energy-momentum tensor must be diagonal and have equal spatial components in order to be compatible with the symmetries of the universe. The simplest realization of such an energy-momentum tensor is that of a perfect fluid. In its rest frame it reads
\begin{equation}
 T_{\mu\nu}=
 \begin{pmatrix}
  \rho & 0 & 0 & 0 \\
   0 & p & 0 & 0 \\
   0 & 0 & p & 0 \\
   0 & 0 & 0 & p
 \end{pmatrix},
\end{equation}
where $\rho$ and $p$ are, respectively, the energy density and the isotropic pressure of the fluid. For arbitrary four-velocities $v^\mu$ of the perfect fluid this expression can be generalized to the form
\begin{equation}
 T_{\mu\nu}=\left( \rho+p\right) v_\mu v_\nu-p\,g_{\mu\nu}.
\end{equation}

Solving Einstein's equations with the above assumptions results in the Friedmann equation~\cite{Friedmann:1924bb}
\begin{equation}
 H^2\equiv\left( \frac{\dot{a}}{a}\right) ^2=\frac{8\,\pi\,G_N}{3}\sum_i\rho_i-\frac{k}{a^2}+\frac{\Lambda}{3}
\end{equation}
and the acceleration equation 
\begin{equation}
 \dot{H}+H^2=\frac{\ddot{a}}{a}=-\frac{4\,\pi\,G_N}{3}\sum_i\left( \rho_i+3\,p_i\right) +\frac{\Lambda}{3}\,,
\end{equation}
where a dot denotes a derivative with respect to time. These equations determine the evolution of the scale factor and thus the dynamics of the universe. In these equations we introduced the Hubble parameter $H=\dot{a}/a$ that characterizes the expansion rate of the universe.\footnote{The Hubble parameter is named after Edwin Hubble, who first observed the expansion of the universe in a study of the relation of distance and redshift of galaxies~\cite{Hubble:1929ig}.} The energy of photons and other relativistic particles decreases during their propagation through the expanding universe and thus the spectra of distant astrophysical objects appear redshifted. This is used to describe the expansion of the universe in terms of the redshift parameter $z$ that is defined as
\begin{equation}
 1+z\equiv\frac{\lambda(t_0)}{\lambda(t_e)}=\frac{a(t_0)}{a(t_e)}\,,
 \label{redshift}
\end{equation}
where $\lambda(t_e)$ and $a(t_e)$ are, respectively, the wavelength and scale factor at emission, and $\lambda(t_0)$ and $a(t_0)$ are their present-day values. The present-day scale factor is usually chosen as $a(t_0)=1$.

From the Friedmann equation one can derive a critical density $\rho_c$ that corresponds to the energy density of a flat universe:
\begin{equation}
 \rho_c=\frac{3\,H^2}{8\,\pi\,G_N}\simeq 1.05\times 10^{-5}\,h^2\,\text{GeV}\,\text{cm}^{-3},
\end{equation}
where $h$ is the scaled Hubble parameter that is defined by
\begin{equation}
 H\equiv100\,h\,\text{km}\,\text{s}^{-1}\,\text{Mpc}^{-1}.
\end{equation}
The critical density can then be used to define a density parameter,
\begin{equation}
 \Omega_i=\frac{\rho_i}{\rho_c}\,,
\end{equation}
that can be employed to rewrite the Friedmann equation in the form of a sum rule:
\begin{equation}
 1=\sum_i\Omega_i-\frac{k}{a^2H^2}\equiv\Omega_\text{tot}-\frac{k}{a^2H^2}\,,
\end{equation}
where the cosmological constant is considered as a part of the total energy density. Written in this form, the notions of an open, a flat and a closed universe correspond, respectively, to $\Omega_\text{tot}<1$, $\Omega_\text{tot}=1$ and $\Omega_\text{tot}>1$. The energy content of the universe can be composed of several forms of energy, characterized by the equation of state
\begin{equation}
 p_i=w_i\,\rho_i\,.
\end{equation}
Relativistic particles and radiation have a pressure that contributes with one third of their energy density, thus $w_r=1/3$, while non-relativistic particles have negligible pressure and therefore $w_m=0$. The cosmological constant $\Lambda$ can be described as an energy component with negative pressure: $w_{\Lambda}=-1$.

From the Friedmann equation and the acceleration equation one can derive the continuity equation
\begin{equation}
 \dot{\rho}_i+3\,H\left( \rho_i+p_i\right) =\dot{\rho}_i+3\,\frac{\dot{a}}{a}\,\rho_i\left( 1+w_i\right) =0\,,
\end{equation}
which is equivalent to the covariant conservation of the energy-momentum tensor. By integration of the continuity equation one can derive the dependence of the energy density on the redshift parameter:
\begin{equation}
 \rho_i\propto a^{-3\left( 1+w_i\right) }=\left( 1+z\right) ^{3\left( 1+w_i\right) }. 
\end{equation}
The energy density of non-relativistic matter decreases with $\left( 1+z\right) ^3$ due to the dilution of its number density during the expansion of the universe. By contrast, the energy density of relativistic matter decreases with an additional factor of $\left( 1+z\right) $ because of the energy redshift in the expanding universe. The cosmological constant is equivalent to an intrinsic energy of the vacuum and is independent of the dynamics of the universe. These different behaviors lead to a change of the fraction that the individual energy forms contribute to the total energy density with time. Thus the early universe is dominated by radiation, while at later times the universe is dominated by matter and finally by vacuum energy.\footnote{The existence of a significant vacuum energy contribution to the total energy density of the universe at late times was first deduced from observations of supernovae indicating an accelerated expansion of our universe~\cite{Riess:1998cb,Perlmutter:1998np}.}

\subsubsection*{The History of the Thermal Universe}

During the phase of radiation domination the universe is mainly filled with relativistic particles in thermal equilibrium. From thermodynamics we know that the total energy density is then given by
\begin{equation}
 \rho=\frac{\pi^2}{30}\,g_*T^4,
\end{equation}
where $g_*$ is the effective number of relativistic degrees of freedom, being defined as
\begin{equation}
 g_*=\sum_{\text{bosons}}g_i\left( \frac{T_i}{T}\right) ^4+\frac{7}{8}\sum_{\text{fermions}}g_i\left( \frac{T_i}{T}\right) ^4,
\end{equation}
and the factors $g_i$ are the numbers of degrees of freedom of the individual relativistic boson and fermion species. This energy density determines the Hubble parameter and thus also the expansion rate of the early universe. In an adiabatically expanding universe the conserved quantity is the comoving entropy density:
\begin{equation}
 \frac{d(s\,a^3)}{dt}=0\,,\qquad\text{where}\qquad s=\frac{\rho+p}{T}=\frac{2\,\pi^2}{45}\,g_{*S}T^3
\end{equation}
for relativistic particles. In the definition of the entropy density $g_{*S}$ is the effective number of relativistic degrees of freedom with respect to entropy:
\begin{equation}
 g_{*S}=\sum_{\text{bosons}}g_i\left( \frac{T_i}{T}\right) ^3+\frac{7}{8}\sum_{\text{fermions}}g_i\left( \frac{T_i}{T}\right) ^3.
\end{equation}
One can then derive that in the thermal universe the temperature is inversely proportional to the scale factor as long as $g_{*S}$ remains constant:
\begin{equation}
 T\propto g_{*S}^{-1/3}\,a^{-1}=g_{*S}^{-1/3}\left( 1+z\right) \,.
\end{equation}
Whenever a particle species vanishes from the thermal plasma due to annihilation or decay processes, its entropy is transferred to the remaining particles in the thermal bath and their temperature is slightly increased. However, this effect is typically an $\mathcal{O}(1)$ correction. Therefore, one can approximate the characteristic temperature of the thermal plasma using the expression
\begin{equation}
 T=T_0\left( 1+z\right) ,
\end{equation}
where $T_0$ is the present-day temperature of the cosmic background radiation. Thus the temperature of photons and other relativistic particles decreases with the expansion of the universe according to their redshift.

\begin{figure}[t]
 \centering
 \includegraphics[scale=0.65]{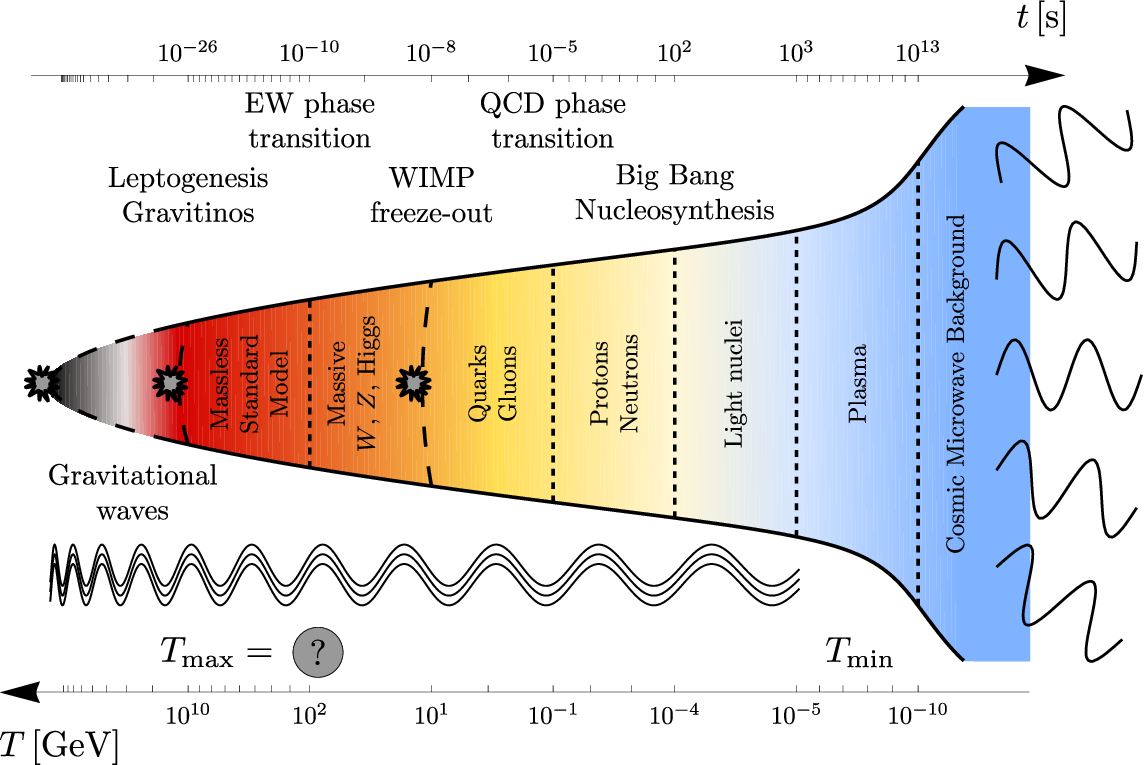}
 \caption[Timeline of the thermal early universe.]{Timeline of the thermal early universe starting from its beginning at the end of inflation. Several stages of the evolution are highlighted, for instance the processes of leptogenesis and gravitino production right after the reheating of the universe, and the nucleosynthesis of light elements as well as the emission of the cosmic microwave background that give strong observational constraints on the cosmological scenario. Courtesy of Kai Schmitz.}
 \label{timeline}
\end{figure}

In the following we want to discuss a few important stages during the early universe, namely inflation, baryogenesis, big bang nucleosynthesis and the emission of the cosmic microwave background. A timeline summarizing the history of the thermal universe is shown in Figure~\ref{timeline}.

\paragraph{Inflation}

A cosmological model starting with a hot thermal phase after the big bang requires very specific initial conditions in order to explain the observed flatness and homogeneity of the universe. In an expanding universe dominated by radiation or matter the initial value of the total energy density must be extremely fine-tuned in order to obtain the observed value of $\Omega_{\text{tot}}\simeq1$ today. This dilemma is called the flatness problem. In addition, observations of the cosmic microwave background indicate that the universe was highly isotropic before structure formation although the observed cosmic microwave background sky is many orders of magnitude larger than the causal horizon at the time of photon decoupling. Thus the homogeneity of the temperature could not be achieved by physical interactions. Instead, it could only be achieved by extremely fine-tuned initial conditions. This dilemma is known as the horizon problem.

Both problems can be solved by the introduction of an inflationary phase in the early universe where $w_{\text{eff}}\simeq-1$ and therefore the universe is expanding exponentially~\cite{Guth:1980zm,Albrecht:1982wi,Linde:1983gd}: During inflation, $\Omega_{\text{tot}}=1$ becomes an attractor solution that is approached regardless of the initial conditions. After the inflationary phase the universe is so close to being flat that the observed value of the total energy density is very close to one even during the phases of radiation and matter domination. The isotropy of the cosmic microwave background sky can also be explained by this mechanism: The entire observed universe had initially been contained in a small causally connected region that expanded tremendously during the phase of inflation.

Such an inflationary phase can be realized by a scalar inflaton field that enters a so-called slow-roll phase. Apart from solving the above issues, inflation theories predict large-scale density perturbations that arise from quantum fluctuations of the inflaton field. These are observed in the form of temperature anisotropies in the CMB and finally lead to the formation of structures like galaxies and stars in the universe.

After inflation, the density of all particles that initially filled the universe is diluted. However, the decay of the inflaton field at the end of the inflationary phase produces a hot thermal plasma of elementary particles. This process of entropy production is known as the reheating of the universe, and the equilibrium temperature of the thermal plasma right after inflation is therefore called the reheating temperature. After this phase the universe is described by standard thermal cosmology.

\paragraph{Baryogenesis}

A fundamental problem in cosmology and particle physics is the origin of the baryon asymmetry, \textit{i.e.} the question why there is more matter than antimatter in the universe. Any initial baryon asymmetry in the universe is diluted by the exponential expansion during inflation. Thus it is necessary to generate a baryon asymmetry dynamically after inflation. However, this process of baryogenesis is only possible when the three Sakharov conditions are satisfied~\cite{Sakharov:1967dj}:
\begin{itemize}
 \item baryon number ($B$) violation,
 \item $C$-symmetry and $CP$-symmetry violation,
 \item departure from thermal equilibrium.
\end{itemize}
Several models have been proposed for baryogenesis: The first model was the production of a baryon asymmetry in the out-of-equilibrium decays of superheavy particles in grand unified theories (GUTs)~\cite{Yoshimura:1978ex,Ignatiev:1978uf}. However, it was found out that the generated baryon asymmetry is washed out by non-perturbative sphaleron processes that are effective at temperatures above the electroweak scale and violate the linear combination $B+L$ of baryon and lepton number but conserve the combination $B-L$~\cite{'tHooft:1976up,Manton:1983nd}.

This observation led to the proposal of electroweak baryogenesis, where the baryon asymmetry is generated in $B+L$-violating sphaleron processes~\cite{Kuzmin:1985mm}. However, this scenario is also disfavored since it only works if the electroweak phase transition is of first order which is not the case for the standard model of particle physics and most supersymmetric extensions. One viable model of baryogenesis is based on the coherent production of baryons in the decay of scalar supersymmetric partners of leptons and baryons. This model is known as Affleck--Dine baryogenesis~\cite{Affleck:1984fy}.

The currently favored model to solve the problem of baryon asymmetry is baryogenesis via thermal leptogenesis~\cite{Fukugita:1986hr}. In this model a lepton asymmetry is generated in $B-L$-violating out-of-equilibrium decays of heavy Majorana neutrinos that is then partly transferred into a baryon asymmetry via sphaleron processes. This mechanism is closely related to the problem of neutrino masses, since heavy Majorana neutrinos can also explain small nonvanishing masses for the light neutrinos via the seesaw mechanism~\cite{Yanagida:1980xy,GellMann:1980vs}. The observation of neutrino oscillations and thus nonvanishing neutrino masses strongly supports the existence of heavy Majorana neutrinos and therefore also the mechanism of thermal leptogenesis. One drawback of this model might be that a high reheating temperature of $T_R\gtrsim 10^9\,$GeV is required in the early universe in order to produce the observed amount of baryon asymmetry~\cite{Davidson:2002qv, Buchmuller:2004nz}.

\paragraph{Primordial Nucleosynthesis}

Big Bang nucleosynthesis (BBN), \textit{i.e.} the production of light elements during the first minutes of the cosmological evolution, is one of the best tools to study the early universe since it takes place at temperatures $T\simeq 1$--0.1\,MeV and is thus based on well-understood standard model physics~\cite{Alpher:1948ve,Wagoner:1966pv}. Therefore, up to now BBN provides the deepest reliable probe of the early universe.

At temperatures above roughly 1\,MeV neutrons and protons are in thermal equilibrium due to weak interactions. When the temperature falls below this value, the neutrons leave chemical equilibrium and the ratio of the neutron and proton number densities is fixed due to the Boltzmann factor at a value of
\begin{equation}
 \frac{n_n}{n_p}=\exp\left( -\frac{m_n-m_p}{T_{\text{fr}}}\right) \simeq \frac{1}{6}\,.
\end{equation}
The exact value of the freeze-out time $T_{\text{fr}}$ depends on the number of relativistic degrees of freedom $g_*(T_{\text{fr}})$ and is therefore sensitive, for instance, to the number of relativistic neutrino species. After chemical decoupling, protons and neutrons could start to form deuterium. However, as a result of photo-dissociation processes due to the large number of photons in the thermal plasma, the deuterium production becomes only efficient at temperatures below 0.1\,MeV. This delay of the production of light elements is known as the deuterium bottleneck. Once deuterium is efficiently produced, virtually all neutrons combine with protons to form $^4\text{He}$ almost independent of the nuclear reaction rates. However, by that time the neutron-to-proton ratio has slightly decreased to about $1/7$ due to neutron decay. The relative abundance of $^4$He by weight can then easily be estimated: 
\begin{equation}
 Y_p\equiv\frac{\rho_{^4\!\text{He}}}{\rho_{p}+\rho_{^4\!\text{He}}}\simeq\frac{2\,n_n}{n_p+n_n}=\frac{2\,n_n/n_p}{1+n_n/n_p}\approx 25\,\%\,.
\end{equation}
By contrast, the calculation of the other light element abundances depends on the details of nuclear interactions and in particular on the value of the baryon-to-photon ratio $\eta\equiv n_b/n_{\gamma}$ that determines at which temperature the process of nucleosynthesis can start after the deuterium bottleneck. The latter is practically the only free parameter of the theory of BBN.

\begin{figure}[t]
 \centering
 \includegraphics[scale=0.782]{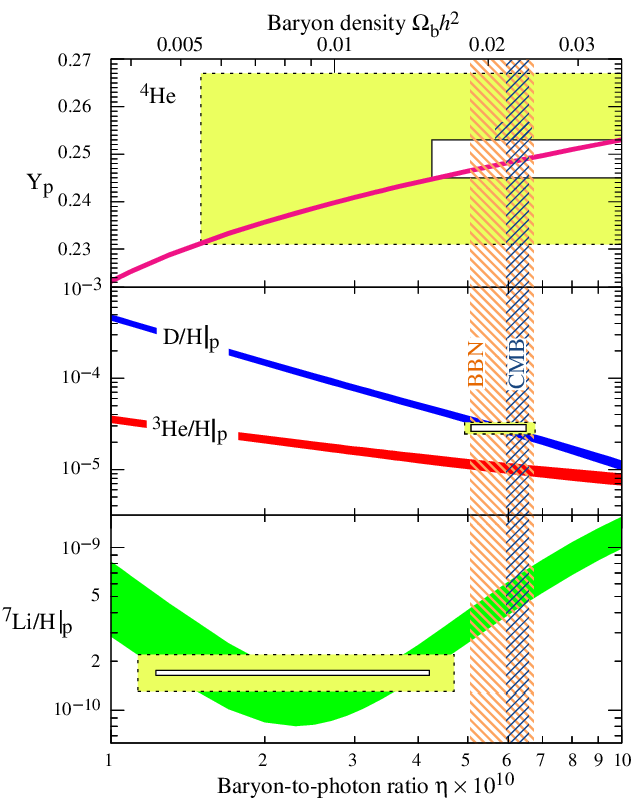}
 \hspace{5mm}
 \includegraphics[scale=0.6,bb=0 -4 262 474]{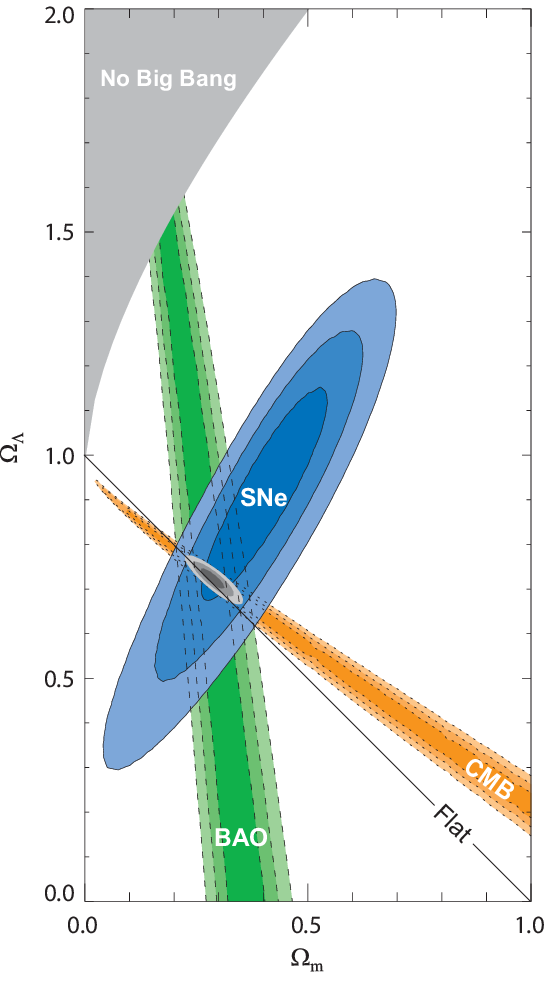}
 \caption[BBN predictions of the light element abundances and $\Omega_{\Lambda}-\Omega_m$ plane of the concordance model of cosmology.]{\textit{Left:} BBN predictions for the abundances of $^4$He, deuterium, $^3$He and $^7$Li as a function of the baryon-to-photon ratio compared to values deduced from astrophysical observations (boxes). There is a remarkable coincidence of the values for the baryon-to-photon ratio derived from BBN and CMB observations. Figure taken from~\cite{Nakamura:2010zzi}. \textit{Right:} $\Omega_{\Lambda}-\Omega_m$ plane of the concordance model of cosmology. The observations of the CMB, of supernovae and of baryon acoustic oscillations overlap, and their combination suggests a flat universe with a dark energy density of $\Omega_{\Lambda}\simeq 0.74$ and a matter density of $\Omega_m\simeq 0.26$. Figure taken from~\cite{Kowalski:2008ez}.}
 \label{ConcordanceBBN}
\end{figure}

The abundances of the light elements predicted for the case of the standard model of particle physics are in good agreement with data from astrophysical observations of the $^4$He abundance in low-metallicity H\,\textsc{ii} regions and of the deuterium abundance in quasar absorption spectra for a value of the baryon-to-photon ratio in the range of $5.1\times 10^{-10}<\eta<6.5\times 10^{-10}$, corresponding to an energy density in baryons of $0.019\leq\Omega_bh^2\leq0.024$. This value is in remarkable agreement with the value determined from observations of the cosmic microwave background (see left panel in Figure~\ref{ConcordanceBBN}) and thus strongly constrains deviations imposed by physics beyond the standard model. The only drawback of this theory is the significant deviation of observations from the calculation of the lithium abundance which is not understood so far.

\paragraph{Cosmic Microwave Background}

The observation of the cosmic microwave background (CMB) radiation~\cite{Penzias:1965wn} is the most compelling evidence for a hot thermal phase in the early universe~\cite{Dicke:1965zz}. Actually, it is a relic from the time of last scattering, when photons decoupled from the thermal plasma of electrons and light elements at a temperature of $T\simeq 0.25\,\text{eV}$ corresponding to a redshift $z\simeq 1100$. The Cosmic Background Explorer (COBE) satellite mission found that the CMB is practically isotropic and corresponds to an almost perfect black body radiation spectrum with a temperature of $T_0=2.725\,\text{K}$~\cite{Mather:1993ij,Mather:1998gm}. In addition, the COBE satellite observed slight temperature anisotropies in the CMB sky map at the level of $\delta T/T\sim10^{-5}$~\cite{Smoot:1992td}.

\begin{figure}[t]
 \centering
 \includegraphics[scale=0.21,bb=0 -60 1024 562]{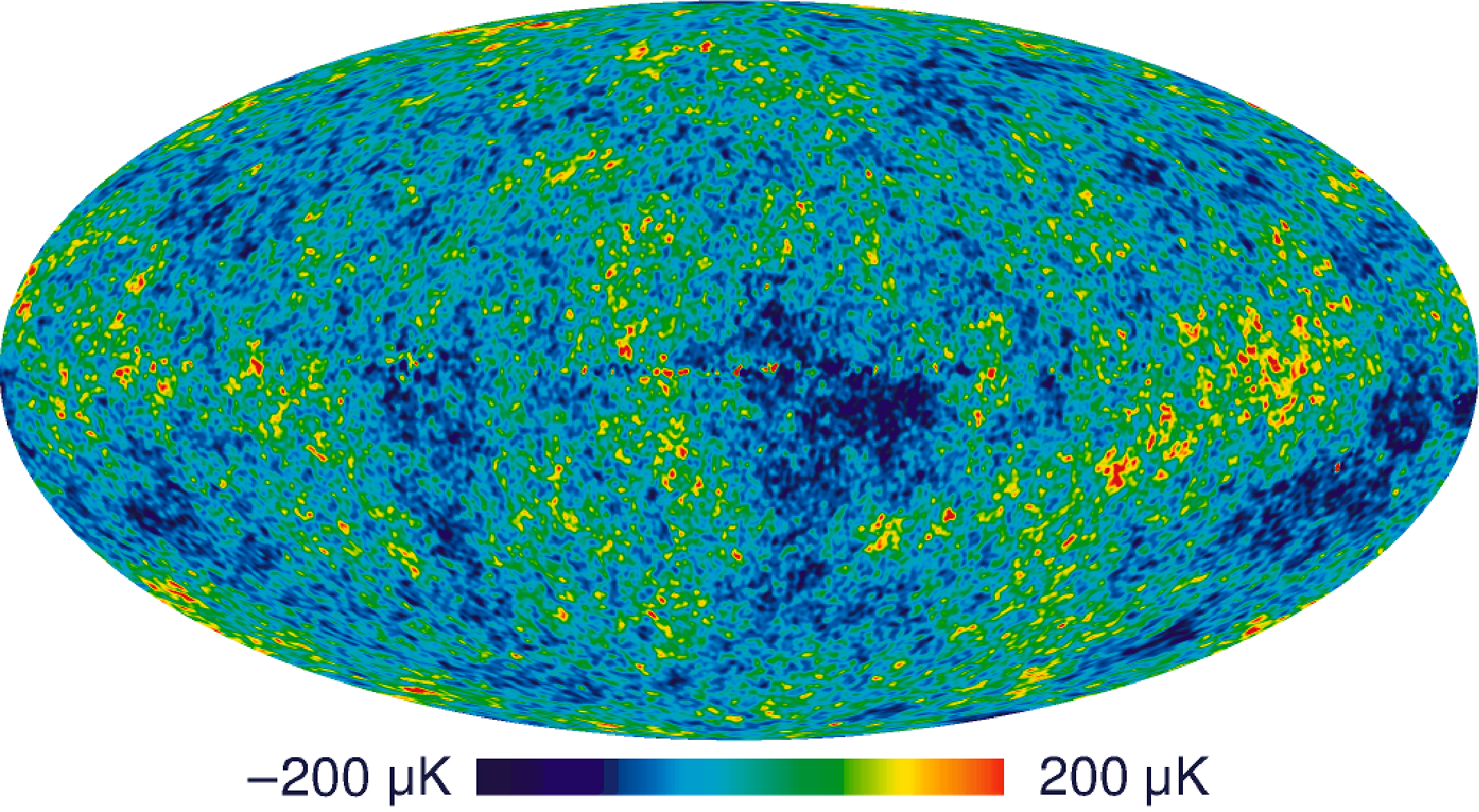}
 \hfill
 \includegraphics[scale=0.88,bb=4 5 256 180]{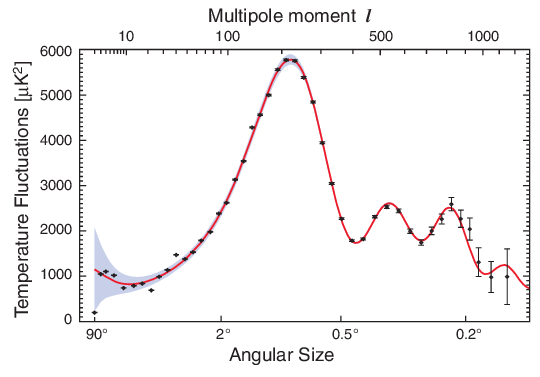}
 \caption[WMAP 7-year temperature anisotropy map and angular power spectrum.]{WMAP 7-year temperature anisotropy map \textit{(left)} and the angular power spectrum of temperature fluctuations from the decomposition of the temperature map into spherical harmonics \textit{(right)}~\cite{Larson:2010gs}. Image credit: NASA / WMAP Science Team.}
 \label{CMB}
\end{figure}
These anisotropies were observed in detail by the Wilkinson Microwave Anisotropy Probe (WMAP) satellite mission and are probably the most valuable probe of the big bang theory (see left panel of Figure~\ref{CMB}). Expansion of the temperature anisotropy map into spherical harmonics 
\begin{equation}
 \frac{\delta T}{T}(\theta,\,\phi)=\sum_{l=2}^{\infty}\sum_{m=-l}^la_{lm}\,Y_{lm}(\theta,\,\phi)
\end{equation}
results in the CMB power spectrum $l\,(l+1)\,C_l/(2\,\pi)$ in terms of multipole moments $l$ (roughly corresponding to an angular size $\theta\sim180^\circ/l$) with
\begin{equation}
 C_l\equiv\left\langle \abs{a_{lm}}^2\right\rangle =\frac{1}{2\,l+1}\sum_{m=-l}^l\abs{a_{lm}}^2.
\end{equation}
The structure of the CMB power spectrum (see right panel in Figure~\ref{CMB}) is mainly determined by four effects: Fluctuations in the photon density at the time of last scattering correspond to temperature fluctuations in the CMB sky. Moreover, photons emitted from a potential well are redshifted on the way to the observer. The combination of intrinsic temperature fluctuations and gravitational redshift is called Sachs--Wolfe effect. In addition, the observed photon temperature is affected by a Doppler shift due to movements of the thermal plasma and by varying potential wells that are crossed by the CMB photons. The latter effect is known as integrated Sachs--Wolfe effect. On scales smaller than roughly $1^\circ$, corresponding to the size of the causal horizon at the time of last scattering, one can observe acoustic oscillations of the thermal plasma that are driven by the gravitational potential of matter and the pressure of photons.

All these effects are very sensitive to cosmological parameters. One particular example is the position of the first acoustic peak which shows that we are living in a universe that is practically spatially flat. The derived cosmological parameters are in good agreement among various observations and thus one speaks of the concordance model of cosmology (see also Figure~\ref{ConcordanceBBN}).

\section{Evidence for the Existence of Dark Matter}
\label{DMevidence}

Various astrophysical observations suggest the existence of dark matter. In this section we want to summarize the evidence from observations on galactic scales up to cosmological scales. All evidence is based on the gravitational effect of dark matter. By now, no evidence for dark matter has been found on microscopic scales. For more extensive reviews on this topic see for instance~\cite{Bertone:2004pz,Einasto:2009zd}.

\subsection*{Galactic Scales}

\paragraph{Rotation Curves of Spiral Galaxies}

\begin{figure}[t]
 \centering
 \includegraphics[scale=0.35,bb=30 157 588 711]{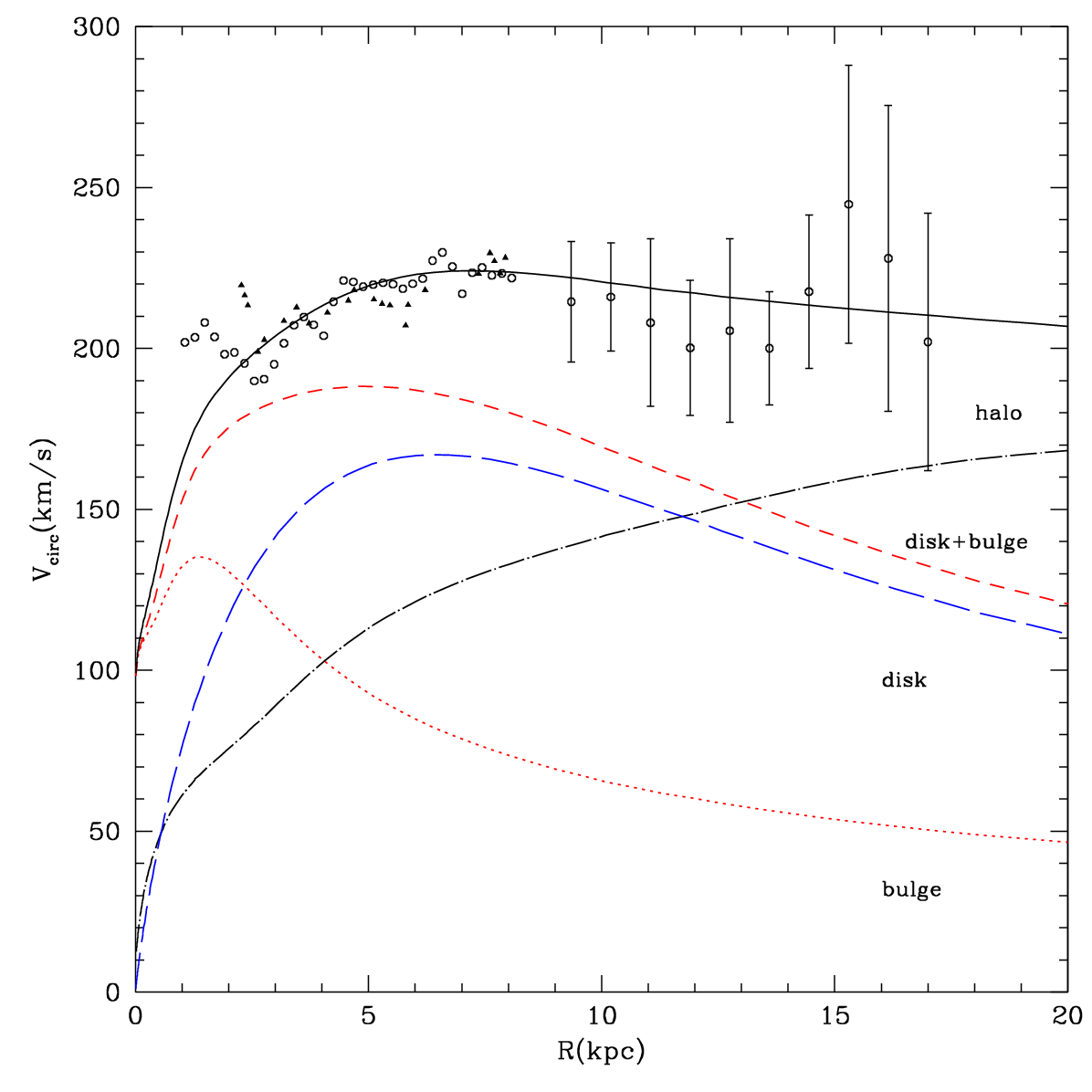}
 \includegraphics[scale=0.4,bb=0 -40 555 373]{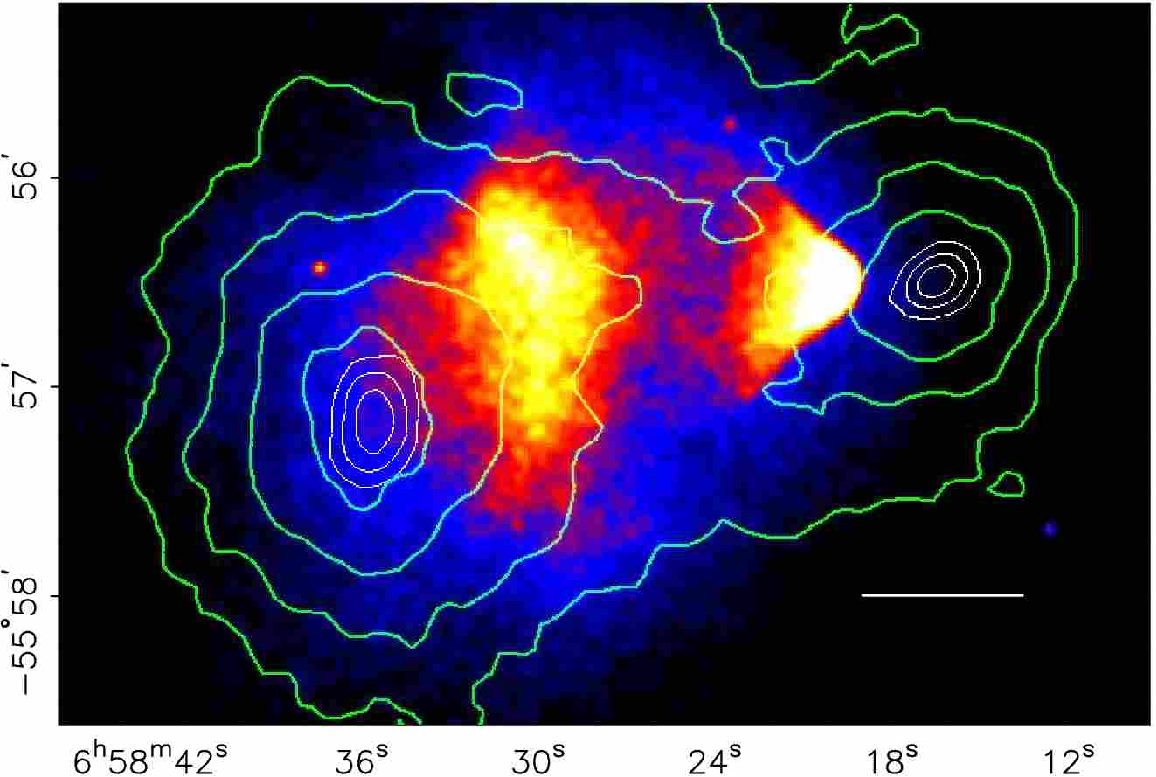}
 \caption[Rotation curve of the Milky Way and the Bullet cluster.]{\textit{Left:} Rotation curve of the Milky Way. The solid line shows a fit to data obtained by the observation of hydrogen gas and the long-dashed, dotted and short-dashed lines show the expected contributions from the luminous disk, the galactic bulge (\textit{i.e.} the very dense central group of stars in a spiral galaxy) and their sum, respectively. In addition, the dash-dotted line shows the required contribution of a dark halo. Figure taken from~\cite{Klypin:2001xu}. \textit{Right:} X-ray image of two galaxy clusters that passed through each other, the smaller of the two being called the Bullet cluster. The hot gas which makes up the dominant contribution of the luminous cluster mass was slowed down in the collision and is clearly displaced from the gravitational potential lines as mapped by weak gravitational lensing. It is concluded that the cluster masses are dominated by collisionless dark matter. Figure borrowed from~\cite{Clowe:2006eq}.}
 \label{rotationBullet}
\end{figure}

A very convicing evidence for dark matter on galactic scales comes from the observation of rotation curves of spiral galaxies, \textit{i.e.} the circular velocity distribution of stars and gas as a function of the distance to the galactic center. In spiral galaxies, stars and gas move on almost circular orbits around the center of their host galaxy. From Newtonian dynamics we know that their circular velocity is given by
\begin{equation}
 v_c(r)=\sqrt{\frac{G_N\,M(r)}{r}}\,,
\end{equation}
where $ M(r)=4\pi\int_0^rdr'\,r'^2\,\rho(r')$ is the total mass inside the sphere with radius $r$. In the outer regions of a galaxy, beyond the visible disc, one would then expect the velocity to fall off as $v\propto1/\sqrt{r}$. However, observations of galactic rotation curves show that the velocity remains constant even far beyond the luminous disk (see Figure~\ref{rotationBullet}).\footnote{A flat rotation curve extending to radii larger than the visible disc of stars was first observed in 1970 when Vera Rubin and Kent Ford studied the velocities of H\,\textsc{ii} regions in the Andromeda galaxy~\cite{Rubin:1970zz}.} This observation can be explained by the existence of a spherical dark halo with a density profile $\rho_{\text{halo}}\propto1/r^2$ in the outer regions. Although the existence of a spherical halo of dark matter is firmly established due to these observations, the situation in the inner parts of galaxies is much less clear. Inside the galactic disc the density is typically dominated by stars and gas and thus the shape of the density profile of the dark component cannot be traced well in these regions.

So far, the outer boundary of galactic halos has not been observed and thus the total mass of galaxies is unknown. Therefore, only a lower limit on the total amount of dark matter in galaxies can be inferred. This translates to a lower limit $\Omega_{\text{DM}}\gtrsim0.1$ on the cosmological dark matter density. It has been proposed that part of the galactic dark halos is composed of non-luminous ordinary matter in the form of Massive Compact Halo Objects (MACHOs)~\cite{Paczynski:1985jf,Griest:1990vu}. There have been searches for these objects through the microlensing effect, finding that MACHOs can only contribute a subdominant part of the galactic dark halo. Thus non-baryonic dark matter is needed to explain the galactic dynamics.

\paragraph{Velocity Dispersion in Galaxies}

A method particularly useful to search for dark matter effects in dwarf galaxies is the observation of their velocity dispersion. Assuming hydrodynamical equilibrium and spherical symmetry of the system, the velocity dispersion is related to the mass of the galaxy. Typically, very high mass-to-light ratios are observed in dwarf spheroidal galaxies, indicating that these systems are dominated by a halo of dark matter.

\subsection*{Scale of Galaxy Clusters}

Also in galaxy clusters a significant discrepancy between the observed amount of luminous matter in the form of stars or gas and the total cluster mass is observed. In the following we will list a few methods that are used to determine the total cluster mass.

\paragraph{Virial Theorem}

From the observation of the velocity dispersion of galaxy clusters their mass can be determined using the virial theorem which relates the average kinetic energy with the average gravitational potential. This method actually gave the very first evidence for the existence of dark matter: Fritz Zwicky found in 1933 that the velocities of galaxies in the Coma galaxy cluster are almost a factor of ten larger than expected from the mass of the luminous galaxies and concluded that the system must be dominated by a dark matter component~\cite{Zwicky:1933gu}. Modern observations of this type are consistent with a total matter density in the range $\Omega_m\approx0.2$--0.3, thereby clearly exceeding the amount of baryonic matter in the universe as derived from big bang nucleosynthesis.

\paragraph{Gravitational Lensing}

Another method to determine the mass of galaxy clusters is based on the effect of gravitational lensing. A massive galaxy cluster can act as a lens that bends the light of distant astrophysical sources in the cluster direction by the gravitational effect of its potential well. One distinguishes between strong lensing, where the effect of the gravitational lens leads to visible distortions of the image of a background source, and weak lensing, where the lensing effect leads to slight distortions of the shapes of background galaxies that can only be observed by a statistical analysis of a large number of these galaxies. However, this method works very well and allows to reconstruct the spatial mass distribution inside the galaxy cluster.

\paragraph{X-Ray Observation of Hot Gas in Galaxy Clusters}

Yet another method to determine the mass of a galaxy cluster is based on X-ray observations of the hot gas contained inside the cluster. Under the assumption of hydrostatical equilibrium there exists a relation between the temperature of the gas and the cluster mass. This method in addition allows to map the distribution of matter in the cluster. In general one finds that the gas temperature is much higher than expected from the amount of luminous matter observed in clusters. Therefore, a dark component is needed to explain the high temperature of the contained gas. The results for the dark matter density obtained by the latter two methods are generally in good agreement with the result deduced from the velocity dispersion of galaxy clusters.

\paragraph{Bullet Cluster}

An even more direct evidence for the existence of dark matter is found in the Bullet cluster~\cite{Clowe:2006eq} (see right panel in Figure~\ref{rotationBullet}). This system actually consists of two galaxy clusters that passed through each other. Although the hot gas, which makes up the dominant contribution of the luminous cluster mass, is displaced from the galaxies due to the collision, the gravitational potential follows the distribution of galaxies. This can only be understood if the cluster mass is dominated by a practically collisionless component of dark matter. This observation also strongly disfavors alternative theories of gravity like Modified Newtonian Dynamics (MOND)~\cite{Milgrom:1983ca} that were proposed to explain the galactic rotation curves without invoking a dark matter component.

\subsection*{Cosmological Scales}

\paragraph{Large Scale Structure}

Although the universe appears to be homogeneous and isotropic on the largest scales, a lot of structure is observed on small and large scales. This structure is expected to originate from quantum fluctuations that were amplified during the stage of inflation. These density perturbations can only start to grow efficiently with the beginning of the matter-dominated phase of the universe. However, as baryons are coupled to the plasma of photons whose pressure counteracts the gravitational collaps, their density perturbations can only start to grow after the decoupling of photons, \textit{i.e.} after the emission of the CMB. In this case the density fluctuations at photon decoupling would have needed to be already quite large, corresponding to temperature fluctuations in the CMB of the order of $\delta T/T\sim10^{-3}$ in order to explain the observed structure in the present universe.

This value is much larger than the observed level of fluctuations ($\delta T/T\sim10^{-5}$) and thus dark matter is needed to explain the amount of structure that we observe today. Since dark matter particles were already decoupled from the thermal plasma, their density perturbations could start to grow from the time of matter-radiation equality (redshift $z\sim3000$) onwards.

The effect of baryon acoustic oscillations can also be observed in the large-scale distribution of galaxies and allows to derive the total density of matter in the universe~\cite{Eisenstein:2005su}: $\Omega_m\simeq0.27$.
Together with the value for the density of baryonic matter from primordial nucleosynthesis this is a strong evidence for dark matter.

\paragraph{Cosmic Microwave Background}

\begin{figure}[t]
 \centering
 \includegraphics[scale=0.08]{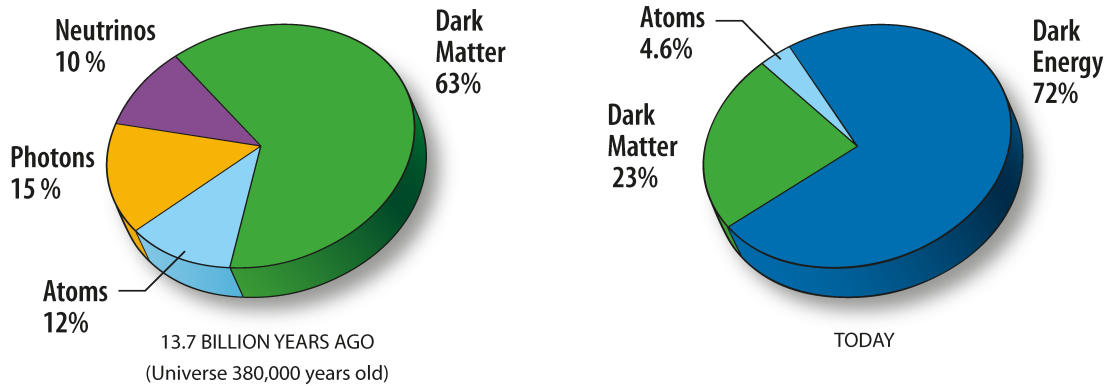}
 \caption[Energy content of the universe at the time of the CMB emission and today.]{Contributions of radiation (\textit{i.e.} photons and neutrinos), non-relativistic matter (atoms and dark matter) and dark energy to the total energy content of the universe at the time of the emission of the cosmic microwave background (\textit{left}) and today (\textit{right}). Image credit: NASA / WMAP Science Team.}
 \label{energycontent}
\end{figure}
As discussed before the temperature fluctuations in the CMB correspond to the acoustic oscillations of the thermal plasma of photons and baryons. The power spectrum of temperature anisotropies has been measured in particular detail by the WMAP experiment. From the shape of the acoustic peaks the densities of baryons and matter as well as a number of other cosmological parameters can be derived. The derived values for the energy densities are shown in Figure~\ref{energycontent} and clearly require the existence of dark matter (as well as dark energy) in the universe.

\section{Constraints on Dark Matter Properties}

In this section we want to discuss shortly the constraints on the properties of dark matter from cosmological and astrophysical considerations. A more thorough discussion of this topic for particle dark matter candidates can be found for instance in~\cite{Taoso:2007qk}.

In principal one could explain the observations of dark matter on the scale of galaxies and clusters with ordinary matter in the form of non-luminous astrophysical objects like MACHOS. In that case the dark matter would basically be composed of charged and colored elementary particles. However, the results from big bang nucleosynthesis and the observations of the large scale structure and the cosmic microwave background point to a strong discrepancy between the amount of baryonic matter and the total amount of matter in the universe. Actually, it turns out that ordinary atoms only amount to one sixth of the total matter density (see Figure~\ref{energycontent}) and hence it is clear that the dominant amount of dark matter must be nonbaryonic.

For this reason, currently the best candidates for the dark matter in the universe are new elementary particles. In this case, however, there exist stringent bounds on their charges with respect to electromagnetic and strong interactions. Electromagnetically interacting dark matter particles should mix with ordinary matter and should then also abundantly exist on Earth in the form of electromagnetically bound states~\cite{Ellis:1983ew}. These particles should then appear in searches for anomalously heavy hydrogen atoms~\cite{Wolfram:1978gp,Yamagata:1993jq}. Since all these searches have been negative so far, one expects dark matter particles to be neutral with respect to electromagnetic interactions.\footnote{Actually, it has been proposed that dark matter could be formed of charged massive particles (CHAMPS)~\cite{DeRujula:1989fe,Dimopoulos:1989hk} but these models are strongly constrained.} Similarly, colored dark matter particles are expected to form color-neutral bound states together with ordinary matter~\cite{Ellis:1983ew}. These heavy hadrons could then be identified in searches for heavy nucleons~\cite{Wolfram:1978gp,Dover:1979sn,Yamagata:1993jq,Javorsek:2001yv}. Also in this case all searches have been negative so far and thus one expects dark matter particles to be color-neutral.

The requirement of a successful structure formation introduces another important constraint. Dark matter is necessary to explain the observed growth of density perturbations after the time of radiation domination but it also needs to have the correct properties in order to match observations. In fact, the observation of small-scale structures constrains the free-streaming length of dark matter particles: they must be sufficiently non-relativistic in order not to wash out small-scale density perturbations. Thus the dominant part of the dark matter in the universe must be in the form of cold dark matter particles, \textit{i.e.} particles that became non-relativistic well before the beginning of structure formation at the time of matter-radiation equality. In any case, however, there is also a subdominant contribution of hot dark matter, namely the background of light standard model neutrinos.

Another important constraint comes from the requirement that the relic density of the dark matter particle as predicted by the particle physics model needs to be in accord with the observed dark matter density in the universe. The current best-fit value from the combination of the seven-year data of WMAP, observations of baryon acoustic oscillations and determinations of the present-day Hubble parameter is given by~\cite{Komatsu:2010fb}
\begin{equation}
  \Omega_{\text{DM}}h^2=0.1126(36)\,.
\end{equation}
Of course, an additional constraint is that the dark matter particle candidate is either perfectly stable or sufficiently long-lived to explain the dark matter observations. Cosmological observations alone constrain the dark matter lifetime to be at least one order of magnitude above the current age of the universe ($t_0\simeq13.7\,\text{Gyr}=4.3\times10^{17}\,$s)~\cite{DeLopeAmigo:2009dc}. We will see later in this work that astrophysical observations are able to set much more stringent constraints on the dark matter lifetime.

There are additional constraints for many particle dark matter candidates from the requirement that the particle physics model leads to a consistent cosmological scenario that is not in conflict with any cosmological observation. In particular, big bang nucleosynthesis is very sensitive to deviations from the standard model particle content at an early epoch of the cosmological evolution. For instance, the presence of charged particles at the time of BBN or the decay of metastable relics during or after the time of BBN can significantly alter the predicted abundances of light elements and thus lead to conflicts with observations.\footnote{For a recent review of constraints from big bang nucleosynthesis and further references see~\cite{Jedamzik:2009uy}.}

In particular the latter point typically provides strong constraints for supergravity theories containing a gravitino in the particle spectrum. We will give an overview of these constraints and how they can be circumvented in Section~\ref{gravitinocosmo}.

\chapter{Supersymmetry and Supergravity}
\label{susysugra}

Supersymmetry (SUSY)~\cite{Wess:1974tw} is a generalization of the space-time symmetries of quantum field theory that relates bosonic and fermionic degrees of freedom. It introduces new fermionic generators $Q$ that transform fermions into bosons and vice versa:
\begin{equation*}
 Q\left| \text{boson}\right\rangle \simeq\left| \text{fermion}\right\rangle ,\qquad Q\left| \text{fermion}\right\rangle \simeq\left| \text{boson}\right\rangle . 
\end{equation*}
This represents a nontrivial extension to the Poincar\'e symmetry of ordinary quantum field theory and its structure is highly constrained by the theorem of Haag, Lopuszanski and Sohnius~\cite{Haag:1974qh} which is a generalization of the Coleman--Mandula theorem~\cite{Coleman:1967ad}.

The introduction of SUSY predicts the existence of supersymmetric partners for the standard model particles relating each fermionic degree of freedom with a bosonic one. If SUSY were an exact symmetry of nature, particles and their superpartners would be degenerate in mass. However, since no superpartners have been observed yet, SUSY must be a broken symmetry.

Although not yet confirmed experimentally, there are several theoretical motivations for interest in this additional symmetry. The first is the hierarchy problem of the standard model. This problem stems from the huge difference of the electroweak scale ($\mathcal{O}(100)\,$GeV) and the (reduced) Planck scale
\begin{equation}
 \MP=\frac{1}{\sqrt{8\,\pi\,G_N}}\simeq2.4\times10^{18}\,\text{GeV}\,,
\end{equation}
where gravitational interactions become comparable in magnitude to gauge interactions. The only scalar particle in the standard model, the Higgs boson, receives quadratic radiative corrections to its mass due to fermion loops. These quadratic divergences are conveniently cancelled by the contributions of the bosonic superpartners to the radiative corrections. This exact cancellation holds only for unbroken SUSY. However, as long as SUSY is broken softly, no quadratic divergences are reintroduced and the hierarchy problem can still be solved~\cite{Girardello:1981wz}.

\begin{figure}
 \centering
 \includegraphics[scale=0.52]{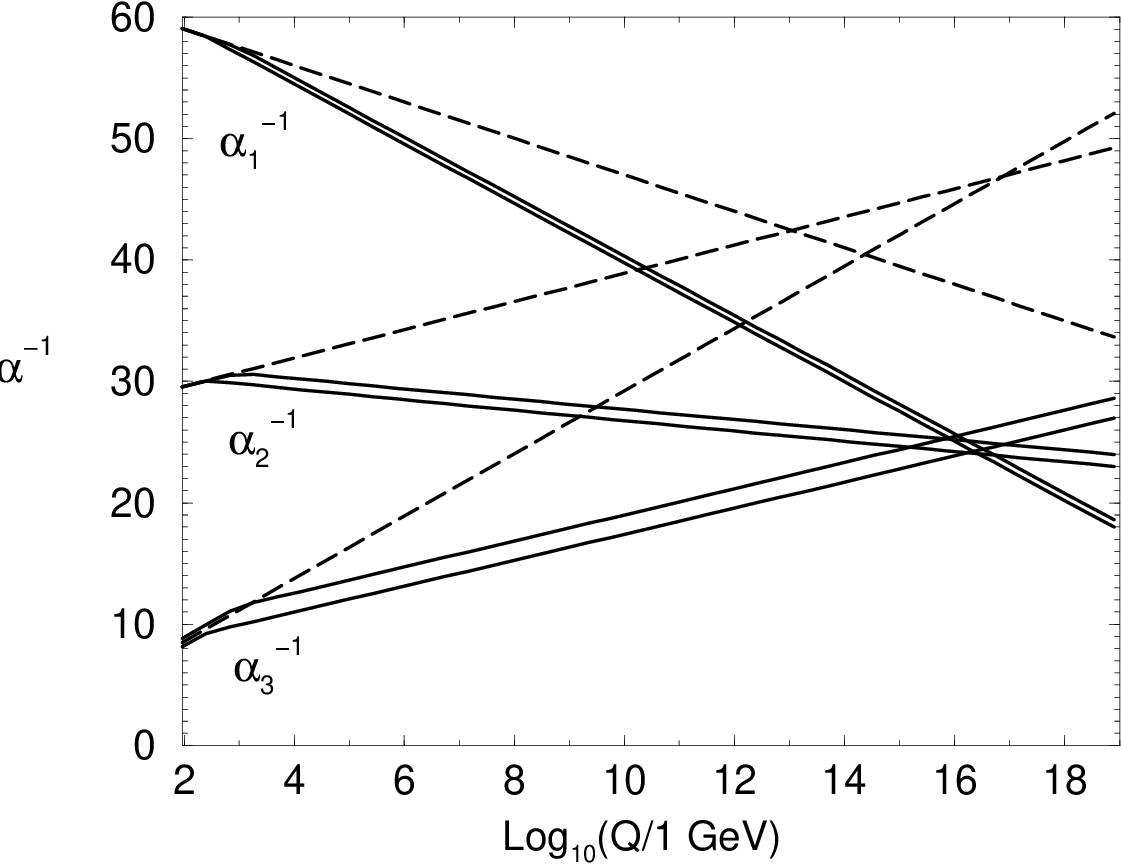} 
 \caption[Renormalization group evolution of the inverse gauge couplings in the standard model and in the minimal supersymmetric standard model.]{Renormalization group evolution of the inverse gauge couplings in the standard model (dashed) and in the MSSM (solid). The bands correspond to assumed masses for the supersymmetric particles in the range from 250\,GeV to 1\,TeV. Figure borrowed from~\cite{Martin:1997ns}.}
 \label{Unification}
\end{figure}

A second motivation is the unification of gauge couplings $\alpha_{\alpha}=g_{\alpha}^2/4\pi $, $\alpha=1,\,2,\,3$, where $g_1=\sqrt{5/3}\,g'$ and $g_2=g$ are the electroweak coupling constants, and $g_3=g_s$ is the strong coupling constant of the unbroken standard model gauge group
\begin{equation*}
 SU(3)_c\times SU(2)_L\times U(1)_Y\,.
\end{equation*}
The evolution of the gauge couplings is determined by renormalization group equations that depend on the particle content of the theory. It is observed that the gauge couplings do not unify in the standard model, while the altered particle content of a supersymmetric theory at the TeV scale leads to a unification of the gauge couplings at a scale $M_{\text{GUT}}\simeq2\times10^{16}\,\text{GeV}$ (\textit{cf.} Figure~\ref{Unification}).

A third motivation for supersymmetry is that it provides a promising candidate for particle dark matter: the lightest supersymmetric particle (LSP). With conserved \textit{R} parity the LSP is absolutely stable and therefore a natural candidate for the dark matter in the universe.

For a general introduction to the topic of supersymmetry see for instance~\cite{Martin:1997ns}.

\section{Supergravity}

If supersymmetry is promoted to a local symmetry, \textit{i.e.} the parameter in SUSY transformations becomes coordinate-dependent, the theory must incorporate gravity. This is because in order to achieve invariance under local SUSY transformations one has to add a new supermultiplet to the theory: the gravity supermultiplet, which consists of the spin-2 graviton and the spin-3/2 gravitino (see Table~\ref{Gravitymultiplet}). The resulting locally supersymmetric theory is therefore called supergravity (SUGRA)~\cite{Freedman:1976xh,Deser:1976eh}. A short introduction to supergravity can be found for instance in~\cite{Cerdeno:1998hs}. For more exhaustive discussions of SUGRA see references therein.
\begin{table}
 \centering
 \begin{tabular}{lccc}
  \toprule 
  Name & Bosons & Fermions & $ \left( SU(3)_c\,,\,SU(2)_L\right) _Y $ \\
  \midrule 
  Graviton, gravitino & $ g_{\mu\nu} $ & $ \psi_{\mu} $ & $ \left( \mathbf{1},\,\mathbf{1}\right) _0 $ \\
  \bottomrule 
 \end{tabular}
 \caption[The gravity supermultiplet of locally supersymmetric theories.]{The gravity supermultiplet of locally supersymmetric theories. Listed are the particles together with their quantum numbers with respect to the unbroken standard model gauge group.}
 \label{Gravitymultiplet}
\end{table}

\subsubsection*{The Supergravity Lagrangian}

Supergravity theories are in general described by three functions of the scalar fields in the theory: The K\"ahler potential $K(\phi,\phi^*)$, which is a real function of the scalar fields, the superpotential $W(\phi)$, which is a holomorphic function of the scalar fields, and the gauge kinetic function $f_{ab}(\phi)$, which also is a holomorphic function of the scalar fields.

The bosonic part of the supergravity Lagrangian is of the form~\cite{Cerdeno:1998hs}
\begin{equation}
 \begin{split}
  \mathscr{L}&=\sqrt{-g}\Bigg[ -\frac{\MP^2}{2}\,R-G_{i\bar{\jmath}}(\phi,\phi^*)\left( D_\mu\,\phi^i\right) \left( D^\mu\phi^{*\,\bar{\jmath}}\right) -V(\phi,\phi^*) \\
  &\qquad\qquad\qquad\qquad-\frac{1}{4}\,(\RE f_{ab})F_{\mu\nu}^aF^{b\,\mu\nu}+\frac{i}{4}\,(\IM f_{ab})F_{\mu\nu}^a\tilde{F}^{b\,\mu\nu}\Bigg] ,
 \end{split}
\end{equation}
where $g=\det g^{\mu\nu}$ is the determinant of the space-time metric. The first part with the Ricci scalar $R$ is the usual Einstein--Hilbert term from general relativity. The second part is the kinetic term of the scalar fields that in general is not of canonical form. However, the non-linear sigma model of scalars is not arbitrary since invariance under supergravity transformations requires that the field space of scalars is a K\"ahler manifold. The metric of this K\"ahler manifold is given by a second derivative of the K\"ahler potential:
\begin{equation}
  G_{i\bar{\jmath}}(\phi,\phi^*)=\frac{\partial}{\partial\phi^i}\,\frac{\partial}{\partial\phi^{*\,\bar{\jmath}}}\,K(\phi,\phi^*)\,.
\end{equation}
In addition, the covariant derivative of the scalars is of the form
\begin{equation}
  D_\mu\phi^i=\partial_\mu\phi^i-gA_\mu^a\,X^{i\,a}=\partial_\mu\phi^i+i\,gA_\mu^a\,G^{i\bar{\jmath}}\,\frac{\partial D^a}{\partial\phi^{*\bar{\jmath}}}\,,
\end{equation}
where the $X^{i\,a}$ are holomorphic Killing vector fields corresponding to isometries of the K\"ahler metric $G_{i\bar{\jmath}}$ and the $D^a$ are the associated Killing potentials. The third term in the Lagrangian is the scalar potential that is given by
\begin{equation}
  V=\exp\left( \frac{K}{\MP^2}\right) \left( (\nabla_iW)\,G^{i\bar{\jmath}}\,(\nabla_{\bar{\jmath}}W^*)-3\,\frac{\abs{W}^2}{\MP^2}\right) +\frac{1}{2}\,g^2(\RE f_{ab})^{-1}D^aD^b,
\end{equation}
where the K\"ahler covariant derivative $\nabla_iW$ and the real Killing potential $D^a$ are, respectively, given by
\begin{equation}
  \nabla_iW=\frac{\partial W}{\partial\phi^i}+\frac{1}{\MP^2}\,\frac{\partial K}{\partial\phi^i}\,W\qquad\text{and}\qquad D^a=-i\,\frac{\partial K}{\partial\phi^i}\,X^{i\,a}.
\end{equation}
The last two terms are the kinetic terms of gauge bosons expressed in form of the field strength tensor and its dual. These terms are proportional to the gauge kinetic function $f_{ab}$. Due to the appearance of non-canonical kinetic terms, supergravity is a nonrenormalizable theory. In addition, the full Lagrangian will contain interaction terms with couplings of negative mass dimension. However, it is usually assumed that supergravity is an appropriate low-energy approximation of a more general theory like string theory.

The most general supergravity Lagrangian including all fermionic terms is usually derived using the superspace formalism. A convenient starting point for phenomenological studies of four-dimensional $N=1$ supergravity models is the complete supergravity Lagrangian in component fields as given in the book of Wess and Bagger~\cite{Wess:1992cp}. Here we want to quote the Lagrangian in a simplified case where $f_{ab}=\delta_{ab}$. In the notation of Wess and Bagger -- but restoring the dependence on the reduced Planck mass -- the Lagrangian reads:
\begin{align}
  \frac{\mathscr{L}}{\sqrt{-g}} &=-\frac{\MP^2}{2}\,R-G_{i\bar{\jmath}}\left( D_\mu\,\phi^i\right) \left( D^\mu\phi^{*\,\bar{\jmath}}\right) -\frac{1}{2}\,g^2D^aD^a-\frac{1}{4}\,F_{\mu\nu}^aF^{a\,\mu\nu}-i\bar{\lambda}^a\bar{\sigma}^\mu D_\mu\,\lambda^a \nonumber\\
  &\qquad-i\,G_{i\bar{\jmath}}\,\bar{\chi}^{\bar{\jmath}}\,\bar{\sigma}^\mu D_\mu\,\chi^i+\varepsilon^{\mu\nu\rho\sigma}\bar{\psi}_\mu\bar{\sigma}_\nu D_\rho\,\psi_\sigma+i\sqrt{2}\,g\,\frac{\partial D^a}{\partial\phi^i}\,\bar{\chi}^i\lambda^a-i\sqrt{2}\,g\,\frac{\partial D^a}{\partial\phi^{*\bar{\jmath}}}\,\bar{\chi}^{\bar{\jmath}}\,\bar{\lambda}^a \nonumber\\
  &\qquad-\frac{1}{\MP}\bigg(\frac{g}{2}\,D^a\psi_\mu\,\sigma^\mu\bar{\lambda}^a-\frac{g}{2}\,D^a\bar{\psi}_\mu\,\bar{\sigma}^\mu\lambda^a+\frac{1}{\sqrt{2}}\,G_{i\bar{\jmath}}\left( D_\mu\,\phi^{*\,\bar{\jmath}}\right) \chi^i\sigma^\mu\,\bar{\sigma}^\nu\psi_\mu \nonumber\\
  &\qquad\qquad+\frac{1}{\sqrt{2}}\,G_{i\bar{\jmath}}\left( D_\mu\,\phi^i\right) \bar{\chi}^{\bar{\jmath}}\,\bar{\sigma}^\nu\sigma^\mu\bar{\psi}_\nu-\frac{i}{4}\left( \psi_\mu\,\sigma^{\rho\sigma}\sigma^\mu\bar{\lambda}^a+\bar{\psi}_\mu\,\bar{\sigma}^{\rho\sigma}\bar{\sigma}^\mu\lambda^a\right) \!\left( F_{\rho\sigma}^a+\widehat{F}_{\rho\sigma}^a\right) \!\!\bigg)\nonumber\\
  &\qquad+\frac{1}{\MP^2}\bigg(\frac{1}{4}\,G_{i\bar{\jmath}}\left( i\,\varepsilon^{\mu\nu\rho\sigma}\psi_\mu\sigma_\nu\bar{\psi}_\rho+\psi_\rho\,\sigma^\sigma\bar{\psi}^\rho\right) \chi^i\sigma_\sigma\bar{\chi}^{\bar{\jmath}}-\frac{3}{16}\,\lambda^a\sigma^\mu\bar{\lambda}^a\lambda^b\sigma_\mu\bar{\lambda}^b \nonumber\\
  &\qquad\qquad-\frac{1}{8}\left( G_{i\bar{\jmath}}\,G_{k\bar{l}}-2\,\MP^2\,R_{i\bar{\jmath}k\bar{l}}\right) \chi^i\,\chi^k\,\bar{\chi}^{\bar{\jmath}}\,\bar{\chi}^{\bar{l}}+\frac{1}{8}\,G_{i\bar{\jmath}}\,\bar{\chi}^{\bar{\jmath}}\,\bar{\sigma}^\mu\,\chi^i\,\bar{\lambda}^a\,\bar{\sigma}_\mu\lambda^a\bigg) \\
  &\qquad-\exp\left( \frac{K}{2\,\MP^2}\right) \!\bigg(\frac{1}{\MP^2}\Big(W^*\psi_\mu\,\sigma^{\mu\nu}\psi_\nu+W\bar{\psi}_\mu\,\bar{\sigma}^{\mu\nu}\bar{\psi}_\nu\Big)+\frac{1}{\MP}\Big(\frac{i}{\sqrt{2}}\,(\nabla_iW)\,\chi^i\sigma^\mu\bar{\psi}_\mu \nonumber\\
  &\qquad\qquad\qquad\qquad+\frac{i}{\sqrt{2}}\,(\nabla_{\bar{\imath}}W^*)\,\bar{\chi}^{\bar{\imath}}\,\bar{\sigma}^\mu\psi_\mu\Big)+\frac{1}{2}\,D_i\nabla_jW\chi^i\,\chi^j+\frac{1}{2}\,D_{\bar{\imath}}\nabla_{\bar{\jmath}}W^*\bar{\chi}^{\bar{\imath}}\,\bar{\chi}^{\bar{\jmath}}\bigg) \nonumber\\
  &\qquad-\exp\left( \frac{K}{\MP^2}\right) \!\left( (\nabla_iW)\,G^{i\bar{\jmath}}\,(\nabla_{\bar{\jmath}}W^*)-3\,\frac{|W|^2}{\MP^2}\right) ,\nonumber
\end{align}
where
\begin{equation}
  \widehat{F}_{\mu\nu}^a\equiv F_{\mu\nu}^a-\frac{i}{2\,\MP}\left( \psi_\mu\,\sigma_\nu\,\bar{\lambda}^a+\bar{\psi}_\mu\,\bar{\sigma}_\nu\,\lambda^a-\psi_\nu\,\sigma_\mu\,\bar{\lambda}^a-\bar{\psi}_\nu\,\bar{\sigma}_\mu\,\lambda^a\right) 
\end{equation}
and
\begin{equation}
  \sigma^{\mu\nu}\equiv\frac{1}{4}\left( \sigma^\mu\bar{\sigma}^\nu-\sigma^\nu\bar{\sigma}^\mu\right) .
\end{equation}
The curvature tensor and the connection of the K\"ahler manifold are, respectively, given by~\cite{Wess:1992cp}
\begin{equation}
  R_{i\bar{\jmath}k\bar{l}}=G_{m\bar{l}}\,\frac{\partial\Gamma_{ik}^m}{\partial\phi^{*\bar{\jmath}}}\qquad\text{and}\qquad\Gamma_{ij}^k=G^{k\bar{l}}\,\frac{\partial G_{j\bar{l}}}{\partial\phi^i}\,.
\end{equation}
The covariant derivative of the scalars and the K\"ahler covariant derivative have been introduced before. The remaining derivatives are given by~\cite{Wess:1992cp}:
\begin{align}
  D_\mu\,\chi^i &=\partial_\mu\chi^i+\chi^i\omega_\mu+\Gamma_{jk}^i\left( D_\mu\,\phi^j\right) \chi^k+i\,gA_\mu^a\,\frac{\partial}{\partial\phi^k}\left( G^{i\bar{\jmath}}\,\frac{\partial D^a}{\partial \phi^{*\bar{\jmath}}}\right) \chi^j \nonumber\\
  &\qquad\qquad-\frac{1}{4\,\MP^2}\left( \frac{\partial K}{\partial\phi^j}\,D_\mu\,\phi^j-\frac{\partial K}{\partial\phi^{*\bar{\jmath}}}\,D_\mu\,\phi^{*\bar{\jmath}}\right) \chi^i-\frac{i}{2\,\MP^2}\,gA_\mu^a\IM F^a\chi^i, \\
  D_\mu\,\lambda^a &=\partial_\mu\lambda^a+\lambda^a\omega_\mu-gf^{abc}A_\mu^b\lambda^c \nonumber\\
  &\qquad\qquad+\frac{1}{4\,\MP^2}\left( \frac{\partial K}{\partial\phi^j}\,D_\mu\,\phi^j-\frac{\partial K}{\partial\phi^{*\bar{\jmath}}}\,D_\mu\,\phi^{*\bar{\jmath}}\right) \lambda^a+\frac{i}{2\,\MP^2}\,gA_\mu^b\IM F^b\lambda^a, \\
  D_\mu\,\psi_\nu &=\partial_\mu\psi_\nu+\psi_\nu\,\omega_\mu+\frac{1}{4\,\MP^2}\left( \frac{\partial K}{\partial\phi^j}\,D_\mu\,\phi^j-\frac{\partial K}{\partial\phi^{*\bar{\jmath}}}\,D_\mu\,\phi^{*\bar{\jmath}}\right) \psi_\nu \nonumber\\
  &\qquad\qquad+\frac{i}{2\,\MP^2}\,gA_\mu^a\IM F^a\psi_\nu\,, \\
  D_i\nabla_jW &=\frac{\partial^2 W}{\partial\phi^i\,\partial\phi^j}+\frac{1}{\MP^2}\left( \frac{\partial^2 K}{\partial\phi^i\,\partial\phi^j}\,W+\frac{\partial K}{\partial\phi^i}\,\nabla_jW+\frac{\partial K}{\partial\phi^j}\,\nabla_iW\right) \nonumber\\
  &\qquad\qquad-\frac{1}{\MP^4}\,\frac{\partial K}{\partial\phi^i}\,\frac{\partial K}{\partial\phi^j}\,W-\Gamma_{ij}^k\nabla_kW\,.
\end{align}
In these expressions $\omega_\mu$ is the spin connection and
\begin{equation}
  F^a\equiv-i\,G^{i\bar{\jmath}}\,\frac{\partial D^a}{\partial \phi^{*\bar{\jmath}}}\,\frac{\partial K}{\partial\phi^i}+iD^a.
\end{equation}
If the K\"ahler potential has a minimal form, \textit{i.e.} $K=\phi_i\,\phi^{*\bar{\imath}}$, the K\"ahler metric becomes trivial ($G_{i\bar{\jmath}}=\delta_{i\bar{\jmath}}$) and canonical kinetic terms for the scalars are recovered. In this case the K\"ahler connection and curvature are vanishing and the Killing potentials coincide with the $D$-terms of the globally supersymmetric theory:
\begin{equation}
  D^a=\phi^{*i}\,T_{ij}^a\,\phi^j.
\end{equation}

In~\cite{Pradler:2007ne} the supergravity Lagrangian was rewritten up to $\mathcal{O}(\MP^{-1})$ in four-component spinor notation which is convenient for the calculation of matrix elements. Useful references on the translation from two-component to four-component notation are for instance~\cite{Dreiner:2008tw} and~\cite{Haber:1984rc}. In the conventions of~\cite{Pradler:2007ne} the four-component spinors of matter fermions, gauginos and gravitinos are defined as
\begin{equation}
  \chi_L^i=
  \begin{pmatrix}
  \chi^i \\
  0
  \end{pmatrix},\qquad
  \lambda^a=
  \begin{pmatrix}
  -i\lambda^a \\
  i\bar{\lambda}^a
  \end{pmatrix}\qquad\text{and}\qquad
  \psi_\mu=
  \begin{pmatrix}
  -i\,\psi_\mu \\
  i\,\bar{\psi}_\mu
  \end{pmatrix}.
\end{equation}
A definition of the gravitino 4-spinor without the factors of $i$ would lead to a change of the relative signs of the gaugino and gravitino masses that might be observable in the interference terms of different Feynman diagrams.

\subsubsection*{Supersymmetry Breaking and the Super-Higgs Mechanism}

As mentioned in the beginning of this chapter, supersymmetry must be spontaneously broken by soft parameters to explain the mass differences of standard model particles and their superpartners. However, there are no realistic models of spontaneously broken supersymmetry where the SUSY breaking arises from the particle interactions of the minimal supersymmetric standard model. Therefore, one usually assumes the existence of a hidden sector where supersymmetry is broken. The effect of supersymmetry breaking is then mediated to the observable sector via suppressed interactions.

Analogous to the Higgs mechanism of electroweak symmetry breaking, in supergravity there exists a super-Higgs mechanism of supersymmetry breaking~\cite{Deser:1977uq}. Since the SUSY generators are fermionic, the breaking of supersymmetry generates a massless goldstone fermion, the goldstino. However, since the Lagrangian of a spontaneously broken supergravity theory contains a gravitino--goldstino mixing mass term, the goldstino is absorbed into the gravitino, which thereby acquires its longitudinal (helicity $\pm1/2$) components and becomes massive. The gravitino mass is then given by~\cite{Chung:2003fi}
\begin{equation}
 m_{3/2}\sim\frac{\left\langle F\right\rangle }{\MP}\,,
\end{equation}
where $\left\langle F\right\rangle $ is the vacuum expectation value of the hidden sector auxiliary field that is responsible for the spontaneous breaking of supersymmetry. The value of the gravitino mass depends on the particular scheme of SUSY breaking mediation and can range from the eV scale to scales beyond many TeV.

Let us shortly mention the two best-known mechanisms for the mediation of supersymmetry breaking.\footnote{For a discussion of various mechanisms of supersymmetry breaking mediation see \textit{e.g.}~\cite{Martin:1997ns,Chung:2003fi}.}

\paragraph{Gravity-Mediated Supersymmetry Breaking}

In gravity mediation it is assumed that supersymmetry is broken spontaneously in a hidden sector and mediated to the observable sector via non-renormalizable interactions (see for instance~\cite{Nilles:1983ge}). For dimensional reasons, in gravity mediation one typically expects soft terms of the order
\begin{equation}
  m_{\text{soft}}\sim\frac{\left\langle F\right\rangle }{\MP}\sim m_{3/2}\,,
\end{equation}
as the soft masses need to vanish in the limits $\left\langle F\right\rangle \rightarrow0$ and $\MP\rightarrow\infty$.
Therefore, the gravitino mass is required to be somewhere around the electroweak scale up to the TeV scale in order to solve the hierarchy problem. The scale of supersymmetry breaking is rather high in this case: $\sqrt{\left\langle F\right\rangle }\sim10^{10}$--$10^{11}\,$GeV.

\paragraph{Gauge-Mediated Supersymmetry Breaking}

In models of gauge mediation~\cite{Dine:1995ag,Dine:1994vc}\footnote{For a review on theories of gauge mediation see also~\cite{Giudice:1998bp}.} the breaking of supersymmetry is mediated to the visible sector via a sector of messenger particles that couple to the MSSM particles via gauge interactions. In these models the soft mass parameters for scalars are generated via loop effects. By dimensional analysis one expects:
\begin{equation}
  m_{\text{soft}}\sim \frac{\alpha}{4\,\pi}\,\frac{\left\langle F\right\rangle }{M_{\text{mess}}}\,,
\end{equation}
where $\alpha/(4\,\pi)\sim\mathcal{O}(1/100)$ is a loop factor. As the messenger scale is typically considerably below the Planck scale, supersymmetry must be broken at a lower scale compared to the case of gravity mediation to achieve soft masses between the electroweak scale and the TeV scale. This leads subsequently to a gravitino mass that is in general much smaller than in the case of gravity mediation. Then, the gravitino is always the lightest supersymmetric particle.

Although the contributions to the soft parameters from gravity mediation are also present in scenarios of gauge-mediated supersymmetry breaking, they are in general negligible due to the suppression of the lower supersymmetry breaking scale by the Planck scale.

\section{The Minimal Supersymmetric Standard Model}

\begin{figure}[t]
 \centering
 \includegraphics[scale=0.22]{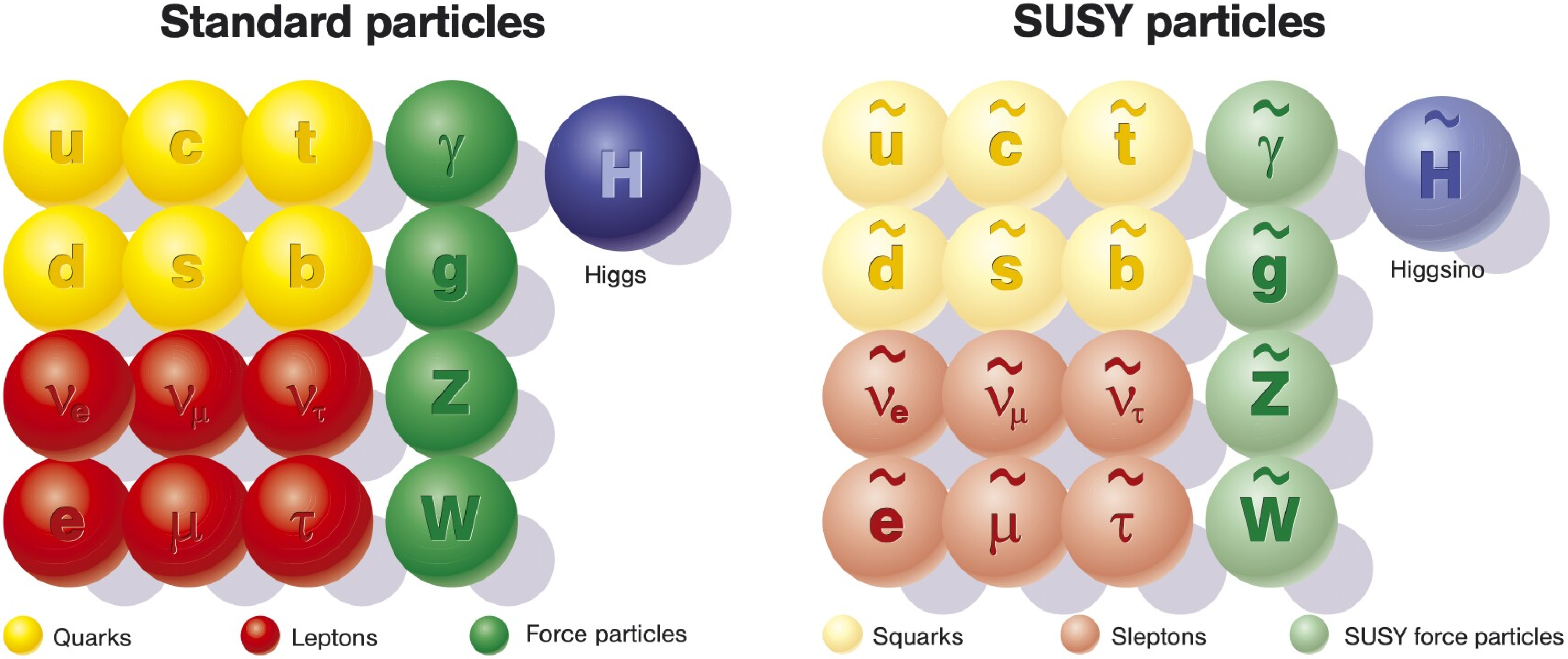} 
 \caption[Schematic particle content of the minimal supersymmetric standard model.]{Particle content of the minimal supersymmetric standard model. There are three generations of quarks and leptons along with their scalar superpartners. Similarly, there are gauge bosons corresponding to the standard model gauge group generators (here shown after electroweak symmetry breaking) and their fermionic superpartners. The Higgs sector is represented by a Higgs boson and its fermionic superpartner, the higgsino. Image Credit: DESY.}
 \label{particles}
\end{figure}
The minimal supersymmetric extension of the standard model (MSSM) introduces superpartners to the standard model matter fermions and gauge bosons, and extends the Higgs sector to include two $SU(2)$ doublets and their superpartners (see Figure~\ref{particles}). The standard model particles and their superpartners are combined into supermultiplets. The gauge or vector supermultiplets consist of a spin-1 vector boson $A_{\mu}^a$, a spin-1/2 Majorana fermion $\lambda^a$ and a scalar auxiliary field $D^a$, where $a$ labels the gauge group generators. In particular, we have the electroweak gauge bosons and the corresponding fermionic gauginos as well as the gluons and the fermionic gluinos (see Table~\ref{GaugeMSSM}). The table gives the notation of the particles in Lagrangians and Feynman rules, and their transformation properties under the standard model gauge group.
\begin{table}[t]
 \centering
 \begin{tabular}{lccc}
  \toprule 
  Name & Gauge bosons & Gauginos & $ \left( SU(3)_c\,,\,SU(2)_L\right) _Y $ \\
  \midrule 
  $ B $ boson, bino & $ A_{\mu}^{(1)}=B_{\mu} $ & $ \lambda^{(1)}=\tilde{B} $ & $ \left( \mathbf{1},\,\mathbf{1}\right) _0 $ \\
  $ W $ bosons, winos & $ A_{\mu}^{(2)\, a}=W_{\mu}^a $ & $ \lambda^{(2)\, a}=\tilde{W}^a $ & $ \left( \mathbf{1},\,\mathbf{3}\right) _0 $ \\
  gluons, gluinos & $ A_{\mu}^{(3)\, a}=G_{\mu}^a $ & $ \lambda^{(3)\, a}=\tilde{g}^a $ & $ \left( \mathbf{8},\,\mathbf{1}\right) _0 $ \\
  \bottomrule 
 \end{tabular}
 \caption[The gauge supermultiplets of the minimal supersymmetric standard model.]{The gauge supermultiplets of the minimal supersymmetric standard model. Listed are the different particles together with their quantum numbers with respect to the unbroken standard model gauge group.}
 \label{GaugeMSSM}
\end{table}

Chiral supermultiplets consist of one complex scalar $\phi$, a two-component chiral fermion $\chi$ and an auxiliary scalar field $F$. There are three generations of left-handed and right-handed leptons and quarks, the corresponding scalar sleptons and squarks, and the respective antiparticles. Additionally, there are two Higgs doublets, the fermionic higgsinos and their antiparticles (see Table~\ref{ChiralMSSM}). The enlarged Higgs sector is needed to guarantee the cancellation of anomalies from the introduction of the higgsino superpartners. In addition, the two Higgs doublets are needed to generate masses for `up'- and `down'-type quarks as well as charged leptons after electroweak symmetry breaking. 
\begin{table}[t]
 \centering
 \begin{tabular}{lccc}
  \toprule 
  Name & Scalars $ \phi^i $ & Fermions $ \chi_L^i $ & $ \left( SU(3)_c\,,\,SU(2)_L\right) _Y $ \\
  \midrule 
  Sleptons, leptons & $ \tilde{\ell}^i=
  \begin{pmatrix}
  \tilde{\nu}_L^i \\
  \tilde{e}_L^i
  \end{pmatrix} $ & $ \ell^i=
  \begin{pmatrix}
  \nu_L^i \\
  e_L^i
  \end{pmatrix} $ & $ \left( \mathbf{1},\,\mathbf{2}\right) _{-\frac{1}{2}} $ \\
  & $ \tilde{e}^{*\,i}=\tilde{e}_R^{*\,i} $ & $ e^{c\,i}=e_R^{c\,i} $ & $ \left( \mathbf{1},\,\mathbf{1}\right) _{+1} $ \\[2pt]
  Squarks, quarks & $ \tilde{q}_h^i=
  \begin{pmatrix}
  \tilde{u}_{L,\,h}^i \\
  \tilde{d}_{L,\,h}^i
  \end{pmatrix} $ & $ q_h^i=
  \begin{pmatrix}
  u_{L,\,h}^i \\
  d_{L,\,h}^i
  \end{pmatrix} $ & $ \left( \mathbf{3},\,\mathbf{2}\right) _{+\frac{1}{6}} $ \\
  & $ \tilde{u}_h^{*\, i}=\tilde{u}_{R,\,h}^{*\, i} $ & $ u_h^{c\, i}=u_{R,\,h}^{c\, i} $ & $ \left( \bar{\mathbf{3}},\,\mathbf{1}\right) _{-\frac{2}{3}} $ \\
  & $ \tilde{d}_h^{*\, i}=\tilde{d}_{R,\,h}^{*\, i} $ & $ d_h^{c\, i}=d_{R,\,h}^{c\, i} $ & $ \left( \bar{\mathbf{3}},\,\mathbf{1}\right) _{+\frac{1}{3}} $ \\[2pt]
  Higgs, higgsinos & $ H_d=
  \begin{pmatrix}
  H_d^0 \\
  H_d^-
  \end{pmatrix} $ & $ \tilde{H}_d=
  \begin{pmatrix}
  \tilde{H}_d^0 \\
  \tilde{H}_d^- 
  \end{pmatrix} $ & $ \left( \mathbf{1},\,\mathbf{2}\right) _{-\frac{1}{2}} $ \\
  & $ H_u=
  \begin{pmatrix}
  H_u^+ \\
  H_u^0
  \end{pmatrix} $ & $ \tilde{H}_u=
  \begin{pmatrix}
  \tilde{H}_u^+ \\
  \tilde{H}_u^0 
  \end{pmatrix} $ & $ \left( \mathbf{1},\,\mathbf{2}\right) _{+\frac{1}{2}} $ \\
  \bottomrule 
 \end{tabular}
 \caption[The chiral supermultiplets of the minimal supersymmetric standard model.]{The chiral supermultiplets of the minimal supersymmetric standard model. Listed are the different particles and their transformation properties under the unbroken standard model gauge group. Here, $ i=1,\,2,\,3 $ is the generation index of (s)leptons and (s)quarks and $h=r,\,g,\,b$ is the color index of (s)quarks.}
 \label{ChiralMSSM}
\end{table}

The Lagrangian of a globally supersymmetric theory is determined by the superpotential $W$, which is a function of the scalar fields. In the MSSM with conserved $R$ parity the superpotential is given by~\cite{Barbier:2004ez}
\begin{equation}
 W_{\text{MSSM}}=\mu H_uH_d+\lambda_{ij}^eH_d\tilde{\ell}_i\tilde{e}_j^*+\lambda_{ij}^dH_d\tilde{q}_i\tilde{d}_j^*-\lambda_{ij}^uH_u\tilde{q}_i\tilde{u}_j^*\,.
 \label{MSSMW}
\end{equation}
In this expression we implicitly sum over generation indices $i,\,j=1,\,2,\,3$ and the suppressed $SU(2)$ gauge indices. For instance we have
\begin{equation}
  \lambda_{ij}^e\,H_d\,\tilde{\ell}_i\,\tilde{e}_j^*=\sum_{i,\,j\,=\,1}^3\sum_{\alpha,\,\beta\,=1}^2\lambda_{ij}^e\,\varepsilon^{\alpha\beta}H_{d\alpha}\,\tilde{\ell}_{i\beta}\,\tilde{e}_j^*=\sum_{i,\,j\,=\,1}^3\lambda_{ij}^e\left( H_d^0\tilde{e}_{Li}-H_d^-\tilde{\nu}_{Li}\right) \tilde{e}_{Rj}^*\,.
\end{equation}
The first term in the superpotential is a supersymmetric mass term for the Higgs fields and the remaining terms are Yukawa couplings between the Higgs fields and the matter particles. 
Additionally, the breaking of supersymmetry introduces soft terms for the scalars and the gauginos that are of the following generic form~\cite{Barbier:2004ez}:
\begin{align}
  -\mathscr{L}_\text{MSSM}^\text{soft} &=m_{H_u}^2|H_u|^2+m_{H_d}^2|H_d|^2+\left( BH_uH_d+\text{h.c.}\right) +m_{\tilde{q}_{ij}}^2\tilde{q}_i^*\tilde{q}_j+m_{\tilde{u}_{ij}}^2\tilde{u}_i^*\tilde{u}_j+m_{\tilde{d}_{ij}}^2\tilde{d}_i^*\tilde{d}_j \nonumber\\
  &\qquad+m_{\tilde{\ell}_{ij}}^2\tilde{\ell}_i^*\tilde{\ell}_j+m_{\tilde{e}_{ij}}^2\tilde{e}_i^*\tilde{e}_j+\left( A_{ij}^eH_d\tilde{\ell}_i\tilde{e}_j^*+A_{ij}^dH_d\tilde{q}_i\tilde{d}_j^*-A_{ij}^uH_u\tilde{q}_i\tilde{u}_j^*+\text{h.c.}\right) \nonumber\\
  &\qquad+\frac{1}{2}\,M_1\bar{\tilde{B}}\tilde{B}+\frac{1}{2}\,M_2\bar{\tilde{W}}^3\tilde{W}^3+M_2\bar{\tilde{W}}^+\tilde{W}^++\frac{1}{2}\,M_3\bar{\tilde{g}}^a\tilde{g}^a. \label{MSSMsoft}
\end{align}
The MSSM Lagrangian is then given as~\cite{Martin:1997ns}
\begin{equation}
 \begin{split}
  \mathscr{L}_\text{MSSM} &=-\left( D_\mu\phi^{*i}\right) \left( D^\mu\phi^i\right) -\frac{1}{4}\,F_{\mu\nu}^aF^{a\,\mu\nu}-i\bar{\chi}^i\bar{\sigma}^\mu\partial_\mu\chi^i-i{\lambda^a}^\dagger\bar{\sigma}^\mu D_\mu\lambda^a \\
  &\qquad\qquad-\frac{1}{2}\left( \frac{\partial^2W}{\partial\phi^i\partial\phi^j}\,\chi^i\chi^j+\frac{\partial^2W^*}{\partial\phi^{*i}\partial\phi^{*j}}\,\bar{\chi}^i\bar{\chi}^j\right) \\
  &\qquad\qquad-\sqrt{2}\,g\left( \phi^*T^a\chi\right) \lambda^a-\sqrt{2}\,g\bar{\lambda}^a\left( \bar{\chi}\,T^a\phi\right) -V(\phi,\phi^*)\,,
 \end{split}
\end{equation}
where the scalar potential has the form
\begin{equation}
  V(\phi,\phi^*)=F^iF^{*i}+\frac{1}{2}\,D^aD^a+\mathscr{L}^\text{soft}=\frac{\partial W}{\partial\phi^i}\,\frac{\partial W^*}{\partial\phi^{*i}}+\frac{1}{2}\sum_ag^2\left( \phi^*T^a\phi\right) ^2+\mathscr{L}^\text{soft}.
  \label{scalarpot}
\end{equation}
Let us now turn to the effects of electroweak symmetry breaking.

\subsubsection*{Electroweak Symmetry Breaking}

As we have stated before, in the MSSM there are two complex Higgs doublets. Electroweak symmetry is broken down to electromagnetism,
\begin{equation*}
 SU(2)_L\times U(1)_Y\rightarrow U(1)_{em}\,,
\end{equation*}
dynamically through radiative corrections to the soft Higgs masses $ m_{H_u} $ and $ m_{H_d} $. The neutral Higgs fields then acquire vacuum expectation values $ \left\langle H_u^0\right\rangle =v_u $ and $ \left\langle H_d^0\right\rangle =v_d $. The ratio of the Higgs VEVs is usually denoted as
\begin{equation}
 \tan{\beta}\equiv\frac{v_u}{v_d}\,.
\end{equation}
The VEVs of the Higgs doublets are related to the standard model Higgs VEV $ v\simeq 174\usk\unit{GeV} $ in the following way:
\begin{equation}
  v^2=v_u^2+v_d^2\,,\qquad v_u=v\sin{\beta}\qquad\text{and}\qquad v_d=v\cos{\beta}\,.
  \label{higgsVEV}
\end{equation}
The new mass eigenstates after electroweak symmetry breaking are the electrically neutral photon $A_\mu$ and the electrically neutral $Z$ boson, defined as
\begin{align}
 \begin{pmatrix}
  A_{\mu} \\
  Z_{\mu}
 \end{pmatrix}
 &=\begin{pmatrix}
  \cos{\theta_W} & \sin{\theta_W} \\
  -\sin{\theta_W} & \cos{\theta_W}
 \end{pmatrix}
 \begin{pmatrix}
  B_{\mu} \\
  W_{\mu}^3
 \end{pmatrix}, \label{neutraleigenstates} \\
\intertext{as well as the electrically charged $W^{\pm}$ bosons, defined as}
 W_{\mu}^{\pm} &=\frac{1}{\sqrt{2}}\left( W_{\mu}^1\mp iW_{\mu}^2\right). \label{chargedeigenstates}
\end{align}
In these expressions the weak mixing angle $\theta_W$ is defined as
\begin{equation}
  \sin{\theta_W}=\frac{g'}{\sqrt{g^2+g'^2}}\qquad\text{and}\qquad\cos{\theta_W}=\frac{g}{\sqrt{g^2+g'^2}}\,,
  \label{weakangle}
\end{equation}
with the $ SU(2)_L $ gauge coupling $ g $ and the $ U(1)_Y $ gauge coupling $ g' $. The gauge coupling of the residual symmetry group $U(1)_{em}$ is the elementary charge
\begin{equation}
  e=\sqrt{4\,\pi\,\alpha}=g\sin\theta_W=g'\cos\theta_W\,.
\end{equation}
The photon $A_\mu$ turns out to be massless, while the other electroweak gauge bosons acquire masses through the Higgs mechanism. They absorb the three massless degrees of freedom of the two complex Higgs doublets: the Goldstone bosons $G$ and $G^{\pm}$. These become the longitudinal modes of the massive $Z$ and $W^{\pm}$ gauge bosons whose masses are then given by
\begin{equation}
  m_W=\frac{g\,v}{\sqrt{2}}\qquad\text{and}\qquad m_Z=\frac{g\,v}{\sqrt{2}\cos{\theta_W}}\,.
  \label{WZmass}
\end{equation}
The Higgs gauge eigenstates can be expressed in terms of the Higgs mass eigenstates: The neutral gauge eigenstates are decomposed as
\begin{equation}
 \begin{pmatrix}
  H_u^0 \\
  H_d^0
 \end{pmatrix}=
 \begin{pmatrix}
  v_u \\
  v_d
 \end{pmatrix}+\frac{1}{\sqrt{2}}
 \begin{pmatrix}
  c_{\alpha} & s_{\alpha} \\
  -s_{\alpha} & c_{\alpha}
 \end{pmatrix}
\begin{pmatrix}
  h \\
  H
 \end{pmatrix}+\frac{i}{\sqrt{2}}
 \begin{pmatrix}
  s_{\beta_0} & c_{\beta_0} \\
  -c_{\beta_0} & s_{\beta_0}
 \end{pmatrix}
 \begin{pmatrix}
  G \\
  A
 \end{pmatrix}
 \label{HiggsStates}
\end{equation}
and the charged gauge eigenstates read
\begin{equation}
 \begin{pmatrix}
  H_u^+ \\
  H_d^{-\,*}
 \end{pmatrix}=
 \begin{pmatrix}
  s_{\beta_{\pm}} & c_{\beta_{\pm}} \\
  -c_{\beta_{\pm}} & s_{\beta_{\pm}}
 \end{pmatrix}
 \begin{pmatrix}
  G^+ \\
  H^+
 \end{pmatrix}.
\end{equation}
Here and in the following parts we use the abbreviations $s_{\alpha}\equiv\sin{\alpha} $and $c_{\alpha}\equiv\cos{\alpha}$ for the mixing angles. In the tree-level approximation one has $\beta_0=\beta_{\pm}=\beta$ and the masses of the Higgs mass eigenstates are given by
\begin{equation}
 \begin{split}
  m_{h,\,H}^2 &=\frac{1}{2}\left( m_A^2+m_Z^2\mp\sqrt{\left( m_A^2-m_Z^2\right) ^2+4\,m_Z^2\,m_A^2\sin^22\,\beta}\right), \\
  m_A^2 &=2\abs{\mu}^2+m_{H_u}^2+m_{H_d}^2\,, \\
  m_{H^{\pm}}^2 &=m_A^2+m_W^2\,.
 \end{split}
\end{equation}
In this case, the mixing angle $\alpha$ is determined by the conditions
\begin{equation}
 \frac{\sin{2\,\alpha}}{\sin{2\,\beta}}=-\left( \frac{m_H^2+m_h^2}{m_H^2-m_h^2}\right) \!\!\qquad\text{and}\qquad\frac{\tan{2\,\alpha}}{\tan{2\,\beta}}=\left( \frac{m_A^2+m_Z^2}{m_A^2-m_Z^2}\right) ,
 \label{alphaangle}
\end{equation}
and is usually chosen to be negative.

In the decoupling limit~\cite{Gunion:2002zf}, \textit{i.e.} for $m_A\gg m_Z$, corresponding to a large $\mu$ parameter, the Higgs particles $H$, $A$ and $H^{\pm}$ are very heavy and decouple from the low-energy phenomenology. Only the lightest Higgs particle $h$ stays at the electroweak scale. In this case, using the relations (\ref{alphaangle}), the mixing angle becomes $\alpha\simeq\beta-\pi/2$ and the lightest Higgs boson $h$ obtains the couplings of the ordinary standard model Higgs boson.

\subsubsection*{Neutralinos and Charginos}

Gauginos and higgsinos mix with each other due to electroweak symmetry breaking. The neutral gauginos and the neutral higgsinos combine to form the four mass eigenstates called neutralinos $\tilde{\chi}_{\alpha}^0\,$, while the charged gauginos and the charged higgsinos mix to form the two mass eigenstates called charginos $\tilde{\chi}_{\alpha}^{\pm}$.

In the gauge eigenstate basis $\psi^0=(-i\tilde{B},\,-i\tilde{W}^3,\,\tilde{H}_u^0,\,\tilde{H}_d^0)^T$ the neutralino mass part of the Lagrangian is given by
\begin{equation}
 \mathscr{L}_{\text{neutralino mass}}=-\frac{1}{2}\,{\psi^0}^TM_N\,\psi^0+\text{h.c.}
\end{equation}
The entries of the neutralino mass matrix $M_N$ can be directly read off from the MSSM Lagrangian. The matrix turns out to be symmetric and reads
\begin{equation}
 M_N=
 \begin{pmatrix}
  M_1 & 0 & \frac{g'v_u}{\sqrt{2}} & -\frac{g'v_d}{\sqrt{2}} \\
  0 & M_2 & -\frac{g\,v_u}{\sqrt{2}} & \frac{g\,v_d}{\sqrt{2}} \\
  \frac{g'v_u}{\sqrt{2}} & -\frac{g\,v_u}{\sqrt{2}} & 0 & -\mu \\
  -\frac{g'v_d}{\sqrt{2}} & \frac{g\,v_d}{\sqrt{2}} & -\mu & 0
 \end{pmatrix}.
 \label{neutralinomatrix}
\end{equation}
Using an orthogonal matrix that includes equation (\ref{neutraleigenstates}), we can change the basis from the gauge eigenstates to the supersymmetric partners of the massive gauge bosons ${\psi^0}^\prime=(-i\tilde{\gamma},\,-i\tilde{Z},\,\tilde{H}_u^0,\,\tilde{H}_d^0)^T$:
\begin{equation}
 \begin{pmatrix}
  -i\tilde{\gamma} \\
  -i\tilde{Z} \\
  \tilde{H}_u^0 \\
  \tilde{H}_d^0
 \end{pmatrix}=R
 \begin{pmatrix}
  -i\tilde{B} \\
  -i\tilde{W}^3 \\
  \tilde{H}_u^0 \\
  \tilde{H}_d^0
 \end{pmatrix},\qquad
 R=
 \begin{pmatrix}
  c_W & s_W & 0 & 0 \\
  -s_W & c_W & 0 & 0 \\
  0 & 0 & 1 & 0 \\
  0 & 0 & 0 & 1 
 \end{pmatrix}.
\end{equation}
The neutralino mass term can then be rewritten in the form
\begin{equation}
 \mathscr{L}_{\text{neutralino mass}}=-\frac{1}{2}\,{\psi^0}^{\prime T}M_N'\,{\psi^0}^\prime+\text{h.c.}\,, 
\end{equation}
where the neutralino mass matrix in the new basis can be written as
\begin{equation}
  M_N'=R\,M_N\,R^T=
  \begin{pmatrix}
   M_1c_W^2+M_2\,s_W^2 & \left( M_2-M_1\right) s_W\,c_W & 0 & 0 \\
   \left( M_2-M_1\right) s_W\,c_W & M_1s_W^2+M_2\,c_W^2 & -m_Z\,s_{\beta} & m_Z\,c_{\beta} \\
   0 & -m_Z\,s_{\beta} & 0 & -\mu \\
   0 & m_Z\,c_{\beta} & -\mu & 0
  \end{pmatrix}.\!\!\!\!\!
\end{equation}
In this expression we used the relations (\ref{higgsVEV}), (\ref{weakangle}) and (\ref{WZmass}) to rewrite the contributions from Yukawa couplings in terms of the $Z$ boson mass. For the mixing angles we used the abbreviations $s_W\equiv\sin{\theta_W}$, $c_W\equiv\cos{\theta_W}$, $s_\beta\equiv\sin{\beta}$ and $c_\beta\equiv\cos{\beta}$.

In a similar way the basis can be changed to the neutralino mass eigenstates. Symmetric matrices can always be diagonalized by a unitary matrix $N$ and its transposed matrix in the following way:
\begin{equation}
 N^*M_N'N^{\dagger}=
 \begin{pmatrix}
  m_{\tilde{\chi}_1^0} & 0 & 0 & 0 \\
  0 & m_{\tilde{\chi}_2^0} & 0 & 0 \\
  0 & 0 & m_{\tilde{\chi}_3^0} & 0 \\
  0 & 0 & 0 & m_{\tilde{\chi}_4^0}
 \end{pmatrix}. 
\end{equation}
The neutralinos are then obtained by the rotation
\begin{equation}
 \begin{pmatrix}
  \tilde{\chi}_1^0 \\
  \tilde{\chi}_2^0 \\
  \tilde{\chi}_3^0 \\
  \tilde{\chi}_4^0 
 \end{pmatrix}=N
 \begin{pmatrix}
  -i\tilde{\gamma} \\
  -i\tilde{Z}^0 \\
  \tilde{H}_u^0 \\
  \tilde{H}_d^0
 \end{pmatrix}.
\end{equation}

The chargino mass term in the gauge eigenstate basis $\psi^-=(-i\tilde{W}^-,\,\tilde{H}_d^-)^T$ and $\psi^+=(-i\tilde{W}^+,\,\tilde{H}_u^+)^T$ reads
\begin{equation}
 \mathscr{L}_{\text{chargino mass}}=-{\psi^-}^TM_C\,\psi^++\text{h.c.}\,,
\end{equation}
where the chargino mass matrix $M_C$ is given by
\begin{equation}
 M_C=
 \begin{pmatrix}
  M_2 & g\,v_u \\
  g\,v_d & \mu
 \end{pmatrix}=
 \begin{pmatrix}
  M_2 & \sqrt{2}\,m_W\,s_{\beta} \\
  \sqrt{2}\,m_W\,c_{\beta} & \mu
 \end{pmatrix}.
 \label{charginomatrix}
\end{equation}
For the last equality we employed the relations (\ref{higgsVEV}) and (\ref{WZmass}). Due to the asymmetric form of the mass matrix, two unitary matrices $U$ and $V$ are required to diagonalize it. This can be done in the following way:
\begin{equation}
 U^*M_CV^{\dagger}=VM_C^{\dagger}U^T=
 \begin{pmatrix}
  m_{\tilde{\chi}_1^{\pm}} & 0 \\
  0 & m_{\tilde{\chi}_2^{\pm}}
 \end{pmatrix}.
\end{equation}
The matrices $U$ and $V$ also rotate the gauge eigenstates into the basis of left- and right-handed chargino mass eigenstates:
\begin{equation}
 \begin{pmatrix}
  \tilde{\chi}_1^- \\
  \tilde{\chi}_2^-
 \end{pmatrix}=U
 \begin{pmatrix}
  -i\tilde{W}^- \\
  \tilde{H}_d^- 
 \end{pmatrix}
 \qquad\text{and}\qquad
 \begin{pmatrix}
  \tilde{\chi}_1^+ \\
  \tilde{\chi}_2^+
 \end{pmatrix}=V
 \begin{pmatrix}
-i  \tilde{W}^+ \\
  \tilde{H}_u^+
 \end{pmatrix}.
\end{equation}
The chargino masses are given as the positive roots of the eigenvalues of $M_C^{\dagger}M_C$, since
\begin{equation}
 VM_C^{\dagger}M_CV^{\dagger}=U^*M_CM_C^{\dagger}U^T=
 \begin{pmatrix}
  m_{\tilde{\chi}_1^{\pm}}^2 & 0 \\
  0 & m_{\tilde{\chi}_2^{\pm}}^2
 \end{pmatrix},
\end{equation}
and can be given in an analytical form:
\begin{equation}
 \begin{split}
  m_{\tilde{\chi}_{1,2}^{\pm}}^2 &=\frac{1}{2}\bigg( \abs{M_2}^2+\abs{\mu}^2+2\,m_W^2 \\
  &\qquad\qquad\mp\left.\sqrt{\left( \abs{M_2}^2+\abs{\mu}^2+2\,m_W^2\right) ^2-4\abs{\mu\,M_2-m_W^2\sin{2\,\beta}}^2}\right) .
 \end{split}
\end{equation}
One should keep in mind, though, that our treatment of the neutralino and chargino mass matrices is only valid at tree level and typically gets corrections from higher-order contributions.

\section[Bilinear Breaking of \texorpdfstring{$R$}{R} Parity]{Bilinear Breaking of \boldmath$R$ Parity}
\label{Rbreaking}

\subsubsection*{\boldmath$R$ Parity}

In the standard model of particle physics the baryon number $B$ and the lepton number $L$ are accidentally conserved quantities. In particular, the difference $B-L$ is even conserved in nonperturbative sphaleron processes coming from the chiral anomaly of the standard model. In the MSSM this conservation of $B$ and $L$ does not generically hold. As a consequence one usually introduces an additional parity by hand in order to forbid processes that lead to rapid proton decay~\cite{Fayet:1977yc,Farrar:1978xj}. This parity is called $R$ parity and is described by a new multiplicative quantum number
\begin{equation}
 R_p=(-1)^{3(B-L)+2s}
 \label{Rparity}
\end{equation}
that is assigned to all particles in the MSSM plus the gravity multiplet according to their baryon and lepton numbers as well as their spin $s$. It turns out that all standard model particles have even ($R_p=+1$) and their supersymmetric partners odd $R$ parity ($R_p=-1$). This imposes that supersymmetric particles can only be produced pairwise and that they cannot decay into standard model particles only.

Thus, in theories with conserved $R$ parity the lightest supersymmetric particle is absolutely stable and provides a natural particle candidate for dark matter as long as it is neutral with respect to electromagnetic and strong interactions. Depending on the mechanism of supersymmetry breaking, it turns out that in many cases the lightest of the neutralinos discussed above is the lightest supersymmetric particle and thus it has been thoroughly studied as the prototype candidate for dark matter in the form of weakly interacting massive particles (WIMPs)~\cite{Jungman:1995df}.

\subsubsection*{\boldmath$R$-Parity Breaking Operators}

Despite the typical assumption of a conserved $R$ parity, in general supersymmetric theories contain $R$-violating terms~\cite{Hall:1983id}. In particular, the most general MSSM superpotential contains additional $R$-parity violating terms that are renormalizable and compatible with all gauge symmetries~\cite{Barbier:2004ez}:
\begin{equation}
 W_{\slashed{R}_p}=\mu_iH_u\tilde{\ell}_i+\frac{1}{2}\lambda_{ijk}\tilde{\ell}_i\tilde{\ell}_j\tilde{e}_k^*+\lambda'_{ijk}\tilde{\ell}_i\tilde{q}_j\tilde{d}_k^*+\frac{1}{2}\lambda_{ijk}^{\prime\prime}\tilde{u}_i^*\tilde{d}_j^*\tilde{d}_k^*\,.
\end{equation}
As in the case of the superpotential with conserved $R$ parity, summation over the generation indices $i,\,j,\,k=1,\,2,\,3$ and the suppressed gauge indices is assumed. Gauge invariance requires that $\lambda_{ijk}$ is antisymmetric with respect to its first two indices while $\lambda_{ijk}^{\prime\prime}$ must be antisymmetric in the last two indices. The factors of one half are typically introduced to avoid factors of two in the calculation of matrix elements. The $R$-parity breaking mass parameters $\mu_i$, and the Yukawa couplings $\lambda_{ijk}$ and $\lambda'_{ijk}$ violate lepton number, while the Yukawa couplings $\lambda_{ijk}^{\prime\prime}$ violate baryon number. We want to stress again that there is in principle no underlying symmetry of the theory that forbids these terms.

Experimentally, however, the strength of these interactions is strongly constrained: The observed lower limit of the proton lifetime of $\tau_{p}>2.1\times10^{29}\,\text{years}$~\cite{Nakamura:2010zzi} demands, for instance, that either the $L$-violating or the $B$-violating couplings vanish, or that all the couplings are extremely suppressed. In addition, the requirement that a baryon asymmetry produced in the early universe is not erased before the electroweak phase transition implies constraints on the order of $\lambda_{ijk},\,\lambda_{ijk}'<10^{-7}$~\cite{Campbell:1990fa, Fischler:1990gn, Dreiner:1992vm,Endo:2009cv}.

In addition to the terms in the superpotential, the most general Lagrangian of a softly broken supersymmetric theory contains $R$-parity violating soft terms~\cite{Barbier:2004ez}:
\begin{equation}
  -\mathscr{L}_{\slashed{R}_p}^\text{soft}=B_iH_u\tilde{\ell}_i+m_{H_d\ell_i}^2H_d^*\tilde{\ell}_i+\frac{1}{2}\,A_{ijk}\tilde{\ell}_i\tilde{\ell}_j\tilde{e}_k^*+\frac{1}{2}\,A_{ijk}'\tilde{\ell}_i\tilde{q}_j\tilde{d}_k^*+\frac{1}{2}\,A_{ijk}''\tilde{u}_i^*\tilde{d}_j^*\tilde{d}_k^*+\text{h.c.}
\end{equation}
From the theoretical point of view, $R$-parity violation could for instance be connected to the spontaneous breaking of a $B-L$ symmetry. A model where $R$-parity breaking is realized in this way was studied in~\cite{Buchmuller:2007ui}.

\subsubsection*{Bilinear \boldmath$R$-Parity Violation}

In this work we want to restrict ourselves to the case of bilinear $R$-parity breaking, \textit{i.e.} only the $R$-parity violating mass terms $\mu_i$, $B_i$ and $m_{H_d\ell_i}^2$ are considered in addition to the MSSM case. In these scenarios proton stability is generically guaranteed as the listed $R$-parity breaking terms violate only lepton number while baryon number is still a conserved quantity.

Once $R$ parity is broken, the sneutrinos acquire a typically nonvanishing vacuum expectation value $v_i$. In order to determine its value we need to minimize the scalar potential that is given by equation~(\ref{scalarpot}). Taking into account the extended superpotential and the additional soft terms, and replacing the neutral components of the scalar $SU(2)$ doublets with their respective VEVs we find the following expression for the minimum of the potential:
\begin{equation}
 \begin{split}
  V_\text{min} &=|\mu|^2\left( v_d^2+v_u^2\right) +\left( \mu\,\mu_i^*+\mu^*\mu_i\right) v_d\,v_i+|\mu_i|^2\left( v_u^2+v_i^2\right) +m_{H_d}^2v_d^2+m_{H_u}^2v_u^2 \\
  &\qquad-\left( B+B^*\right) v_uv_d+m_{\tilde{\ell}_{ij}}^2v_i^2-\left( B_i+B_i^*\right) v_uv_i+\left( m_{H_d\ell_i}^2+m_{H_d\ell_i}^{*2}\right) v_d\,v_i \\
  &\qquad+\frac{1}{8}\left( g^2+g'^2\right) \left( v_u^2-v_d^2-v_i^2\right) ^2
 \end{split}
\end{equation}
The parameters $B$ and $B_i$ can always be chosen real by a suitable choice of the phase of the slepton fields~\cite{Hajer:2010zz}. As long as the $R$-parity violating parameters are small, the Higgs VEVs are practically the same as in the $R$-parity conserving case. The VEVs for the sneutrino fields can then be found by looking at the minimum of the scalar potential in the sneutrino directions:
\begin{equation}
 \begin{split}
  0\stackrel{!}{=}\left. \frac{\partial V}{\partial\tilde{\nu}_i^*}\right| _\text{min}= &\;\mu\,\mu_i^*v_d+|\mu_i|^2v_i+m_{\tilde{\ell}_{ij}}^2v_j-B_i\,v_u+m_{H_d\ell_i}^{*2}v_d \\
  &\qquad+\frac{1}{4}\left( g^2+g'^2\right) \left( v_d^2-v_u^2+v_i^2\right) v_i\,.
 \end{split}
 \label{potmin}
\end{equation}
The term $|\mu_i|^2v_i$ can be safely neglected since it is quadratic in the small parameter $\mu_i$. In addition, $v_i^2$ does not contribute significantly to the sum of squared VEVs and can be neglected in the $D$-term part:
\begin{equation}
  \frac{1}{4}\left( g^2+g'^2\right) \left( v_d^2-v_u^2+v_i^2\right) \simeq\frac{1}{2}\,m_Z^2\left( \cos^2\beta-\sin^2\beta\right) =\frac{1}{2}\,m_Z^2\cos2\,\beta\,.
\end{equation}
After rearranging equation~(\ref{potmin}) we find the following expression for the VEVs of the sneutrino fields:
\begin{equation}
 \frac{v_i}{v_d}\simeq\frac{B_i\tan\beta-m_{H_d\ell_i}^{*2}-\mu\mu_i^*}{m_{\tilde{\ell}_{ij}}^2+\frac{1}{2}\,m_Z^2\cos2\,\beta}\,.
\end{equation}

Now we want to perform a supersymmetry-preserving rotation of the down-type Higgs and lepton $SU(2)$ doublets. After the field redefinitions
\begin{equation}
  H_d=H_d'-\epsilon_i\tilde{\ell}_i'\,,\quad\tilde{H}_d=\tilde{H}_d'-\epsilon_i\ell_i'\,,\quad\tilde{\ell}_i=\tilde{\ell}_i'+\epsilon_iH_d'\,,\quad\text{and}\quad\ell_i=\ell_i'+\epsilon_i\tilde{H}_d'\,,
\end{equation}
where $\epsilon_i=\mu_i/\mu$, the bilinear $R$-parity breaking mass term vanishes from the superpotential, \textit{i.e.} $\mu_i'=0$. However, new $R$-parity violating Yukawa couplings are generated in this way:
\begin{equation}
  W'=\mu H_uH_d'+\lambda_{ij}^eH_d'\tilde{\ell}_i'\tilde{e}_j^*+\lambda_{ij}^dH_d'\tilde{q}_i\tilde{d}_j^*-\lambda_{ij}^uH_u\tilde{q}_i\tilde{u}_j^*\mu+\frac{1}{2}\,\lambda_{ijk}\tilde{\ell}_i'\tilde{e}_j^*\tilde{\ell}_k'+\lambda_{ijk}'\tilde{q}_i\tilde{d}_j^*\tilde{\ell}_k'\,,
\end{equation}
where
\begin{equation}
  \lambda_{ijk}=\epsilon_k\,\lambda_{ij}^e-\epsilon_i\,\lambda_{kj}^e\qquad\text{and}\qquad\lambda_{ijk}'=\epsilon_k\,\lambda_{ij}^d\,.
\end{equation}
It is important to note that although this rotation generates new trilinear couplings it does not lead to baryon number violation and therefore proton stability is still guaranteed. In the soft scalar Lagrangian only the $R$-parity violating mass mixings change at first order in the small $R$-parity breaking parameters:
\begin{equation}
  B_i'=B_i-\epsilon_i\,B\qquad\text{and}\qquad m_{H_d\ell_i}'^2=m_{H_d\ell_i}^2+\epsilon_i\left( m_{\tilde{\ell}_{ij}}^2-m_{H_d}^2\right) .
\end{equation}
In this new basis the VEV of the sneutrino fields turns out to be
\begin{equation}
 \frac{v_i}{v_d}\simeq\frac{B_i\tan\beta-m_{H_d\ell_i}^{*2}}{m_{\tilde{\ell}_{ij}}^2+\frac{1}{2}\,m_Z^2\cos2\,\beta}\,,
 \label{sneutrinoVEV}
\end{equation}
where we dropped the primes on the rotated parameters. This expression for the vacuum expectation values of the sneutrino fields has been used in several works on gravitino dark matter to parametrize the effect of bilinear $R$-parity violation, see for instance~\cite{Buchmuller:2007ui,Ishiwata:2008cu,Covi:2008jy,Grefe:2008zz}. In some of these works, though, the $D$-term contribution in the denominator has been neglected.

In a different parametrization of bilinear $R$-parity breaking one can perform an additional rotation of the fields in the scalar sector such that the sneutrino VEV vanishes~\cite{Bobrovskyi:2010ps}. In that case all effects of bilinear $R$-parity violation are encoded in the form of $R$-parity breaking Yukawa couplings. Of course, there should be no physical consequences of using a different parametrization of the theory.

\subsubsection*{Neutralino--Neutrino and Chargino--Charged Lepton Mixing}

In models with bilinear $R$-parity breaking the distinction between down-type Higgs and lepton supermultiplets is lost. Since lepton number is not a conserved quantity anymore, the left-handed neutrinos mix with the neutralinos to form new mass eigenstates. Similarly, the charged leptons mix with the charginos. These mixings lead to decays of the lightest supersymmetric particle and they are therefore a crucial ingredient for the calculation of gravitino dark matter decay channels in this work.

The $4\times4$ neutralino mixing matrix in equation~(\ref{neutralinomatrix}) is extended to a $7\times7$ matrix that also includes mixings with the three flavors of the light neutrinos. In its most general form this neutralino--neutrino mixing matrix can be written as~\cite{Barbier:2004ez}
\begin{equation}
 M_N^7=
 \begin{pmatrix}
  M_1 & 0 & \frac{g'v_u}{\sqrt{2}} & -\frac{g'v_d}{\sqrt{2}} & \frac{g'v_j}{\sqrt{2}} \\
  0 & M_2 & -\frac{g\,v_u}{\sqrt{2}} & \frac{g\,v_d}{\sqrt{2}} & -\frac{g\,v_j}{\sqrt{2}} \\
  \frac{g\,v_d}{\sqrt{2}} & -\frac{g'v_d}{\sqrt{2}} & 0 & -\mu & -\mu_j \\
  -\frac{g\,v_u}{\sqrt{2}} & \frac{g'v_u}{\sqrt{2}} & -\mu & 0 & 0 \\
  \frac{g'v_i}{\sqrt{2}} & -\frac{g\,v_i}{\sqrt{2}} & -\mu_i & 0 & 0
 \end{pmatrix},
\end{equation}
where the basis is given by $\psi_i^0=(-i\tilde{B},\,-i\tilde{W}^3,\,\tilde{H}_u^0,\,\tilde{H}_d^0,\,\nu_i)^T$. As mentioned in the previous section, we want to present our results in a parametrization where the bilinear $R$-parity violating terms in the superpotential are vanishing ($\mu_i'=0$). Rotating the bino and wino fields into the photino--zino basis and expressing the Higgs VEVs in terms of the $Z$ mass, the neutralino--neutrino mixing matrix then reads
\begin{equation}
  M_N^{7\prime}=
  \begin{pmatrix}
   M_1c_W^2+M_2\,s_W^2 & \left( M_2-M_1\right) s_W\,c_W & 0 & 0 & 0 \\
   \left( M_2-M_1\right) s_Wc_W & M_1s_W^2+M_2\,c_W^2 & -m_Z\,s_\beta & m_Z\,c_\beta & -m_Z\,\xi_j \\
   0 & -m_Z\,s_\beta & 0 & -\mu & 0 \\
   0 & m_Z\,c_\beta & -\mu & 0 & 0 \\
   0 & -m_Z\,\xi_i & 0 & 0 & 0
  \end{pmatrix}.
\end{equation}
In this expression we introduced the ratio of the sneutrino VEVs and the Higgs VEV $\xi_i=v_i/v$ for a dimensionless parametrization of $R$-parity violation. This mixing matrix is then diagonalized by a unitary matrix $N^7$ and the mass eigenstates are obtained by the rotation
\begin{equation}
 \begin{pmatrix}
  \tilde{\chi}_1^0 \\
  \tilde{\chi}_2^0 \\
  \tilde{\chi}_3^0 \\
  \tilde{\chi}_4^0 \\
  \tilde{\chi}_{4+i}^0
 \end{pmatrix}=N^7
 \begin{pmatrix}
  -i\tilde{\gamma} \\
  -i\tilde{Z}^0 \\
  \tilde{H}_u^0 \\
  \tilde{H}_d^0 \\
  \nu_i
 \end{pmatrix}.
\end{equation}
Due to the very small mixing via the sneutrino VEV the neutrino eigenstate is practically not changed by the rotation. Thus we have
\begin{equation}
  \nu_i\simeq\tilde{\chi}_{4+i}^0
\end{equation}
and mixings between the fields $\psi_i^0$ and the neutrinos can be simply calculated as
\begin{equation}
  \nu_i=N_{4+i\,j}^7\,\psi_j^0=N_{\nu_i\,j}^7\,\psi_j^0\,.
\end{equation}
If one considers neutrinos and antineutrinos as separate particles one should, however, employ an additional chirality projector in the calculation of matrix elements:
\begin{equation}
  \nu_{Li}=P_L\,\nu_i=N_{\nu_i\,j}^7\,P_L\,\psi_j^0\,.
\end{equation}
Another consequence of the fact that the mixing of neutralinos with the neutrinos is a very small perturbation to the neutralino mass matrix is that it can be described using the method of mass insertions that is frequently used in studies of small flavor-violating effects\cite{Hall:1985dx,Gabbiani:1996hi}. Since this method was used in previous treatments of gravitino decays via neutrino--neutralino and charged lepton-chargino mass mixings~\cite{Ishiwata:2008cu,Covi:2008jy} we will also present the connection to the mixing parameters used in those works. In that case the mixing of neutralinos with the neutrinos can be expressed in terms of a mixing of the fields $\psi_i^0$ into the zino in the standard neutralino sector multiplied with the coupling of the zino to the neutrinos:
\begin{equation}
  \begin{split}
    N_{4+i\,1}^7&=N_{\nu_i\,\tilde{\gamma}}^7\simeq-\xi_i\,U_{\tilde{\gamma}\tilde{Z}}\,, \\
    N_{4+i\,2}^7&=N_{\nu_i\,\tilde{Z}}^7\simeq-\xi_i\,U_{\tilde{Z}\tilde{Z}}\,, \\
    N_{4+i\,3}^7&=N_{\nu_i\,\tilde{H}_u^0}^7\simeq-\xi_i\,U_{\tilde{H}_u^0\tilde{Z}}\,, \\
    N_{4+i\,4}^7&=N_{\nu_i\,\tilde{H}_d^0}^7\simeq-\xi_i\,U_{\tilde{H}_d^0\tilde{Z}}\,.
  \end{split}
\end{equation}
In these expressions the photino--zino, zino--zino and higgsino--zino mixing parameters are defined in the following way:
\begin{align}
  U_{\tilde{\gamma}\tilde{Z}}&=m_Z\sum_{i=1}^4\frac{N^*_{i\tilde{\gamma}}N_{i\tilde{Z}}}{m_{\tilde{\chi}_{i}^0}}\,, \label{UgammaZ}\\
  U_{\tilde{Z}\tilde{Z}}&=m_Z\sum_{i=1}^4\frac{N^*_{i\tilde{Z}}N_{i\tilde{Z}}}{m_{\tilde{\chi}_{i}^0}}\,, \label{UZZ}\\
  U_{\tilde{H}_u^0\tilde{Z}}&=m_Z\sum_{i=1}^4\frac{N^*_{i\tilde{H}_u^0}N_{i\tilde{Z}}}{m_{\tilde{\chi}_{i}^0}}\,, \label{UHuZ}\\
  U_{\tilde{H}_d^0\tilde{Z}}&=m_Z\sum_{i=1}^4\frac{N_{i\tilde{H}_d^0}^*N_{i\tilde{Z}}}{m_{\tilde{\chi}_{i}^0}}\,. \label{UHdZ}
\end{align}
Following the treatment in~\cite{Bobrovskyi:2010ps} we now perform a perturbative diagonalization of the mixing matrix. Neglecting the off-diagonal elements proportional to the $Z$ mass and the sneutrino VEV the matrix is diagonalized by
\begin{equation}
  N^7=
  \begin{pmatrix}
  c_W & -s_W & 0 & 0 & 0 \\
  s_W & c_W & 0 & 0 & 0 \\
  0 & 0 & \frac{1}{\sqrt{2}} & \frac{1}{\sqrt{2}} & 0 \\
  0 & 0 & -\frac{1}{\sqrt{2}} & \frac{1}{\sqrt{2}} & 0 \\
  0 & 0 & 0 & 0 & 1 
  \end{pmatrix},\quad
  N^7\,M_N^{7\prime}\,{N^7}^T=
  \begin{pmatrix}
  M_1 & 0 & 0 & 0 & 0 \\
  0 & M_2 & 0 & 0 & 0 \\
  0 & 0 & -\mu & 0 & 0 \\
  0 & 0 & 0 & \mu & 0 \\
  0 & 0 & 0 & 0 & 0
  \end{pmatrix}.
\end{equation}
Starting from this zeroth order approximation, we expand the diagonalizing matrix $N^7$ to leading order in $m_Z$ and $\xi_i$.\footnote{A Mathematica package to perform perturbative matrix diagonalizations was kindly provided by Jonas Schmidt.} From the result of this perturbative expansion we find useful approximate formulae for the mixing parameters:
\begin{align}
  U_{\tilde{\gamma}\tilde{Z}}&\simeq -m_Z\sin\theta_W\cos\theta_W\,\frac{M_2-M_1}{M_1\,M_2}\,, \label{UgammaZapx}\\
  U_{\tilde{Z}\tilde{Z}}&\simeq m_Z\left( \frac{\sin^2\theta_W}{M_1}+\frac{\cos^2\theta_W}{M_2}\right) , \label{UZZapx}\\
  U_{\tilde{H}_u^0\tilde{Z}}&\simeq m_Z^2\cos\beta\,\frac{M_1\cos^2\theta_W+M_2\sin^2\theta_W}{M_1\,M_2\,\mu}\,, \label{UHuZapx}\\
  U_{\tilde{H}_d^0\tilde{Z}}&\simeq -m_Z^2\sin\beta\,\frac{M_1\cos^2\theta_W+M_2\sin^2\theta_W}{M_1\,M_2\,\mu}\,. \label{UHdZapx}
\end{align}
Comparing these relations to results from numerical diagonalizations we find that the accuracy of these formulae is at the percent level as long as $M_1, M_2, \mu\gtrsim m_Z$. Hence they are more accurate than the approximations for the mixing parameters previously presented in~\cite{Grefe:2008zz}. In addition, we now also have approximate formulae for higgsino--zino mixing that were not available before.

Similar to the case of the neutralino mass matrix, the $2\times2$ chargino mixing matrix in equation~(\ref{charginomatrix}) is extended to a $5\times5$ matrix that also includes mixings with the three flavors of charged leptons. In its most general form this chargino--lepton mixing matrix can be written as~\cite{Barbier:2004ez}
\begin{equation}
 M_C^5=
 \begin{pmatrix}
  M_2 & g\,v_u & 0 \\
  g\,v_d & \mu & -\lambda_{ij}^e\,v_i \\
  g\,v_i & \mu_i & \lambda_{ij}^e\,v_d
 \end{pmatrix},
\end{equation}
where the basis vectors are $\psi^-=(-i\tilde{W}^-,\,\tilde{H}_d^-,\,\ell_i^-)^T$ and $\psi^+=(-i\tilde{W}^+,\,\tilde{H}_u^+,\,e_i^{c\,+})^T$. Also in this case we rotate away the $\mu_i$ and express the Higgs VEVs in terms of the $W$ mass and and the lepton mass:
\begin{equation}
 M_C^{5\prime}=
 \begin{pmatrix}
  M_2 & \sqrt{2}\,m_W\,s_\beta & 0 \\
  \sqrt{2}\,m_W\,c_\beta & \mu & -m_{\ell_{ij}}\,\xi_i\,c_\beta \\
  \sqrt{2}\,m_W\,\xi_i & 0 & m_{\ell_{ij}}
 \end{pmatrix}.
\end{equation}
This mass matrix is diagonalized by two unitary matrices $U^5$ and $V^5$ and the mass eigenstates are obtained by the rotations
\begin{equation}
 \begin{pmatrix}
  \tilde{\chi}_1^- \\
  \tilde{\chi}_2^- \\
  \tilde{\chi}_{2+i}^-
 \end{pmatrix}=U^5
 \begin{pmatrix}
  -i\tilde{W}^- \\
  \tilde{H}_d^- \\
  \ell_i^-
 \end{pmatrix}
 \qquad\text{and}\qquad
 \begin{pmatrix}
  \tilde{\chi}_1^+ \\
  \tilde{\chi}_2^+ \\
  \tilde{\chi}_{2+i}^+
 \end{pmatrix}=V^5
 \begin{pmatrix}
  -i\tilde{W}^+ \\
  \tilde{H}_u^+ \\
  e_i^{c\,+}
 \end{pmatrix}.
\end{equation}
Due to the very small mixing via the sneutrino VEV the left- and the right-handed part of the charged lepton eigenstates are practically not changed by the rotation. Thus we have
\begin{equation}
  \ell_i^-\simeq\tilde{\chi}_{2+i}^-\qquad\text{and}\qquad e_i^{c\,+}\simeq\tilde{\chi}_{2+i}^+\,,
\end{equation}
and mixings between the fields $\psi_i^\pm$ and the charged leptons can be simply calculated as
\begin{equation}
  \ell_i^-=U_{2+i\,j}^5\,\psi_j^-=U_{\ell_i\,j}^5\,\psi_j^-\qquad\text{and}\qquad e_i^{c\,+}=V_{2+i\,j}^5\,\psi_j^+=V_{e_i^c\,j}^5\,\psi_j^+\,.
\end{equation}
In the calculation of matrix elements with massive leptons as external particles this should be written in the form:
\begin{equation}
  \ell_i^-=P_L\,U_{\ell_i\,j}^5\,\psi_j^-+P_R\,V_{e_i^c\,j}^{5*}\,\bar{\psi}_j^+\,.
\end{equation}
The mixing of charginos to the left-handed leptons is a very small perturbation to the mass matrix induced by the tiny $R$-parity violating coupling of the left-handed leptons to the right-handed wino. Therefore, using the method of mass insertions it decouples into a mixing of the fields $\psi_i^-$ into the right-handed wino in the standard chargino sector multiplied with the coupling of the right-handed wino to the charged leptons:
\begin{equation}
  \begin{split}
    U_{2+i\,1}^5&=U_{\ell_i\,\tilde{W}}^5\simeq-\sqrt{2}\,\xi_i\,U_{\tilde{W}\tilde{W}}\,, \\
    U_{2+i\,2}^5&=U_{\ell_i\,\tilde{H}_d^-}^5\simeq-\sqrt{2}\,\xi_i\,U_{\tilde{H}_d^-\tilde{W}}\,.
  \end{split}
\end{equation}
In principle the situation for the right-handed leptons is similar. Their mixing with the charginos is induced by a tiny $R$-parity violating coupling to the down-type higgsino. However, in addition it mixes via a detour over the left-handed leptons. In both cases the coupling is proportional to the lepton mass and thus both mixings contribute at the same order. In this case the mixing in the standard chargino sector does not decouple from the mixing with the right-handed leptons and thus in cannot be described in a simple form as written above for the left-handed leptons.\footnote{Actually, also for the left-handed leptons there is an additional contribution to the mixing with the charginos via a detour over the right-handed leptons. However, this contribution is suppressed by $\mathcal{O}(m_{\ell_i}^2/m_W^2)$ compared to the dominant contribution and can therefore safely be neglected.} The wino--wino and higgsino--wino mixing parameters are defined in the following way:
\begin{align}
  U_{\tilde{W}\tilde{W}}&=m_W\sum_{i=1}^2\frac{U_{i\tilde{W}^-}V^*_{i\tilde{W}^+}}{m_{\tilde{\chi}_{i}^\pm}}\,, \label{UWW}\\
  U_{\tilde{H}_d^-\tilde{W}}&=m_W\sum_{i=1}^2\frac{U_{i\tilde{H}_d^-}V^*_{i\tilde{W}^+}}{m_{\tilde{\chi}_{i}^\pm}}\,. \label{UHW}
\end{align}
Similar to the case of neutralino--neutrino mixing we can perform a perturbative diagonalization of the mixing matrix in order to find approximate formulae for these mixing parameters. Neglecting the off-diagonal elements proportional to the $W$ mass and the sneutrino VEV the matrix is diagonalized by the identity matrix. Starting from this zeroth order approximation, we expand the diagonalizing matrices $U^5$ and $V^5$ to leading order in $m_W$ and $\xi_i$ and find the relations:
\begin{align}
  U_{\tilde{W}\tilde{W}}&\simeq \frac{m_W}{M_2}\,, \label{UWWapx}\\
  U_{\tilde{H}_d^-\tilde{W}}&\simeq -\frac{\sqrt{2}\,m_W^2\sin\beta}{M_2\,\mu}\,. \label{UHWapx}
\end{align}
Also these formulae are accurate at the percent level as long as $M_2, \mu\gtrsim m_W$. The first relation has already been found in~\cite{Grefe:2008zz} while the higgsino--wino mixing formula was not available before. As mentioned before, the mixing of the charginos with the right-handed leptons is proportional to the lepton mass. Therefore, it turns out that it is suppressed compared to the mixing of the left-handed leptons by $\mathcal{O}(m_{\ell_i}/m_W)$. Hence we will neglect the mixing of the charginos with the right-handed leptons due to $R$-parity breaking in this work and only consider the dominant mixing of the charginos with the left-handed leptons.

\subsubsection*{Higgs--Sneutrino Mixing}

In addition to the extended mixing in the fermionic sector $R$-parity violation also introduces new mixings in the scalar sector, in particular between the Higgs boson and lepton doublets. In principle there are mixings in the neutral part of the doublets as well as in the charged part. Here, however, we want to restrict ourselves to the mixing between the lightest Higgs boson and the sneutrinos in the decoupling limit. The mixing matrix has the form
\begin{equation}
 \mathscr{L}_{h\tilde{\nu}_i}=-\left( h\;\tilde{\nu}_i^*\right) 
 \begin{pmatrix}
  m_h^2 & \frac{1}{\sqrt{2}}\left( B_j\,s_\beta-m_{H_d\ell_i}^2\,c_\beta\right) \\
  \frac{1}{\sqrt{2}}\left( B_i^*\,s_\beta-m_{H_d\ell_i}^{*2}\,c_\beta\right) & m_{\tilde{\ell}_{ij}}^2+\frac{1}{2}\,m_Z^2\,\xi_i^2
 \end{pmatrix}
 \begin{pmatrix}
  h \\
  \tilde{\nu}_j
 \end{pmatrix}.
\end{equation}
Since we only consider small values for the $R$-parity breaking parameters, the masses of the lightest Higgs boson and the sneutrinos are practically not affected by the mixing. The mixing between the lightest Higgs particle and the sneutrinos can then be expressed in terms of the sneutrino VEV:
\begin{equation}
  \mathscr{L}_{h\tilde{\nu}_i}\simeq m_h^2\,h^2+m_{\tilde{\ell}_{ij}}^2\abs{\tilde{\nu}_i}^2+\left( \frac{\xi_i}{\sqrt{2}}\,\left( m_{\tilde{\ell}_{ij}}^2+\frac{1}{2}\,m_Z^2\cos2\,\beta\right) h\tilde{\nu}_i+\text{h.c}\right) .
\end{equation}

\chapter{Gravitino Decays}
\label{gravitino}

In this chapter we discuss the effects of adding the gravitino to the supersymmetric particle spectrum. After an introduction of the field-theoretical treatment of the massive gravitino and its interactions, we review the implications of the presence of the gravitino in the early universe. In the remaining part of this chapter we will present a main result of this thesis: the calculation of the decay widths of the gravitino lightest supersymmetric particle in theories with bilinear $R$-parity breaking including decay processes with three particles in the final state. In addition, we discuss the gravitino branching ratios and their dependence on the parameters of the theory. Finally, we present the spectra of stable final state particles produced in different gravitino decay channels.

\section{The Massive Gravitino and its Interactions}

The massive gravitino is described by the following Lagrangian~\cite{Bolz:2000fu}:
\begin{equation}
 \mathscr{L}_{3/2}=-\frac{1}{2}\,\varepsilon^{\mu\nu\rho\sigma}\bar{\psi}_{\mu}\,\gamma^{5}\gamma_{\nu}\,\partial_{\rho}\psi_{\sigma}-\frac{1}{4}\,m_{3/2}\,\bar{\psi}_{\mu}\left[ \gamma^{\mu},\,\gamma^{\nu}\right] \psi_{\nu}+\mathscr{L}_{\text{int}}\,. 
\end{equation}
The free gravitino obeys the equations of motion
\begin{equation}
 \frac{\partial\mathscr{L}_{3/2}}{\partial\bar{\psi}_{\mu}}-\partial_{\nu}\frac{\partial\mathscr{L}_{3/2}}{\partial(\partial_{\nu}\bar{\psi}_{\mu})}=-\frac{1}{2}\,\varepsilon^{\mu\nu\rho\sigma}\gamma^{5}\gamma_{\nu}\,\partial_{\rho}\psi_{\sigma}-\frac{1}{4}\,m_{3/2}\left[ \gamma^{\mu},\,\gamma^{\nu}\right] \psi_{\nu}=0\,
  \label{GravitinoEoM}
\end{equation}
that finally lead to the Rarita--Schwinger equations~\cite{Rarita:1941mf} for the massive gravitino field: 
\begin{equation}
 \gamma^{\mu}\psi_{\mu}(x)=0\qquad\text{and}\qquad\left( i\,\slashed{\partial}-m_{3/2}\right) \psi_{\mu}(x)=0\,.
\end{equation}
These equations additionally imply the constraint $\partial^{\mu}\psi_{\mu}(x)=0\,.$ Going to momentum space there are positive and negative frequency solutions for the Rarita--Schwinger equations:
\begin{equation}
 \psi_{\mu}^s(x)=\psi_{\mu}^{+\,s}(p)\,e^{-ip\cdot x}\qquad\text{and}\qquad\psi_{\mu}^s(x)=\psi_{\mu}^{-\,s}(p)\,e^{ip\cdot x},\quad s=\pm\frac{3}{2}, \pm\frac{1}{2}\,, 
\end{equation}
where the mode functions $ \psi_{\mu}^+ $ and $ \psi_{\mu}^- $ have to obey the constraints
\begin{alignat}{2}
 \gamma^{\mu}\psi_{\mu}^{+\,s}(p) &=0\,, &\gamma^{\mu}\psi_{\mu}^{-\,s}(p) &=0\,, \nonumber\\
 \left( \slashed{p}-m_{3/2}\right) \psi_{\mu}^{+\,s}(p) &=0\,, &\qquad\text{and}\qquad\left( \slashed{p}+m_{3/2}\right) \psi_{\mu}^{-\,s}(p) &=0\,, \label{modeRarita}\\
 p^{\mu}\psi_{\mu}^{+\,s}(p) &=0 &p^{\mu}\psi_{\mu}^{-\,s}(p) &=0\,. \nonumber
\end{alignat}
For the calculation of unpolarized matrix elements we will need the gravitino polarization tensors
\begin{equation}
 P_{\mu\nu}^{\pm}(p)=\sum_s\psi_{\mu}^{\pm\,s}(p)\,\bar{\psi}_{\nu}^{\pm\,s}(p), 
\end{equation}
where the sum is performed over the four helicity states $s=\pm\frac{3}{2}, \pm\frac{1}{2} $ of the spin-3/2 gravitino. The polarization tensors for a gravitino with four-momentum $p$ are given by 
\begin{equation}
 P_{\mu\nu}^+(p)=-\left( \slashed{p}+m_{3/2}\right) \left\lbrace \Pi_{\mu\nu}(p)-\frac{1}{3}\,\Pi_{\mu\sigma}(p)\,\Pi_{\nu\lambda}(p)\,\gamma^{\sigma}\gamma^{\lambda}\right\rbrace =-\left( \slashed{p}+m_{3/2}\right) \Phi_{\mu\nu}(p)
 \label{PolTensor+}
\end{equation}
for the positive frequency mode functions and 
\begin{equation}
 P_{\mu\nu}^-(p)=-\left( \slashed{p}-m_{3/2}\right) \left\lbrace \Pi_{\mu\nu}(p)-\frac{1}{3}\,\Pi_{\mu\sigma}(p)\,\Pi_{\nu\lambda}(p)\,\gamma^{\sigma}\gamma^{\lambda}\right\rbrace =-\left( \slashed{p}-m_{3/2}\right) \Phi_{\mu\nu}(p)
 \label{PolTensor-}
\end{equation}
for the negative frequency mode functions. In the above expressions we use 
\begin{equation}
 \Pi_{\mu\nu}(p) =\left( g_{\mu\nu}-\frac{p_{\mu}\,p_{\nu}}{m_{3/2}^2}\right) \quad\text{and}\quad\Phi_{\mu\nu}(p)=\Pi_{\mu\nu}(p)-\frac{1}{3}\,\Pi_{\mu\sigma}(p)\,\Pi_{\nu\lambda}(p)\,\gamma^{\sigma}\gamma^{\lambda}.
\end{equation}
For a derivation of these polarization tensors see for instance~\cite{Grefe:2008zz}. Since the gravitino field is a solution of the Rarita--Schwinger equations of motion, the polarization tensors obey the relations
\begin{alignat}{2}
 \gamma^{\mu}P_{\mu\nu}^{\pm}(p) &=0\,, &P_{\mu\nu}^{\pm}(p)\,\gamma^{\nu} &=0\,, \nonumber\\
 p^{\mu}P_{\mu\nu}^{\pm}(p) &=0\,, &\qquad\text{and}\qquad\qquad\qquad P_{\mu\nu}^{\pm}(p)\,p^{\nu} &=0\,, \label{polarizationIDs}\\
 \left( \slashed{p}\mp m_{3/2}\right) P_{\mu\nu}^{\pm}(p) &=0 &P_{\mu\nu}^{\pm}(p)\left( \slashed{p}\mp m_{3/2}\right) &=0\,. \nonumber
\end{alignat}
These constraints -- as well as those in equation~(\ref{modeRarita}) -- can be used to significantly simplify the calculation of matrix elements including gravitinos.

After this short discussion of the gravitino equations of motion we want to introduce interactions between the gravitino and the MSSM particles. The interaction part of the gravitino Lagrangian reads~\cite{Bolz:2000fu,Pradler:2007ne}
\begin{equation}
 \begin{split}
  \mathscr{L}_{\text{int}}= &-\frac{i}{\sqrt{2}\,\MP}\left[ \left( D_{\mu}^*\phi^{i*}\right) \bar{\psi}_{\nu}\gamma^{\mu}\gamma^{\nu}P_L\chi^i-\left( D_{\mu}\phi^i\right) \bar{\chi}^iP_R\gamma^{\nu}\gamma^{\mu}\psi_{\nu}\right] \\
  &-\frac{i}{8\MP}\,\bar{\psi}_{\mu}\left[ \gamma^{\nu},\,\gamma^{\rho}\right] \gamma^{\mu}\lambda^{(\alpha)\,a}F_{\nu\rho}^{(\alpha)\,a}+\mathcal{O}(\MP^{-2})\,. 
 \label{interaction}
 \end{split}
\end{equation}
In this expression the covariant derivative of scalar fields is given as~\cite{Pradler:2007ne}
\begin{align}
 D_{\mu}\phi_i &=\partial_{\mu}\phi_i+i\sum_{\alpha=1}^3g_{\alpha}A_{\mu}^{(\alpha)\, a}T_{a,\, ij}^{(\alpha)}\phi_j 
 \intertext{and the field strength tensor for the gauge bosons reads}
 F_{\mu\nu}^{(\alpha)\, a} &=\partial_{\mu}A_{\nu}^{(\alpha)\, a}-\partial_{\nu}A_{\mu}^{(\alpha)\, a}-g_{\alpha}f^{(\alpha)\, abc}A_{\mu}^{(\alpha)\, b}A_{\nu}^{(\alpha)\, c}. 
\end{align}
The $T_{a,\, ij}^{(\alpha)}$ with $\alpha=1,\,2,\,3$ are the generators of the standard model gauge groups:
\begin{equation}
  T_{a,\, ij}^{(1)}=Y_i\,\delta_{ij}\,,\qquad T_{a,\, ij}^{(2)}=\frac{1}{2}\,\sigma_{a,\, ij}\quad\text{and}\quad T_{a,\, ij}^{(3)}=\frac{1}{2}\,\lambda_{a,\, ij}\,,
 \label{generators}
\end{equation}
where $Y_i$ is the hypercharge of the chiral supermultiplets as listed in Table~\ref{ChiralMSSM}. The Pauli matrices $\sigma_a$ are given in equation~(\ref{PauliMatrix}) and $\lambda_a$ are the eight Gell-Mann matrices which will not be needed in this work. The $f^{(\alpha)\, abc}$ are the totally antisymmetric structure constants of the corresponding gauge group.

The gravitino Feynman rules have been extracted from the interaction Lagrangian for instance in~\cite{Pradler:2007ne}. Since the gravitino and the gauginos are Majorana fields for which exist Wick contractions different from those of Dirac fermions, amplitudes will contain charge conjugation matrices and there may arise ambiguities concerning the relative sign of interfering diagrams. Therefore we employ a notation that introduces a continuous fermion flow~\cite{Denner:1992vza}. The direction of this fermion flow in a process can be chosen arbitrarily if the corresponding Feynman rules are used. Amplitudes are then written down in the direction opposite to the continuous fermion flow. This method avoids the appearance of charge conjugation matrices and gives the correct relative signs of different Feynman diagrams contributing to a single process. A set of Feynman rules that are relevant for this work is provided in Appendix~\ref{feynmanrules}.

\newpage

\section{Gravitino Cosmology}
\label{gravitinocosmo}

The addition of the gravitino to the supersymmetric particle spectrum leads to several strong constraints from the requirement of a consistent cosmological scenario. These generic problems are therefore called the cosmological gravitino problems.

It was early understood that a population of gravitinos in thermal equilibrium with the hot plasma in the early universe can have an unacceptably high abundance, by far exceeding the critical density of the universe if the gravitino is stable and not lighter than $\mathcal{O}(1)\,$keV~\cite{Pagels:1981ke}. In addition, an unstable gravitino may affect the successful predictions of primordial nucleosynthesis if it is present during or after the time of BBN~\cite{Weinberg:1982zq}. Since the gravitino interactions are suppressed by the Planck scale, its lifetime is expected to be of the order
\begin{equation}
\tau_{\text{3/2}}\sim\frac{\MP^2}{m_{3/2}^3}\approx 3\,\text{years}\left( \frac{100\,\text{GeV}}{m_{3/2}}\right) ^3.
\end{equation}
This led to the conclusion that an unstable gravitino should be heavier than $\mathcal{O}(10)\,$TeV in order to decay before the time of BBN~\cite{Weinberg:1982zq}. Another constraint comes from the overproduction of the stable lightest supersymmetric particle in gravitino decays~\cite{Krauss:1983ik}.

These problems can partly be circumvented in an inflationary universe, since during inflation any initial abundance of gravitinos is diluted due to the exponential expansion of the universe~\cite{Ellis:1982yb}. However, gravitinos can also be abundantly produced after a phase of inflation, again leading to strong cosmological constraints~\cite{Nanopoulos:1983up,Khlopov:1984pf}.

In particular, gravitinos are produced in scattering processes in the hot thermal plasma after the reheating of the universe (\textit{cf.} Figure~\ref{timeline}). Thermal gravitino production in a hot plasma is enhanced due to the contribution of the less suppressed interactions of the goldstino component. The gravitino relic density in that case is proportional to the reheating temperature $T_R$~\cite{Ellis:1984eq,Moroi:1993mb} and is given by~\cite{Bolz:2000fu,Pradler:2006qh,Rychkov:2007uq}
\begin{equation}
\Omega_{3/2}h^2\simeq 0.27\left( \frac{T_R}{10^{10}\,\text{GeV}}\right) \left( \frac{100\,\text{GeV}}{m_{3/2}}\right) \left( \frac{m_{\tilde{g}}}{1\,\text{TeV}}\right) ^2, 
\end{equation}
where $m_{\tilde{g}}$ is the gluino mass. Late entropy production as well as late electromagnetic and/or hadronic cascades from the decay of a metastable gravitino during the time of BBN can significantly alter the predictions of the light element abundances. In consequence this leads to strong constraints on the reheating temperature~\cite{Kawasaki:1994af,Kawasaki:2008qe,Ellis:1984eq,Ellis:1984er,Reno:1987qw,Dimopoulos:1988ue,Cyburt:2002uv,Kohri:2005wn}.

If the gravitino is assumed to be stable, the requirement that its abundance from thermal production does not exceed the critical density of the universe gives, depending on the gravitino mass, also strong constraints on the reheating temperature~\cite{Moroi:1993mb}. On the other hand, for reasonable values of the gluino mass and a gravitino mass in the $\mathcal{O}(100)\,$GeV range, thermally produced gravitinos can amount to the observed dark matter density, \textit{i.e.} $\Omega_{3/2}\simeq\Omega_{\text{DM}}$. This constrains the reheating temperature to be $T_R\approx\mathcal{O}(10^{10})\,\text{GeV}$, which is compatible with the constraint $T_R\gtrsim10^9\,\text{GeV}$ from thermal leptogenesis (\textit{cf.} Chapter~\ref{cosmology}). As the gravitino is neutral with respect to all standard model gauge interactions, a gravitino with a mass roughly on the order of the electroweak scale thus appears to be a natural candidate for the cold dark matter in the universe that is also consistent with the production of the baryon asymmetry via thermal leptogenesis.

In this case an additional contribution to the gravitino abundance arises from the decay of the next-to-lightest supersymmetric particle (NLSP)~\cite{Covi:1999ty,Feng:2003xh}:\footnote{It has been proposed that gravitinos could also be produced in non-thermal processes at the end of inflation~\cite{Kallosh:1999jj,Giudice:1999yt,Kawasaki:2006gs,Endo:2007sz}. Another recently proposed gravitino production mechanism is the freeze-in mechanism~\cite{Cheung:2011nn}. As all these mechanisms are hypothetical, we will not further consider them in this work.}
\begin{equation}
  \Omega_{3/2}h^2=\frac{m_{3/2}}{m_\text{NLSP}}\,\Omega_\text{NLSP}h^2.
\end{equation}
However, since its decay rate into the gravitino and standard model particles is suppressed by the Planck scale, the NLSP is expected to be long-lived~\cite{Martin:1997ns}:
\begin{equation}
  \tau_{\text{NLSP}}\simeq\frac{48\,\pi\,\MP^2\,m_{3/2}^2}{m_{\text{NLSP}}^5}\approx 9\,\text{days}\left( \frac{m_{3/2}}{10\,\text{GeV}}\right) ^2\left( \frac{150\,\text{GeV}}{m_{\text{NLSP}}}\right) ^5.
\end{equation}
Thus the late NLSP decays may -- depending on its particle nature -- spoil the predictions of standard BBN~\cite{Kawasaki:2008qe,Jedamzik:2006xz}. For instance, the hadronic decays of a neutralino NLSP typically dissociate the primordial light elements~\cite{Feng:2004zu,Cerdeno:2005eu,Buchmuller:2006nx,Covi:2009bk}, whereas a long-lived stau NLSP can form a bound state with $^4$He and catalyze the production of $^6$Li~\cite{Pospelov:2006sc,Cyburt:2006uv,Hamaguchi:2007mp,Kawasaki:2007xb}.\footnote{It has also been proposed that the catalyzed production of $^7$Li in specific scenarios might explain the discrepancy between the observed lithium abundance and that predicted by BBN~\cite{Jedamzik:2005dh}.} Stop NLSPs are also strongly constrained~\cite{DiazCruz:2007fc}, while sneutrino NLSPs affect the standard BBN predictions much less and could lead to a viable scenario~\cite{Kanzaki:2006hm}.

Several solutions to the NLSP decay problem have been proposed in the literature. For instance, the gravitino could be very light, thus leading to a sufficiently short NLSP lifetime~\cite{Moroi:1993mb}. Other options are that the NLSP has additional allowed decay channels to hidden sector particles~\cite{Cheung:2010qf} or that the number density of the NLSP is diluted by a sufficient amount of entropy production before the onset of BBN~\cite{Hamaguchi:2007mp,Buchmuller:2006tt,Hasenkamp:2010if}.

However, there is another solution to the gravitino problem: The introduction of a small amount of $R$-parity violation at a level $\lambda,\lambda'\gtrsim10^{-14}$ suffices to cause the NLSP to decay into standard model particles before the onset of BBN. The requirement that the baryon asymmetry is not washed out imposes an upper limit $\lambda,\lambda'\lesssim10^{-7}$ (\textit{cf.} Section~\ref{Rbreaking}), thus leaving a large allowed range for the $R$-parity breaking parameters. Due to the double suppression of the gravitino couplings to standard model particles by the Planck mass and a small $R$-parity violating parameter, the gravitino remains very long-lived with a lifetime exceeding the age of the universe by several orders of magnitude~\cite{Takayama:2000uz}. Since the NLSPs in this scenario decay mainly via the $R$-parity violating interactions, no contribution to the gravitino abundance from NLSP decays is expected and the abundance is given by the thermal production rate.

It is highly nontrivial that the combination of constraints from the requirement of a successful baryogenesis via leptogenesis and a sufficiently short NLSP lifetime predict a scenario where the gravitino naturally has the correct relic density and a sufficiently long lifetime to constitute the dark matter of the universe. For this reason a slight breaking of $R$ parity is a very attractive solution to the NLSP decay problem.\footnote{An additional motivation for models with a slight violation of $R$ parity is that these scenarios can also relax cosmological constraints on the axion multiplet and thus lead to a more natural solution for the strong $CP$ problem~\cite{Hasenkamp:2011xh}.}

Let us shortly summarize the observations: Supergravity theories encounter strong constraints due to the impact of the gravitino on the cosmological scenario. For instance, in scenarios where the lightest neutralino is the dark matter candidate strong upper limits on the reheating temperature arise from the requirement that the late gravitino decay is not in conflict with the predictions of primordial nucleosynthesis. In this case thermal leptogenesis cannot be the mechanism responsible for the generation of the baryon asymmetry in the universe.

On the other hand, gravitino dark matter leads to cosmological scenario that is naturally consistent with big bang nucleosynthesis and baryogenesis via thermal leptogenesis for gravitino masses roughly above 5\,GeV and for a small amount of $R$-parity violation in the range~\cite{Buchmuller:2007ui}
\begin{equation}
  10^{-14}\lesssim\lambda,\lambda'\lesssim10^{-7}.
\end{equation}
A similar allowed range is found for the bilinear $R$-parity breaking parameter $\xi$ from the requirement that the NLSP decays before the time of BBN and that the contribution to the light neutrino masses from $R$-parity breaking is consistent with observations~\cite{Ishiwata:2008cu}:
\begin{equation}
  10^{-11}\lesssim\xi\lesssim10^{-7}.
\end{equation}
Several models for $R$-parity breaking have been discussed in the literature that lead to scenarios with very long-lived gravitino dark matter~\cite{Buchmuller:2007ui,Ji:2008cq,Endo:2009by,FileviezPerez:2009gr,Choi:2009ng}. An appealing consequence is that these scenarios lead to an interesting and testable phenomenology at colliders~\cite{Buchmuller:2004rq,Ellis:2006vu,Bobrovskyi:2010ps} and in indirect detection experiments~\cite{Bertone:2007aw,Ibarra:2008qg,Ishiwata:2008cu,Covi:2008jy}. This is particularly intriguing as the gravitino is usually considered to be one of the most elusive particles with respect to experimental detection.

In the main part of this work we will confront the predicted signals from unstable gravitino dark matter with observations of indirect detection experiments in all possible cosmic-ray channels. For this purpose we will start with a study of the decay channels of the unstable gravitino assuming that $R$ parity is violated by small bilinear terms without discussing a specific underlying model.

\newpage

\section{Gravitino Decay Channels}
\label{gravdecay}

In this section we want to discuss the decay channels of the gravitino LSP via bilinear $R$-parity violating terms as introduced in Section~\ref{Rbreaking}.\footnote{Discussions of gravitino decays via trilinear operators can be found for instance in~\cite{Lola:2008bk,Moreau:2001sr,Lola:2007rw}.} These can be used subsequently to calculate the branching ratios of different decay channels and the spectra of final state particles in gravitino decays. With these calculations we quantitatively connect the lifetime of gravitino dark matter to the underlying parameters of supergravity and bilinear $R$-parity violation. In addition, the spectra predicted for the stable final state particles are a crucial input for all phenomenological studies concerning the indirect detection of gravitino dark matter.

\subsection{Two-Body Decays}

Calculations for the typically dominating two-body gravitino decays have been presented in several works. The tree-level decay of a gravitino into a photon and a neutrino via $R$-parity violating photino--neutrino mixing is described by the Feynman diagram
\begin{equation*}
 \parbox{5.5cm}{
  \begin{picture}(172,162) (92,-96)
    \SetWidth{0.5}
    \SetColor{Black}
    \Line(211,-55.999)(215,-44.001)\Line(207.001,-48)(218.999,-52)
    \Line[arrow,arrowpos=0.5,arrowlength=3.75,arrowwidth=1.5,arrowinset=0.2](213,-50)(232,-84)
    \Line(194,-17)(213,-51)
    \Photon(194,-17)(232,51){-5}{4}
    \Photon(194,-17)(213,-51){-5}{2}
    \Line[double,sep=4](117,-17)(194,-17)
    \Vertex(194,-17){2.5}
    \Text(104,-17)[]{\normalsize{\Black{$\psi_{3/2}$}}}
    \Text(237,58)[]{\normalsize{\Black{$\gamma$}}}
    \Text(191,-39)[]{\normalsize{\Black{$\tilde{\gamma}$}}}
    \Text(237,-91)[]{\normalsize{\Black{$\nu_i$}}}
  \end{picture}
 }
\end{equation*}
and was first calculated in~\cite{Takayama:2000uz}. Above the threshold for the production of $W$ and $Z$ bosons two more decay channels become kinematically allowed: the decay of a gravitino into a $W$ boson and a charged lepton via wino--charged lepton mixing as well as that into a $Z$ boson and a neutrino via zino--neutrino mixing. The contributions of the tree-level Feynman diagrams
\begin{equation*}
 \parbox{5.5cm}{
  \begin{picture}(172,162) (92,-96)
    \SetWidth{0.5}
    \SetColor{Black}
    \Line(211,-55.999)(215,-44.001)\Line(207.001,-48)(218.999,-52)
    \Line[arrow,arrowpos=0.5,arrowlength=3.75,arrowwidth=1.5,arrowinset=0.2](213,-50)(232,-84)
    \Line(194,-17)(213,-51)
    \Photon(194,-17)(232,51){-5}{4}
    \Photon(194,-17)(213,-51){-5}{2}
    \Line[double,sep=4](117,-17)(194,-17)
    \Vertex(194,-17){2.5}
    \Text(104,-17)[]{\normalsize{\Black{$\psi_{3/2}$}}}
    \Text(237,58)[]{\normalsize{\Black{$W^+$}}}
    \Text(191,-39)[]{\normalsize{\Black{$\tilde{W}^-$}}}
    \Text(237,-91)[]{\normalsize{\Black{$\ell_i^-$}}}
  \end{picture}
 }
 \text{and}\qquad
 \parbox{5.5cm}{
  \begin{picture}(172,162) (92,-96)
    \SetWidth{0.5}
    \SetColor{Black}
    \Line(211,-55.999)(215,-44.001)\Line(207.001,-48)(218.999,-52)
    \Line[arrow,arrowpos=0.5,arrowlength=3.75,arrowwidth=1.5,arrowinset=0.2](213,-50)(232,-84)
    \Line(194,-17)(213,-51)
    \Photon(194,-17)(232,51){-5}{4}
    \Photon(194,-17)(213,-51){-5}{2}
    \Line[double,sep=4](117,-17)(194,-17)
    \Vertex(194,-17){2.5}
    \Text(104,-17)[]{\normalsize{\Black{$\psi_{3/2}$}}}
    \Text(237,58)[]{\normalsize{\Black{$Z$}}}
    \Text(191,-39)[]{\normalsize{\Black{$\tilde{Z}$}}}
    \Text(237,-91)[]{\normalsize{\Black{$\nu_i$}}}
  \end{picture}
 }
\end{equation*}
were first discussed in~\cite{Ibarra:2007wg} finding that these channels dominate once they are kinematically accessible. However, there are two more tree-level Feynman diagrams contributing to these decay channels:
\begin{equation*}
 \parbox{5.5cm}{
  \begin{picture}(172,162) (92,-96)
    \SetWidth{0.5}
    \SetColor{Black}
    \Line[arrow,arrowpos=0.5,arrowlength=3.75,arrowwidth=1.5,arrowinset=0.2](194,-17)(232,-84)
    \Photon(194,-17)(232,51){-5}{4}
    \Line[double,sep=4](117,-17)(194,-17)
    \Vertex(194,-17){2.5}
    \Text(104,-17)[]{\normalsize{\Black{$\psi_{3/2}$}}}
    \Text(237,58)[]{\normalsize{\Black{$W^+$}}}
    \Text(182,-40)[]{\normalsize{\Black{$v_i$}}}
    \Text(237,-91)[]{\normalsize{\Black{$\ell_i^-$}}}
    \Line[dash,dashsize=4,arrow,arrowpos=0.5,arrowlength=3.75,arrowwidth=1.5,arrowinset=0.2](185,-34)(195,-17)
  \end{picture}
 }
 \text{and}\qquad
 \parbox{5.5cm}{
  \begin{picture}(172,162) (92,-96)
    \SetWidth{0.5}
    \SetColor{Black}
    \Line[arrow,arrowpos=0.5,arrowlength=3.75,arrowwidth=1.5,arrowinset=0.2](194,-17)(232,-84)
    \Photon(194,-17)(232,51){-5}{4}
    \Line[double,sep=4](117,-17)(194,-17)
    \Vertex(194,-17){2.5}
    \Text(104,-17)[]{\normalsize{\Black{$\psi_{3/2}$}}}
    \Text(237,58)[]{\normalsize{\Black{$Z$}}}
    \Text(182,-40)[]{\normalsize{\Black{$v_i$}}}
    \Text(237,-91)[]{\normalsize{\Black{$\nu_i$}}}
    \Line[dash,dashsize=4,arrow,arrowpos=0.5,arrowlength=3.75,arrowwidth=1.5,arrowinset=0.2](185,-34)(195,-17)
  \end{picture}
 }.
\end{equation*}
These contributions were first calculated in~\cite{Ishiwata:2008cu} taking also into account the possible decay into the lightest Higgs boson via neutral Higgs--sneutrino mixing and neutral higgsino--neutrino mixing described by the tree-level Feynman diagrams
\begin{equation*}
 \parbox{5.5cm}{
  \begin{picture}(172,162) (92,-96)
    \SetWidth{0.5}
    \SetColor{Black}
    \Line(211,22)(215,10)\Line(207,14)(219,18)
    \Line[arrow,arrowpos=0.5,arrowlength=3.75,arrowwidth=1.5,arrowinset=0.2](194,-17)(232,-84)
    \Line[dash,dashsize=4,arrow,arrowpos=0.5,arrowlength=3.75,arrowwidth=1.5,arrowinset=0.2](213,17)(194,-17)
    \Line[dash,dashsize=4](213,17)(232,51)
    \Line[double,sep=4](117,-17)(194,-17)
    \Vertex(194,-17){2.5}
    \Text(104,-17)[]{\normalsize{\Black{$\psi_{3/2}$}}}
    \Text(237,58)[]{\normalsize{\Black{$h$}}}
    \Text(193,5)[]{\normalsize{\Black{$\tilde{\nu}_i^*$}}}
    \Text(237,-91)[]{\normalsize{\Black{$\nu_i$}}}
  \end{picture}
 }
 +\qquad
 \parbox{5.5cm}{
  \begin{picture}(172,162) (92,-96)
    \SetWidth{0.5}
    \SetColor{Black}
    \Line(211,-55.999)(215,-44.001)\Line(207.001,-48)(218.999,-52)
    \Line[arrow,arrowpos=0.5,arrowlength=3.75,arrowwidth=1.5,arrowinset=0.2](213,-51)(232,-84)
    \Line[dash,dashsize=4](194,-17)(232,51)
    \Line[double,sep=4](117,-17)(194,-17)
    \Line(194,-17)(213,-51)
    \Vertex(194,-17){2.5}
    \Text(104,-17)[]{\normalsize{\Black{$\psi_{3/2}$}}}
    \Text(237,58)[]{\normalsize{\Black{$h$}}}
    \Text(193,-39)[]{\normalsize{\Black{$\tilde{h}$}}}
    \Text(237,-91)[]{\normalsize{\Black{$\nu_i$}}}
  \end{picture}
 }.
\end{equation*}
It was found that the additional channels further decrease the branching ratio for the decay into a photon and a neutrino, thereby reducing the strength of a monoenergetic signal of gamma rays from gravitino decays in indirect detection experiments. Assuming a standard model-like lightest Higgs boson, the decay into the Higgs plus neutrino channel contributes at the same level as the $Z$ boson plus neutrino channel if the gravitino mass is sufficiently above the production threshold.

The calculation of all tree-level gravitino decay channels mentioned above has been revised in detail in~\cite{Grefe:2008zz} leading to the following results for the decay widths of the gravitino LSP decay channels with two particles in the final state:
\begin{equation}
 \begin{split}
  \Gamma\left( \psi_{3/2}\rightarrow\gamma\,\nu_i\right) &\simeq\frac{\xi_i^2\,m_{3/2}^3}{64\,\pi\,\MP^2}\abs{U_{\tilde{\gamma}\tilde{Z}}}^2, \\
  \Gamma\left( \psi_{3/2}\rightarrow Z\nu_i\right) &\simeq\frac{\xi_i^2\,m_{3/2}^3\,\beta_Z^2}{64\,\pi\,\MP^2}\left\lbrace \,U_{\tilde{Z}\tilde{Z}}^2f_Z-\frac{8}{3}\frac{m_Z}{m_{3/2}}\,U_{\tilde{Z}\tilde{Z}}\,j_Z+\frac{1}{6}\,h_Z\right\rbrace , \\
  \Gamma\left( \psi_{3/2}\rightarrow W^+\ell_i^-\right) &\simeq\frac{\xi_i^2\,m_{3/2}^3\,\beta_W^2}{32\,\pi\,\MP^2}\left\lbrace \,U_{\tilde{W}\tilde{W}}^2f_W-\frac{8}{3}\frac{m_W}{m_{3/2}}\,U_{\tilde{W}\tilde{W}}\,j_W+\frac{1}{6}\,h_W\right\rbrace , \\
  \Gamma\left( \psi_{3/2}\rightarrow h\,\nu_i\right) &\simeq\frac{\xi_i^2\,m_{3/2}^3\,\beta_h^4}{384\,\pi\,\MP^2}\abs{\frac{m_{\tilde{\nu}_i}^2}{m_{\tilde{\nu}_i}^2-m_h^2}+\sin{\beta}\,U_{\tilde{H}_u^0\tilde{Z}}+\cos{\beta}\,U_{\tilde{H}_d^0\tilde{Z}}}^2,
 \end{split}
\end{equation}
where $\xi_i\equiv v_i/v$ is the dimensionless number parametrizing the strength of bilinear $R$-parity breaking and the mixing parameters $U_{\tilde{\gamma}\tilde{Z}}$, $U_{\tilde{Z}\tilde{Z}}$, $U_{\tilde{W}\tilde{W}}$, $U_{\tilde{H}_u^0\tilde{Z}}$ and $U_{\tilde{H}_d^0\tilde{Z}}$ are those defined in Section~\ref{Rbreaking}. In these expressions the kinematic functions $\beta_X, f_X, j_X$ and $h_X$ are given by
\begin{alignat}{2}
  \beta_X &=1-\frac{m_X^2}{m_{3/2}^2}\,, &\qquad\qquad f_X &=1+\frac{2}{3}\,\frac{m_X^2}{m_{3/2}^2}+\frac{1}{3}\,\frac{m_X^4}{m_{3/2}^4}\,, \nonumber\\
  j_X &=1+\frac{1}{2}\,\frac{m_X^2}{m_{3/2}^2}\,, &\qquad\qquad h_X &=1+10\,\frac{m_X^2}{m_{3/2}^2}+\frac{m_X^4}{m_{3/2}^4}\,. \label{kinematicfunctions}
\end{alignat}
The same results hold for the conjugate processes $\psi_{3/2}\rightarrow\gamma\,\bar{\nu}_i$, $\psi_{3/2}\rightarrow Z\,\bar{\nu}_i$, $\psi_{3/2}\rightarrow W^-\ell_i^+$ and $\psi_{3/2}\rightarrow h\,\bar{\nu}_i$. Indeed, these decay channels represent the complete set of all possible two-body final states in the decay of the gravitino LSP in theories with bilinear $R$-parity violation according to the interaction Lagrangian from equation~(\ref{interaction}). All other two-particle final states that one could imagine are kinematically forbidden: Decays into sleptons, squarks, neutralinos, charginos and gluinos are obviously forbidden since the gravitino represents the lightest supersymmetric particle. But also the decays into the heavier neutral Higgs bosons and the charged Higgs bosons are impossible since their masses are in general larger than the corresponding higgsino masses.

Although there are no additional two-particle final states that have been neglected in previous calculations, it turns out that there are two more tree-level Feynman diagrams contributing to the gravitino decay into a $W$ boson and a charged lepton, and that into a $Z$ boson and a neutrino. These contributions come from 4-vertex diagrams with a Higgs VEV at one of the legs:
\begin{equation*}
 \parbox{5.5cm}{
  \begin{picture}(172,162) (92,-96)
    \SetWidth{0.5}
    \SetColor{Black}
    \Line(211,-55.999)(215,-44.001)\Line(207.001,-48)(218.999,-52)
    \Line[arrow,arrowpos=0.5,arrowlength=3.75,arrowwidth=1.5,arrowinset=0.2](213,-51)(232,-84)
    \Line(194,-17)(213,-51)
    \Photon(194,-17)(232,51){-5}{4}
    \Line[double,sep=4](117,-17)(194,-17)
    \Vertex(194,-17){2.5}
    \Text(104,-17)[]{\normalsize{\Black{$\psi_{3/2}$}}}
    \Text(237,58)[]{\normalsize{\Black{$W^+$}}}
    \Text(219,-28)[]{\normalsize{\Black{$\tilde{H}_{d}^-$}}}
    \Text(182,-40)[]{\normalsize{\Black{$v_{d}$}}}
    \Text(237,-91)[]{\normalsize{\Black{$\ell_i^-$}}}
    \Line[dash,dashsize=4,arrow,arrowpos=0.5,arrowlength=3.75,arrowwidth=1.5,arrowinset=0.2](185,-34)(195,-17)
  \end{picture}
 }
 \text{and}\qquad
 \parbox{5.5cm}{
  \begin{picture}(172,162) (92,-96)
    \SetWidth{0.5}
    \SetColor{Black}
    \Line(211,-55.999)(215,-44.001)\Line(207.001,-48)(218.999,-52)
    \Line[arrow,arrowpos=0.5,arrowlength=3.75,arrowwidth=1.5,arrowinset=0.2](213,-51)(232,-84)
    \Line(194,-17)(213,-51)
    \Photon(194,-17)(232,51){-5}{4}
    \Line[double,sep=4](117,-17)(194,-17)
    \Vertex(194,-17){2.5}
    \Text(104,-17)[]{\normalsize{\Black{$\psi_{3/2}$}}}
    \Text(237,58)[]{\normalsize{\Black{$Z$}}}
    \Text(219,-28)[]{\normalsize{\Black{$\tilde{H}_{u,\,d}^0$}}}
    \Text(182,-40)[]{\normalsize{\Black{$v_{u,\,d}$}}}
    \Text(237,-91)[]{\normalsize{\Black{$\nu_i$}}}
    \Line[dash,dashsize=4,arrow,arrowpos=0.5,arrowlength=3.75,arrowwidth=1.5,arrowinset=0.2](185,-34)(195,-17)
  \end{picture}
 }.
\end{equation*}
For the calculation of the contribution of these diagrams we refer to Appendix~\ref{gravitinodecay}. In cases where the $\mu$ parameter is large compared to the gaugino mass parameters (as in the discussions in~\cite{Ishiwata:2008cu,Covi:2008jy,Grefe:2008zz}) these additional diagrams do not contribute significantly since the higgsino--gaugino mixing parameters are suppressed in that case (\textit{cf.} equations~(\ref{UHuZapx}), (\ref{UHdZapx}) and (\ref{UHWapx})). Since these contributions are of the same form as the diagrams with a sneutrino VEV, the final results can be altered in a simple way to include this correction. The complete set of tree-level two-body decay widths is then given by
\begin{align}
  \Gamma\left( \psi_{3/2}\rightarrow\gamma\,\nu_i\right) &\simeq\frac{\xi_i^2\,m_{3/2}^3}{64\,\pi\,\MP^2}\abs{U_{\tilde{\gamma}\tilde{Z}}}^2, \nonumber\\
  \Gamma\left( \psi_{3/2}\rightarrow Z\nu_i\right) &\simeq\frac{\xi_i^2\,m_{3/2}^3\,\beta_Z^2}{64\,\pi\,\MP^2}\left\lbrace \,U_{\tilde{Z}\tilde{Z}}^2f_Z+\frac{1}{6}\abs{1+s_\beta\,U_{\tilde{H}_u^0\tilde{Z}}-c_\beta\,U_{\tilde{H}_d^0\tilde{Z}}}^2h_Z\right. \nonumber\\
    &\left. \quad\qquad-\frac{8}{3}\frac{m_Z}{m_{3/2}}\,U_{\tilde{Z}\tilde{Z}}\left( 1+s_\beta \RE U_{\tilde{H}_u^0\tilde{Z}}-c_\beta \RE U_{\tilde{H}_d^0\tilde{Z}}\right) j_Z \right\rbrace, \label{gravitinowidths}\\
  \Gamma\left( \psi_{3/2}\rightarrow W^+\ell_i^-\right) &\simeq\frac{\xi_i^2\,m_{3/2}^3\,\beta_W^2}{32\,\pi\,\MP^2}\left\lbrace \,U_{\tilde{W}\tilde{W}}^2f_W+\frac{1}{6}\abs{1-\sqrt{2}\,c_\beta\,U_{\tilde{H}_d^-\tilde{W}}}^2h_W\right. \nonumber\\
    &\left. \quad\qquad-\frac{8}{3}\frac{m_W}{m_{3/2}}\,U_{\tilde{W}\tilde{W}}\left( 1-\sqrt{2}\,c_\beta \RE U_{\tilde{H}_d^-\tilde{W}}\right) j_W \right\rbrace , \nonumber\\
  \Gamma\left( \psi_{3/2}\rightarrow h\,\nu_i\right) &\simeq\frac{\xi_i^2\,m_{3/2}^3\,\beta_h^4}{384\,\pi\,\MP^2}\abs{\frac{m_{\tilde{\nu}_i}^2+\frac{1}{2}\,m_Z^2\cos2\beta}{m_h^2-m_{\tilde{\nu}_i}^2}+2\sin{\beta}\,U_{\tilde{H}_u^0\tilde{Z}}+2\cos{\beta}\,U_{\tilde{H}_d^0\tilde{Z}}}^2. \nonumber
\end{align}
In the result for the Higgs channel we also included a correction since we neglected the $D$-term contribution to the sneutrino VEV in our previous calculation in~\cite{Grefe:2008zz}. This affected also the mixing that was assumed between the lightest Higgs particle and the sneutrinos (see Section~\ref{Rbreaking}).

\subsection{Three-Body Decays}
\label{threebody}

Recently it has been pointed out that below the threshold for the production of the massive electroweak gauge bosons the three-body decays involving virtual gauge bosons have to be taken into account~\cite{Choi:2010xn}. The decay widths of the channels $\psi_{3/2}\rightarrow Z^*\,\nu\rightarrow f\,\bar{f}\,\nu$ and $\psi_{3/2}\rightarrow {W^+}^*\,\ell^-\rightarrow f\,\bar{f'}\,\ell^-$ were first calculated in~\cite{Choi:2010jt}. In this thesis we revise this calculation in detail and in addition calculate the contributions from virtual photons and virtual lightest Higgs bosons, \textit{i.e.} contributions from the three-body decay diagrams $\psi_{3/2}\rightarrow \gamma^*\,\nu\rightarrow f\,\bar{f}\,\nu$ and $\psi_{3/2}\rightarrow h^*\,\nu\rightarrow f\,\bar{f}\,\nu$.

\subsection*{\boldmath$\psi_{3/2}\rightarrow \gamma^*/Z^*\,\nu\rightarrow f\,\bar{f}\,\nu$}

At tree level there are five diagrams contributing to the decay of a gravitino into a fermion-antifermion pair and a neutrino via an intermediate photon or $Z$ boson:
\begin{equation*}
 \parbox{5.5cm}{
  \begin{picture}(172,162) (92,-96)
    \SetWidth{0.5}
    \SetColor{Black}
    \Line(211,-55.999)(215,-44.001)\Line(207.001,-48)(218.999,-52)
    \Line[arrow,arrowpos=0.5,arrowlength=3.75,arrowwidth=1.5,arrowinset=0.2](213,-50)(232,-84)
    \Line(194,-17)(213,-51)
    \Photon(194,-17)(213,17){-5}{2}
    \Photon(194,-17)(213,-51){-5}{2}
    \Line[arrow,arrowpos=0.5,arrowlength=3.75,arrowwidth=1.5,arrowinset=0.2](213,17)(232,51)
    \Line[arrow,arrowpos=0.5,arrowlength=3.75,arrowwidth=1.5,arrowinset=0.2](233,-17)(214,17)
    \Line[double,sep=4](117,-17)(194,-17)
    \Vertex(194,-17){2.5}
    \Vertex(213,17){2.5}
    \Text(104,-17)[]{\normalsize{\Black{$\psi_{3/2}$}}}
    \Text(237,58)[]{\normalsize{\Black{$f$}}}
    \Text(237,-24)[]{\normalsize{\Black{$\bar{f}$}}}
    \Text(188,6)[]{\normalsize{\Black{$\gamma, Z$}}}
    \Text(188,-39)[]{\normalsize{\Black{$\tilde{\gamma}, \tilde{Z}$}}}
    \Text(237,-91)[]{\normalsize{\Black{$\nu_i$}}}
  \end{picture}
 }\qquad
 +\qquad
 \parbox{5.5cm}{
  \begin{picture}(172,162) (92,-96)
    \SetWidth{0.5}
    \SetColor{Black}
    \Line[arrow,arrowpos=0.5,arrowlength=3.75,arrowwidth=1.5,arrowinset=0.2](194,-17)(232,-84)
    \Photon(194,-17)(213,17){-5}{2}
    \Line[arrow,arrowpos=0.5,arrowlength=3.75,arrowwidth=1.5,arrowinset=0.2](213,17)(232,51)
    \Line[arrow,arrowpos=0.5,arrowlength=3.75,arrowwidth=1.5,arrowinset=0.2](233,-17)(214,17)
    \Line[double,sep=4](117,-17)(194,-17)
    \Vertex(194,-17){2.5}
    \Vertex(213,17){2.5}
    \Text(104,-17)[]{\normalsize{\Black{$\psi_{3/2}$}}}
    \Text(237,58)[]{\normalsize{\Black{$f$}}}
    \Text(237,-24)[]{\normalsize{\Black{$\bar{f}$}}}
    \Text(193,6)[]{\normalsize{\Black{$Z$}}}
    \Text(182,-40)[]{\normalsize{\Black{$v_i$}}}
    \Text(237,-91)[]{\normalsize{\Black{$\nu_i$}}}
    \Line[dash,dashsize=4,arrow,arrowpos=0.5,arrowlength=3.75,arrowwidth=1.5,arrowinset=0.2](185,-34)(195,-17)
  \end{picture}
 }
\end{equation*}
and in addition the diagrams coming from the 4-vertex with a Higgs VEV that were also discussed for the two-body decay:
\begin{equation*}
 \parbox{5.5cm}{
  \begin{picture}(172,162) (92,-96)
    \SetWidth{0.5}
    \SetColor{Black}
    \Line(211,-55.999)(215,-44.001)\Line(207.001,-48)(218.999,-52)
    \Line[arrow,arrowpos=0.5,arrowlength=3.75,arrowwidth=1.5,arrowinset=0.2](213,-51)(232,-84)
    \Photon(194,-17)(213,17){-5}{2}
    \Line[arrow,arrowpos=0.5,arrowlength=3.75,arrowwidth=1.5,arrowinset=0.2](213,17)(232,51)
    \Line[arrow,arrowpos=0.5,arrowlength=3.75,arrowwidth=1.5,arrowinset=0.2](233,-17)(214,17)
    \Line(194,-17)(213,-51)
    \Line[double,sep=4](117,-17)(194,-17)
    \Vertex(194,-17){2.5}
    \Vertex(213,17){2.5}
    \Text(104,-17)[]{\normalsize{\Black{$\psi_{3/2}$}}}
    \Text(237,58)[]{\normalsize{\Black{$f$}}}
    \Text(237,-24)[]{\normalsize{\Black{$\bar{f}$}}}
    \Text(193,6)[]{\normalsize{\Black{$Z$}}}
    \Text(219,-28)[]{\normalsize{\Black{$\tilde{H}_{u,\,d}^0$}}}
    \Text(182,-40)[]{\normalsize{\Black{$v_{u,\,d}$}}}
    \Text(237,-91)[]{\normalsize{\Black{$\nu_i$}}}
    \Line[dash,dashsize=4,arrow,arrowpos=0.5,arrowlength=3.75,arrowwidth=1.5,arrowinset=0.2](185,-34)(195,-17)
  \end{picture}
 }.
\end{equation*}
Although there are in principle additional interference effects from the virtual exchange of the lightest Higgs boson, we discuss that process separately. This is justified since the Higgs exchange process is strongly suppressed below the threshold by the small Higgs--fermion couplings. In addition, there exists an interference for decay into the specific final state $\ell_i^+\ell_i^-\nu_i$ since it can proceed via the photon, $Z$ and Higgs channels, but also via the $W$ channel. In the calculation presented here, however, we did not take into account this exceptional case.

We present the calculation of the decay width corresponding to the Feynman diagrams above in detail in Appendix~\ref{gravitinodecay} of this thesis. Here we directly show the result for the differential decay width with respect to $s$ and $t$, where $s$ is the invariant mass of the two fermions $f$ and $\bar{f}$ and $t$ is the invariant mass of the neutrino and one of the fermions. Neglecting the masses of the final state particles we find\footnote{In the case of virtual photon exchange the effect of nonvanishing final state fermion masses cannot be neglected since that could lead to a divergent propagator. Therefore, we will keep track of fermion masses for the pure photon channel.}
\begin{align}
  &\frac{d\Gamma\left( \psi_{3/2}\rightarrow \gamma^*/Z^*\,\nu_i\rightarrow f\,\bar{f}\,\nu_i\right) }{ds\,dt}\simeq\frac{\xi_i^2}{1536\,\pi^3\,m_{3/2}^5\,\MP^2}\,\Bigg[\frac{e^2\,Q^2}{s}\abs{U_{\tilde{\gamma}\tilde{Z}}}^2 \nonumber\\
  &\qquad\qquad\times\bigg( 3\,m_{3/2}^6\left( 1+2\,\frac{m_f^2}{s}\right) +m_{3/2}^2\left( s^2+8\,s\,t+6\,t^2+6\,m_f^2\left( m_f^2-s-2\,t\right) \right) \nonumber\\
  &\qquad\qquad\qquad-3\,m_{3/2}^4\,(s+2\,t)-s\left( s^2+2\,s\,t+2\,t^2+2\,m_f^2\,(m_f^2-2\,t)\right) \bigg) \nonumber\\
  &\qquad\quad+\frac{g_Z}{\left( s-m_Z^2\right) ^2+m_Z^2\,\Gamma_Z^2}\,\bigg\lbrace g_Z\,s\,U_{\tilde{Z}\tilde{Z}}^2\Big((C_V-C_A)^2\left( 3\,m_{3/2}^4-s^2\right) \nonumber\\
  &\qquad\qquad\quad\times\left( m_{3/2}^2-s-2\,t\right) +\left( C_V^2+C_A^2\right) \left( m_{3/2}^2\left( 2\,s^2+8\,s\,t+6\,t^2\right) -2\,s\,(s+t)^2 \right) \Big) \nonumber\\
  &\qquad\qquad-2\,g_Z\,s\,m_{3/2}\,m_Z\,U_{\tilde{Z}\tilde{Z}}\left( 1+s_\beta \RE U_{\tilde{H}_u^0\tilde{Z}}-c_\beta \RE U_{\tilde{H}_d^0\tilde{Z}}\right) \nonumber\\
  &\qquad\qquad\quad\times\Big(-2\,C_V\,C_A\left( 3\,m_{3/2}^2-s\right) \left( m_{3/2}^2-s-2\,t\right) \nonumber\\
  &\qquad\qquad\qquad+\left( C_V^2+C_A^2\right) \left( 3\,m_{3/2}^4-2\,m_{3/2}^2\,(s+t)-s^2+2\,s\,t+2\,t^2\right) \!\Big) \label{3bodygammaZnufull}\\
  &\qquad\qquad+2\,g_Z\,m_Z^2\abs{1+s_\beta\,U_{\tilde{H}_u^0\tilde{Z}}-c_\beta\,U_{\tilde{H}_d^0\tilde{Z}}}^2\Big(-2\,m_{3/2}^2\,s\,C_V\,C_A\left( m_{3/2}^2-s-2\,t\right) \nonumber\\
  &\qquad\qquad\quad+\left( C_V^2+C_A^2\right) \left( m_{3/2}^4\,(2\,s+t)-m_{3/2}^2\left( 2\,s^2+2\,s\,t+t^2\right) +s\,t\,(s+t)\right) \Big) \nonumber\\
  &\qquad\qquad-2\,e\,Q\left( \RE U_{\tilde{\gamma}\tilde{Z}}\left( m_Z^2-s\right) +\IM U_{\tilde{\gamma}\tilde{Z}}\,m_Z\,\Gamma_Z\right) \nonumber\\
  &\qquad\qquad\quad\times\bigg(U_{\tilde{Z}\tilde{Z}}\Big(C_V\big( 3\,m_{3/2}^6-3\,m_{3/2}^4\,(s+2\,t)+m_{3/2}^2\left( s^2+8\,s\,t+6\,t^2\right) \nonumber\\
  &\qquad\qquad\qquad\quad-s\left( s^2+2\,s\,t+2\,t^2\right) \big)-C_A\left( 3\,m_{3/2}^4-s^2\right) \left( m_{3/2}^2-s-2\,t\right) \Big) \nonumber\\
  &\qquad\qquad\qquad-m_{3/2}\,m_Z\left( 1+s_\beta \RE U_{\tilde{H}_u^0\tilde{Z}}-c_\beta \RE U_{\tilde{H}_d^0\tilde{Z}}\right) \nonumber\\
  &\qquad\qquad\qquad\quad\times\Big(C_V\left( 3\,m_{3/2}^4-2\,m_{3/2}^2\,(s+t)-s^2+2\,s\,t+2\,t^2\right) \nonumber\\
  &\qquad\qquad\qquad\qquad-C_A\left( 3\,m_{3/2}^2-s\right) \left( m_{3/2}^2-s-2\,t\right) \Big) \bigg) \bigg\rbrace \Bigg] ,\nonumber
\end{align}
where $g_Z\equiv g/\cos\theta$ is the gauge coupling of the $Z$ boson, $Q$ is the charge of the final state fermions, and $C_V$ and $C_A$ are the coefficients of the $V-A$ structure of the $Z$ boson vertex with two fermions:
\begin{equation}
  C_V=\frac{1}{2}\,T^3-Q\sin^2\theta_W\qquad\text{and}\qquad C_A=-\,\frac{1}{2}\,T^3\,.
  \label{CVCA}
\end{equation}
After integrating over the kinematically allowed range of $t$, $0\lesssim t\lesssim m_{3/2}^2-s$, we find the following differential decay rate:\footnote{For the pure photon channel we integrate over the range of $t$ given by equation~(\ref{m13range}), only neglecting the neutrino mass.}
\begin{align}
  &\frac{d\Gamma\left( \psi_{3/2}\rightarrow \gamma^*/Z^*\,\nu_i\rightarrow f\,\bar{f}\,\nu_i\right) }{ds} \nonumber\\
  &\qquad\simeq\frac{\xi_i^2\,m_{3/2}^3\,\beta_s^2}{768\,\pi^3\MP^2}\Bigg[ \frac{e^2\,Q^2}{s}\abs{U_{\tilde{\gamma}\tilde{Z}}}^2f_s\,\sqrt{1-4\,\frac{m_f^2}{s}}\left( 1+2\,\frac{m_f^2}{s}\right) \nonumber\\
  &\qquad\qquad+\frac{g_Z}{\left( s-m_Z^2\right) ^2+m_Z^2\,\Gamma_Z^2}\,\bigg\lbrace\,g_Z\,U_{\tilde{Z}\tilde{Z}}^2\,s\left( C_V^2+C_A^2\right) f_s \nonumber\\
  &\qquad\quad\qquad-\frac{8}{3}\,\frac{m_Z}{m_{3/2}}\,g_Z\,U_{\tilde{Z}\tilde{Z}}\left( 1+s_\beta \RE U_{\tilde{H}_u^0\tilde{Z}}-c_\beta \RE U_{\tilde{H}_d^0\tilde{Z}}\right) s\left( C_V^2+C_A^2\right) j_s \label{3bodygammaZnu}\\
  &\qquad\quad\qquad+\frac{1}{6}\,g_Z\,m_Z^2\abs{1+s_\beta\,U_{\tilde{H}_u^0\tilde{Z}}-c_\beta\,U_{\tilde{H}_d^0\tilde{Z}}}^2\left( C_V^2+C_A^2\right) h_s \nonumber\\
  &\qquad\quad\qquad+e\,Q\left( \RE U_{\tilde{\gamma}\tilde{Z}}\left( m_Z^2-s\right) +\IM U_{\tilde{\gamma}\tilde{Z}}\,m_Z\,\Gamma_Z\right) C_V \nonumber\\
  &\qquad\qquad\qquad\times\left( 2\,U_{\tilde{Z}\tilde{Z}}\,f_s +\frac{8}{3}\frac{m_Z}{m_{3/2}}\left( 1+s_\beta \RE U_{\tilde{H}_u^0\tilde{Z}}-c_\beta \RE U_{\tilde{H}_d^0\tilde{Z}}\right) j_s\right) \bigg\rbrace \Bigg] . \nonumber
\end{align}
In this expression the kinematic functions $\beta_s, f_s, j_s$ and $h_s$ are defined corresponding to the kinematic functions in the two-body decays (see equation(\ref{kinematicfunctions})):
\begin{alignat}{2}
  \beta_s &=1-\frac{s}{m_{3/2}^2}\,, &\qquad\qquad f_s &=1+\frac{2}{3}\,\frac{s}{m_{3/2}^2}+\frac{1}{3}\,\frac{s^2}{m_{3/2}^4}\,, \nonumber\\
  j_s &=1+\frac{1}{2}\,\frac{s}{m_{3/2}^2}\,, &\qquad\qquad h_s &=1+10\,\frac{s}{m_{3/2}^2}+\frac{s^2}{m_{3/2}^4}\,.
  \label{kinematicfunc}
\end{alignat}
The total decay width can be obtained by integrating the above differential decay width over the invariant mass range $0\leq s\leq m_{3/2}^2$.\footnote{To avoid a divergent propagator in the case of virtual photon exchange one should integrate over the range $4\,m_f^2\leq s\leq m_{3/2}^2$ in that case (see equation~(\ref{m12range})).}

Although the decay width $\psi_{3/2}\rightarrow \gamma^*/Z^*\,\nu\rightarrow f\,\bar{f}\,\nu$ is dominated by virtual photon exchange at low gravitino masses, it turns out that virtual photon exchange never significantly contributes to the total gravitino decay width. This is because the virtual photon exchange contribution is always suppressed compared to the two-body decay into a real photon by $\mathcal{O}(\alpha)$. In this respect the approximation in~\cite{Choi:2010jt} to neglect the photon-mediated diagram in the decay width calculation is justified. However, one should not forget that -- depending on the mixing parameters -- this diagram could give the dominant contribution to cosmic-ray antimatter fluxes at gravitino masses below the threshold for $W$ production. Neglecting the photon-mediated diagram our results simplify to
\begin{align}
  &\frac{d\Gamma\left( \psi_{3/2}\rightarrow Z^*\,\nu_i\rightarrow f\,\bar{f}\,\nu_i\right) }{ds\,dt}\simeq\frac{g_Z^2\,\xi_i^2}{1536\,\pi^3\,m_{3/2}^5\,\MP^2\left( \left( s-m_Z^2\right) ^2+m_Z^2\,\Gamma_Z^2\right) } \nonumber\\
  &\qquad\quad\times\bigg\lbrace s\,U_{\tilde{Z}\tilde{Z}}^2\Big((C_V-C_A)^2\left( 3\,m_{3/2}^4-s^2\right) \left( m_{3/2}^2-s-2\,t\right) \nonumber\\
  &\qquad\qquad\quad+\left( C_V^2+C_A^2\right) \left( m_{3/2}^2\left( 2\,s^2+8\,s\,t+6\,t^2\right) -2\,s\,(s+t)^2 \right) \Big) \nonumber\\
  &\qquad\qquad-2\,s\,m_{3/2}\,m_Z\,U_{\tilde{Z}\tilde{Z}}\left( 1+s_\beta \RE U_{\tilde{H}_u^0\tilde{Z}}-c_\beta \RE U_{\tilde{H}_d^0\tilde{Z}}\right) \nonumber\\
  &\qquad\qquad\quad\times\Big(-2\,C_V\,C_A\left( 3\,m_{3/2}^2-s\right) \left( m_{3/2}^2-s-2\,t\right) \label{3bodyZnufull}\\
  &\qquad\qquad\qquad+\left( C_V^2+C_A^2\right) \left( 3\,m_{3/2}^4-2\,m_{3/2}^2\,(s+t)-s^2+2\,s\,t+2\,t^2\right) \Big) \nonumber\\
  &\qquad\qquad+2\,m_Z^2\abs{1+s_\beta\,U_{\tilde{H}_u^0\tilde{Z}}-c_\beta\,U_{\tilde{H}_d^0\tilde{Z}}}^2\Big(-2\,m_{3/2}^2\,s\,C_V\,C_A\left( m_{3/2}^2-s-2\,t\right) \nonumber\\
  &\qquad\qquad\quad+\left( C_V^2+C_A^2\right) \left( m_{3/2}^4\,(2\,s+t)-m_{3/2}^2\left( 2\,s^2+2\,s\,t+t^2\right) +s\,t\,(s+t)\right) \Big) \bigg\rbrace \Bigg] ,\nonumber
\end{align}
including the complete dependence on kinematic variables, and
\begin{align}
  &\frac{d\Gamma\left( \psi_{3/2}\rightarrow Z^*\,\nu\rightarrow f\,\bar{f}\,\nu\right) }{ds} \nonumber\\
  &\qquad\simeq\frac{g_Z^2\,\xi_i^2\,m_{3/2}^3\,\beta_s^2\left( C_V^2+C_A^2\right) }{768\,\pi^3\MP^2\left( \left( s-m_Z^2\right) ^2+m_Z^2\,\Gamma_Z^2\right) }\bigg( s\,U_{\tilde{Z}\tilde{Z}}^2\,f_s+\frac{1}{6}\,m_Z^2\abs{1+s_\beta\,U_{\tilde{H}_u^0\tilde{Z}}-c_\beta\,U_{\tilde{H}_d^0\tilde{Z}}}^2h_s \nonumber\\
  &\qquad\qquad\quad-\frac{8}{3}\,\frac{m_Z}{m_{3/2}}\,s\,U_{\tilde{Z}\tilde{Z}}\left( 1+s_\beta \RE U_{\tilde{H}_u^0\tilde{Z}}-c_\beta \RE U_{\tilde{H}_d^0\tilde{Z}}\right) j_s\bigg) \label{3bodyZnu}
\end{align}
after integration over the invariant mass $t$.

\subsection*{\boldmath$\psi_{3/2}\rightarrow W^*\,\ell\rightarrow f\,\bar{f}'\,\ell$}

At tree level there are three Feynman diagrams contributing to the decay of a gravitino into two fermions and a charged lepton:
\begin{equation*}
 \parbox{5.5cm}{
  \begin{picture}(172,162) (92,-96)
    \SetWidth{0.5}
    \SetColor{Black}
    \Line(211,-55.999)(215,-44.001)\Line(207.001,-48)(218.999,-52)
    \Line[arrow,arrowpos=0.5,arrowlength=3.75,arrowwidth=1.5,arrowinset=0.2](213,-50)(232,-84)
    \Line(194,-17)(213,-51)
    \Photon(194,-17)(213,17){-5}{2}
    \Photon(194,-17)(213,-51){-5}{2}
    \Line[arrow,arrowpos=0.5,arrowlength=3.75,arrowwidth=1.5,arrowinset=0.2](213,17)(232,51)
    \Line[arrow,arrowpos=0.5,arrowlength=3.75,arrowwidth=1.5,arrowinset=0.2](233,-17)(214,17)
    \Line[double,sep=4](117,-17)(194,-17)
    \Vertex(194,-17){2.5}
    \Vertex(213,17){2.5}
    \Text(104,-17)[]{\normalsize{\Black{$\psi_{3/2}$}}}
    \Text(237,58)[]{\normalsize{\Black{$f$}}}
    \Text(237,-24)[]{\normalsize{\Black{$\bar{f}'$}}}
    \Text(191,6)[]{\normalsize{\Black{$W^+$}}}
    \Text(191,-39)[]{\normalsize{\Black{$\tilde{W}^-$}}}
    \Text(237,-91)[]{\normalsize{\Black{$\ell_i^-$}}}
  \end{picture}
 }\qquad
 +\qquad
 \parbox{5.5cm}{
  \begin{picture}(172,162) (92,-96)
    \SetWidth{0.5}
    \SetColor{Black}
    \Line[arrow,arrowpos=0.5,arrowlength=3.75,arrowwidth=1.5,arrowinset=0.2](194,-17)(232,-84)
    \Photon(194,-17)(213,17){-5}{2}
    \Line[arrow,arrowpos=0.5,arrowlength=3.75,arrowwidth=1.5,arrowinset=0.2](213,17)(232,51)
    \Line[arrow,arrowpos=0.5,arrowlength=3.75,arrowwidth=1.5,arrowinset=0.2](233,-17)(214,17)
    \Line[double,sep=4](117,-17)(194,-17)
    \Vertex(194,-17){2.5}
    \Vertex(213,17){2.5}
    \Text(104,-17)[]{\normalsize{\Black{$\psi_{3/2}$}}}
    \Text(237,58)[]{\normalsize{\Black{$f$}}}
    \Text(237,-24)[]{\normalsize{\Black{$\bar{f}'$}}}
    \Text(191,6)[]{\normalsize{\Black{$W^+$}}}
    \Text(182,-40)[]{\normalsize{\Black{$v_i$}}}
    \Text(237,-91)[]{\normalsize{\Black{$\ell_i^-$}}}
    \Line[dash,dashsize=4,arrow,arrowpos=0.5,arrowlength=3.75,arrowwidth=1.5,arrowinset=0.2](185,-34)(195,-17)
  \end{picture}
 }
\end{equation*}
and in addition the diagram coming from the 4-vertex with a down-type Higgs VEV that was also discussed for the two-body decay:
\begin{equation*}
 \parbox{5.5cm}{
  \begin{picture}(172,162) (92,-96)
    \SetWidth{0.5}
    \SetColor{Black}
    \Line(211,-55.999)(215,-44.001)\Line(207.001,-48)(218.999,-52)
    \Line[arrow,arrowpos=0.5,arrowlength=3.75,arrowwidth=1.5,arrowinset=0.2](213,-51)(232,-84)
    \Photon(194,-17)(213,17){-5}{2}
    \Line[arrow,arrowpos=0.5,arrowlength=3.75,arrowwidth=1.5,arrowinset=0.2](213,17)(232,51)
    \Line[arrow,arrowpos=0.5,arrowlength=3.75,arrowwidth=1.5,arrowinset=0.2](233,-17)(214,17)
    \Line(194,-17)(213,-51)
    \Line[double,sep=4](117,-17)(194,-17)
    \Vertex(194,-17){2.5}
    \Vertex(213,17){2.5}
    \Text(104,-17)[]{\normalsize{\Black{$\psi_{3/2}$}}}
    \Text(237,58)[]{\normalsize{\Black{$f$}}}
    \Text(237,-24)[]{\normalsize{\Black{$\bar{f}'$}}}
    \Text(193,6)[]{\normalsize{\Black{$W^+$}}}
    \Text(219,-28)[]{\normalsize{\Black{$\tilde{H}_{d}^-$}}}
    \Text(182,-40)[]{\normalsize{\Black{$v_{d}$}}}
    \Text(237,-91)[]{\normalsize{\Black{$\ell_i^-$}}}
    \Line[dash,dashsize=4,arrow,arrowpos=0.5,arrowlength=3.75,arrowwidth=1.5,arrowinset=0.2](185,-34)(195,-17)
  \end{picture}
 }.
\end{equation*}
For the detailed calculation of the decay width corresponding to the Feynman diagrams above we refer to Appendix~\ref{gravitinodecay} of this thesis. Here we directly show the result for the differential decay width with respect to $s$ and $t$, where $s$ is the invariant mass of the two fermions $f$ and $\bar{f}'$ and $t$ is the invariant mass of the charged lepton and one of the fermions. Neglecting the masses of the final state particles we find
\begin{equation}
 \begin{split}
  &\frac{d\Gamma\left( \psi_{3/2}\rightarrow {W^+}^*\,\ell_i^-\rightarrow f\,\bar{f'}\,\ell_i^-\right) }{ds\,dt} \\
  &\qquad\simeq\frac{g^2\,\xi_i^2}{1536\,\pi^3\,m_{3/2}^5\,\MP^2\left( \left( s-m_W^2\right) ^2+m_W^2\,\Gamma_W^2\right) } \\
  &\qquad\qquad\times\Big[\,s\,U_{\tilde{W}\tilde{W}}^2\left( m_{3/2}^2-t\right) \left( 3\,m_{3/2}^4-3\,m_{3/2}^2\left( s+t\right) +s\,t\right) \\
  &\qquad\qquad\qquad-2\,s\,m_{3/2}\,m_W\,U_{\tilde{W}\tilde{W}}\left( 1-\sqrt{2}\,c_\beta\RE U_{\tilde{H}_d^-\tilde{W}}\right) \\
  &\qquad\qquad\qquad\quad\times\left( 3\,m_{3/2}^4-m_{3/2}^2\left( 3\,s+4\,t\right) +t\left( 2\,s+t\right) \right) \\
  &\qquad\qquad\qquad+m_W^2\abs{1-\sqrt{2}\,c_\beta\,U_{\tilde{H}_d^-\tilde{W}}}^2 \\
  &\qquad\qquad\qquad\quad\times\left( m_{3/2}^4\,(3\,s+t)-m_{3/2}^2\left( 3\,s^2+4\,s\,t+t^2\right) +s\,t\,(s+t)\right) \Big]\,.
 \end{split}
 \label{3bodyWlfull}
\end{equation}
After integrating over the kinematically allowed range of $t$ we find the following differential decay rate:
\begin{align}
  &\frac{d\Gamma\left( \psi_{3/2}\rightarrow {W^+}^*\,\ell_i^-\rightarrow f\,\bar{f'}\,\ell_i^-\right) }{ds} \nonumber\\
  &\qquad\simeq\frac{g^2\,\xi_i^2\,m_{3/2}^3\,\beta_s^2}{1536\,\pi^3\MP^2\left( \left( s-m_W^2\right) ^2+m_W^2\,\Gamma_W^2\right) }\bigg( s\,U_{\tilde{W}\tilde{W}}^2\,f_s \label{3bodyWl}\\
  &\qquad\qquad-\frac{8}{3}\,\frac{m_W}{m_{3/2}}\,s\,U_{\tilde{W}\tilde{W}}\left( 1-\sqrt{2}\,c_\beta\RE U_{\tilde{H}_d^-\tilde{W}}\right) j_s+\frac{1}{6}\,m_W^2\abs{1-\sqrt{2}\,c_\beta\,U_{\tilde{H}_d^-\tilde{W}}}^2h_s\bigg)\,, \nonumber
\end{align}
where the kinematic functions $\beta_s, f_s, j_s$ and $h_s$ are those given in equation~(\ref{kinematicfunc}).

\subsection*{\boldmath$\psi_{3/2}\rightarrow h^*\,\nu\rightarrow f\,\bar{f}\,\nu$}

At tree level there are two Feynman diagrams contributing to the decay of a gravitino into a fermion--antifermion pair and a neutrino via an intermediate lightest Higgs boson:
\begin{equation*}
 \parbox{5.5cm}{
  \begin{picture}(172,162) (92,-96)
    \SetWidth{0.5}
    \SetColor{Black}
    \Line(201,5)(205,-7)\Line(197,-3)(209,1)
    \Line[arrow,arrowpos=0.5,arrowlength=3.75,arrowwidth=1.5,arrowinset=0.2](194,-17)(232,-84)
    \Line[dash,dashsize=4,arrow,arrowpos=0.75,arrowlength=3.75,arrowwidth=1.5,arrowinset=0.2](213,17)(194,-17)
    \Line[arrow,arrowpos=0.5,arrowlength=3.75,arrowwidth=1.5,arrowinset=0.2](213,17)(232,51)
    \Line[arrow,arrowpos=0.5,arrowlength=3.75,arrowwidth=1.5,arrowinset=0.2](233,-17)(214,17)
    \Line[double,sep=4](117,-17)(194,-17)
    \Vertex(194,-17){2.5}
    \Vertex(213,17){2.5}
    \Text(104,-17)[]{\normalsize{\Black{$\psi_{3/2}$}}}
    \Text(237,58)[]{\normalsize{\Black{$f$}}}
    \Text(237,-24)[]{\normalsize{\Black{$\bar{f}$}}}
    \Text(198,15)[]{\normalsize{\Black{$h$}}}
    \Text(212,-15)[]{\normalsize{\Black{$\tilde{\nu}_i^*$}}}
    \Text(237,-91)[]{\normalsize{\Black{$\nu_i$}}}
  \end{picture}
 }\qquad
 +\qquad
 \parbox{5.5cm}{
  \begin{picture}(172,162) (92,-96)
    \SetWidth{0.5}
    \SetColor{Black}
    \Line(211,-55.999)(215,-44.001)\Line(207.001,-48)(218.999,-52)
    \Line[arrow,arrowpos=0.5,arrowlength=3.75,arrowwidth=1.5,arrowinset=0.2](213,-51)(232,-84)
    \Line[dash,dashsize=4](194,-17)(213,17)
    \Line[arrow,arrowpos=0.5,arrowlength=3.75,arrowwidth=1.5,arrowinset=0.2](213,17)(232,51)
    \Line[arrow,arrowpos=0.5,arrowlength=3.75,arrowwidth=1.5,arrowinset=0.2](233,-17)(214,17)
    \Line[double,sep=4](117,-17)(194,-17)
    \Line(194,-17)(213,-51)
    \Vertex(194,-17){2.5}
    \Vertex(213,17){2.5}
    \Text(104,-17)[]{\normalsize{\Black{$\psi_{3/2}$}}}
    \Text(237,58)[]{\normalsize{\Black{$f$}}}
    \Text(237,-24)[]{\normalsize{\Black{$\bar{f}$}}}
    \Text(193,6)[]{\normalsize{\Black{$h$}}}
    \Text(193,-39)[]{\normalsize{\Black{$\tilde{h}$}}}
    \Text(237,-91)[]{\normalsize{\Black{$\nu_i$}}}
  \end{picture}
 }.
\end{equation*}
As for the other decay channels we present the detailed calculation in Appendix~\ref{gravitinodecay}. Neglecting the masses of the final state particles the differential decay width of this process is found to be
\begin{equation}
  \begin{split}
  &\frac{d\Gamma\left( \psi_{3/2}\rightarrow h^*\,\nu\rightarrow f\,\bar{f}\,\nu\right) }{ds\,dt} \\
  &\qquad\simeq\frac{\xi_i^2\,m_{3/2}\,m_f^2\,\beta_s^3\,s\abs{\frac{m_{\tilde{\nu}_i}^2+\frac{1}{2}\,m_Z^2\cos2\beta}{s-m_{\tilde{\nu}_i}^2}+2\sin{\beta}\,U_{\tilde{H}_u^0\tilde{Z}}+2\cos{\beta}\,U_{\tilde{H}_d^0\tilde{Z}}}^2}{6144\,\pi^3\MP^2\,v^2\left( \left( s-m_h^2\right) ^2+m_h^2\,\Gamma_h^2\right) }\,.
  \end{split}
  \label{3bodyhnufull}
\end{equation}
The differential decay width is independent of the invariant mass $t$ and thus the integration over $t$ just results in an additional factor of $(m_{3/2}^2-s)=m_{3/2}^2\,\beta_s$ leading to
\begin{equation}
  \begin{split}
  &\frac{d\Gamma\left( \psi_{3/2}\rightarrow h^*\,\nu\rightarrow f\,\bar{f}\,\nu\right) }{ds} \\
  &\qquad\simeq\frac{\xi_i^2\,m_{3/2}^3\,m_f^2\,\beta_s^4\,s\abs{\frac{m_{\tilde{\nu}_i}^2+\frac{1}{2}\,m_Z^2\cos2\beta}{s-m_{\tilde{\nu}_i}^2}+2\sin{\beta}\,U_{\tilde{H}_u^0\tilde{Z}}+2\cos{\beta}\,U_{\tilde{H}_d^0\tilde{Z}}}^2}{6144\,\pi^3\MP^2\,v^2\left( \left( s-m_h^2\right) ^2+m_h^2\,\Gamma_h^2\right) }\,.
  \end{split}
  \label{3bodyhnu}
\end{equation}
The conjugate processes $\psi_{3/2}\rightarrow\gamma^*/Z^*\,\bar{\nu}_i\rightarrow \bar{f}\,f\,\bar{\nu}_i$, $\psi_{3/2}\rightarrow {W^-}^*\,\ell_i^+\rightarrow \bar{f}\,f'\,\ell_i^+$ and $\psi_{3/2}\rightarrow h^*\,\bar{\nu}_i\rightarrow \bar{f}\,f\,\bar{\nu}_i$ have identical decay widths.

\subsection*{Discussion}

We observe that our results for the three-body decay widths presented in equations~(\ref{3bodyZnufull}) and (\ref{3bodyWlfull}) (neglecting the contributions from higgsino--neutrino mixing) do not coincide with those presented in~\cite{Choi:2010jt}. Therefore, we carefully repeated our calculations and additionally verified that above the thresholds for $W^\pm$, $Z$ and $h$ boson production our results for the decay widths coincide with the two-body decay results of equation~(\ref{gravitinowidths}) in the narrow-width approximation (NWA), \textit{i.e.} we replaced the boson propagator by the limiting value
\begin{equation}
  \lim_{\Gamma_X/m_X\rightarrow 0}\frac{1}{\left( s-m_X^2\right) ^2+m_X^2\Gamma_X^2}=\frac{\pi}{m_X\,\Gamma_X}\,\delta(s-m_X^2)\,
  \label{narrowwidth}
\end{equation}
and verified that
\begin{equation}
 \begin{split}
  \Gamma_{\text{NWA}}\left( \psi_{3/2}\rightarrow Z^*\nu_i\rightarrow f\,\bar{f}\,\nu_i\right) &=\Gamma\left( \psi_{3/2}\rightarrow Z\,\nu_i\right) \times\BR\left( Z\rightarrow f\,\bar{f}\right) , \\
  \Gamma_{\text{NWA}}\left( \psi_{3/2}\rightarrow {W^+}^*\ell_i^-\rightarrow f\,\bar{f}'\,\ell_i^-\right) &=\Gamma\left( \psi_{3/2}\rightarrow W^+\ell_i^-\right) \times\BR\left( W^+\rightarrow f\,\bar{f}'\right) .
 \end{split}
\end{equation}
Thus we are confident that our calculations are correct. The same holds for the three-body decay via virtual Higgs exchange, where we checked that
\begin{equation}
  \Gamma_{\text{NWA}}\left( \psi_{3/2}\rightarrow h^*\nu_i\rightarrow f\,\bar{f}\,\nu_i\right) =\Gamma\left( \psi_{3/2}\rightarrow h\,\nu_i\right) \times\BR\left( h\rightarrow f\,\bar{f}\right) .
\end{equation}
For a derivation of the results in the narrow-width approximation we refer to Appendix~\ref{gravitinodecay}.

It might thus well be that the details of the final state particle spectra presented in~\cite{Choi:2010jt} are not completely correct, although their conclusions for phenomenological studies are not expected to change qualitatively.

Let us shortly comment on additional possible Feynman diagrams for gravitino decays into three final state particles. One class of these additional contributions are decays via trilinear $R$-parity breaking Yukawa couplings as induced by the field redefinitions in Section~\ref{Rbreaking}. As these decays always involve off-shell propagators of sleptons or squarks we expect their contribution to be negligible compared to the resonantly enhanced two-body decay channels.

In addition, at sufficiently large gravitino masses, the gravitino can decay into an on-shell three-body final state including gauge bosons and leptons, namely $\psi_{3/2}\rightarrow W^+W^-\nu_i$ and $\psi_{3/2}\rightarrow Z\,W^+\ell_i^-$. These processes can proceed via the nonabelian 4-vertex including two massive gauge bosons and a gaugino that mixes with the neutrinos or charged leptons. In addition, various other diagrams involving off-shell propagators contribute to these processes.\footnote{The diagrams contributing to the process $\psi_{3/2}\rightarrow W^+W^-\nu_i$ have been discussed in a different context in~\cite{Luo:2010he}.} However, also in this case we expect the additional contributions to the total gravitino decay width to be negligible compared to the resonantly enhanced two-body decay channels.

To conclude this section we want to give an estimate of the dependence of the gravitino lifetime on the amount of $R$-parity breaking. For this reason we give a numerical result for the prefactor of the lifetime in two-body decays into a photon and a neutrino or an antineutrino:
\begin{equation}
  \tau_{3/2}^{\gamma\,\nu_i/\bar{\nu_i}}=\frac{32\,\pi\,\MP^2}{\xi_i^2\,m_{3/2}^3\abs{U_{\tilde{\gamma}\tilde{Z}}}^2}\simeq3.8\times10^{26}\,\text{s}\left( \frac{10^{-9}}{\xi_i}\right) ^2\left( \frac{100\,\text{GeV}}{m_{3/2}}\right) ^3\frac{1}{\abs{U_{\tilde{\gamma}\tilde{Z}}}^2}\,.
\end{equation}
Thus a gravitino with a mass around the electroweak scale naturally has a lifetime of the order of $10^{26}$--$10^{27}$\,s for an $R$-parity violating parameter in the center of the cosmologically favored range. This is exactly the order of magnitude of lifetimes leading to interesting prospects for indirect dark matter searches (see Chapter~\ref{indirectdetection}).

\subsection{Gravitino Branching Ratios}
\label{branchingratios}

In order to calculate the cosmic-ray signals from gravitino decays we need to know the branching ratios of the different gravitino decay channels. These can easily be obtained from the decay widths we calculated before by dividing the partial decay widths by the total gravitino decay width:
\begin{equation}
 \BR\left( X\right) =\frac{\Gamma\left( \psi_{3/2}\rightarrow X\right) }{\Gamma_{\text{tot}}}\,,\qquad\text{where}\qquad\Gamma_{\text{tot}}=\sum_X\Gamma\left( \psi_{3/2}\rightarrow X\right) .
\end{equation}
In the calculation of the gravitino branching ratios the $R$-parity violating parameter $\xi_i$ drops out. Therefore, the branching ratios present a general result that can be obtained without knowledge of the exact size of $\xi_i$. By contrast, there is a significant dependence on the mixing parameters that enter in the calculation of decay widths.

Assuming that the gaugino masses are at the same order ($M_1\sim M_2\sim M_{1/2}$), we find the following proportionalities for the mixing parameters from the approximate formulae in equations~(\ref{UgammaZapx}) to (\ref{UHdZapx}), (\ref{UWWapx}) and~(\ref{UHWapx}):
\begin{equation}
   U_{\tilde{\gamma}\tilde{Z}}\,,\;U_{\tilde{Z}\tilde{Z}}\propto\frac{m_Z}{M_{1/2}}\qquad\text{and}\qquad U_{\tilde{H}_u^0\tilde{Z}}\,,\;U_{\tilde{H}_d^0\tilde{Z}}\propto\frac{m_Z^2}{M_{1/2}\,\mu}
\end{equation}
for the neutralino--neutrino mixing parameters, and
\begin{equation}
   U_{\tilde{W}\tilde{W}}\propto\frac{m_W}{M_2}\qquad\text{and}\qquad U_{\tilde{H}_d^-\tilde{W}}\propto\frac{m_W^2}{M_2\,\mu}
\end{equation}
for the chargino--charged lepton mixing parameters.\footnote{The chargino--charged lepton mixing parameters only depend on the wino mass parameter $M_2$.} This means that for large gravitino masses, where the gaugino and Higgs mass parameters are necessarily much larger than the gauge boson masses, the Feynman diagrams involving a mass mixing are suppressed compared to those that do not include mixing parameters. Therefore, the two-body decay into a photon and a neutrino will always be suppressed above the threshold for electroweak gauge boson production. The other decay channels then have the asymptotic decay widths:
\begin{equation}
  \frac{1}{2}\,\Gamma\left( \psi_{3/2}\rightarrow W^+\ell_i^-\right) =\Gamma\left( \psi_{3/2}\rightarrow Z\nu_i\right) =\Gamma\left( \psi_{3/2}\rightarrow h\,\nu_i\right) =\frac{\xi_i^2\,m_{3/2}^3}{384\,\pi\,\MP^2}\,.
\end{equation}
In addition, all higgsino mixing parameters are suppressed compared to the gaugino mixings by an additional factor of $\mathcal{O}(m_{Z/W}/\mu)$. Therefore, they typically contribute even less. In particular, in the decoupling limit, which we employ as our standard setup, these mixings are negligible for all gravitino masses.

We want to fix now a standard choice of parameters for the following phenomenological studies in this thesis. From mediation mechanisms of supersymmetry breaking like gravity mediation it is expected that the soft mass parameters are proportional to the gravitino mass. Therefore, we choose the gaugino masses to be given by $M_1=1.5\,m_{3/2}$ and $M_2=1.89\,M_1$. The latter relation corresponds to the typical assumption of universal gaugino masses at a high unification scale. Similarly, we choose $m_{\tilde{\nu}_i}=2\,m_{3/2}$ for the masses of sneutrinos. This choice is not very critical as the dependence of the decay widths on the sneutrino mass is very mild. In addition, we take $\tan\beta=10$ and a large $\mu$ parameter. Technically, we set $\mu=\text{Max}\left( 10\,m_Z,10\,m_{3/2}\right) $ in order to guarantee that we are in the decoupling limit for all choices of the gravitino mass. For the mass of the standard model-like lightest Higgs boson we assume a low value of $m_h=115\,$GeV, which is right above the LEP limit and favored by electroweak precision data~\cite{Nakamura:2010zzi}.

\begin{figure}[t]
 \centering
 \includegraphics[scale=0.8]{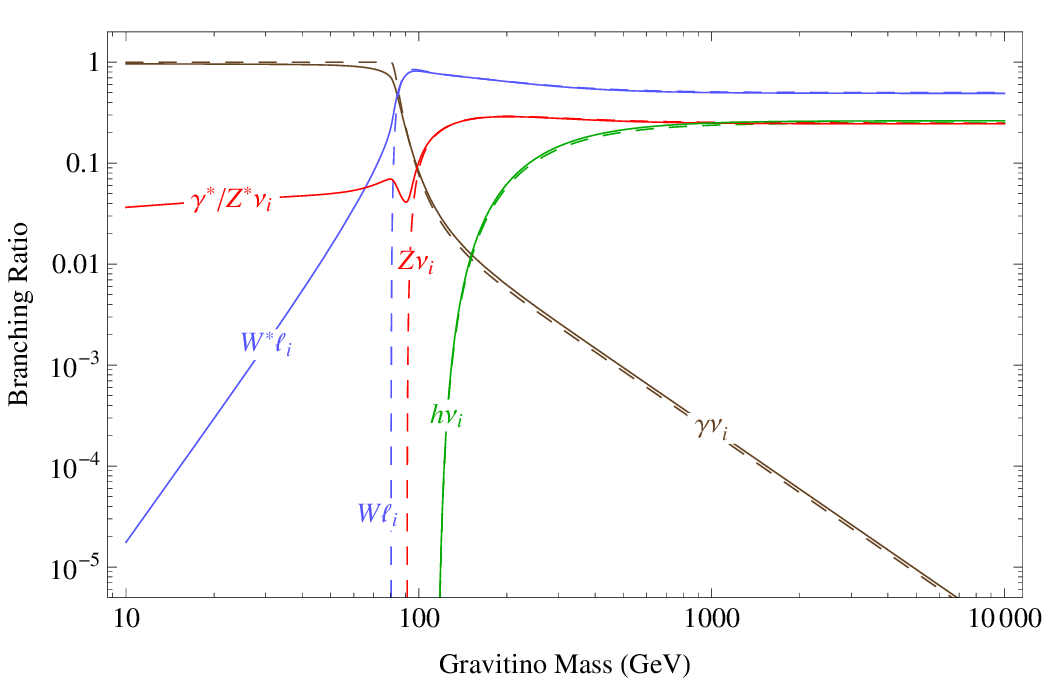} 
 \caption[Branching ratios of the gravitino decay channels in the decoupling limit.]{Branching ratios of the different gravitino decay channels as a function of the gravitino mass in the decoupling limit for our standard choice of parameters. The dashed lines show the result taking into account only the two-body decays of the gravitino, while the solid lines show the full result including three-body decays.}
 \label{BRplot}
\end{figure}
\begin{table}[t]
 \centering
 \begin{tabular}{rccccc}
  \toprule
  $m_{3/2}\quad$ & $\BR(\gamma\,\nu_i)$ & $\BR(W^*\ell_i)$ & $\BR(\gamma^*/Z^*\nu_i)$ & $\BR(h^*\nu_i)$ \\
  \midrule
  30\,GeV & 95\,\% & 1.6\,\% & 4.5\,\% & --- \\
  80\,GeV & 71\,\% & 22\,\% & 7.0\,\% & --- \\
  85\,GeV & 43\,\% & 51\,\% & 5.4\,\% & --- \\
  100\,GeV & 8.3\,\% & 81\,\% & 10\,\% & --- \\
  150\,GeV & 1.3\,\% & 70\,\% & 27\,\% & 1.1\,\% \\
  300\,GeV & 0.26\,\% & 57\,\% & 28\,\% & 15\,\% \\
  1\,TeV & 0.024\,\% & 50\,\% & 25\,\% & 25\,\% \\
  3\,TeV & 0.0026\,\% & 49\,\% & 25\,\% & 26\,\% \\
  10\,TeV & 0.0002\,\% & 49\,\% & 25\,\% & 26\,\% \\
  \bottomrule
 \end{tabular}
 \caption[Branching ratios of the gravitino decay channels in the decoupling limit.]{Branching ratios of the different gravitino decay channels for several gravitino masses in the decoupling limit for our standard choice of parameters. We only present values larger than $10^{-6}$.}
 \label{BRtable}
\end{table}
The result for the gravitino branching ratios using this set of parameters is presented in Figure~\ref{BRplot} as a function of the gravitino mass and in Table~\ref{BRtable} for several exemplary values of the gravitino mass. As expected from the calculation in the narrow-width approximation, the branching ratios derived from the three-body decay widths agree well with the two-body decay results above the thresholds for on-shell production of $W$, $Z$ and Higgs bosons. In addition, we observe the asymptotic behavior of the branching ratios at large gravitino masses as expected from the discussion of the mixing parameters above.

Below the thresholds for on-shell boson production we see the effects of the off-shell processes. In particular the virtual $W$ channel can play a role at gravitino masses slightly below the $W$ production threshold. At lower masses, however, it drops fast in this scenario. In that region actually the virtual photon exchange channel is not completely negligible. It dominates the decay rate in the channel $\psi_{3/2}\rightarrow\gamma^*/Z^*\,\nu_i\rightarrow f\,\bar{f}\,\nu_i$ in this setup and is then the only source of different final state particles than the monoenergetic photon and neutrino. The virtual Higgs channel plays no role at all below the threshold since only light fermions are kinematically allowed in the final state and their couplings to the Higgs boson are suppressed by their small masses.

In~\cite{Choi:2010jt} it was found that the two-body decay into photon and neutrino could be strongly suppressed compared to the three-body decays via virtual gauge bosons. This was achieved by setting the gaugino masses to a high scale such that all mass mixings are suppressed. In that case the diagrams which do not involve mixing parameters become dominant, \textit{i.e.} the 4-vertex diagrams with a sneutrino VEV at one of the legs. Then, typically the virtual $W$ channel dominates in a certain range below the threshold. The relative importance of the two-body decay channel and the virtual $W$ and $Z$ channels strongly depends on the choice of the gaugino masses.

We want to propose here another scenario where only the diagram with photon exchange is suppressed. From the analytical approximation in equation~(\ref{UgammaZapx}) we observe that the photino--neutrino mixing is vanishing for degenerate gaugino masses. This is not an artifact of the approximation as also in the full numerical calculation the mixing is highly suppressed. In this case, the photon decay channel does not contribute at all while all other mixing parameters remain at the same order of magnitude. Although one could argue that this scenario is a bit fine-tuned it can serve as an exemplary scenario where no gamma-ray line is expected from gravitino dark matter decays.

\begin{figure}[t]
 \centering
 \includegraphics[scale=0.8]{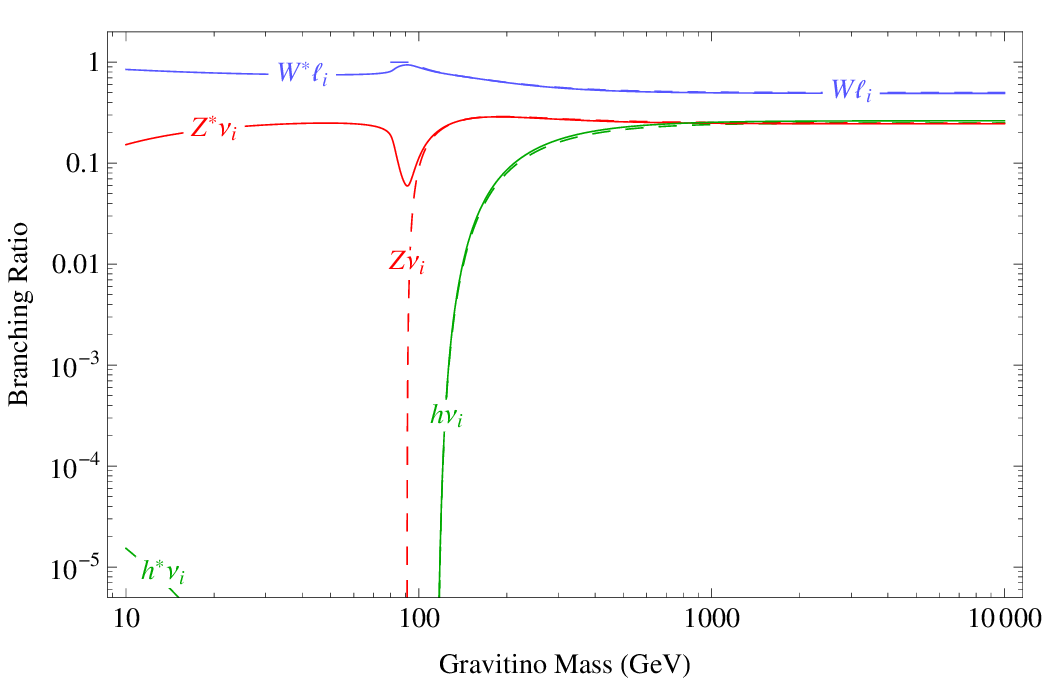} 
 \caption[Branching ratios of the gravitino decay channels in the decoupling limit for the case of a suppressed photon channel.]{Branching ratios of the different gravitino decay channels as a function of the gravitino mass in the decoupling limit for the case of a suppressed photon channel. The dashed lines show the result taking into account only the two-body decays of the gravitino, while the solid lines show the full result including three-body decays.}
 \label{BRplotWOline}
\end{figure}
In Figure~\ref{BRplotWOline} we present the branching ratios for the choice $M_1=M_2=1.5\,m_{3/2}$ and all other parameters as in our standard setup. In this case the photon exchange diagram does not contribute to the gravitino decay. While the branching ratios for the other channels are practically unchanged above the threshold for on-shell production, they now dominate also for low gravitino masses. The ratio between the $W$ and the $Z$ channel remains at the same level for all gravitino masses except for the suppression of the $Z$ channel for gravitino masses close to the $Z$ boson mass.

Due to the strong dependence of the branching ratios on the mixing parameters for low gravitino masses, the phenomenology of the light gravitino becomes much more model-dependent than that of heavier gravitinos. Let us discuss this further in the derivation of the spectra of final state particles produced in gravitino decays.

\section{Spectra of Final State Particles from Gravitino Decays}
\label{gravspectra}

In this section we want to discuss the energy spectra of the final state particles from gravitino decay as this is a crucial input for indirect searches for gravitino dark matter. As a first step we want to determine the spectra of the particles directly produced in gravitino decays. The spectrum of final state particles from a decay is in general given by the expression
\begin{equation}
  \frac{dN}{dE}=\frac{1}{\Gamma}\,\frac{d\Gamma}{dE}\,.
\end{equation}
The simplest spectra are observed for the photon and the neutrino in the two-body gravitino decay as those correspond simply to a monochromatic line at an energy corresponding to one half of the gravitino mass:
\begin{equation}
  \frac{dN_{\gamma/\nu}}{dE}=\delta\left( E-\frac{m_{3/2}}{2}\right) .
\end{equation}
We can also extract the spectra of the particles directly produced in the gravitino three-body decays from the differential decay widths in equations~(\ref{3bodygammaZnu}), (\ref{3bodyWl}) and (\ref{3bodyhnu}). For the neutrino and the charged lepton produced in association with a (virtual) boson we find
\begin{equation}
  \frac{dN_{\nu/\ell}}{dE}=\frac{1}{\Gamma_{X^*\nu/\ell}}\,\frac{d\Gamma_{X^*\nu/\ell}}{dE}=\frac{2\,m_{3/2}}{\Gamma_{X^*\nu/\ell}}\,\frac{d\Gamma_{X^*\nu/\ell}}{ds}\bigg|_{s\,=\,m_{3/2}^2-2\,m_{3/2}\,E}\,,
  \label{nulspec}
\end{equation}
where $X$ can stand for a $\gamma$, a $Z$, a $W$ or a $h$ boson. For gravitino masses below the production threshold for the massive bosons, one should, of course, correctly take into account the interference between the photon and the $Z$ boson diagram and thus use the differential decay width of the complete channel.\footnote{As mentioned before, the interference contribution of the Higgs diagram can be neglected due to the small Higgs--fermion coupling.} For gravitino masses above the production threshold for the massive bosons, the neutrino and charged lepton spectra become a narrow line centered around the energy
\begin{equation}
  E=\frac{m_{3/2}}{2}\left( 1-\frac{m_X^2}{m_{3/2}^2}\right) .
  \label{lineenergy}
\end{equation}
The shape of this line is practically independent of the mixing parameters and can be described by the approximate expression
\begin{equation}
  \frac{dN_{\nu/\ell}}{dE}\simeq\frac{8\,E^2/m_{3/2}}{\left( m_{3/2}^2-2\,m_{3/2}\,E-m_X^2\right) ^2+m_X^2\,\Gamma_X^2}\left( \int_0^{m_{3/2}^2}\!\!ds\,\frac{\left( 1-s/m_{3/2}^2\right) ^2}{\left( s-m_X^2\right) ^2+m_X^2\,\Gamma_X^2}\right) ^{-1}\!\!.
\end{equation}
In a similar way we can extract the spectra of the other two fermions in the three-body decays from equations~(\ref{3bodygammaZnufull}), (\ref{3bodyWlfull}) and (\ref{3bodyhnufull}). Since the diagrams are symmetric with respect to these fermions, the spectra of both fermions are identical\footnote{In the case of the $W$-mediated diagram this is only approximately true since in this case the final state fermions have different masses. However, this effect is negligible in many situations.} and we find:
\begin{equation}
  \frac{dN_f}{dE}=\frac{dN_{\bar{f}^{(\prime)}}}{dE}=\frac{1}{\Gamma_{X^*\nu/\ell}}\,\frac{d\Gamma_{X^*\nu/\ell}}{dE}=\frac{2\,m_{3/2}}{\Gamma_{X^*\nu/\ell}}\int_0^{m_{3/2}^2-t}\!\!ds\,\frac{d\Gamma_{X^*\nu/\ell}}{ds\,dt}\bigg|_{t\,\simeq\,m_{3/2}^2-2\,m_{3/2}\,E}\,.
  \label{ffspec}
\end{equation}
However, many of the fermions produced in the gravitino decay are unstable and therefore these analytical expressions do not yet give us the spectra for all stable final state particles. Muons will decay via a virtual $W$ boson and produce a softer spectrum of electrons as well as electron and muon neutrinos. Tau leptons also decay via a virtual $W$ boson (thus producing tau neutrinos), but in this case also hadronic final states are kinematically allowed. These lead to a soft spectrum of photons, electrons and muon neutrinos as well as electron neutrinos coming from pion decays. Quark final states from the three-body decay also finally lead to soft spectra of photons, electrons and neutrinos coming from the decays of pions produced in hadronization processes. An important addition is, however, that also protons and neutrons can be formed during the fragmentation of quark final states. Most neutrons subsequently decay into protons but there is a small probability that protons and neutrons form a bound state: a deuteron.

\paragraph{Deuteron and Antideuteron Formation}

A usual approach to model the formation of deuterons and antideuterons in particle physics processes is the coalescence model~\cite{Csernai:1986qf}. In this prescription a deuteron is formed when a proton and a neutron come sufficiently close in momentum space, \textit{i.e.} when the absolute value of the difference of their four-momenta is below a threshold coalescence momentum:
\begin{equation}
  \abs{p_p-p_n}<p_0\,.
\end{equation}
We adopt here a value for the coalescence momentum of $p_0=160\,$MeV that was extracted by the authors of~\cite{Kadastik:2009ts} to match the data on deuteron and antideuteron production in hadronic $Z$ decays of the ALEPH experiment at the LEP collider~\cite{Schael:2006fd}.

A common approximation for deuteron formation is that the distributions of neutrons and protons in momentum space are spherically symmetric and uncorrelated. This leads to an energy distribution of deuterons that can be directly calculated from the spectra of neutrons and protons~\cite{Kadastik:2009ts,Brauninger:2009pe}:
\begin{equation}
  \frac{dN_d}{dT_d}=\frac{2}{3}\,p_0^3\,\frac{m_d}{m_p\,m_n}\,\frac{1}{\sqrt{T_d^2+2\,m_d\,T_d}}\,\frac{dN_p}{dT_p}\,\frac{dN_n}{dT_n}\,,
\end{equation}
where $T_p=T_n=T_d/2$ are the kinetic energies of protons, neutrons and deuterons. As pointed out in~\cite{Kadastik:2009ts}, this approximation is qualitatively wrong and significantly underestimates the deuteron yield in high-energetic processes. This is due to the fact, that the distributions of neutrons and protons are actually neither spherically symmetric nor uncorrelated. For instance, in the decay of a gravitino into a $Z$ boson and a neutrino the probability for the formation of a deuteron in the fragmentation of the $Z$ boson should be independent of the gravitino mass. This is due to the fact that the $Z$ boson fragmentation process is always the same as viewed from the $Z$ boson rest frame. However, the spherical approximation of the coalescence model gives a lower yield of deuterons for larger gravitino masses, since the protons and neutrons are distributed over a larger phase space for higher injection energies. Another qualitatively wrong behavior of this approximation is the possibility that protons and neutrons from distinct gravitino decays form a deuteron. This is due to the fact that the simply the spectra of protons and neutrons are multiplied, while in principle the coalescence condition on the four-momenta of protons and neutrons should be applied on an event-by-event basis.

This can be achieved in a Monte Carlo simulation of the decay process. For instance, using an event generator like PYTHIA one can simulate the hadronization of massive gauge and Higgs bosons explicitly requiring the neutrons not to decay. One can then apply event by event the coalescence condition on the protons and neutrons. This method leads to plausible results, \textit{e.g.} the deuteron yield in the gravitino decay to final states including $W$, $Z$ or Higgs bosons is independent of the gravitino mass. This method, however, requires a lot of computing time to generate smooth spectra as only one deuteron or antideuteron is produced in $\mathcal{O}(10^4)$ decay processes.

Let us now turn to the discussion of final state particle spectra from the decay of gravitinos with masses above the threshold for the on-shell production of massive gauge and Higgs bosons.

\paragraph{Two-Body Decay Spectra}

In cases where the gravitino mass is large enough to produce on-shell massive bosons, the decay process is well described by a two-body decay with subsequent fragmentation of the boson (\textit{cf.} the discussion in the last section). Since in this case the phase-space distribution is independent of the structure of the squared matrix element, the whole decay process including the hadronization can be mimicked using the event generator PYTHIA 6.4~\cite{Sjostrand:2006za} although the gravitino is not included in that program. We employ a similar treatment of the implementation of this process as described in~\cite{Grefe:2008zz} and simulate $\mathcal{O}(10^5)$ events per decay channel for several values of the gravitino mass to generate the spectra of photons, electrons, positrons, electron neutrinos and antineutrinos, and muon neutrinos and antineutrinos. In order to generate the spectra of the less abundant protons, antiprotons, and tau neutrinos and antineutrinos, we use a higher statistics of $\mathcal{O}(10^6)$ events. For the generation of the spectra of deuterons and antideuterons an even higher statistics of $\mathcal{O}(10^7)$ events is used.\footnote{The spectra of deuterons and antideuterons from gravitino decay were kindly provided by Gilles Vertongen.}

\begin{table}
 \centering
 \begin{tabular}{cccccc}
  \toprule
  Particle type & $Z\,\nu_i$ & $We$ & $W\mu$ & $W\tau$ & $h\,\nu_i$ \\
  \midrule
  $\gamma$ & 17.0 & 15.0 & 15.0 & 16.1 & 31.5 \\
  $e^-+e^+$ & 16.4 & 15.7 & 15.7 & 16.0 & 28.8 \\
  $p+\bar{p}$ & 1.67 & 1.60 & 1.60 & 1.60 & 2.98 \\
  $\nu_e+\bar{\nu}_e$ & 16.3 & 14.5 & 15.5 & 15.8 & 29.5 \\
  $\nu_\mu+\bar{\nu}_\mu$ & 29.7 & 26.5 & 27.5 & 28.8 & 54.1 \\
  $\nu_\tau+\bar{\nu}_\tau$ & 0.23 & 0.22 & 0.22 & 1.22 & 0.23 \\
  $d+\bar{d}$ & $1.1\times10^{-4}$ & $1.1\times10^{-4}$ & $1.1\times10^{-4}$ & $1.1\times10^{-4}$ & $3.6\times10^{-4}$ \\
  \bottomrule
 \end{tabular}
 \caption[Multiplicities of stable final state particles from gravitino decay.]{Multiplicities of stable final state particles from gravitino decays after the fragmentation of the different on-shell intermediate particles simulated with PYTHIA.}
 \label{particlenumbers}
\end{table}
The resulting particle multiplicities per single gravitino decay for the decay channels $\psi_{3/2}\rightarrow Z\,\nu_i$, $We$, $W\mu$, $W\tau$ and $h\,\nu_i$ are summarized in Table~\ref{particlenumbers}. Note that the multiplicities are independent of the gravitino mass. We observe that similar amounts of particles are produced in the $Z\,\nu_i$ and $W\ell_i$ decay channels while the numbers are roughly twice as high for the decay channel into $h\,\nu_i$. In addition, the amount of photons, electrons and electron neutrinos is similar for every single decay channel. The number of muon neutrinos is larger by a factor of two compared to the other particles. By contrast, the number of protons is lower by one order of magnitude and the number of tau neutrinos is low in all channels. As discussed above, the probability for deuteron formation is extremely low.

Several features of these numbers can be understood qualitatively or even quantitatively: Electrons, electron neutrinos and muon neutrinos are mainly produced in pion and subsequent muon decays and therefore the number of muon neutrinos is a factor of two higher. We expect a factor of two less neutral pions compared to charged pions, thus explaining that the number of photons is at the same level as the number of electrons. Tau neutrinos are practically only produced in direct decays of gauge bosons and tau leptons. A smaller contribution comes also from $B$ meson decays. The larger particle multiplicities in the $h\,\nu_i$ channel are probably due to the dominant Higgs decay into pairs of bottom quarks. Therefore, we expect more hadronic activity compared to the other channels leading to a larger number of final state particles.

The differences between the three $W\ell_i$ channels exclusively come from the decay of the hard leptons in the respective decay channels. While the electron only contributes a single electron to the final spectrum, the muon decay produces a muon neutrino, an electron neutrino and an electron. The tau decay is more complicated, but we expect an additional tau neutrino as well as $\sim1\,e,\nu_e$, $\sim2\,\nu_\mu$ and $\sim1\,\gamma$ which is in good agreement with the simulation.

\begin{figure}
 \centering
 \includegraphics[scale=0.44]{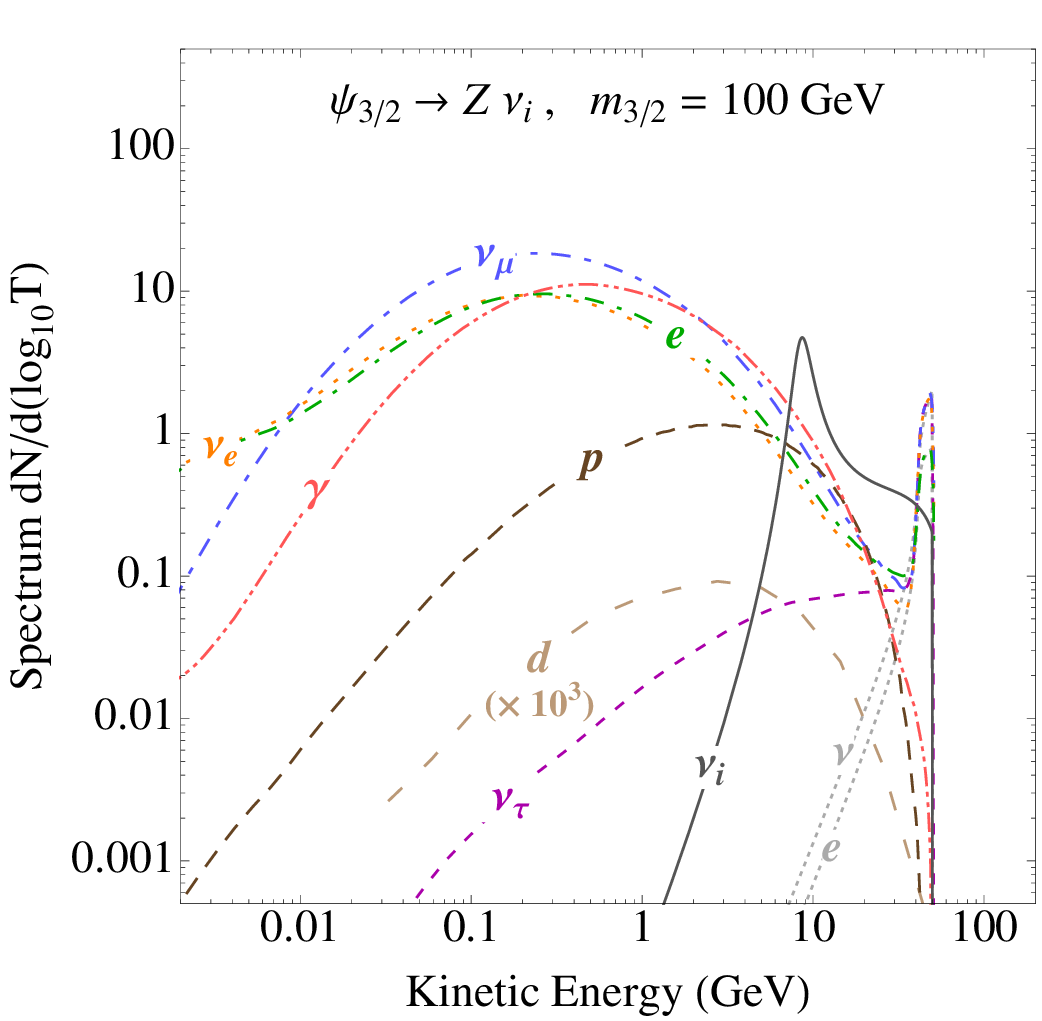}
 \hfill
 \includegraphics[scale=0.44]{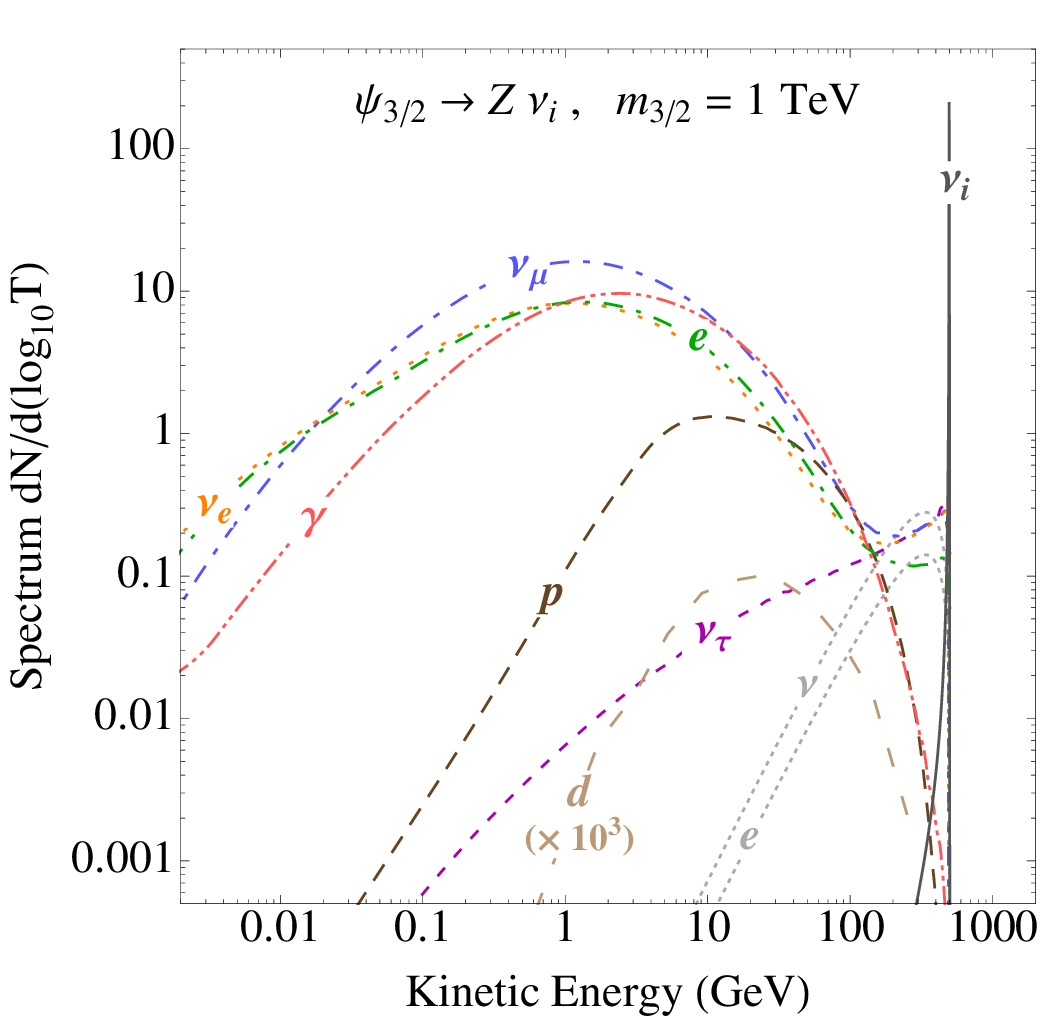}
 \caption[Spectra of stable final state particles from gravitino decay in the channel $\psi_{3/2}\rightarrow Z\nu_i$.]{Spectra of stable final state particles from the decay of a gravitino with a mass of 100\,GeV (\textit{left}) or 1\,TeV (\textit{right}) in the channel $\psi_{3/2}\rightarrow Z\nu_i$. The features of these spectra are discussed in the text.}
 \label{Znu}
\end{figure}
\begin{figure}
 \centering
 \includegraphics[scale=0.44]{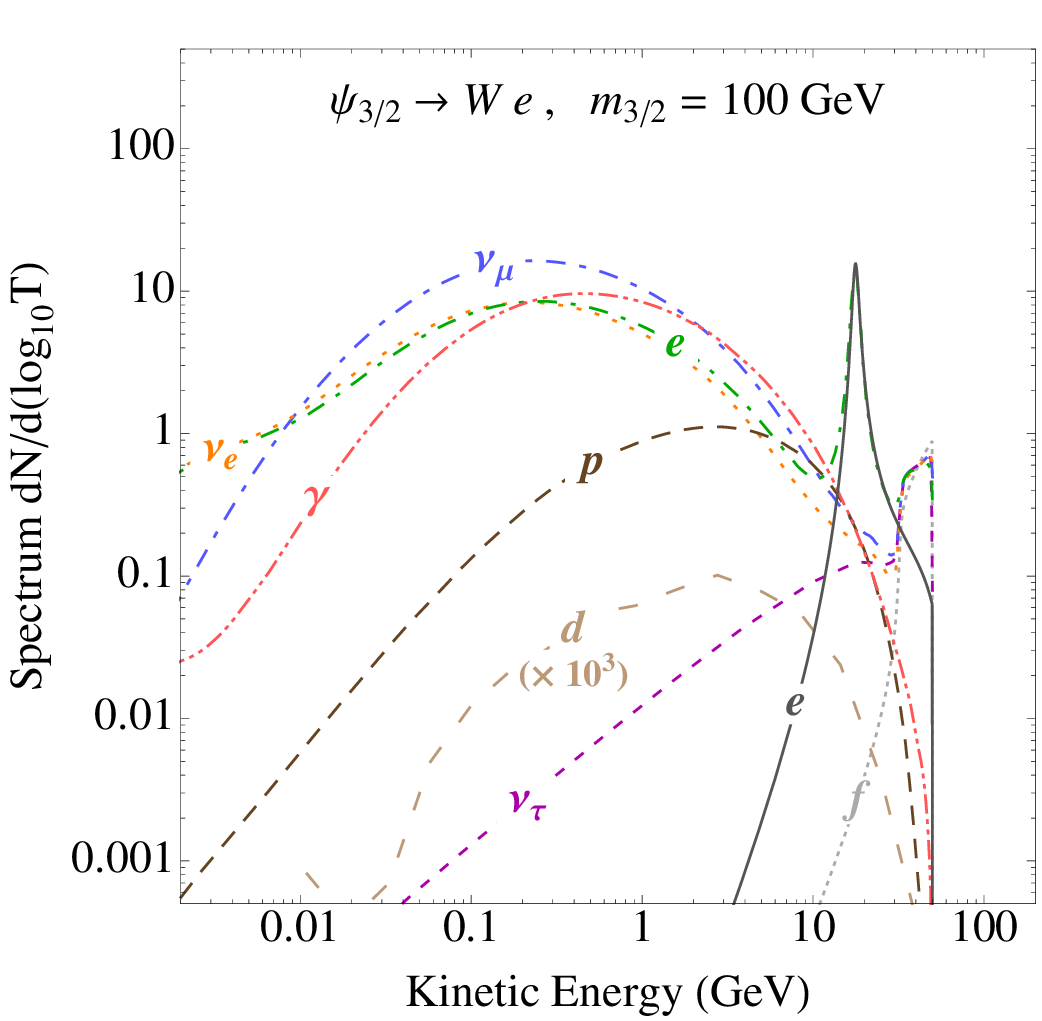}
 \hfill
 \includegraphics[scale=0.44]{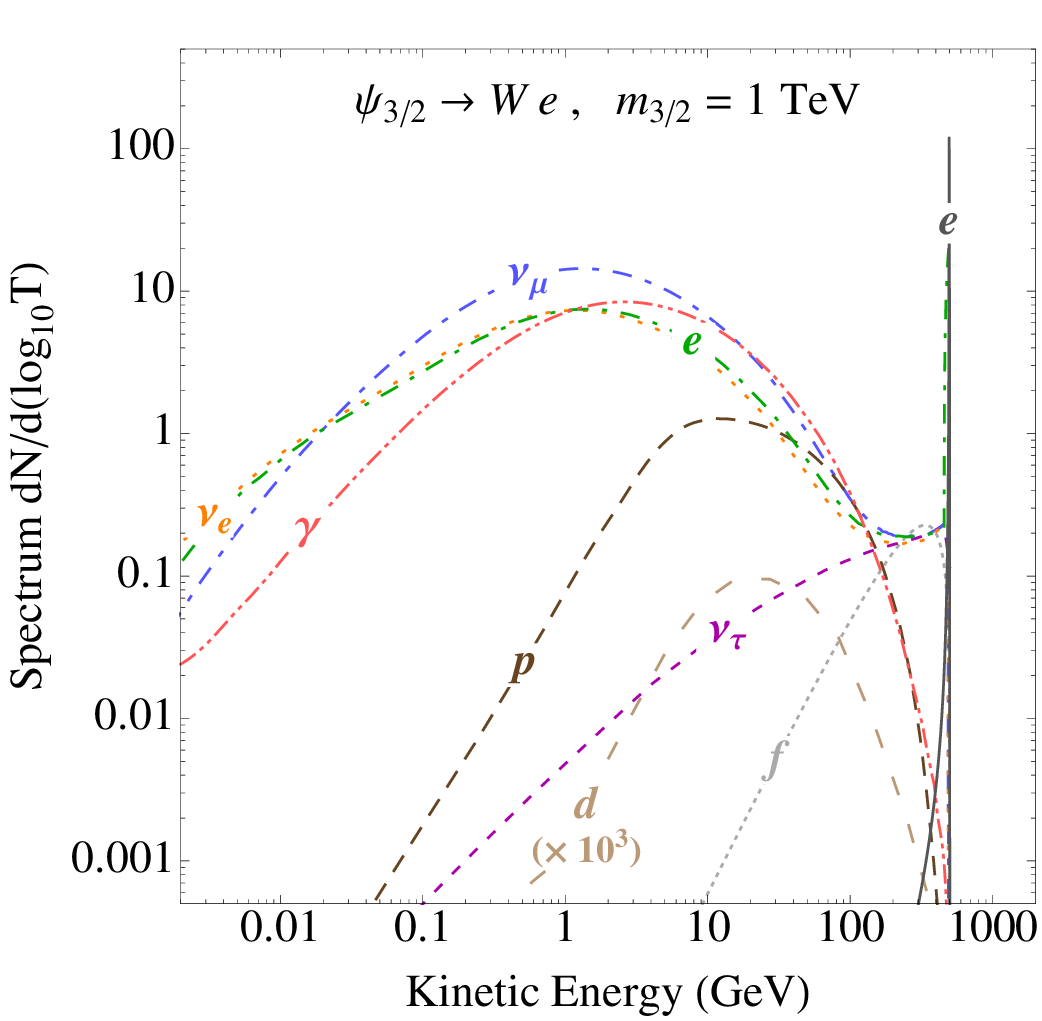}
 \caption[Spectra of stable final state particles from gravitino decay in the channel $\psi_{3/2}\rightarrow We$.]{Spectra of stable final state particles from the decay of a gravitino with a mass of 100\,GeV (\textit{left}) or 1\,TeV (\textit{right}) in the channel $\psi_{3/2}\rightarrow We$. The features of these spectra are discussed in the text.}
 \label{We}
\end{figure}
\begin{figure}
 \centering
 \includegraphics[scale=0.44]{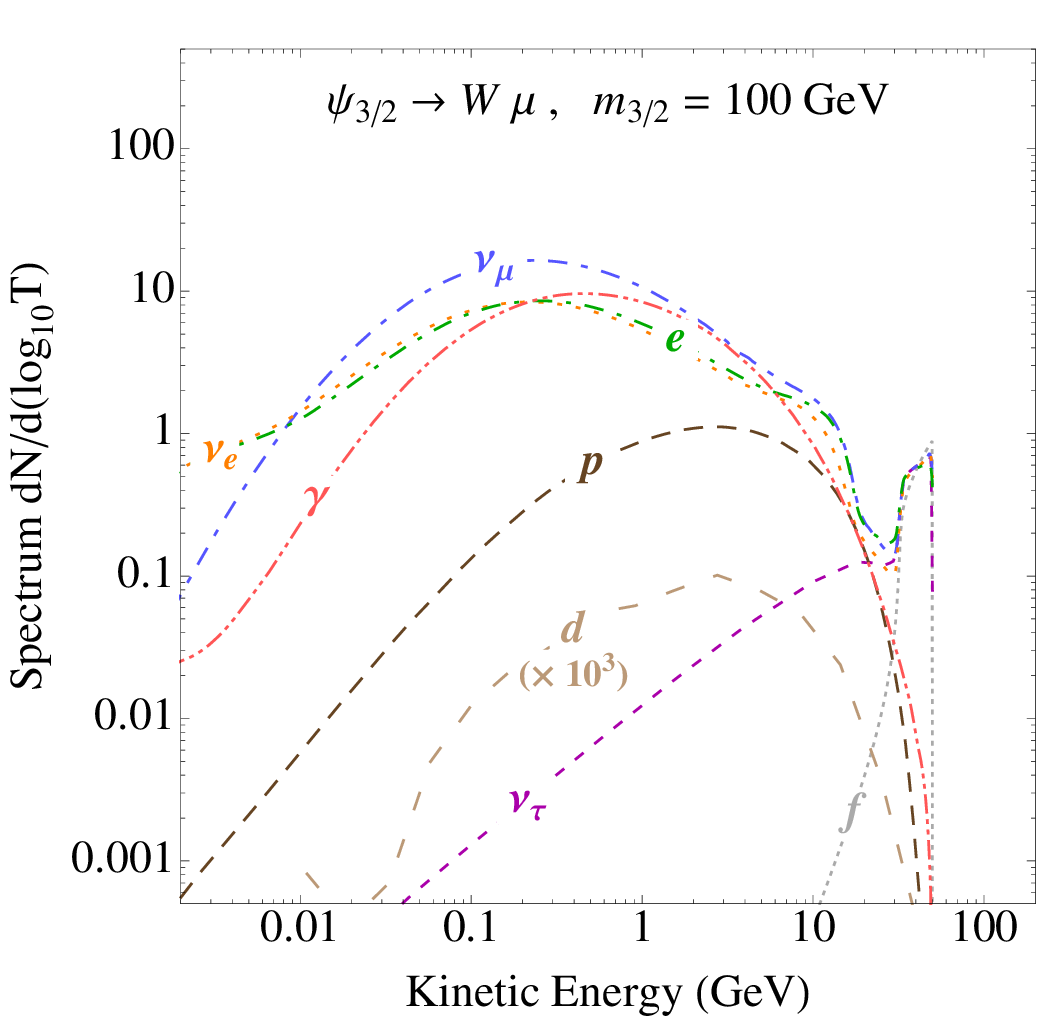}
 \hfill
 \includegraphics[scale=0.44]{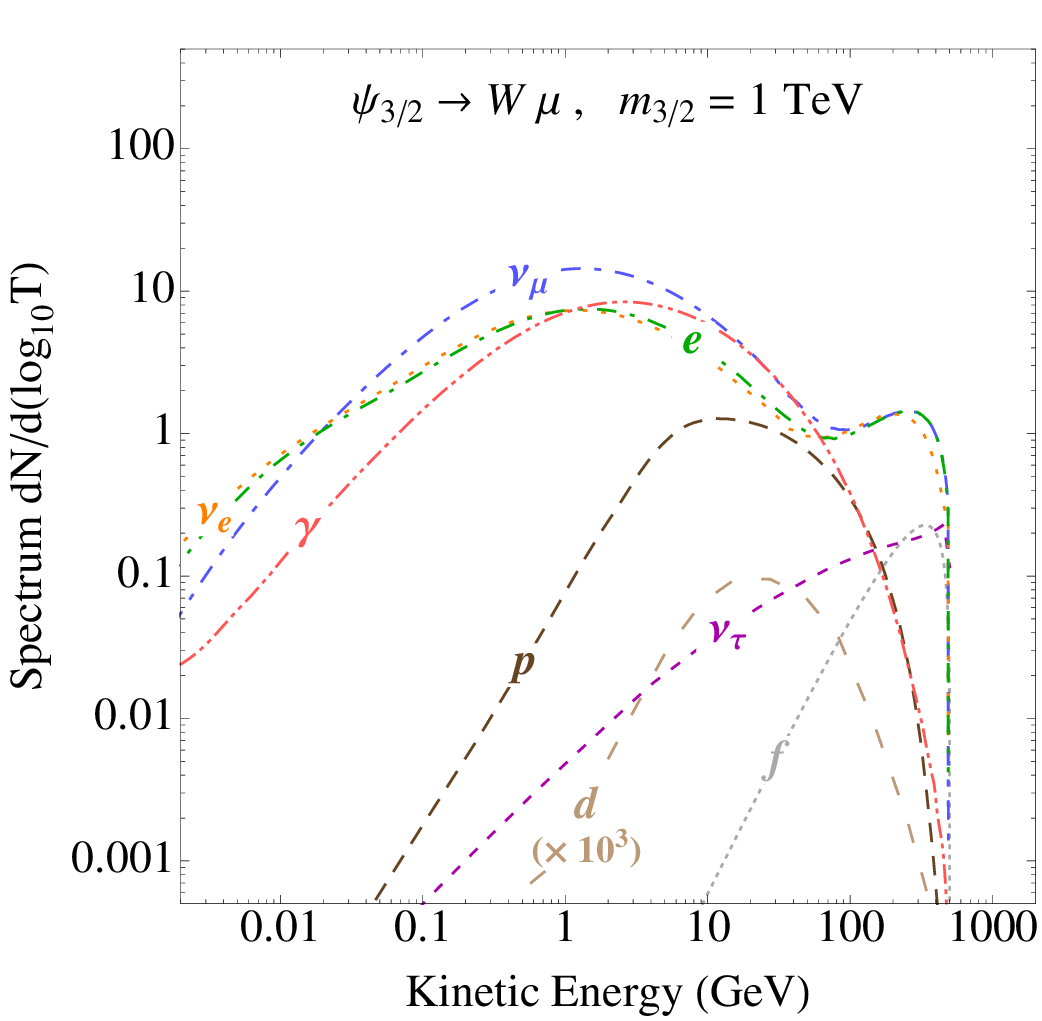}
 \caption[Spectra of stable final state particles from gravitino decay in the channel $\psi_{3/2}\rightarrow W\mu$.]{Spectra of stable final state particles from the decay of a gravitino with a mass of 100\,GeV (\textit{left}) or 1\,TeV (\textit{right}) in the channel $\psi_{3/2}\rightarrow W\mu$. The features of these spectra are discussed in the text.}
 \label{Wmu}
\end{figure}
\begin{figure}
 \centering
 \includegraphics[scale=0.44]{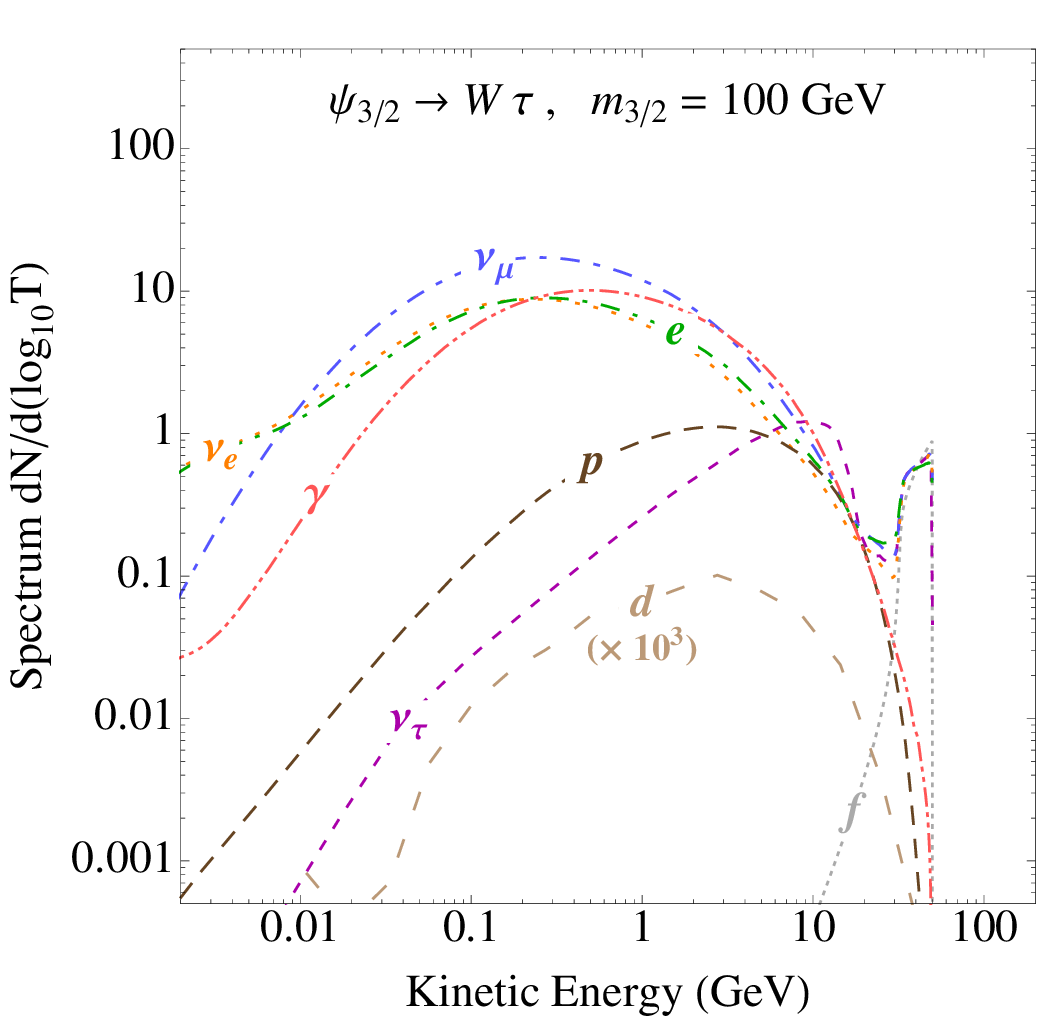}
 \hfill
 \includegraphics[scale=0.44]{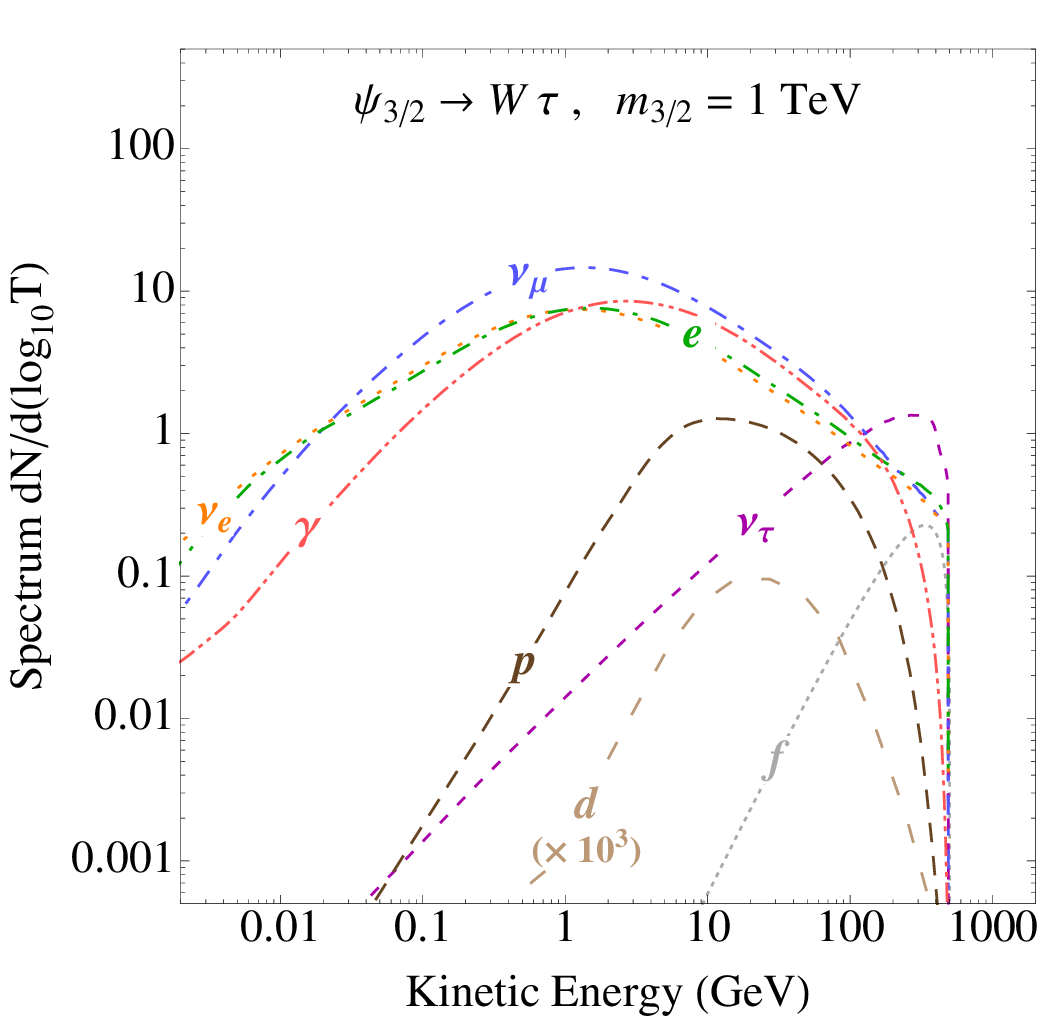}
 \caption[Spectra of stable final state particles from gravitino decay in the channel $\psi_{3/2}\rightarrow W\tau$.]{Spectra of stable final state particles from the decay of a gravitino with a mass of 100\,GeV (\textit{left}) or 1\,TeV (\textit{right}) in the channel $\psi_{3/2}\rightarrow W\tau$. The features of these spectra are discussed in the text.}
 \label{Wtau}
\end{figure}
\begin{figure}
 \centering
 \includegraphics[scale=0.44]{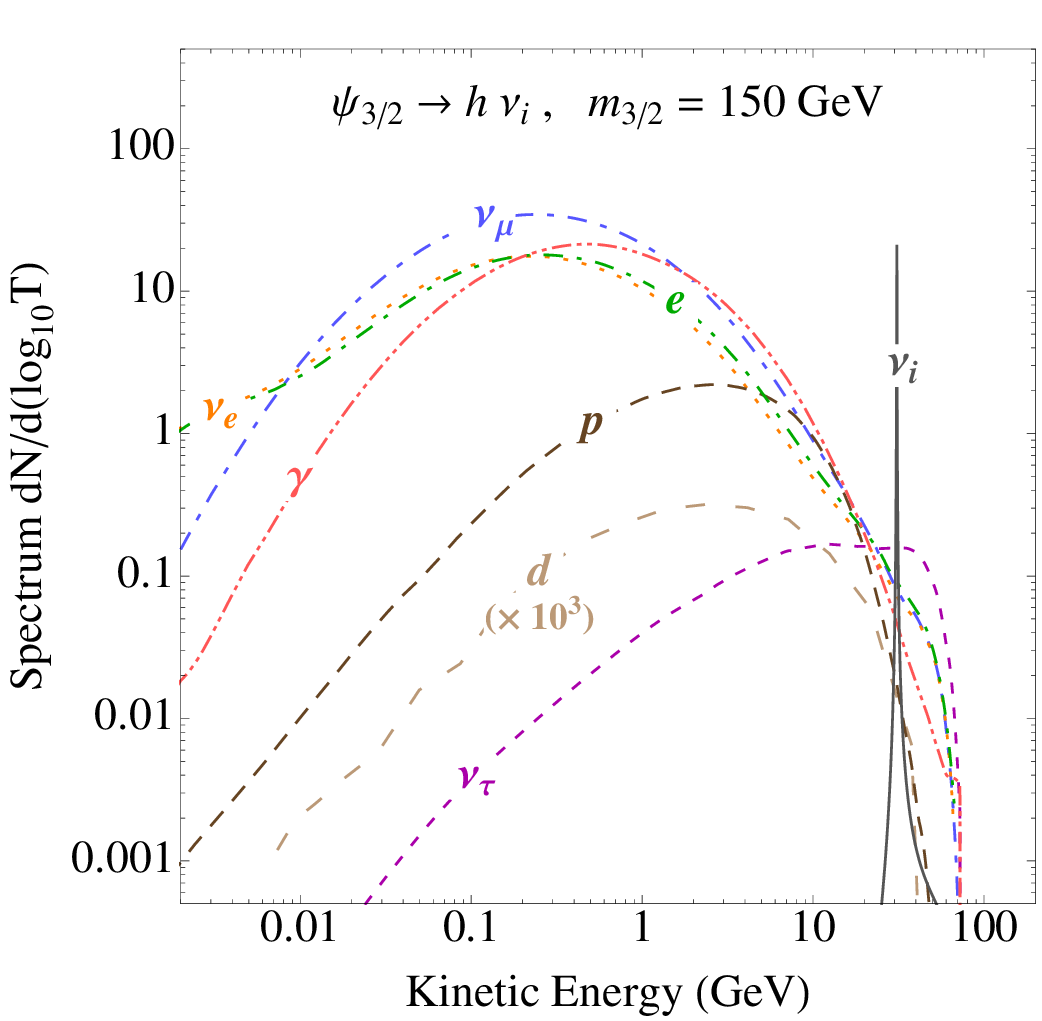}
 \hfill
 \includegraphics[scale=0.44]{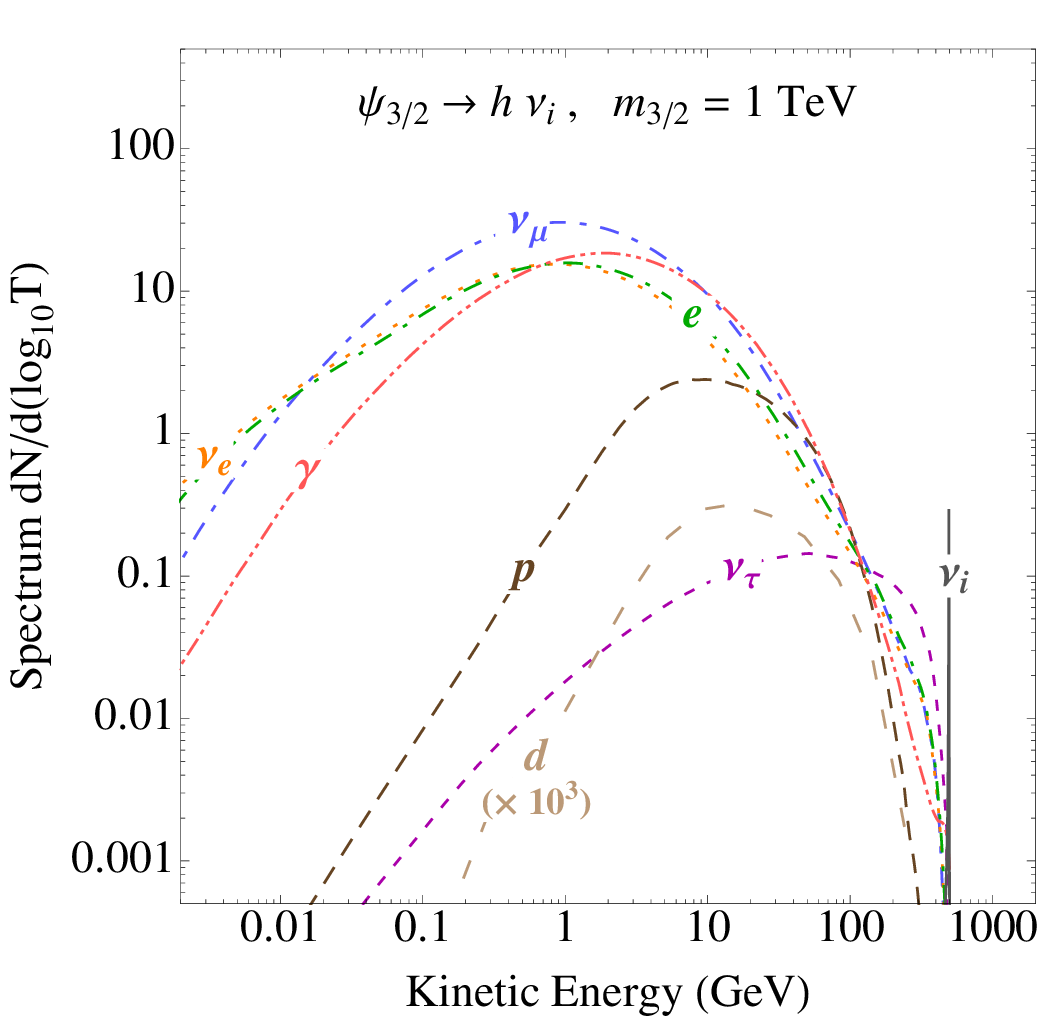}
 \caption[Spectra of stable final state particles from gravitino decay in the channel $\psi_{3/2}\rightarrow h\nu_i$.]{Spectra of stable final state particles from the decay of a gravitino with a mass of 150\,GeV (\textit{left}) or 1\,TeV (\textit{right}) in the channel $\psi_{3/2}\rightarrow h\,\nu_i$ assuming a mass of $m_h=115\,$GeV for the standard model-like lightest Higgs boson. The features of these spectra are discussed in the text.}
 \label{hnu}
\end{figure}

In Figures~\ref{Znu}--\ref{hnu} we present the energy spectra of photons, electrons, protons, deuterons and the different neutrino flavors for the gravitino two-body decay channels $\psi_{3/2}\rightarrow Z\,\nu_i$, $We$, $W\mu$, $W\tau$ and $h\,\nu_i$ obtained from the PYTHIA simulation. Note that the deuteron spectra are multiplied by a factor of $10^3$ to increase their visibility. It is also important to note that these spectra are independent of the choice of mixing parameters as the phase-space distribution of the final state particles is completely fixed by the on-shell condition. Let us discuss in the following the features of these spectra.

Figure~\ref{Znu} shows the spectra from the decay channel $\psi_{3/2}\rightarrow Z\,\nu_i$ for two exemplary values of the gravitino mass: $m_{3/2}=100\,$GeV and $m_{3/2}=1\,$TeV. For the lower mass, slightly above the production threshold, there are particularly prominent features close to the high-energy end of the spectrum at $E=m_{3/2}/2$. In this range the spectra of electrons and neutrinos are dominated by the direct decay of the $Z$ boson into these particles. For comparison we show the spectra of electrons and neutrinos calculated according to the three-body decay formula in equation~(\ref{ffspec}) as light dotted curves. In addition we show the prominent spectrum of the hard neutrino produced directly in the gravitino decay in association with the $Z$ boson calculated according to equation~(\ref{nulspec}). The soft spectra of photons, electrons and neutrinos are mainly produced from pion decays after the hadronization of quark final states. At the lowest energies there is a contribution to the electron and electron neutrino spectra from neutron decay that leads to a departure from the factor of one half compared to muon neutrinos. Tau neutrinos are exclusively produced in rather hard processes, so the soft part of their spectrum is suppressed compared to the other particles. Protons are mainly produced in the decays of heavy baryons and neutrons. The heavy baryons are typically produced early during hadronization thus producing a harder spectrum of protons compared to those of particles coming from pion decay.

A gravitino mass of 1\,TeV represents the typical spectra when threshold effects are negligible. The hard neutrino forms a narrow line at the end of the spectrum, while the spectra of electrons and neutrinos directly produced in the three-body decay become softer due to the larger available phase-space. The soft part of the spectrum coming from fragmentation processes is roughly scale invariant.

For the decay channel $\psi_{3/2}\rightarrow We$ the situation is similar as the soft part of the spectra is determined by practically the same hadronization processes (see Figure~\ref{We}). The hard part shows prominent features of the hard electron and the electrons and neutrinos directly produced in the three-body decay. For the shape of the electron line we observe a remarkable agreement of the calculation according to equation~(\ref{nulspec}) and the PYTHIA simulation. One comment is in order in this case: The spectra simulated in the described way (and also the analytical calculations) do not take into account the effect of final state radiation. This effect would lead to a softening of the high-energy part of the electron spectrum -- in particular of the line features -- and at the same time harden the high-energetic photon spectrum. We plan to include this effect in a future treatment of these decay processes.

For the decay channels $\psi_{3/2}\rightarrow W\mu$ and $W\tau$ we get a slightly different result coming, respectively, from the decay of the hard muon and the decay of the hard tau lepton (see Figures~\ref{Wmu} and~\ref{Wtau}). The muon decay produces a hard contribution to the spectra of electrons as well as electron and muon neutrinos close to the position of the muon line from the direct decay. For a gravitino mass of 1\,TeV we clearly observe the enhancement of the lepton spectra compared to the contribution directly from the three-body decay as shown by the curve from the analytical calculation according to equation~(\ref{ffspec}). In the case of tau decay a prominent hard contribution of tau neutrinos is generated close to the position of the tau line from the direct decay. Also in this case the enhancement of the tau neutrino spectrum at the high-energy end is clearly visible for $m_{3/2}=1\,$TeV. The effect on the photon, electron, and electron and muon neutrino spectra is also slightly visible.

In Figure~\ref{hnu} we show the final state particle spectra for the decay channel $\psi_{3/2}\rightarrow h\,\nu_i$ for gravitino masses of 150\,GeV and 1\,TeV. In this case the situation is a bit different. Due to the coupling of the Higgs boson to fermion masses, mainly pairs of bottom quarks are produced directly in the three-body decay for $m_h=115\,$GeV. Therefore, the spectra are generally softer at the high-energy end of the spectrum. There is, however,
also a contribution from pairs of tau leptons and virtual $W$ bosons that leads to a relatively hard spectrum of tau neutrinos.

\paragraph{Three-Body Decay Spectra}

For gravitino masses below the threshold for gauge and Higgs boson production, \textit{i.e.} when the boson propagators are off-shell, the situation is significantly more complicated. Although we can calculate the energy spectra of all particles directly produced in the gravitino decay, one needs to employ event generators to simulate the hadronization of colored final state particles.

In contrast to the two-body decays this situation cannot be easily mimicked in an event generator like PYTHIA since the kinematic distributions are not solely determined by phase space but depend explicitly on the matrix element of the process.\footnote{In this respect, it is unclear to us how the gravitino three-body decay spectra were derived in~\cite{Choi:2010jt} using PYTHIA.} In addition, PYTHIA only treats $2\rightarrow1$ and $2\rightarrow2$ processes and is not capable of handling the phase space of generic processes with three or more final state particles. Only a limited number of specific three-body decay processes like muon decay is explicitly included in PYTHIA~\cite{Sjostrand:2006za}.

The usual treatment of processes with multiparticle final states is to calculate the matrix element and the phase-space distribution of the process with a matrix element generator like MadGraph/MadEvent~\cite{Alwall:2007st,Alwall:2011uj} or WHIZARD~\cite{Kilian:2007gr}. These tools produce a table of events where the four-momenta of all  particles from the direct decay are listed according to the kinematic distribution given by the matrix element and the phase space. This output is subsequently handed over to PYTHIA where the hadronization processes and particle decays are treated.

For this method, of course, the particle physics model needs to be implemented in the matrix element generator. This is a bit problematic for the case of the gravitino in models with $R$-parity violation. Although MadGraph and WHIZARD are able to handle spin-3/2 particles,\footnote{In the case of MadGraph a support for gravitinos and goldstinos has only recently been added~\cite{Hagiwara:2010pi,Mawatari:2011jy}. For the current status of gravitino implementation in MadGraph see also~\cite{Mawatari:2011}.} the violation of $R$ parity leads to changes in the behavior of the model that cannot be implemented in an easy way by slightly modifying the existing model files. It is, however, planned to extend the capabilities of Madgraph in this direction in the future~\cite{Mawatari:2010}.

Another approach could be to implement the model using a Feynman rules generator like FeynRules~\cite{Christensen:2008py} that allows to generate model files for matrix element generators like MadGraph and WHIZARD directly from the Lagrangian of the theory. In this case, however, there is no publically available implementation of spin-3/2 particles yet, although the authors are working on this topic~\cite{Duhr:2011}. Thus one can expect that it will be possible to treat the phenomenology of $R$-parity violating gravitino dark matter decays with the help of automated calculations in the near future.

For the time being, however, we see no possibility to reliably calculate the full set of final state particle spectra for the gravitino three-body decays. Therefore, in the following we will only discuss some features of the spectra directly calculated from the gravitino decay and restrict mainly to the case of two-body decays in the phenomenological studies in this work.

\begin{figure}
 \centering
 \includegraphics[scale=0.44]{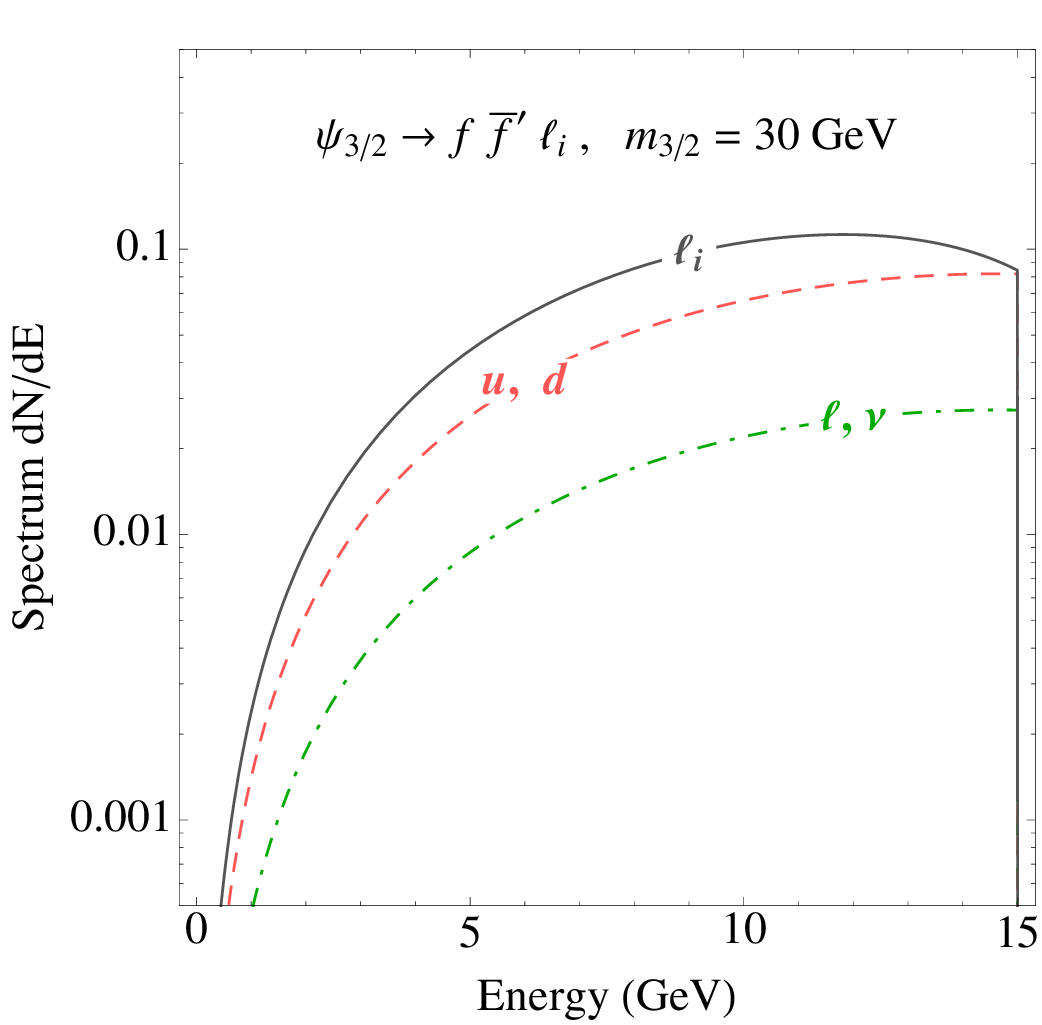}
 \hfill
 \includegraphics[scale=0.44]{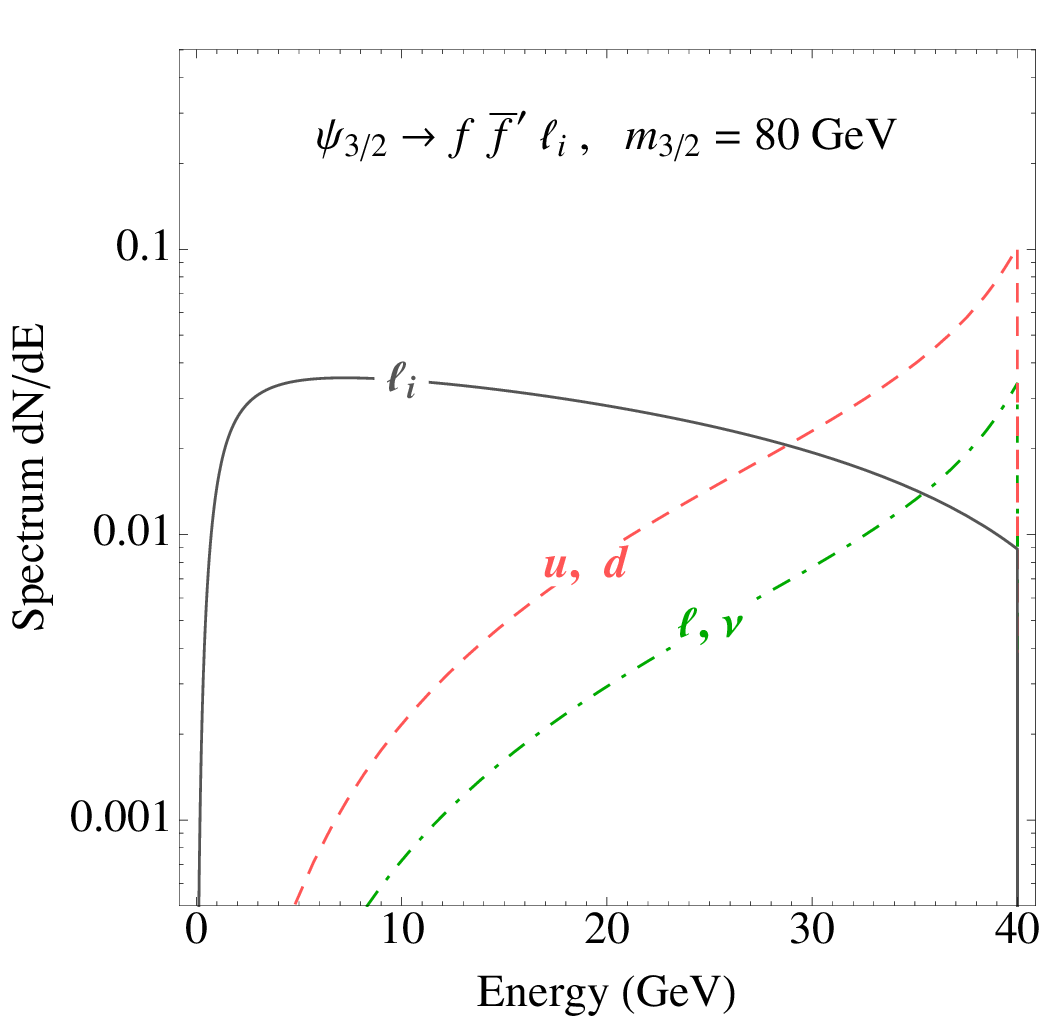}
 \caption[Fermion spectra from gravitino decay in the channel $\psi_{3/2}\rightarrow W^*\ell_i\rightarrow f\,\bar{f}'\,\ell_i$.]{Fermion spectra from the decay of a gravitino with a mass of 30\,GeV (\textit{left}) or 80\,GeV (\textit{right}) in the channel $\psi_{3/2}\rightarrow W^*\ell_i\rightarrow f\,\bar{f}'\,\ell_i$. The features of these spectra are discussed in the text.}
 \label{Wffl}
\end{figure}
Let us start with the channel $\psi_{3/2}\rightarrow W^*\ell_i\rightarrow f\,\bar{f}'\,\ell_i$. The spectra calculated from equations~(\ref{nulspec}) and~(\ref{ffspec}) using the standard set of parameters defined in Section~\ref{branchingratios} are presented in Figure~\ref{Wffl} for two exemplary values of the gravitino mass: $m_{3/2}=80\,$GeV, slightly below the threshold for on-shell $W$ production, and $m_{3/2}=30\,$GeV. As the $W$ boson couples with equal strength to all fermion pairs, their spectra are identical except for the color factor of three for quark final states. However, since the decay into heavy top quarks is forbidden by kinematics, also the production of the associated bottom quarks is strongly suppressed as it can only proceed via mixings of quarks as described by the Cabbibo-Kobayashi-Maskawa (CKM) matrix (see \textit{e.g.}~\cite{Nakamura:2010zzi}). For $m_{3/2}=80\,$GeV the quark and lepton spectra peak at the high-energy end of the spectrum since the propagator favors their invariant mass to be close to the $W$ mass. Further away from the threshold mass this effect is still visible but less prominent. The spectrum for the hard lepton is a bit different. While it resembles the other fermion spectra at low gravitino masses, it peaks at low energies for gravitino masses close to the $W$ production threshold. This is due to the fact that an invariant mass of the fermion pair close to the $W$ mass corresponds to low energies for the hard lepton. As the gravitino mass is increased above the threshold, a strong peak at low energies is developed. This peak is then shifted towards the high-energy end of the spectrum for larger gravitino masses (\textit{cf.} equation~(\ref{lineenergy}) and Figure~\ref{We}). Note that in all cases the spectrum is suppressed as the energy goes to zero since the phase-space factor $\beta_s$ vanishes when the invariant mass of the fermion pair, $s$, equals the gravitino mass.

\begin{figure}
 \centering
 \includegraphics[scale=0.44]{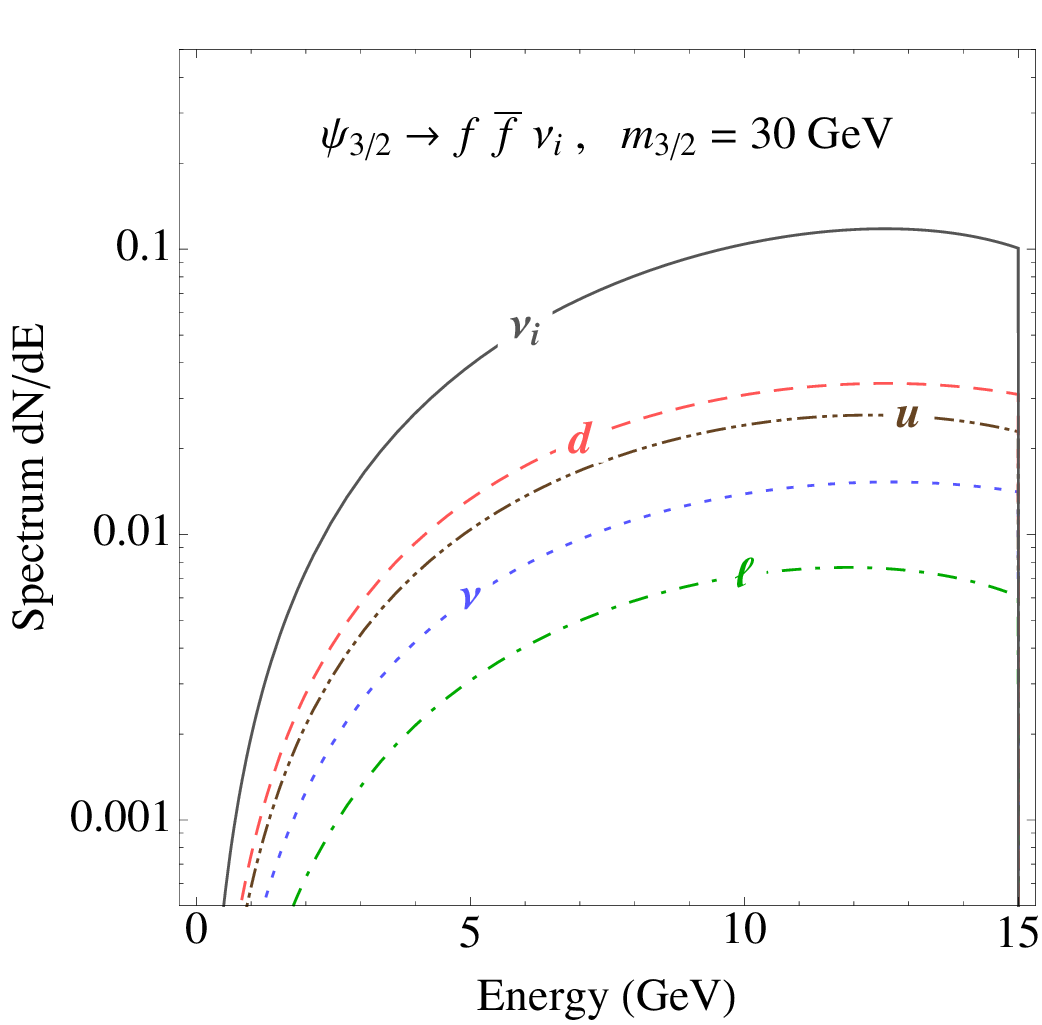}
 \hfill
 \includegraphics[scale=0.44]{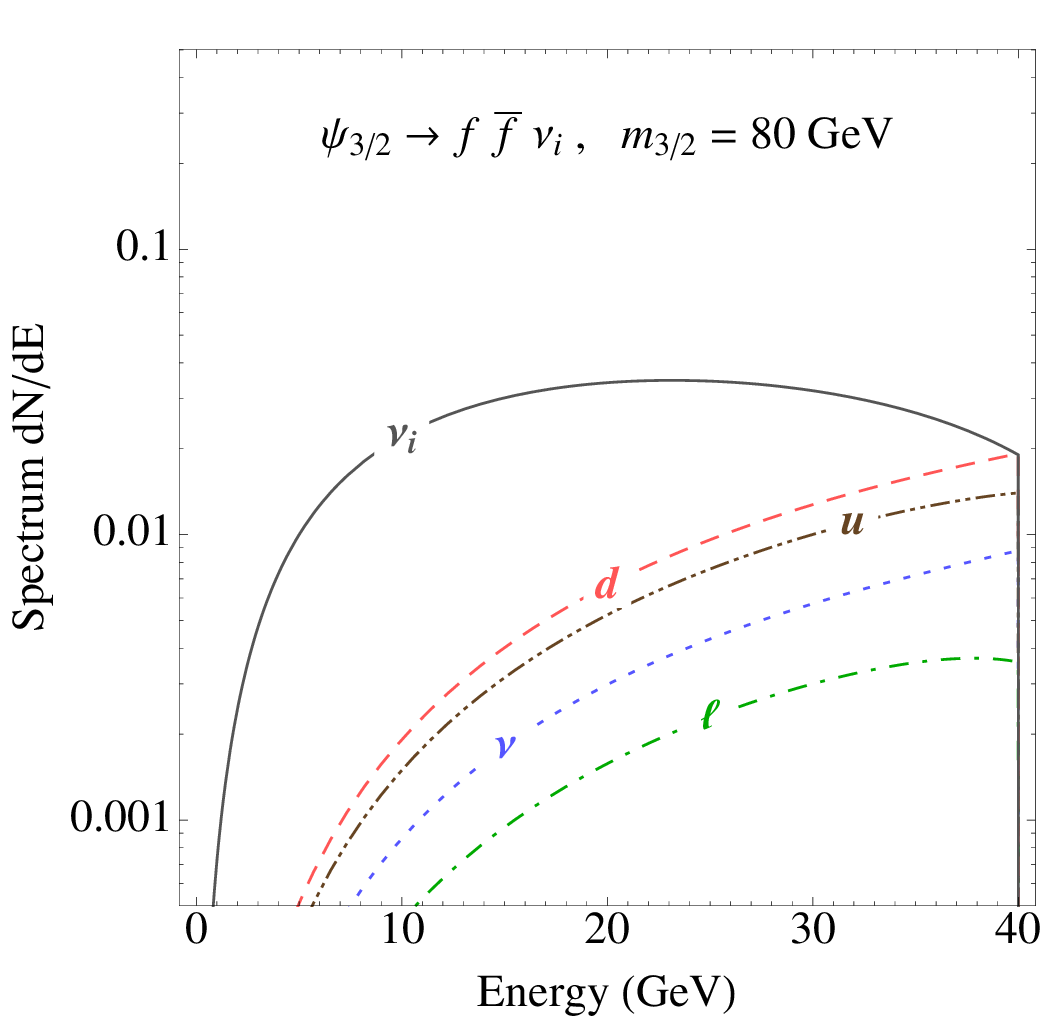}
 \caption[Fermion spectra from gravitino decay in the channel $\psi_{3/2}\rightarrow Z^*\nu_i\rightarrow f\,\bar{f}\,\nu_i$.]{Fermion spectra from the decay of a gravitino with a mass of 30\,GeV (\textit{left}) or 80\,GeV (\textit{right}) in the channel $\psi_{3/2}\rightarrow Z^*\nu_i\rightarrow f\,\bar{f}\,\nu_i$. The features of these spectra are discussed in the text.}
 \label{Zffnu}
\end{figure}
Before we discuss the spectra in the channel $\psi_{3/2}\rightarrow \gamma^*/Z^*\nu_i\rightarrow f\,\bar{f}\,\nu_i$ let us first discuss the case, when the virtual photon contribution is negligible. The corresponding spectra are presented in Figure~\ref{Zffnu}. In principle the features of the spectra are the same as for the previously discussed case, the only difference being that the coupling of the $Z$ boson to fermions is not universal. In fact, it depends on the electromagnetic charge and the third component of the weak isospin of the particles via the coefficients of the $V-A$ structure of the $Z$ coupling (see equation~(\ref{CVCA})). This leads to slightly different spectra for up- and down-type quarks as well as neutrinos and charged leptons. Another difference to the $W$ channel is that the pair production of bottom quarks is not suppressed.

\begin{figure}
 \centering
 \includegraphics[scale=0.445]{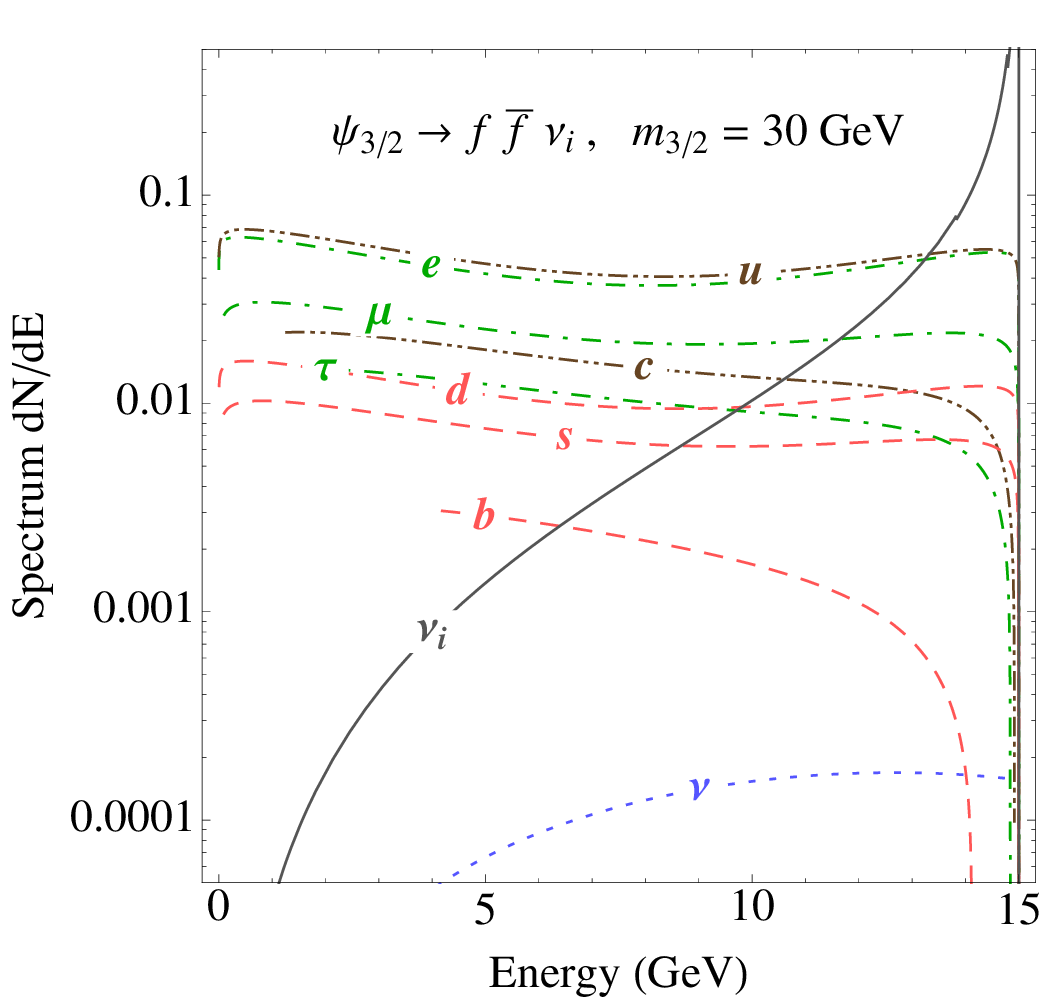}
 \hfill
 \includegraphics[scale=0.435]{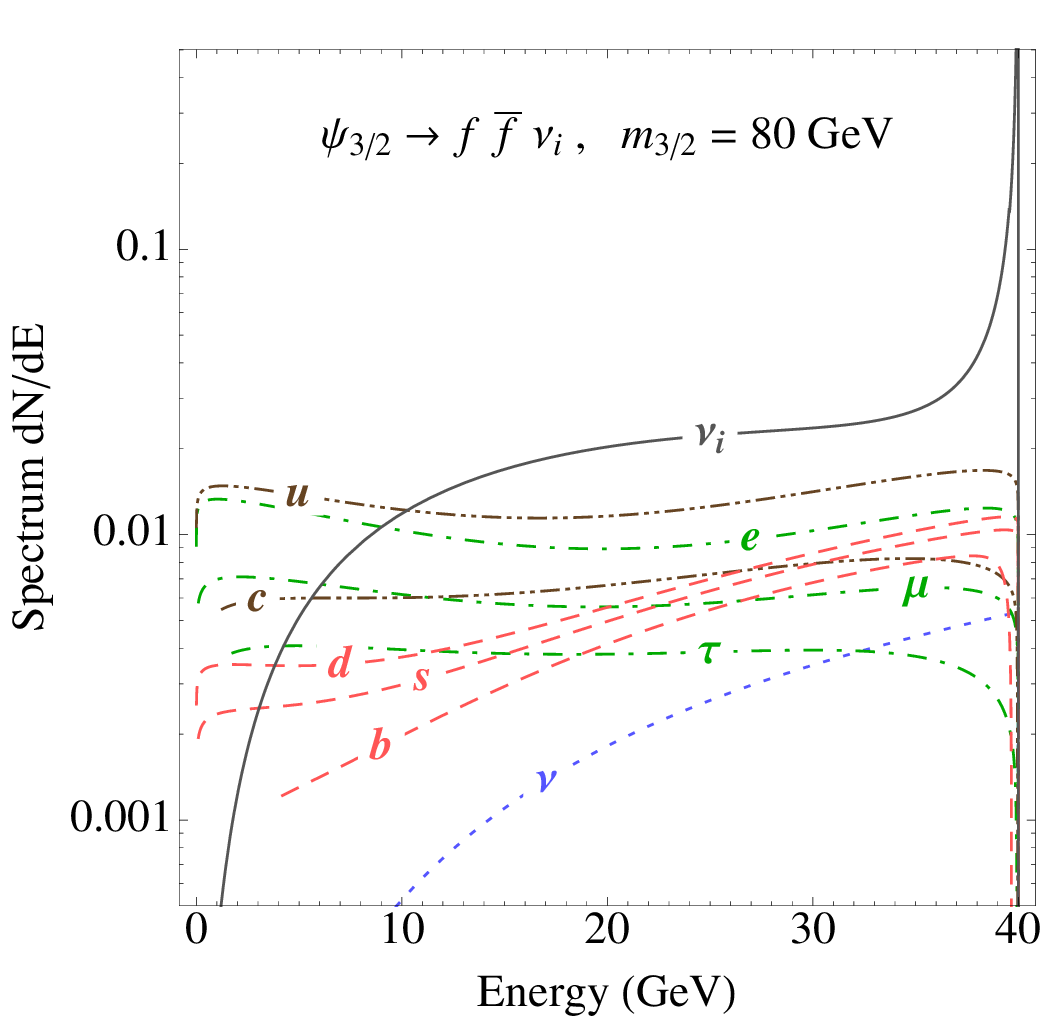}
 \caption[Fermion spectra from gravitino decay in the channel $\psi_{3/2}\rightarrow \gamma^*/Z^*\nu_i\rightarrow f\,\bar{f}\,\nu_i$.]{Fermion spectra from the decay of a gravitino with a mass of 30\,GeV (\textit{left}) or 80\,GeV (\textit{right}) in the channel $\psi_{3/2}\rightarrow \gamma^*/Z^*\nu_i\rightarrow f\,\bar{f}\,\nu_i$. The features of these spectra are discussed in the text.}
 \label{gammaZffnu}
\end{figure}
In the case where the virtual exchange of a photon contributes, the spectra will be significantly more complicated. As the decay width depends on the fermion masses due to the singular behavior of the photon propagator, we expect different spectra for the different fermion generations. The spectra calculated taking into account the dependence on the fermion masses are presented in Figure~\ref{gammaZffnu}. These spectra are a superposition of the photon and $Z$ exchange contributions and their interference. However, at a gravitino mass of 30\,GeV the photon channel strongly dominates for our standard choice of parameters and even at $m_{3/2}=80\,$GeV the $Z$ channel is subdominant. The most prominent feature is the strongly peaked spectrum of the hard neutrino at one half of the gravitino mass, coming from the preference of a close to massless photon in the propagator. In particular for larger gravitino masses, there is a notable contribution from the $Z$ channel to the low-energy tail of the neutrino spectrum.

The reason for the rather flat spectra of the other fermions is less obvious, though the contribution at low energies certainly comes from the fact that a vanishing invariant mass of the fermion system is favored. Since the minimal value of the invariant mass strongly depends on the fermion masses, the spectra of heavier generations are suppressed for all fermion species. Since neutrinos do not couple to the photon, their only contribution -- apart from the hard neutrino spectrum directly produced in the gravitino decay -- comes from the subdominant $Z$ boson exchange and thus their spectrum is strongly suppressed compared to the other particles. For a gravitino mass closer to the $Z$ threshold, one can already see the contribution of the $Z$ channel, in particular for down-type quarks (compare the spectra in Figures~\ref{Zffnu} and~\ref{gammaZffnu}).

The phenomenology of these decay processes for low-energetic gravitinos definitely deserves further treatment in future work, but -- as discussed above -- at the moment we see no option but to concentrate on the phenomenology of the two-body decays of heavier gravitinos.

\paragraph{Gravitino Decay Spectra}

In real situations the final state spectra cannot be observed for the individual decay channels. Thus, in order to study the cosmic-ray signals from gravitino decay in the following chapter, we will need the combined spectrum of a final state particle $X$ from all gravitino decay channels with particle $X$ in the final state. Using the branching ratios of the individual channels the combined spectrum is given by
\begin{equation}
  \frac{dN_X}{dE}=\sum_i\frac{\Gamma_{i}}{\Gamma_{\text{tot}}}\times\frac{1}{\Gamma_{i}}\,\frac{d\Gamma_{i}}{dE_X}=\sum_i\BR_i\times\frac{dN_{i,X}}{dE}\,.
\end{equation}
As the flavor structure of the $R$-parity breaking coupling $\xi_i$ is not known and since we do not employ a model that gives a prediction for it, we will assume throughout the rest of this thesis that the coupling is equivalent for all flavors. This means, in particular, that equal amounts of hard electron, muon and tau neutrinos, or electrons, muons and tau leptons are produced directly in the gravitino decay.

\chapter{Indirect Detection of Gravitino Dark Matter}
\label{indirectdetection}

The method of indirect dark matter detection is based on the observation of cosmic rays. Typically, cosmic rays are either directly produced in astrophysical sources or in spallation processes during the propagation of cosmic rays through the interstellar medium. However, the annihilation or decay of dark matter particles in the galactic halo or extragalactic structures produces an additional contribution to the spectra of cosmic rays that might be observable on top of the astrophysical background. In this introductory part of the chapter we want to review those aspects of indirect dark matter searches that are relevant for the case of unstable gravitino dark matter. We also want to elaborate on the differences between the case of a decaying dark matter candidate, like the gravitino in models with broken $R$-parity, and annihilating dark matter candidates, like the well-studied case of weakly-interacting massive particles (WIMPs). For more general discussions on the topic of indirect dark matter detection we refer the reader to extensive reviews in the literature~\cite{Bergstrom:2000pn,Bertone:2004pz,Feng:2010gw,Salati:2010rc,Porter:2011nv}.

\section{Indirect Searches for Dark Matter}

As we have seen in Section~\ref{DMevidence}, there is compelling evidence that dark matter makes up a significant part of the energy density of the universe, actually at a level of five times the contribution of baryonic matter. That means that a particle dark matter candidate explaining this observation will have an average number density of
\begin{equation}
 n_{\text{DM}}=\frac{\Omega_{\text{DM}}\,\rho_c}{m_{\text{DM}}}\simeq1.1\times10^{-3}\left( \frac{1\,\text{TeV}}{m_{\text{DM}}}\right) \text{m}^{-3}
 \label{DMdensity}
\end{equation}
on cosmological scales. When these dark matter particles decay, they produce intermediate standard model particles that eventually hadronize and/or decay into a set of stable particles: electrons, protons, deuterons, neutrinos and their respective antiparticles as well as gamma rays (\textit{cf.} Section~\ref{gravspectra}). The amount of particles produced per unit volume, time and energy is described by a source term of the form
\begin{equation}
  Q_X(E)\equiv\frac{dN_X}{dVdt\,dE}=\frac{n_{\text{DM}}}{\tau_{\text{DM}}}\,\frac{dN_X}{dE}\,,
\end{equation}
where $dN_X/dE$ is the energy spectrum of final state particles $X$ in the decay of the dark matter particle. These particles will then propagate through the intergalactic medium and might finally be observed in the fluxes of cosmic rays at the Earth. Neutrinos and gamma rays practically do not suffer deflections while traversing the intergalactic medium and thus propagate on straight paths. By contrast, charged cosmic rays in the GeV to TeV energy range considered in this work are strongly affected by magnetic fields and are thus not expected to reach Earth from extragalactic distances (see Section~\ref{antimatter} for a discussion of the propagation of charged cosmic rays).

We can then determine the flux of gamma rays and neutrinos at the position of the Earth coming from dark matter decays at extragalactic distances. With the term 'flux' we refer here to the number of particles per unit area, time and solid angle:\footnote{Later in this work, we will also use a more conventional notion of flux, namely the number of particles per unit area and time. This quantity will then be denoted as $\phi$ in contrast to $\Phi=d\phi/d\Omega$.}
\begin{equation}
 \Phi\equiv\frac{dN}{dA\,dt\,d\Omega}\,.
\end{equation}
Taking into account that the particles are propagating through an expanding universe we obtain for the differential flux of gamma rays and neutrinos the expression~\cite{Grefe:2008zz}
\begin{align}
  \frac{d\Phi_{\gamma/\nu}^{\text{eg}}}{dE} &=\frac{\Omega_{\text{DM}}\,\rho_c}{4\,\pi\,\tau_{\text{DM}}\,m_{\text{DM}}\,H_0\,\Omega_m^{1/2}}\!\int_1^{y_{\text{max}}}\!dy\,\frac{dN_{\gamma/\nu}}{d(y\,E)}\,\frac{y^{-3/2}}{\sqrt{1+\Omega_\Lambda/\Omega_m\,y^{-3}}} \label{EGflux}\\
  &\simeq2.3\times 10^{-4}\,(\text{m}^2\,\text{s}\,\text{sr})^{-1}\left( \frac{10^{26}\,\text{s}}{\tau_{\text{DM}}}\right) \left( \frac{1\,\text{TeV}}{m_{\text{DM}}}\right) \!\int_1^{y_{\text{max}}}\!dy\,\frac{dN_{\gamma/\nu}}{d(y\,E)}\,\frac{y^{-3/2}}{\sqrt{1+\Omega_\Lambda/\Omega_m\,y^{-3}}}\,, \nonumber
\end{align}
where $y\equiv1+z$ and $y_{\text{max}}=1+z_{\text{dec}}$ corresponds to the time when photons/neutrinos decoupled from the thermal plasma and started to propagate freely.\footnote{For photons this statement is not exactly true since they can produce electron pairs in collisions with photons from the CMB or the intergalactic background light. These processes lead to a nonvanishing optical depth for high-energy photons~\cite{Stecker:2005qs}. In addition, the expression for the redshift integral is only valid in the late universe when matter and dark energy dominate the energy density of the universe. However, since the contributions from higher redshifts are practically negligible, we will actually perform the integration up to $y_{\text{max}}=\infty$.} For monochromatic spectra of photons or neutrinos at injection, the redshift integral can be solved analytically and results in~\cite{Buchmuller:2007ui}
\begin{equation}
 \frac{d\Phi_{\gamma/\nu}^{\text{eg}}}{dE}=\frac{\Omega_{\text{DM}}\,\rho_c}{2\,\pi\,\tau_{\text{DM}}\,m_{\text{DM}}^2\,H_0\,\Omega_m^{1/2}}\!\left[ 1+\frac{\Omega_\Lambda}{\Omega_m}\left( \frac{2\,E}{m_{\text{DM}}}\right) ^3\right] ^{-1/2}\!\!\!\left( \frac{2\,E}{m_{\text{DM}}}\right) ^{1/2}\!\!\theta\left( 1-\frac{2\,E}{m_{\text{DM}}}\right) ,
\end{equation}
where $\theta(x)$ denotes the Heaviside step function.

In addition to the extragalactic signal, there exists a strong contribution from a more local source, namely dark matter decays in the Milky Way halo. In contrast to the isotropic extragalactic signal, the halo signal is anisotropic since the source term depends on the dark matter distribution in the galactic halo which is anisotropic as viewed from the Earth's position in the Milky Way:
\begin{equation}
 Q_X(E,\,\vec{l})\equiv\frac{dN_X(\vec{l})}{dVdE\,dt}=\frac{n_{\text{DM}}(\vec{l})}{\tau_{\text{DM}}}\,\frac{dN_X}{dE}=\frac{\rho_{\text{halo}}(\vec{l})}{\tau_{\text{DM}}\,m_{\text{DM}}}\,\frac{dN_X}{dE}\,,
\end{equation}
where $\vec{l}=\left( s,\,b,\,l \right) $ denotes galactic coordinates in terms of the distance $s$ to the Sun, the galactic latitude $b$ and the galactic longitude $l$. The density profile of the dark halo is not precisely known besides the fact that in order to fit galactic rotation curves a behavior $\rho\propto r^{-2}$ at larger radii is needed (\textit{cf.} Section~\ref{DMevidence}). The shape in the inner parts of the galaxy, however, is not completely determined by observations.

\begin{figure}[t]
 \centering
  \includegraphics[scale=0.443]{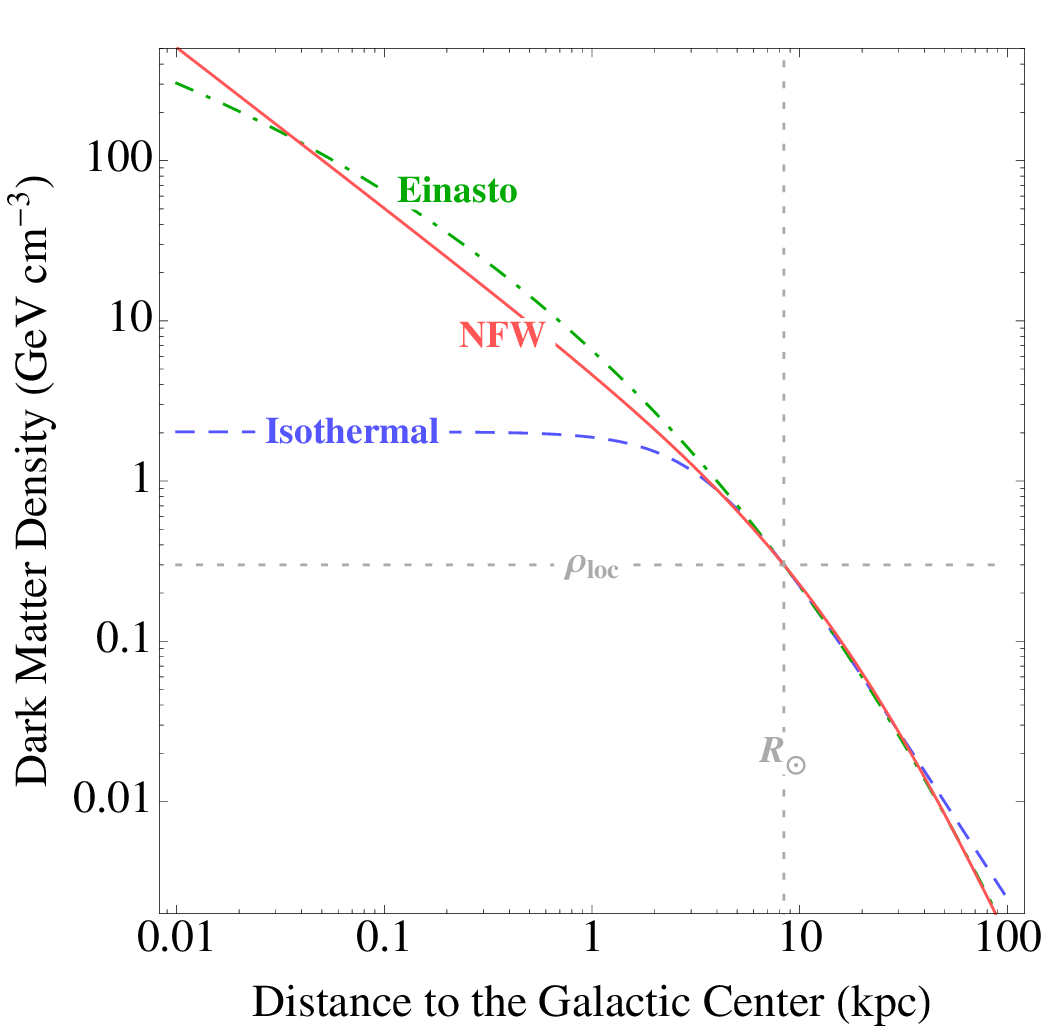}
  \includegraphics[scale=0.438]{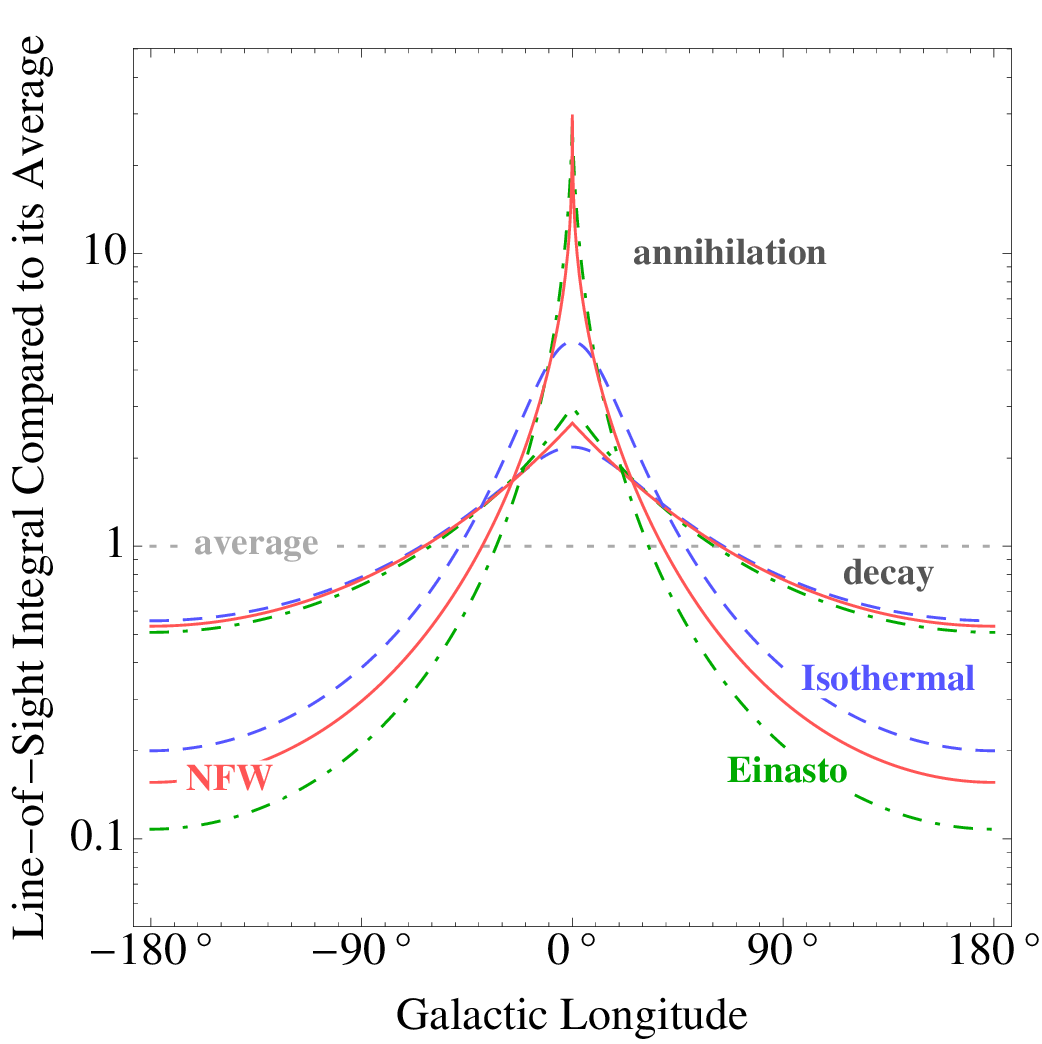}
 \caption[Density profiles for different dark matter halo models and angular dependence of the line-of-sight integral for dark matter annihilations and decays.]{\textit{Left:} Shapes of the dark matter halo density profiles listed in the text. All halo models are normalized to the local dark matter density and reproduce the $\rho\propto r^{-2}$ behavior at larger radii that is needed to explain the observed galactic rotation curve. The Navarro--Frenk--White profile is singular at the galactic center while the Einasto profile finally approaches a constant value. The isothermal profile, by contrast, exhibits an extended flat core. \textit{Right:} Comparison of the angular dependence of the line-of-sight integrals for the cases of dark matter annihilations and decays. The curves are shown for different halo profiles and are normalized to the average value for each individual case.}
 \label{profiles}
\end{figure}

Fits to observational data, though, tend to favor a cored isothermal profile, \textit{i.e.} the density approaches a constant value in the core of the galaxy~\cite{Bertone:2004pz}:
\begin{equation}
 \rho_{\text{iso}}(r)=\rho_{\text{loc}}\,\frac{1+\left( R_{\odot}/r_s\right) ^2}{1+\left( r/r_s\right) ^2}\,,
\end{equation}
where the normalization is given by the local dark matter density\footnote{The determination of the local dark matter density is subject to quite some uncertainty. We will use in this work the standard value given in the text though recent determinations seem to favor a value closer to $0.4\,\text{GeV}\,\text{cm}^{-3}$~\cite{Catena:2009mf}.} $\rho_{\text{loc}}\approx0.3\,\text{GeV}\,\text{cm}^{-3}$, $R_{\odot}=8.4\,\text{kpc}$ is the radius of the solar orbit around the galactic center and we adopt $r_s=3.5\,$kpc for the case of the Milky Way halo. By contrast, results from numerical N-body simulations seem to favor cuspy profiles, \textit{i.e.} halo profiles with a singular behavior at the galactic center. The best-known example is the Navarro--Frenk--White (NFW) profile~\cite{Navarro:1995iw}:
\begin{equation}
 \rho_{\text{NFW}}(r)=\rho_{\text{loc}}\,\frac{\left( R_{\odot}/r_s\right) \left( 1+R_{\odot}/r_s\right) ^2}{\left( r/r_s\right) \left( 1+r/r_s\right) ^2}\,,
\end{equation}
where $r_s=20\,$kpc for the case of the Milky Way. More recent simulations, on the other hand, favor the Einasto profile with a finite central density~\cite{Navarro:2003ew,Einasto:1965}:
\begin{equation}
 \rho_{\text{Ein}}(r)=\rho_{\text{loc}}\exp\left( -\frac{2}{\alpha}\left( \left( \frac{r}{r_s}\right) ^\alpha-\left( \frac{R_{\odot}}{r_s}\right) ^\alpha\right) \right) ,
\end{equation}
where we adopt $\alpha=0.17$ and $r_s=20\,$kpc for the case of the Milky Way. A comparison of the shapes of these halo profiles is presented in the left panel of Figure~\ref{profiles}.
The radius $r$ in the density profiles is given with respect to the galactic center, so we need to express it in terms of galactic coordinates in order to calculate the cosmic-ray fluxes from dark matter decays in the halo:
\begin{equation}
 r(s,\,b,\,l)=\sqrt{s^2+R_{\odot}^2-2\,s\,R_{\odot}\cos{b}\cos{l}}\,.
\end{equation}
Then, the differential gamma-ray and neutrino fluxes from dark matter decays in the galactic halo observed in the direction specified by $b$ and $l$ are given by an integration along the line of sight:
\begin{equation}
 \frac{d\Phi_{\gamma/\nu}^{\text{halo}}}{dE}(b,\,l)=\frac{1}{4\,\pi\,\tau_{\text{DM}}\,m_{\text{DM}}}\,\frac{dN_{\gamma/\nu}}{dE}\int_0^{\infty}ds\,\rho_{\text{halo}}(r(s,\,b,\,l))\,. 
\end{equation}
Due to the linear dependence of the flux on the halo density the anisotropy is not very strong in the case of decaying dark matter. Thus we will mainly work with an averaged signal in the following. For the same reason the dependence on the choice of the halo model is very weak. For definiteness, we will adopt the NFW density profile as our standard choice in this work. In this case we can solve the integration over the halo analytically and obtain the following result for the averaged full-sky flux:
\begin{align}
 \left\langle \frac{d\Phi_{\gamma/\nu}^{\text{halo}}}{dE}\right\rangle  &=\frac{1}{4\,\pi}\int_{-\pi}^\pi dl\int_{-\pi/2}^{\pi/2}\cos b\,db\,\frac{d\Phi_{\gamma/\nu}^{\text{halo}}}{dE}(b,\,l) \nonumber\\
 &=\frac{\rho_{\text{loc}}\,(R_{\odot}+r_s)}{4\,\pi\,\tau_{\text{DM}}\,m_{\text{DM}}}\,\frac{dN_{\gamma/\nu}}{dE}\left(  \operatorname{artanh}\left(\frac{1}{1+\frac{r_s}{R_{\odot}}}\right) -\frac{\ln(\frac{R_{\odot}}{r_s})}{\frac{r_s}{R_{\odot}}-1}-\frac{1}{2}\ln\left(1+2\,\frac{R_{\odot}}{r_s}\right) \right) \nonumber\\
 &\simeq1.3\times 10^{-4}\,(\text{m}^2\,\text{s}\,\text{sr})^{-1}\left( \frac{10^{26}\,\text{s}}{\tau_{\text{DM}}}\right) \left( \frac{1\,\text{TeV}}{m_{\text{DM}}}\right) \frac{dN_{\gamma/\nu}}{dE}\,.
 \label{haloflux}
\end{align}
In the following we want to outline the differences in the indirect detection strategies for the cases of dark matter annihilations and decays.

\paragraph{Annihilating and Decaying Dark Matter}

We have seen before that the flux expected from dark matter decays is proportional to the number density of dark matter particles and inversely proportional to their lifetime. In the case of dark matter annihilations the situation is different: Since a collision of two dark matter particles is required, the resulting cosmic-ray flux is proportional to the square of the dark matter number density. In addition, the annihilation rate depends on the relative velocity of the colliding particles and on their annihilation cross section.

Let us first discuss the dependence on the dark matter density: Since the average present-day dark matter density in the universe is rather low in absolute numbers (\textit{cf.} equation~(\ref{DMdensity})) we do not expect a sizable annihilation signal from diffuse extragalactic sources at low redshifts as in the case of dark matter decays. In fact, the expected signal is very sensitive to the actual distribution of dark matter in the universe since overdense regions significantly enhance the expected total flux. Let us demonstrate this on the expected differential flux of gamma rays and neutrinos from dark matter annihilations in the Milky Way halo. It is given by the following integral along the line of sight:
\begin{equation}
 \frac{d\Phi_{\gamma/\nu}^{\text{halo}}}{dE}(b,\,l)=\frac{\left\langle \sigma v\right\rangle _{\text{DM}}}{8\,\pi\, m_{\text{DM}}^2}\,\frac{dN_{\gamma/\nu}}{dE}\int_0^{\infty}ds\,\rho_{\text{halo}}^2(r(s,\,b,\,l))\,,
\end{equation}
where $\left\langle \sigma v\right\rangle _{\text{DM}}$ is the thermally averaged dark matter annihilation cross section and $dN_{\gamma/\nu}/dE$ is the energy spectrum of photons/neutrinos in the annihilation of two dark matter particles. Due to the dependence on the square of the halo density, the signal from annihilating dark matter is strongly enhanced in the direction towards the galactic center, especially for cuspy halo profiles. In the right panel of Figure~\ref{profiles} we compare the angular dependencies of the signals expected from dark matter annihilations and decays for different halo profiles. Clearly visible is the enhancement of the signal in the galactic center direction where the dark matter density has its maximum. In the case of dark matter annihilations, however, the signal varies over several orders of magnitude depending on the line of sight while in the case of decays the signal varies by less than a factor of ten. Thus the expected signal from dark matter annihilations is much more anisotropic than that from dark matter decays and the observation of the angular dependence of gamma-ray or neutrino signals can be used as a strategy to discriminate annihilating and decaying dark matter particles. In addition, we observe that the expected annihilation signal suffers much larger uncertainties from the choice of the halo profile than the signal expected from decays.

However, not only the galactic center is an important source for signals from dark matter annihilation. Also other dense structures could emit an observable signal. Among these are substructures of the galactic halo that are expected from numerical simulations, the dark matter halos of dwarf galaxies in the neighborhood of the Milky Way and the dark matter concentration in galaxy clusters. But also celestial bodies like the Sun or the Earth could emit an enhanced annihilation signal if dark matter particles are captured inside their centers by energy losses due to weak elastic scattering processes. In this case, however, only neutrinos could leave the dense objects and be observed.

Let us now come to the dependence on the dark matter annihilation cross section: If dark matter exists in the form of weakly interacting massive particles, the observed dark matter density is a relic from the freeze-out of WIMP annihilations in the early universe~\cite{Jungman:1995df}. When the temperature in the early universe dropped below the WIMP mass, these particles left thermal equilibrium and efficiently annihilated with each other. However, due to the expansion of the universe their physical number density was diluted and the annihilation process became inefficient before all WIMPs were destroyed, leaving a relic density of WIMPs that could play the role of dark matter. The annihilations of WIMPs in the present universe are thus determined by the same annihilation cross section that is responsible for the production of the correct relic density in the early universe. This circumstance makes the WIMP scenario very predictive, as the thermally averaged WIMP annihilation cross section is required to be $\left\langle \sigma v\right\rangle _{\text{WIMP}}\approx3\times10^{-26}\,\text{cm}^3\,\text{s}^{-1}$ to match the observed dark matter density~\cite{Jungman:1995df}.

However, as this annihilation cross section is too small to predict observable anomalous contributions in cosmic-ray signals, typically a boost factor is introduced to increase the annihilation signal. A small part of this boost factor could be accommodated by astrophysical effects like halo substructures~\cite{Lavalle:1900wn}, but in general an additional enhancement like the proposed Sommerfeld enhancement of non-relativistic annihilation cross sections is needed (see for instance~\cite{Cirelli:2007xd}).

For decaying dark matter, on the other hand, the value of the lifetime is in general not directly related to the cross section determining the production in the early universe. Therefore, scenarios with decaying dark matter particles are less constrained and in principle less predictive, but they do not rely on the assumption of additional enhancement mechanisms to explain cosmic-ray signatures.

In the specific case of gravitino dark matter no accumulation inside astrophysical objects is expected since the gravitational interaction is too weak. Also, the existence of halo substructures and large scale structures does not significantly change the signal predictions since the average flux is not affected by inhomogeneities in the dark matter distribution. Thus one can work with a gravitino distribution according to the halo density profile for the signal from gravitino decays in the Milky Way halo and with the average cosmological dark matter density for the signal from gravitino decays at extragalactic distances. In addition, there is in general no chance to observe the contribution from gravitino annihilations, since these are suppressed by higher orders of the Planck scale.

In the following sections of this thesis we want to discuss the prospects for the indirect detection of unstable gravitino dark matter in the spectra of gamma rays, charged cosmic rays and neutrinos.

\section[Probing Gravitino Dark Matter with Gamma Rays]{Probing Gravitino DM with Gamma Rays}
\label{gammas}

In this section we discuss the gamma-ray signals coming from the decay of gravitino dark matter. Gamma rays are a very important channel to search for dark matter signals since they contain spectral and directional information that can be well measured. Since the gamma-ray signal expected from gravitino decays is a diffuse flux from all directions with only mild angular dependence, it contributes to the isotropic diffuse gamma-ray background that is often referred to as the extragalactic gamma-ray background (EGB). This is a diffuse flux of high-energetic photons that is thought to have its origin in unresolved sources like active galactic nuclei, blazars, starburst galaxies, gamma-ray bursts or in truly diffuse emission (see for instance~\cite{Dermer:2007fg}). However, it is not clear that this flux is completely of extragalactic nature since exotic sources of diffuse emission in the galactic halo could also contribute.

Another strategy to search for a dark matter signal is to search for gamma-ray lines. Gamma rays of astrophysical origin are typically expected to have a spectrum that follows a power law. Monochromatic signals in the spectrum are a clear sign for a particle physics origin. Therefore, one can use the good energy resolution of gamma-ray observatories to constrain the partial lifetime of dark matter two-body decay channels with at least one photon in the final state.

The signals of gamma rays from the decay of gravitino dark matter in scenarios with bilinear $R$-parity violation have already been studied in several works~\cite{Takayama:2000uz,Buchmuller:2007ui,Bertone:2007aw,Ibarra:2007wg,Ishiwata:2008cu,Buchmuller:2009xv}. In this thesis, however, we want to employ recent data from the Large Area Telescope on the Fermi Gamma-Ray Space Telescope (Fermi LAT) to constrain the parameters of gravitino dark matter.

\subsubsection*{Diffuse Gamma-Ray Flux}

The diffuse extragalactic gamma-ray background is overwhelmed by a diffuse component expected from galactic gamma-ray emission. Thus it has to be extracted in a complicated way from gamma-ray observations by subtracting the galactic contribution as modeled by galactic cosmic-ray propagation models. This subtraction of a dominant component can, however, lead to large systematical uncertainties on the derived flux.

An isotropic diffuse gamma-ray component on top of the galactic emission was first observed by the SAS-2 satellite~\cite{Fichtel:1978} and has since then been measured in more detail by the Energetic Gamma Ray Experiment Telescope (EGRET) on the Compton Gamma Ray Observatory~\cite{Sreekumar:1997un} and by Fermi LAT~\cite{Abdo:2010nz}. To reduce the foreground contamination, the galactic disc (corresponding to galactic latitudes $-10^{\circ}\leq b\leq10^{\circ}$) is usually excluded from the analysis. The extragalactic gamma-ray background amounts to roughly one quarter of the total diffuse flux measured away from the galactic disc.

Here we will compare the expected contribution from gravitino decays to the extragalactic gamma-ray background as recently derived from Fermi LAT data~\cite{Abdo:2010nz}. No anomalous behavior of the spectrum is observed and the data points are well described by a power law:
\begin{equation}
  E^2\,\frac{d\Phi_\gamma^{\text{EGB}}}{dE}\approx5.5\times10^{-3}\left( \frac{E}{\text{GeV}}\right) ^{-0.41}\text{GeV}\,(\text{m}^2\,\text{s}\,\text{sr})^{-1}.
\end{equation}
For comparison we also show previous results obtained from data of the EGRET experiment~\cite{Strong:2004ry,Sreekumar:1997un}. In the original analysis of the EGRET data a power-law behavior of the extragalactic gamma-ray background in the energy range from 50\,MeV--10\,GeV was found~\cite{Sreekumar:1997un}:
\begin{equation}
  E^2\,\frac{d\Phi_\gamma^{\text{EGB}}}{dE}=1.37\times10^{-2}\left( \frac{E}{\text{GeV}}\right) ^{-0.1}\text{GeV}\,(\text{m}^2\,\text{s}\,\text{sr})^{-1}.
\end{equation}
An independent analysis using an optimized cosmic-ray propagation model to match the observations of the galactic diffuse emission~\cite{Strong:2004de} led to a softer power law in the energy range from 50\,MeV--2\,GeV~\cite{Strong:2004ry}:
\begin{equation}
  E^2\,\frac{d\Phi_\gamma^{\text{EGB}}}{dE}=6.8\times10^{-3}\left( \frac{E}{\text{GeV}}\right) ^{-0.32}\text{GeV}\,(\text{m}^2\,\text{s}\,\text{sr})^{-1}.
\end{equation}
However, in this analysis an excess of gamma rays above the power law was observed at energies above 2\,GeV stimulating the interpretation of this anomaly as a signal from dark matter annihilations or decays (see for instance~\cite{Ibarra:2007wg,deBoer:2005tm}). In particular, it was found that the gamma-ray signal from gravitino dark matter could explain the excess in the EGRET data for a gravitino mass of roughly 150\,GeV and a lifetime on the order of $10^{26}\,$s~\cite{Ibarra:2007wg}. However, as mentioned before, although the new Fermi LAT data are consistent with the reanalysis of the EGRET data by Strong \textit{et al.} at low energies they show no evidence for the existence of an excess at higher energies and therefore we will use the data only as an upper limit for the contribution from gravitino decays.

\begin{figure}[t]
 \centering
 \includegraphics[scale=0.9,bb=0 11 500 300]{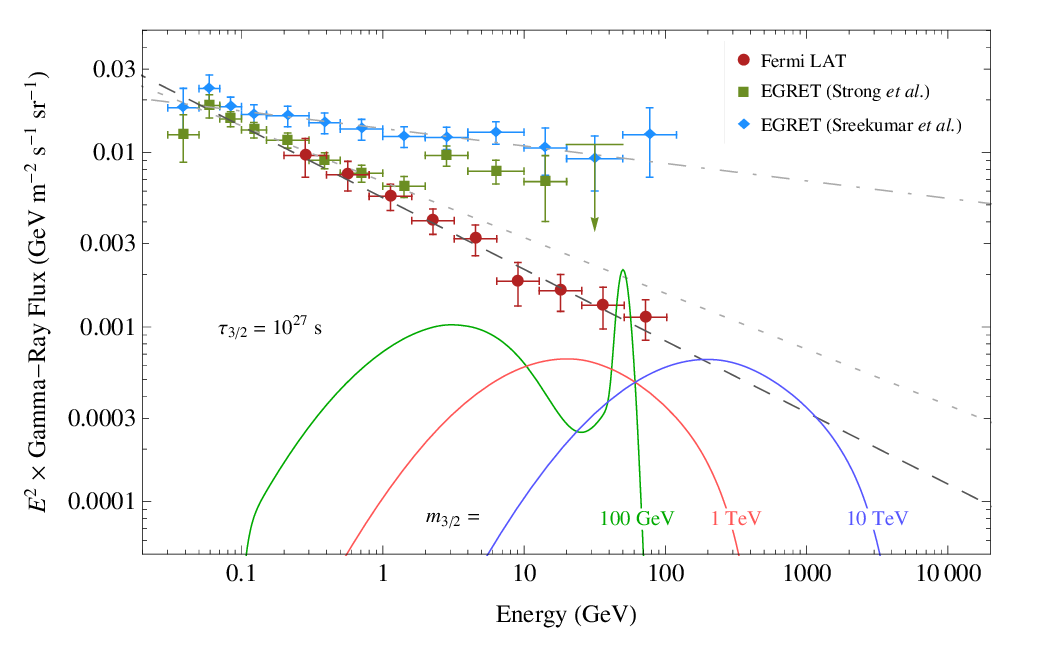} 
 \caption[Diffuse gamma-ray flux from gravitino dark matter decays compared to the measured extragalactic gamma-ray background.]{Diffuse gamma-ray flux from gravitino dark matter decays compared to the extragalactic gamma-ray background as derived from Fermi LAT and EGRET data. The gravitino signal is shown for a lifetime of $10^{27}\,$s and gravitino masses of 100\,GeV, 1\,TeV and 10\,TeV. The dashed, dotted and dash-dotted lines show the power-law fits to the Fermi LAT and EGRET data points.}
 \label{gammaflux}
\end{figure}
The main dark matter contribution to the diffuse gamma-ray flux at high energies comes from photons directly produced in the decays of gravitinos. In addition, there is a contribution from the upscattering of photons due to inverse Compton scattering of electrons and positrons from gravitino decays on the interstellar radiation field (see \textit{e.g.}~\cite{Ishiwata:2009dk,Ibarra:2009dr,Zhang:2009ut}).\footnote{Actually, the contribution from inverse Compton scatterings dominate the low-energy tail of the gamma-ray spectrum for large gravitino masses~\cite{Ibarra:2009dr}.} In this work, however, we will only consider the prompt flux for an order of magnitude estimate of the gamma-ray signal. The averaged gamma-ray flux from gravitino decays in the galactic halo excluding the galactic disk is then calculated as
\begin{equation}
 \begin{split}
   \left\langle \frac{d\Phi_\gamma^{\text{halo}}}{dE}\right\rangle _{\!\abs{b}\,>\,10^\circ} &=\frac{1}{2\,\pi\left( 1-\sin\pi/18\right)}\int_{-\pi}^\pi dl\int_{\pi/18}^{\pi/2}\cos{b}\,db\,\frac{d\Phi_{\gamma}^{\text{halo}}}{dE}(b,\,l) \\
   &\simeq1.2\times 10^{-8}\,(\text{cm}^2\,\text{s}\,\text{sr})^{-1}\left( \frac{10^{26}\,\text{s}}{\tau_{3/2}}\right) \left( \frac{1\,\text{TeV}}{m_{3/2}}\right) \frac{dN_{\gamma}}{dE}\,.
 \end{split}
\end{equation}
In addition, we consider the isotropic extragalactic contribution according to equation~(\ref{EGflux}) so that the total diffuse gamma-ray signal expected from gravitino decays is calculated as
\begin{equation}
 \frac{d\Phi_\gamma}{dE}=\left\langle \frac{d\Phi_\gamma^{\text{halo}}}{dE}\right\rangle _{\!\abs{b}\,>\,10^\circ}+\frac{d\Phi_{\gamma}^{\text{eg}}}{dE}\,.
\end{equation}
In Figure~\ref{gammaflux} we compare the contribution of gamma rays from the decay of gravitino dark matter to the observed extragalactic gamma-ray background as derived from Fermi LAT~\cite{Abdo:2010nz} and EGRET data~\cite{Sreekumar:1997un,Strong:2004ry}. For the gravitino signal we choose a lifetime of $10^{27}\,$s and three exemplary masses: 100\,GeV, 1\,TeV and 10\,TeV. Additionally, we adopt an energy resolution of $\sigma(E)/E=10\,\%$ to estimate the sensitivity of Fermi LAT to spectral features~\cite{Rando:2009yq}. Technically we apply the energy resolution by a convolution of the gamma-ray spectrum with a Gaussian distribution:
\begin{equation}
  \frac{d\Phi_\gamma^{\text{Gauss}}}{dE}=\frac{1}{\sqrt{2\,\pi}\,\sigma(E)}\int_0^{\infty}dE'\,\exp\left( -\frac{1}{2}\left( \frac{E-E'}{\sigma(E)}\right) ^2\right) \frac{d\Phi_\gamma}{dE'}\,.
\end{equation}
We clearly see the gamma-ray line at the high-energy end of the spectrum for low gravitino masses. For higher masses, where the gravitino two-body decay into a photon and a neutrino is suppressed, this contribution vanishes practically completely and the gamma-ray flux is dominated by the soft contributions from gauge and Higgs boson fragmentation. As the Fermi LAT data do not exhibit any spectral features, the gravitino decay contribution to the flux can only be subdominant. In the following we will present limits on the gravitino lifetime obtained from diffuse gamma-ray data and searches for gamma-ray lines.

\subsubsection*{Lifetime Bound from the Diffuse Flux and from Photon Lines}

We estimate a bound on the gravitino lifetime by requiring that the prompt gamma-ray flux from gravitino decays does not overshoot the error bars of the Fermi LAT measurement of the extragalactic gamma-ray background. This bound is rather conservative since we neglect the possible contribution from astrophysical backgrounds in the derivation of the lifetime limit.

\begin{figure}[t]
 \centering
 \includegraphics[scale=0.9,bb=0 11 500 300]{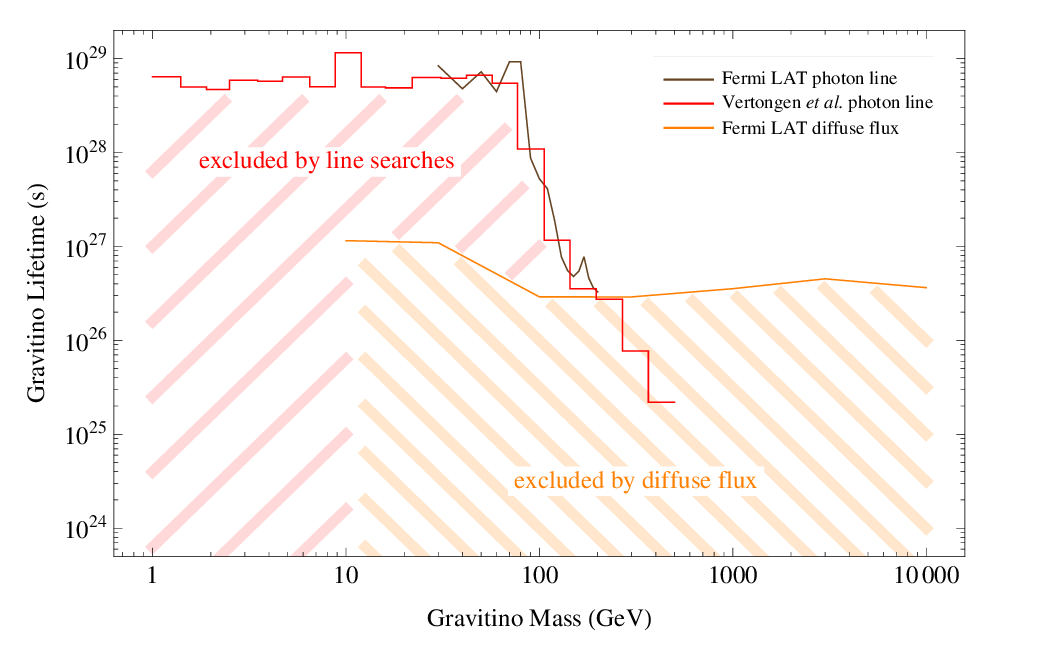} 
 \caption[Bounds on the gravitino lifetime from gamma-ray observations.]{Bounds on the gravitino lifetime from observations of the diffuse extragalactic gamma-ray background and from photon line searches.}
 \label{gammabound}
\end{figure}

In addition, a strong upper limit on the gravitino lifetime can be deduced from a search for photon lines by the Fermi LAT collaboration~\cite{Abdo:2010nc}. Their analysis covers the full sky excluding the galactic disc ($\abs{b}<10^\circ$) but taking also into account a patch of $20^\circ\times20^\circ$ around the galactic center, where a strongly enhanced signal is expected for the case of annihilating dark matter. We take their limits on the decay width into a two-photon final state in the energy range 30--200\,GeV and apply it to the case of the gravitino two-body decay into a photon and a neutrino. In order to calculate a lifetime limit for the gravitino we multiply with the appropriate branching ratio and rescale their lifetime limits by a factor of one half since we have a single-photon final state:
\begin{equation}
  \tau_{3/2}\geq\BR\left( \psi_{3/2}\rightarrow\gamma\,\nu_i\right) \times\frac{\tau_{\gamma\gamma}}{2}\,.
\end{equation}
Recently, in an independent analysis using Fermi LAT data, these limits have been extended to cover the energy range 1--500\,GeV~\cite{Vertongen:2011mu}.\footnote{This energy range has been corrected to 1--300\,GeV in the published version, but that does not affect our discussion.} The authors give separate limits deduced from a full-sky analysis excluding the galactic disc and from an analysis using a region around the galactic center. In addition, the specific case of a decay channel into a photon and a neutrino is discussed. Thus we derive upper limits on the gravitino lifetime in the following way:
\begin{equation}
  \tau_{3/2}\geq\BR\left( \psi_{3/2}\rightarrow\gamma\,\nu_i\right) \times\tau_{\gamma\nu}^{\text{halo}}\qquad\text{and}\qquad\tau_{3/2}\geq\BR\left( \psi_{3/2}\rightarrow\gamma\,\nu_i\right) \times\tau_{\gamma\nu}^{\text{center}}.
\end{equation}
We will only use the bounds from the full-sky analysis since they are stronger than the bounds from the central region of the Milky Way for decaying dark matter particles. The results of the conservative lifetime estimate from the diffuse flux and the limits derived from the photon-line searches are presented in Figure~\ref{gammabound}. For gravitino masses below the $W$ boson mass, the branching fraction for the gamma-ray line is close to 100\,\% in our standard gravitino scenario (\textit{cf.} Section~\ref{branchingratios}) and a very strong lower limit on the gravitino lifetime on the order of $\tau_{3/2}\gtrsim5\times10^{28}\,$s is obtained from line searches. At larger masses the branching ratio for the line drops quickly, reducing the significance of the lifetime limit. The comparison of the continuum signal with the diffuse extragalactic gamma-ray background leads to an estimate for the lower limit of the gravitino lifetime at a constant level around $\tau_{3/2}\gtrsim3\times10^{26}\,$s. In our standard scenario this limit becomes more important than the limit from line searches for gravitino masses above a few hundred GeV.

If the photon line is suppressed for some reason (see Section~\ref{branchingratios}), the low gravitino mass region is significantly less constrained since line searches do not provide a lifetime limit in that case. There will, however, still be a constraint on the continuum gamma-ray flux expected from three-body decays. Thus the lifetime limit derived from the diffuse flux is expected to remain at the same order of magnitude.

\section{Probing Gravitino Dark Matter with Cosmic-Ray Antimatter}
\label{antimatter}

In this section we want to discuss the signals expected from gravitino dark matter decays in the spectra of charged cosmic rays. Also in this case, related studies have been presented before in the literature~\cite{Ibarra:2008qg,Ishiwata:2008cu,Buchmuller:2009xv}.

While cosmic-ray particles are abundantly produced in astrophysical sources, antimatter particles are thought to be practically only produced in spallation processes involving high-energetic cosmic rays impinging on the interstellar medium. Therefore, one typically expects a charge-symmetric exotic contribution from dark matter to appear first as an anomalous component in the spectra of cosmic-ray antiparticles.

As mentioned earlier in this chapter, the propagation of charged cosmic rays through the Milky Way halo is much more complicated than in the case of gamma rays and neutrinos. In order to predict the contributions from gravitino dark matter decays to the spectra of charged cosmic rays one thus needs to have a detailed understanding of the effects of cosmic-ray propagation. In the following we will therefore give a short overview of the propagation of charged particles. For a more extensive discussion and further references we refer to~\cite{Salati:2010rc}.

\paragraph{Propagation of Charged Cosmic Rays}

While charged cosmic rays traverse the galactic halo, they are deflected by Alfv\'en waves, the irregularities of the galactic magnetic field. In the case of the Milky Way the magnetic turbulence is strong and thus the effect of multiple deflections can be described by space diffusion with a coefficient
\begin{equation}
 K(E)=K_0\,\beta\left( \mathcal{R}/\text{GV}\right) ^\delta,
\end{equation}
where $\mathcal{R}=pc/(Ze)$ and $\beta=v/c$ are, respectively, the rigidity and the velocity of the cosmic-ray particle, $\delta$ is a spectral index, and $K_0$ is a normalization coefficient.\footnote{It is usually also assumed that $K(E)$ is constant throughout the diffusive halo, so there is no explicit spatial dependence.} The Alfv\'en waves drift inside the Milky Way with the Alfv\'en velocity $V_a$ which is typically in the range of 20--100\,km\,s$^{-1}$. This drift of the magnetic irregularities leads to a small effect of diffusive reacceleration of cosmic rays due to second-order Fermi acceleration. This effect depends on the strength of space diffusion as well as on the velocity of the cosmic rays and their total energy:
\begin{equation}
 K_{EE}(E)=\frac{2}{9}\,V_a^2\,\frac{E^2\beta}{K(E)}\,.
\end{equation}
In addition, the galactic wind coming from the stars in the galactic disc with a velocity $V_C$ in the range of 5--15\,km\,s$^{-1}$ leads to a drift of cosmic rays away from the galactic disc.

The values for the normalization $K_0$ and the spectral index $\delta$ of the diffusion coefficient, the galactic drift velocity $V_C$, and the Alfv\'en velocity $V_a$ can be constrained by studying the spectra of primary cosmic rays produced in astrophysical sources and secondary cosmic rays produced in spallation processes during propagation. In particular the cosmic-ray boron-to-carbon ratio (B/C) is quite sensitive to the parameters of cosmic-ray transport and is thus usually employed to determine the free parameters of the propagation model~\cite{Maurin:2001sj}.

The transport equation for the cosmic-ray number density per unit kinetic energy $f_X(T)=dn_X/dT$ has the general form~\cite{Berezinskii:1990}
\begin{equation}
  \frac{\partial f_X}{\partial t}=\vec{\nabla}\cdot\left( K(T)\,\vec{\nabla}f_X-\vec{V}_C(\vec{r})\,f_X\right) -\frac{\partial}{\partial T}\left( b(T,\vec{r})\,f_X-K_{EE}(T)\,\frac{\partial f_X}{\partial T}\right) +Q_X(T,\vec{r})\,,
  \label{CRpropagation}
\end{equation}
where $b(T,\vec{r})$ is a coefficient describing energy losses and $Q_X(T,\vec{r})$ is a source term that describes particle production and depletion processes. This transport equation holds for all cosmic-ray species as long as the correct source terms and energy loss coefficients are taken into account. A useful assumption is that the number densities of cosmic-ray particles are in a steady state, \textit{i.e.} $\partial f_X/\partial t=0$.

The transport equation can be solved in a semi-analytical two-zone diffusion model for the galaxy~\cite{Maurin:2001sj}. In this approach the diffusive halo is modeled as a thick disc that matches the circular structure of the Milky Way. The disc of stars and gas with a height of $2\,h=200\,$pc and a radius of 20\,kpc lies in the middle. Above and beneath the disc are confinement layers with a respective height of $L=1$--15\,kpc where cosmic rays are trapped by diffusion.

The solution to the transport equation for cosmic rays originating from gravitino dark matter decay can then be expressed in the following form~\cite{Ibarra:2008qg}:
\begin{equation}
  f_X(T)=\frac{1}{m_{3/2}\,\tau_{3/2}}\int_0^{\infty}dT'\,G(T,T')\,\frac{dN_X}{dT'}\,,
  \label{propGreen}
\end{equation}
where $dN_X/dT$ is the source spectrum of cosmic rays from gravitino decay and $G(T,T')$ is a Green's function that accounts for the propagation through the diffusive halo and includes all astrophysical parameters like the halo density profile. The differential flux of cosmic rays coming from gravitino decays in the galactic halo can then be expressed as
\begin{equation}
  \frac{d\Phi_X^{\text{DM}}}{dT}=\frac{v_X}{4\,\pi}\,f_X(T)\,,
  \label{propFlux}
\end{equation}
where $v_X$ is the velocity of the cosmic-ray particle.

\paragraph{Solar Modulation}

An additional complication for the accurate prediction of charged-cosmic-ray spectra as observed by experiments is the effect of solar modulation. Cosmic-ray particles that enter the heliosphere of the solar system are affected by the solar wind of charged particles. The plasma of charged particles emitted from the sun mainly has the effect of decelerating charged cosmic rays that traverse the solar system on the way to Earth. This effect depends on the intensity of the solar wind and thus changes over the eleven-year cycle of solar activity. We adopt here the charge-independent solar modulation treatment in the force field approximation~\cite{Gleeson:1968zz}. In this model the expected interstellar (IS) flux of charged cosmic rays can be compared to the flux measured by cosmic-ray experiments at the top of the atmosphere (TOA) via the following relation~\cite{Maurin:2001sj,Perko:1987}:
\begin{equation}
  \frac{d\Phi_X^{\text{IS}}}{dT_{\text{IS}}}=\left( \frac{p^{\text{IS}}}{p^{\text{TOA}}}\right) ^2\frac{d\Phi_X^{\text{TOA}}}{dT_{\text{TOA}}}\,.
  \label{solarmod}
\end{equation}
In this expression, $T$ is the kinetic energy and $p$ the momentum of the cosmic-ray particles. The kinetic energy at the top of the atmosphere is shifted to lower energies depending on the value of the Fisk potential $\phi_F$ that characterizes the solar activity:
\begin{equation}
 T_{\text{TOA}}=T_{\text{IS}}-\abs{Z_X}e\,\phi_F\,,
\end{equation}
where $\abs{Z_X}e$ is the absolute value of the electric charge of the cosmic-ray particle. The value of the Fisk potential in a data taking period of an experiment is usually determined by fits on cosmic-ray data. A period of minimal solar activity corresponds to a value of $\phi_F\approx500\,$MV. Due to this rather low value the effect of solar modulation is mainly relevant for low-energetic cosmic rays. At energies above $\mathcal{O}(10)\,$GeV the effect is practically negligible.

\subsection{Positrons and Electrons}

In the case of positrons and electrons the propagation through the diffusive halo of the Milky Way is dominated by space diffusion and energy loss processes. Therefore, the general propagation equation~(\ref{CRpropagation}) can be simplified to the following form:\footnote{As the electron mass is negligible compared to the total energy in the considered energy range we will work with the total positron and electron energies instead of their kinetic energy.}
\begin{equation}
  \frac{\partial f_{e^\pm}}{\partial t}=\vec{\nabla}\cdot\left( K(E)\,\vec{\nabla}f_{e^\pm}\right) -\frac{\partial}{\partial E}\left( b(E,\vec{r})\,f_{e^\pm}\right) +Q_{e^\pm}(E,\vec{r})=0\,.
\end{equation}
In the energy range above a few GeV, which is relevant for dark matter searches, the energy loss of positrons is dominated by the emission of synchrotron radiation in the galactic magnetic fields and by inverse Compton scatterings on stellar light and CMB photons.\footnote{The synchrotron radiation of electrons and positrons from dark matter decays contributes as an exotic component to the photon spectrum at radio frequencies and thus can also be used to search for dark matter signals. It is strongest in the direction of the galactic center region, where the dark matter density and the strength of the galactic magnetic fields are maximal (see for instance~\cite{Ishiwata:2008qy,Zhang:2009pr}). Photons produced in inverse Compton scattering processes contribute to the diffuse gamma-ray flux (\textit{cf.} Section~\ref{gammas}).} Then, the energy loss rate $b(E,\vec{r})$ is typically assumed to be a spatially constant function that can be written as
\begin{equation}
 b(E)=-\frac{E^2}{E_0\,\tau_E}\,,
\end{equation}
where $E_0=1\,$GeV is a reference energy and $\tau_E=10^{16}\,$s is the typical timescale of energy loss processes. Since high-energetic positrons are ultra-relativistic their velocity is practically given by the speed of light and their rigidity is proportional to their total energy. Hence, the space diffusion coefficient $K(E)$ can be written in the form
\begin{equation}
 K(E)=K_0\left( E/\text{GeV}\right) ^\delta.
\end{equation}
Using equations~(\ref{propGreen}) and~(\ref{propFlux}) we find that the interstellar positron and electron fluxes in the vicinity of the solar system coming from gravitino dark matter decays in the galactic halo are given by
\begin{equation}
  \frac{d\Phi_{e^\pm}^{\text{DM}}}{dE}=\frac{c}{4\,\pi\,m_{3/2}\,\tau_{3/2}}\int_0^{\infty}dE'\,G_{e^\pm}(E,E')\,\frac{dN_{e^\pm}}{dE'}\,,
\end{equation}
where the Green's function for the propagation of the positrons can be approximated in the following way~\cite{Ibarra:2008qg}:
\begin{equation}
  G_{e^\pm}(E,E')\simeq\frac{10^{16}}{\epsilon^2}\exp\left( a+b\left( \epsilon^{\delta-1}-\epsilon^{\prime\delta-1}\right) \right) \theta(E'-E)\,\text{cm}^{-3}\,\text{s}\,.
\end{equation}
The energy of electrons and positrons is parametrized in the form $\epsilon=E/$GeV and $\epsilon'=E'/$GeV, and the parameters $a$ and $b$ are given in Table~\ref{positronparameters} for our choice of a Navarro--Frenk--White dark matter halo density profile and different sets of propagation parameters. In practice the upper limit for the integration over the Green's function is given by the maximal energy of positrons in a gravitino decay: $E_{\text{max}}=m_{3/2}/2$. Due to their sizable energy losses during propagation, electrons and positrons observed at Earth are expected to originate from rather local regions of the galactic halo.
\begin{table}[t]
 \centering
 \begin{tabular}{cccccccc}
  \toprule
  Model & $\delta$ & $K_0\,(\text{kpc}^2\!/\text{Myr})$ & $L\,(\text{kpc})$ & $a$ & $b$ \\
  \midrule
  M2 & 0.55 & 0.00595 & 1 & $-0.9716$ & $-10.012$ \\
  MED & 0.70 & 0.0112 & 4 & $-1.0203$ & $-1.4493$ \\
  M1 & 0.46 & 0.0765 & 15 & $-0.9809$ & $-1.1456$ \\
  \bottomrule
 \end{tabular}
 \caption[Cosmic-ray propagation parameters for positrons.]{Parameters of cosmic-ray propagation models that correspond, respectively, to the best fit of cosmic-ray B/C data (MED) as well as the minimal (M2) or maximal (M1) positron flux compatible with cosmic-ray B/C data. Figures taken from~\cite{Ibarra:2008qg,Delahaye:2007fr}.}
 \label{positronparameters}
\end{table}

Let us now discuss the signal spectra expected from gravitino decays and how they compare to expected astrophysical backgrounds and experimental observations.

\paragraph{Positron Fraction}

From the experimental point of view it is convenient to present measurements of cosmic-ray positrons and electrons in form of the positron fraction that is defined as the ratio of the cosmic-ray positron flux and the sum of cosmic-ray electron and positron fluxes:
\begin{equation}
  \frac{e^+}{e^++e^-}\equiv\frac{\Phi_{e^+}}{\Phi_{e^+}+\Phi_{e^-}}\,.
\end{equation}
It is generally assumed that systematic uncertainties connected with the measurement method (for instance the detection efficiency) are similar for positrons and electrons. Therefore, these effects to a large extent cancel if the results are presented in form of the positron fraction (see for instance~\cite{Adriani:2008zr}).

This treatment, however, introduces more uncertainties in the theoretical prediction, since in order to calculate this quantity we will need to take into account the astrophysical background fluxes of positrons and electrons in addition to the contribution from gravitino decays. A background of secondary positrons and electrons is produced by the spallation of the interstellar medium by impinging high-energetic cosmic rays. For the case of electrons one expects an additional primary component directly produced in astrophysical sources. Usually the fluxes of cosmic-ray secondary positrons as well as those of primary and secondary electrons are obtained from numerical simulations. In this work we will employ the parametrization of the background fluxes from~\cite{Baltz:1998xv} for the numerical result found in~\cite{Moskalenko:1997gh}:
\begin{align}
  \frac{d\Phi_{e^+}^{\text{sec}}}{dE} &\simeq\frac{4.5\,\epsilon^{0.7}}{1+650\,\epsilon^{2.3}+1500\,\epsilon^{4.2}}\,(\text{GeV}\,\text{cm}^2\,\text{s}\,\text{sr})^{-1}, \\
  \frac{d\Phi_{e^-}^{\text{prim}}}{dE} &\simeq\frac{0.16\,\epsilon^{-1.1}}{1+11\,\epsilon^{0.9}+3.2\,\epsilon^{2.15}}\,(\text{GeV}\,\text{cm}^2\,\text{s}\,\text{sr})^{-1}, \\
  \frac{d\Phi_{e^-}^{\text{sec}}}{dE} &\simeq\frac{0.70\,\epsilon^{0.7}}{1+110\,\epsilon^{1.5}+600\,\epsilon^{2.9}+580\,\epsilon^{4.2}}\,(\text{GeV}\,\text{cm}^2\,\text{s}\,\text{sr})^{-1},
\end{align}
where $\epsilon=E/$GeV. Using these expressions we can calculate the differential positron fraction of the astrophysical background as
\begin{equation}
  \frac{e^+}{e^++e^-}(E)\bigg|^{\text{bkg}}=\frac{d\Phi_{e^+}^{\text{sec}}/dE}{d\Phi_{e^+}^{\text{sec}}/dE+d\Phi_{e^-}^{\text{prim}}/dE+d\Phi_{e^-}^{\text{sec}}/dE}\,.
\end{equation}
In the case of a signal from gravitino decays the exotic contribution to the positron fraction is calculated as
\begin{equation}
  \frac{e^+}{e^++e^-}(E)\bigg|^{\text{DM}}=\frac{d\Phi_{e^+}^{\text{DM}}/dE}{2\,d\Phi_{e^\pm}^{\text{DM}}/dE+d\Phi_{e^+}^{\text{sec}}/dE+d\Phi_{e^-}^{\text{prim}}/dE+d\Phi_{e^-}^{\text{sec}}/dE}\,,
\end{equation}
where the normalization is modified due to the dark matter contribution to the total flux of cosmic-ray electrons and positrons.

\begin{figure}[t]
 \centering
 \includegraphics[scale=0.9,bb=0 11 500 300]{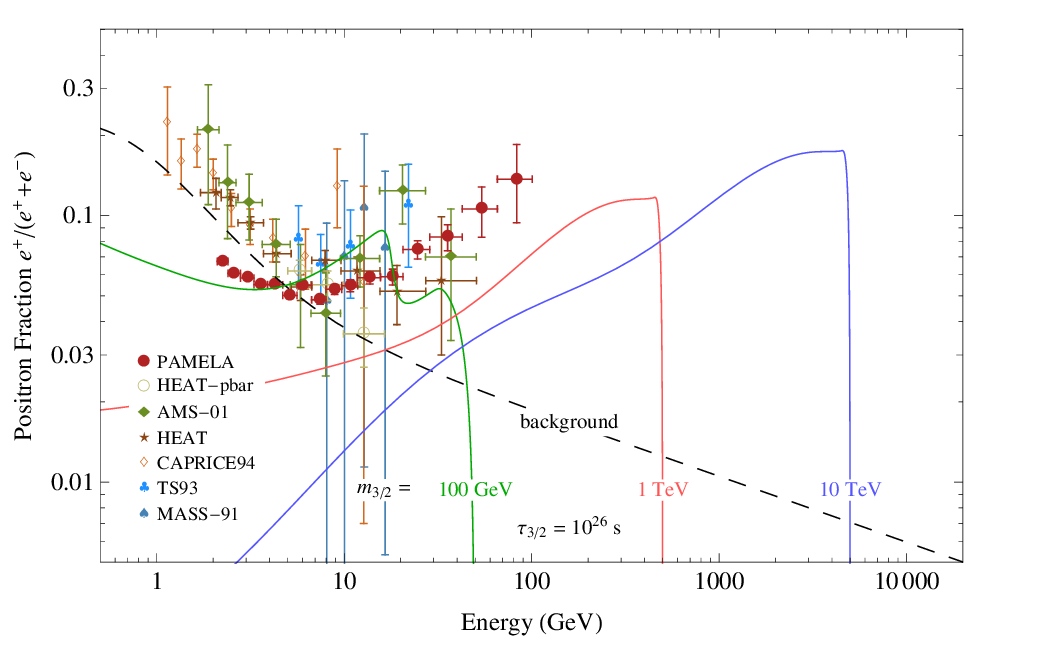} 
 \caption[Contribution to the cosmic-ray positron fraction from gravitino decays compared to measurements and the expectation from astrophysical primary and secondary production.]{Contribution to the cosmic-ray positron fraction from the decay of gravitino dark matter in the MED propagation model compared to measurements of the positron fraction by various experiments and the expectation from astrophysical primary and secondary production. The contribution from gravitino decays is shown for a lifetime of $10^{26}\,$s and gravitino masses of 100\,GeV, 1\,TeV and 10\,TeV. In order to account for effects of solar modulation, the data points were demodulated using values for the Fisk potential quoted by the experimental groups (\textit{cf.} Table~\ref{fiskpotentiale}).}
 \label{posfraction}
\end{figure}
In Figure~\ref{posfraction} we compare the contribution of gravitino dark matter decays to the cosmic-ray positron fraction with the expected astrophysical background and the data of several experiments: the balloon-borne Matter Antimatter Superconducting Spectrometer (MASS-91)~\cite{Grimani:2002yz}, the balloon-borne TS93 experiment~\cite{Golden:1995sq}, the balloon-borne Cosmic AntiParticle Ring Imaging Cherenkov Experiment (CAPRICE94)~\cite{Boezio:2000}, the balloon-borne High-Energy Antimatter Telescope (HEAT)~\cite{Barwick:1997ig}, the balloon-borne HEAT-pbar experiment~\cite{Beatty:2004cy}, the Alpha Magnetic Spectrometer on Space Shuttle flight STS-91 (AMS-01)~\cite{Aguilar:2007yf} and the Payload for Antimatter Matter Exploration and Light-nuclei Astrophysics (PAMELA) on the Resurs-DK1 satellite~\cite{Adriani:2008zr}.\footnote{The data tables were partly extracted from the cosmic-ray database of Strong and Moskalenko~\cite{Strong:2009xp}.}
\begin{table}[t]
 \centering
 \begin{tabular}{cc}
  \toprule
  Experiment & Fisk potential (MV) \\
  \midrule
  PAMELA & 600 \\
  Fermi LAT & 550 \\
  AMS-01 & 650 \\
  HEAT & 710 \\
  \bottomrule
 \end{tabular}
 \hspace{5mm}
 \begin{tabular}{cc}
  \toprule
  Experiment & Fisk potential (MV) \\
  \midrule
  BETS & 550 \\
  CAPRICE94 & 600 \\
  MASS-91 & 550 \\
   & \\
  \bottomrule
 \end{tabular}
 \caption[Values of the Fisk potential to account for the effect of solar modulation in the electron and positron data sets.]{Values of the Fisk potential to account for the effect of solar modulation in the electron and positron data sets. The figures are taken from the publications of the respective data sets (see text for the list of references).}
 \label{fiskpotentiale}
\end{table}

Since the gravitino signal and the astrophysical background calculated above correspond to the interstellar positron fraction, we need to correct the experimental data for the effect of solar modulation according to equation~(\ref{solarmod}). For flux ratios like the positron fraction, however, the rescaling with the squared ratio of the interstellar momentum and the the momentum at the top of the atmosphere cancels and only a shift in the energy scale remains:\footnote{In several works in the literature it is, however, erroneously stated that solar modulation effects completely cancel in the positron fraction.}
\begin{equation}
  \frac{e^+}{e^++e^-}(E^{\text{IS}})\bigg|^{\text{\text{IS}}}=\frac{e^+}{e^++e^-}(E^{\text{IS}}-e\,\phi_F)\Bigg|^{\text{TOA}}.
\end{equation}
The corresponding values of the Fisk potential for the data taking periods of the experiments measuring the positron fraction are listed in Table~\ref{fiskpotentiale}. We observe from Figure~\ref{posfraction} that the data for the positron fraction from different experiments are in approximate agreement with each other and with the expected astrophysical background, but at low energies the PAMELA data significantly deviate from the other experiments, in particular in view of the small error bars of the PAMELA measurement. This raises the question if the effect of solar modulation is really independent of the sign of charge~\cite{Adriani:2008zr}. Due to this source of uncertainty, data at energies below 10\,GeV are usually ignored in the search for exotic contributions.

On the other hand, at energies above 10\,GeV the PAMELA experiment observes a drastic rise of the positron fraction, thereby supporting earlier hints for a rising positron fraction in the HEAT and AMS-01 data.\footnote{The observation of a steep rise of the positron fraction at energies above 10\,GeV is also supported by a recent analysis of Fermi LAT data~\cite{Mitthumsiri:2011}.} This observation cannot be explained by the background of secondary positrons and is a strong indication for a primary source of cosmic-ray positrons. It has been proposed, for instance, that supernova remnants or nearby pulsars might act as a primary source for electron-positron pairs (see \textit{e.g.}~\cite{Hooper:2008kg,Profumo:2008ms,Shaviv:2009bu,Blasi:2009bd,Barger:2009yt}). An intriguing possibility, however, is also that the rise of the positron fraction is a signal of particle dark matter. This has led to a multitude of studies trying to explain the data by an exotic contribution from dark matter annihilations~\cite{Cholis:2008qq,Ishiwata:2008cv,Fox:2008kb,Nezri:2009jd} or decays~\cite{Ibarra:2008qg,Ishiwata:2008cu,Ibarra:2008jk,Nardi:2008ix,Hamaguchi:2008ta,Yin:2008bs,Chen:2009ew,Ibarra:2009bm}.

As we see in Figure~\ref{posfraction} gravitino decays predict a rising contribution in the positron fraction with a cutoff at half of the gravitino mass. For a gravitino mass of 100\,GeV, slightly above the $W$ boson production threshold, the positron line from the direct decay is displaced from the kinematic endpoint and is visible as a second peak. This spectrum clearly cannot explain the data points. Since the rise is measured up to almost 100\,GeV a gravitino of at least 200\,GeV is needed to explain the data. In general, however, it turns out that the gravitino signal does not give a perfect fit to the data as the rise of the sum of signal and background contributions is typically too soft. This is due to the soft contributions to the positron fraction from gravitino decays that comes from the fragmentation of gauge and Higgs bosons and from the decays of muons and tau leptons. This situation is improved if the $R$-parity breaking coupling is only in the electron flavor as the larger amount of positrons directly produced in gravitino decays leads to a harder spectrum (see also~\cite{Buchmuller:2009xv,Ishiwata:2009vx}).

However, all explanations of the rise in the positron fraction by unstable gravitino dark matter require a lifetime of the order of $10^{26}\,$s that is in conflict with bounds from searches for a contribution to the diffuse isotropic gamma-ray flux (\textit{cf.} Figure~\ref{gammabound}).

\paragraph{Electron and Positron Flux}

\begin{figure}[t]
 \centering
 \includegraphics[scale=0.9,bb=0 11 500 300]{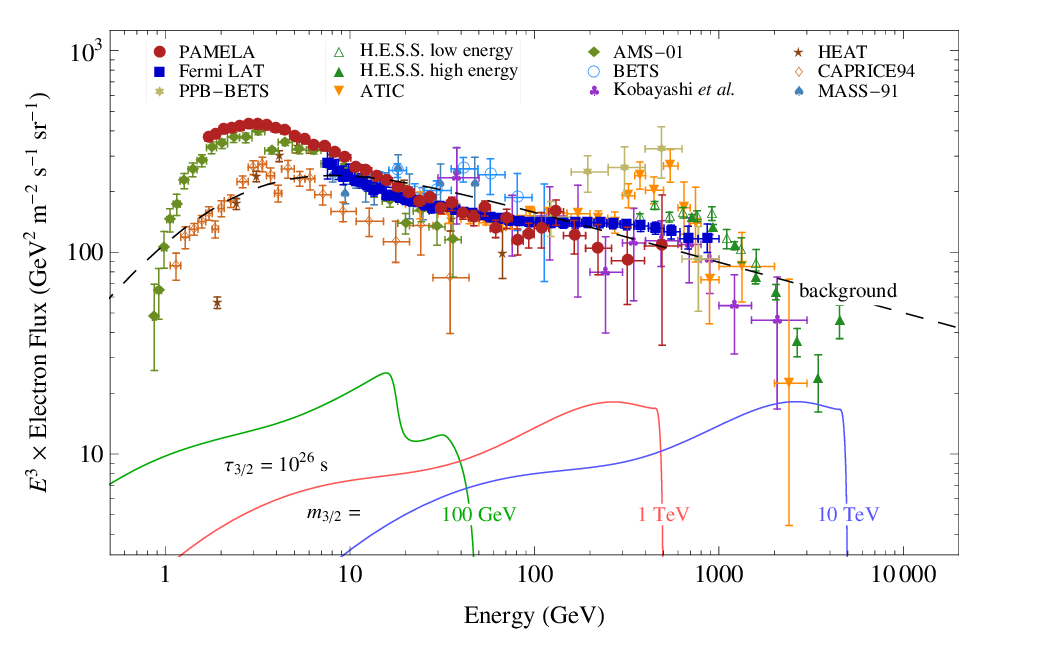}
 \caption[Cosmic-ray electron flux from gravitino decays compared to measurements and the expectation from astrophysical primary and secondary production.]{Cosmic-ray electron flux expected from the decay of gravitino dark matter in the MED propagation model compared to measurements of the electron (+ positron) flux by various experiments and the expectation from astrophysical primary and secondary production. The flux from gravitino decay is shown for a lifetime of $10^{26}\,$s and gravitino masses of 100\,GeV, 1\,TeV and 10\,TeV. In order to account for effects of solar modulation, the data points were demodulated using values for the Fisk potential quoted by the experimental groups (\textit{cf.} Table~\ref{fiskpotentiale}).}
 \label{eflux}
\end{figure}
In addition to the positron fraction many experiments report results for the total flux of cosmic-ray electrons and positrons. In Figure~\ref{eflux} we compare the contribution of gravitino dark matter decays to the cosmic-ray electron spectrum with the expected astrophysical background and the data of MASS-91~\cite{Grimani:2002yz}, CAPRICE94~\cite{Boezio:2000}, HEAT~\cite{DuVernois:2001bb}, a balloon-borne emulsion-chamber experiment by Kobayashi {\it et al.}~\cite{Kobayashi:1999he}, the Balloon-borne Electron Telescope with Scintillating fibers (BETS)~\cite{Torii:2001aw}, AMS-01~\cite{Alcaraz:2000bf}, the balloon-borne Advanced Thin Ionization Calorimeter (ATIC)~\cite{:2008zzr}, the BETS experiment on the Polar Patrol Balloon (PPB-BETS)~\cite{Torii:2008xu}, the High Energy Stereoscopic System (H.E.S.S.)~\cite{Aharonian:2008aa,Aharonian:2009ah}, Fermi LAT~\cite{Ackermann:2010ij} and PAMELA~\cite{Adriani:2011xv}.\footnote{The data tables were partly extracted from the cosmic-ray database of Strong and Moskalenko~\cite{Strong:2009xp}.}

Many of the experiments only use calorimetric information to measure electrons and thus cannot discriminate between electrons and positrons. In particular, the measurements at high energies by Fermi LAT, H.E.S.S., PPB-BETS, ATIC and Kobayashi {\it et al.} show only the combined spectrum of electrons and positrons. We observe that not all the data points agree well with each other. In particular at low energies there are strong discrepancies among the experiments although we already accounted for solar modulation effects. This could be another hint for the existence of charge-dependent solar modulation effects. But also at higher energies the results from ATIC and PPB-BETS strongly deviate from other data and from the expected flux of astrophysical origin. In particular before the publication of the Fermi LAT results this has led to many works trying to explain these observations in terms of exotic contributions from dark matter~\cite{Ishiwata:2009vx,Kyae:2009jt,Cheung:2009si,Fukuoka:2009cu}. However, also the Fermi LAT data are not compatible with the power-law behavior of the expected astrophysical background, although at a smaller level than the ATIC and PPB-BETS data. Also this deviation has been interpreted as a signal from dark matter~\cite{Shirai:2009fq,Ibarra:2009dr} but could also be explained by pulsars~\cite{Grasso:2009ma}.

Also a gravitino with a mass of the order of 1\,TeV can lead to an exotic contribution in the correct energy range. In this work, however, we do not try to fit the exotic contribution from gravitino decays to the Fermi LAT data points as there are strong constraints on this explanation from bounds on the gravitino lifetime from gamma-ray data.

\begin{figure}[t]
 \centering
 \includegraphics[scale=0.9,bb=0 11 500 300]{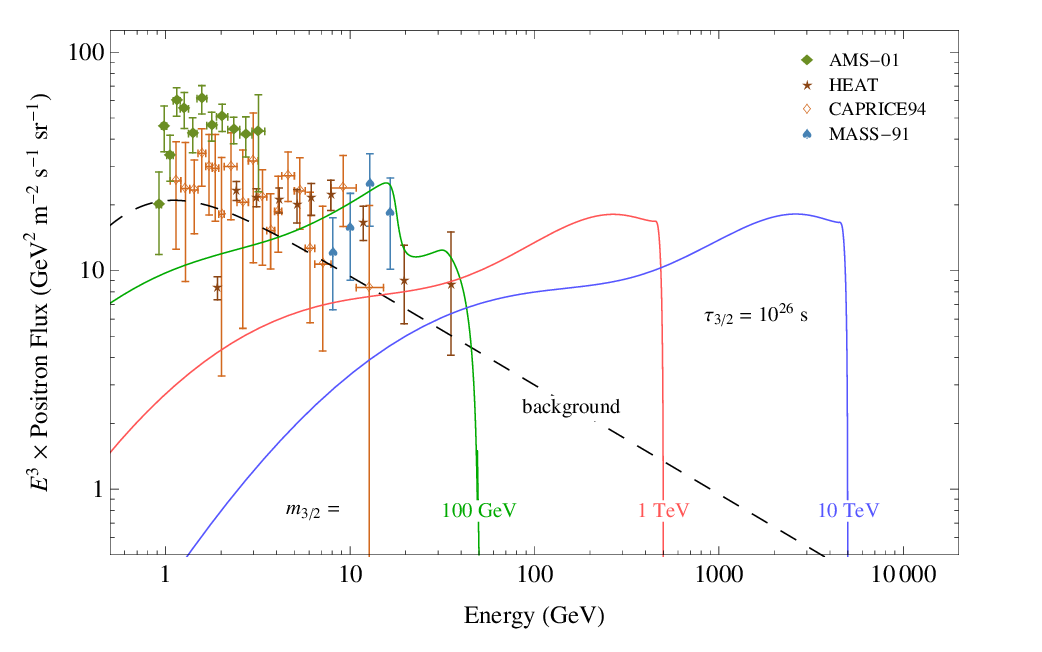}
 \caption[Cosmic-ray positron flux from gravitino decays compared to measurements and the expectation from astrophysical secondary production.]{Cosmic-ray positron flux expected from the decay of gravitino dark matter in the MED propagation model compared to measurements of the absolute positron flux by various experiments and the expectation from astrophysical secondary production. The flux from gravitino decay is shown for a lifetime of $10^{26}\,$s and gravitino masses of 100\,GeV, 1\,TeV and 10\,TeV. In order to account for effects of solar modulation, the data points were demodulated using values for the Fisk potential quoted by the experimental groups (\textit{cf.} Table~\ref{fiskpotentiale}).}
 \label{e+flux}
\end{figure}
Let us also comment on measurements of the positron flux. In Figure~\ref{e+flux} we present the absolute flux of positrons expected from gravitino dark matter decays and compare it to the data of the spectrometric measurements of MASS-91~\cite{Grimani:2002yz}, CAPRICE94~\cite{Boezio:2000}, HEAT~\cite{DuVernois:2001bb} and AMS-01~\cite{Alcaraz:2000bf}. Since the signal from gravitino decay is charge-symmetric the expected positron flux is equivalent to the electron flux presented in Figure~\ref{eflux}. However, the expected astrophysical positron background is significantly lower than the electron background and decreases faster with rising energy. Therefore, the background at TeV energies is lower by two orders of magnitude and a dark matter signal that contributes visibly to the electron spectrum should be clearly observable in the positron spectrum. But also for smaller contributions, like a gravitino signal compatible with gamma-ray observations, a signature above the background could be observable.

Unfortunately, only measurements at rather low energies are available so far, where the effects of solar modulation are important. In addition, the measurements exhibit rather large uncertainties. Looking at the data for the positron fraction, one can, however, expect that a significant deviation from the astrophysical background will be observed. It is expected that the PAMELA collaboration will eventually publish data for the absolute flux of cosmic-ray positrons up to energies of 300\,GeV~\cite{Adriani:2011xv}.

\subsection{Antiprotons}

In contrast to the case of positrons, energy loss processes during propagation are negligible for antiprotons (and protons). Then again, the effect of galactic convection has to be taken into account in this case and there is an additional negative contribution to the source term that comes from the annihilation of antiprotons with hydrogen and helium nuclei in the interstellar medium. Thus the propagation equation can be written in the form:
\begin{equation}
  \frac{\partial f_{\bar{p}}}{dt}=\vec{\nabla}\cdot\left( K(T)\,\vec{\nabla} f_{\bar{p}}-\vec{V}_C(\vec{r})\,f_{\bar{p}}\right) -2\,h\,\delta(z)\,\Gamma_{\bar{p}}^{\text{ann}}\,f_{\bar{p}}+Q_{\bar{p}}(T,\vec{r})=0\,.
\end{equation}
The interstellar antiproton flux at the position of the solar system coming from dark matter decays in the galactic halo is given by
\begin{equation}
  \frac{d\Phi_{\bar{p}}^{\text{DM}}}{dT}=\frac{v_{\bar{p}}}{4\,\pi\,m_{3/2}\,\tau_{3/2}}\int_0^{\infty}dT'\,G_{\bar{p}}(T,T')\,\frac{dN_{\bar{p}}}{dT'}\,,
\end{equation}
where the velocity of antiprotons in terms of their kinetic energy is given by
\begin{equation}
  v_{\bar{p}}=c\,\sqrt{1-\frac{m_{\bar{p}}^2}{(T+m_{\bar{p}})^2}}\,.
\end{equation}
The Green's function for the propagation of antiprotons can be approximated by~\cite{Ibarra:2008qg}
\begin{equation}
  G_{\bar{p}}(T,T')\simeq 10^{14}\exp\left( x+y\,\ln \tau+z\,(\ln\tau)^2\right) \delta(T'-T)\,{\text{cm}}^{-3}\,{\text{s}}\,,
\end{equation}
where $\tau=T/$GeV. The parameters $x$, $y$ and $z$ are given in Table~\ref{antiprotonparameters} for the case of an NFW dark matter halo density profile and different sets of propagation parameters. In practice the upper limit for the integration over the Green's function is given by the maximal kinetic energy of antiprotons in a gravitino decay: $T_{\text{max}}=m_{3/2}/2-m_{\bar{p}}$\,.
\begin{table}[t]
 \centering
 \begin{tabular}{cccccccc}
  \toprule
  Model & $\delta$ & $K_0\,(\text{kpc}^2\!/\text{Myr})$ & $L\,(\text{kpc})$ & $V_C\,(\text{km/s})$ & $x$ & $y$ & $z$ \\
  \midrule
  MIN & 0.85 & 0.0016 & 1 & 13.5 & $-0.0537$ & 0.7052 & $-0.1840$ \\
  MED & 0.70 & 0.0112 & 4 & 12 & 1.8002 & 0.4099 & $-0.1343$ \\
  MAX & 0.46 & 0.0765 & 15 & 5 & 3.3602 & $-0.1438$ & $-0.0403$ \\
  \bottomrule
 \end{tabular}
 \caption[Cosmic-ray propagation parameters for antiprotons.]{Parameters of cosmic-ray propagation models that correspond, respectively, to the best fit of cosmic-ray B/C data (MED) as well as the minimal (MIN) or maximal (MAX) antiproton flux compatible with cosmic-ray B/C data. Figures taken from~\cite{Ibarra:2008qg,Donato:2003xg}.}
 \label{antiprotonparameters}
\end{table}

\paragraph{Antiproton-to-Proton Ratio}

Similar to the case of the cosmic-ray positron fraction, systematic uncertainties of antiproton measurements are reduced if the results are presented in the form of the ratio of antiproton and proton fluxes:
\begin{equation}
  \frac{\bar{p}}{p}=\frac{\Phi_{\bar{p}}}{\Phi_p}\,.
\end{equation}
Also in this case a larger theoretical uncertainty is introduced since the astrophysical background of protons needs to be taken into account in addition to that of antiprotons. Even if there are no primary antiprotons expected from astrophysical objects, a background of secondary antiprotons is produced in spallation processes between impinging primary cosmic rays and the interstellar medium~\cite{Donato:2001ms}. The dominant contribution comes from interactions of cosmic-ray protons and helium nuclei with hydrogen and helium nuclei in the interstellar medium. For the background of cosmic-ray antiprotons at the top of the atmosphere we use the parametrization from~\cite{Nezri:2009jd} for the numerical result of~\cite{Bringmann:2006im}:
\begin{equation}
  \frac{d\Phi_{\bar{p}}^{\text{bkg,TOA}}}{dT}=\frac{0.9\,\tau^{-0.9}}{14+30\,\tau^{-1.85}+0.08\,\tau ^{2.3}}\,(\text{GeV}\,\text{m}^2\,\text{s}\,\text{sr})^{-1},
\end{equation}
where $\tau=T/$GeV. In the simulation of this spectrum the effect of solar modulation has been taken into account assuming a Fisk potential of 500\,MV. The main uncertainty on this background comes from uncertainties in the knowledge of nuclear cross sections and amounts to roughly 25\,\% in the energy range of 100\,MeV--100\,GeV~\cite{Donato:2001ms}.

\begin{figure}[t]
 \centering
 \includegraphics[scale=0.9,bb=0 11 500 300]{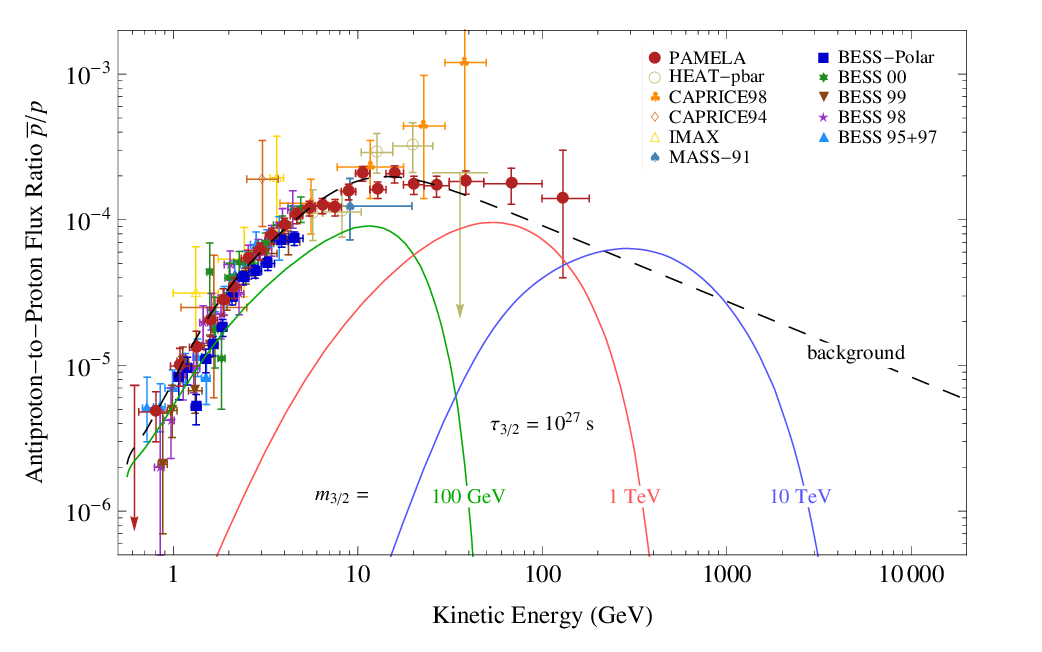} 
 \caption[Cosmic-ray antiproton-to-proton flux ratio from gravitino decays compared to measurements and the expected astrophysical background.]{Cosmic-ray antiproton-to-proton flux ratio expected from the decay of gravitino dark matter in the MED propagation model compared to measurements by various experiments and the expected astrophysical background. The contribution from gravitino decays is shown for a lifetime of $10^{27}\,$s and gravitino masses of 100\,GeV, 1\,TeV and 10\,TeV. In order to account for effects of solar modulation, the data points were demodulated using values for the Fisk potential quoted by the experimental groups (\textit{cf.} Table~\ref{fiskpotentialp}).}
 \label{antipratio}
\end{figure}
\begin{table}[t]
 \centering
 \begin{tabular}{cc}
  \toprule
  Experiment & Fisk potential (MV) \\
  \midrule
  PAMELA & 525 \\
  HEAT-pbar & 1344 \\
  AMS-01 & 650 \\
  CAPRICE98 & 600 \\
  CAPRICE94 & 500 \\
  IMAX & 750 \\
  \bottomrule
 \end{tabular}
 \hspace{5mm}
 \begin{tabular}{cc}
  \toprule
  Experiment & Fisk potential (MV) \\
  \midrule
  BESS-Polar & 850 \\
  BESS 00 & 1344 \\
  BESS 99 & 648 \\
  BESS 98 & 610 \\
  BESS 95+97 & 500 \\
  MASS-91 & 500 \\
  \bottomrule
 \end{tabular}
 \caption[Values of the Fisk potential to account for the effect of solar modulation in the antiproton data sets.]{Values of the Fisk potential to account for the effect of solar modulation in the antiproton data sets. The figures are taken from the publications of the respective data sets (see text for the list of references). The values for MASS-91 and HEAT-pbar are taken from~\cite{Donato:2008jk}.}
 \label{fiskpotentialp}
\end{table}
The background flux of cosmic-ray protons is mainly of primary origin. For the expected flux at the top of the atmosphere we use the parametrization from~\cite{Nezri:2009jd} for the numerical result of~\cite{Lionetto:2005jd}:
\begin{equation}
  \frac{d\Phi_p^{\text{bkg,TOA}}}{dT}=\frac{0.9\,\tau^{-1}}{8+1.1\,\tau^{-1.85}+0.8\,\tau ^{1.68}}\,(\text{GeV}\,\text{cm}^2\,\text{s}\,\text{sr})^{-1},
\end{equation}
where $\tau=T/$GeV. In the generation of this spectrum solar modulation has been taken into account assuming a Fisk potential of 550\,MV. In order to compare the astrophysical background to the expected interstellar flux originating from gravitino decays, we need to account for solar modulation in a similar way as for the case of the positron fraction:
\begin{equation}
  \frac{\bar{p}}{p}(T^{\text{IS}})\bigg|^{\text{\text{IS}}}=\frac{\bar{p}}{p}(T^{\text{IS}}-e\,\phi_F)\Bigg|^{\text{TOA}}.
\end{equation}
In Figure~\ref{antipratio} we compare the contribution of gravitino dark matter decays to the cosmic-ray antiproton-to-proton flux ratio with the expected astrophysical background and the data measured by
MASS-91~\cite{Hof:1996yx}, the balloon-borne Isotope Matter Antimatter Experiment (IMAX)~\cite{Mitchell:1996bi}, CAPRICE94~\cite{Boezio:1997ec}, CAPRICE98~\cite{Boezio:2001ac}, a series of flights of the Balloon-borne Experiment with a Superconducting Spectrometer (BESS) in 1995+1997~\cite{Orito:1999re}, 1998~\cite{Maeno:2000qx}, and 1999+2000~\cite{Asaoka:2001fv}, the balloon-borne BESS-Polar experiment~\cite{Abe:2008sh}, the balloon-borne HEAT-pbar experiment~\cite{Beach:2001ub} and PAMELA~\cite{Adriani:2010rc}.\footnote{The data tables were partly extracted from the cosmic-ray database of Strong and Moskalenko~\cite{Strong:2009xp}.}

In particular in the lower-energetic part of the spectrum we observe an excellent agreement of the various data sets among each other and also with the expected astrophysical background. Hence there is no room for a sizable exotic contribution from dark matter. Only at higher energies the CAPRICE98 and PAMELA data show a tendency to overshoot the background expectation, even though at a level that is compatible with the error bars of the measurements. As the expected flux ratio from gravitino decays with a lifetime of $10^{27}\,$s is only slightly below the measured values, strong limitations are put on gravitino explanations of the cosmic-ray positron and electron excesses also from antiproton data. Note, however, that the uncertainty from the allowed range of propagation parameters amounts to one order of magnitude above and below the presented fluxes in the MED propagation model.

\paragraph{Antiproton Flux}

\begin{figure}[t]
 \centering
 \includegraphics[scale=0.9,bb=0 11 500 300]{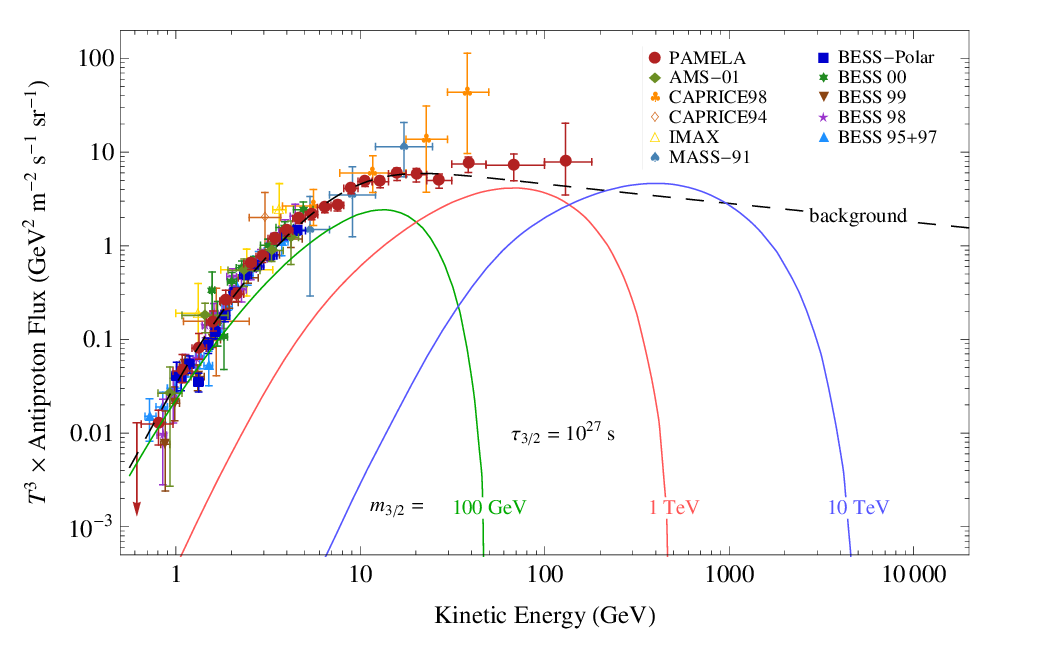} 
 \caption[Cosmic-ray antiproton flux from gravitino decays compared to measurements and the expectation from astrophysical secondary production.]{Cosmic-ray antiproton flux expected from the decay of gravitino dark matter in the MED propagation model compared to measurements by various experiments and the expectation from astrophysical secondary production. The flux from gravitino decay is shown for a lifetime of $10^{27}\,$s and gravitino masses of 100\,GeV, 1\,TeV and 10\,TeV. In order to account for effects of solar modulation, the data points were demodulated using values for the Fisk potential quoted by the experimental groups (\textit{cf.} Table~\ref{fiskpotentialp}).}
 \label{antipflux}
\end{figure}
Also in the case of antiprotons most experimental collaborations have finally also published data for the absolute flux. In Figure~\ref{antipflux} we compare the contribution of gravitino dark matter decays to the absolute cosmic-ray antiproton spectrum with the expected astrophysical background and the data points measured by MASS-91~\cite{Basini:1999hh}, IMAX~\cite{Mitchell:1996bi}, CAPRICE94~\cite{Boezio:1997ec}, CAPRICE98~\cite{Boezio:2001ac}, BESS 95+97~\cite{Orito:1999re}, BESS 98~\cite{Maeno:2000qx}, BESS 99+00~\cite{Asaoka:2001fv}, BESS-Polar~\cite{Abe:2008sh}, AMS-01~\cite{Aguilar:2002ad} and PAMELA~\cite{Adriani:2010rc}.%\footnote{The data tables were partly extracted from the cosmic-ray database of Strong and Moskalenko~\cite{Strong:2009xp}.}
\footnotemark[\value{footnote}]

In this case the theoretical uncertainty of the background prediction is slightly decreased as one does not need to employ the astrophysical background of primary protons. Apart from that, the conclusions from these data sets are practically unchanged compared to those from the data of the antiproton-to-proton flux ratio.

\subsection{Antideuterons}

The use of a signal of antideuterons to search for dark matter was first proposed by Donato \textit{et al.}~\cite{Donato:1999gy}. It is a particularly convenient dark matter detection channel, since the expected astrophysical background at low energies is at a very low level. Therefore, antideuterons allow for an almost background-free search for an exotic dark matter component in certain parameter ranges. In fact, no cosmic-ray antideuterons have been observed so far and there exists only an upper bound on the antideuteron flux from the BESS experiment~\cite{Fuke:2005it}. However, several new experiments are currently planned to improve this situation: The Gaseous Antiparticle Spectrometer (GAPS) is planned to perform two balloon flights, the Long Duration Balloon flight (LDB) and the Ultra-Long Duration Balloon flight (ULDB)~\cite{Fuke:2008zz}. In addition, the recently launched AMS-02 experiment on the International Space Station will greatly improve on the sensitivity to low antideuteron fluxes~\cite{Choutko:2007}.

Several studies on antideuteron fluxes from dark matter annihilations or decays can be found in the literature~\cite{Donato:1999gy,Brauninger:2009pe,Baer:2005tw,Donato:2008yx,Ibarra:2009tn}. All of these studies, however, employ the spherical approximation of the coalescence model for antideuteron formation which in general is insufficient to describe the actual production rate~\cite{Kadastik:2009ts}. Only very recent studies employ the Monte Carlo approach~\cite{Kadastik:2009ts,Cui:2010ud}. As discussed in Section~\ref{gravspectra} we also employ decay spectra obtained by this method.

The propagation equation for antideuterons is given in analogy to that for antiprotons:
\begin{equation}
  \frac{\partial f_{\bar{d}}}{dt}=\vec{\nabla}\cdot\left( K(T)\,\vec{\nabla}f_{\bar{d}}-\vec{V}_C(\vec{r})\,f_{\bar{d}}\right) -2\,h\,\delta(z)\,\Gamma_{\bar{d}}^{\text{ann}}\,f_{\bar{d}}+Q_{\bar{d}}(T,\vec{r})=0\,.
\end{equation}
The interstellar antideuteron flux at the position of the solar system coming from dark matter decays in the galactic halo is then given by
\begin{equation}
  \frac{d\Phi_{\bar{d}}^{\text{DM}}}{d(T/\text{n})}=\frac{v_{\bar{d}}}{4\,\pi\,m_{3/2}\,\tau_{3/2}}\int_0^{\infty}d(T'\!/\text{n})\,G_{\bar{d}}(T/\text{n},T'\!/\text{n})\,\frac{dN_{\bar{d}}}{d(T'\!/\text{n})}\,,
\end{equation}
where the velocity of antideuterons in terms of their kinetic energy is given by
\begin{equation}
  v_{\bar{d}}=c\,\sqrt{1-\frac{m_{\bar{d}}^2}{(T+m_{\bar{d}})^2}}\,.
\end{equation}
We use a numerical approximation of the Green's function that was found in~\cite{Ibarra:2009tn} and is of the same form as the Green's function for antiproton propagation:
\begin{equation}
  G_{\bar{d}}(T/\text{n},T'\!/\text{n})\simeq 10^{14}\exp\left( x+y\,\ln \tau+z\,(\ln\tau)^2\right) \delta(T'\!/\text{n}-T/\text{n})\,{\text{cm}}^{-3}\,{\text{s}}\,,
\end{equation}
where the kinetic energy per nucleon is parametrized in the form $\tau=(T/\text{n})/$(GeV/n) and $\tau'=(T'\!/\text{n})/$(GeV/n). The parameters $x$, $y$ and $z$ are given in Table~\ref{antideuteronparameters} for the case of an NFW halo profile. In practice the upper limit for the integration over the Green's function is given by the maximal kinetic energy per nucleon of antideuterons in a gravitino decay: $(T/\text{n})_{\text{max}}=(m_{3/2}/2-m_{\bar{d}})/2$.
\begin{table}[t]
 \centering
 \begin{tabular}{cccccccc}
  \toprule
  Model & $\delta$ & $K_0\,(\text{kpc}^2\!/\text{Myr})$ & $L\,(\text{kpc})$ & $V_C\,(\text{km/s})$ & $x$ & $y$ & $z$ \\
  \midrule
  MIN & 0.85 & 0.0016 & 1 & 13.5 & $-0.3889$ & 0.7532 & $-0.1788$ \\
  MED & 0.70 & 0.0112 & 4 & 12 & 1.6023 & 0.4382 & $-0.1270$ \\
  MAX & 0.46 & 0.0765 & 15 & 5 & 3.1992 & $-0.1098$ & $-0.0374$ \\
  \bottomrule
 \end{tabular}
 \caption[Cosmic-ray propagation parameters for antideuterons.]{Parameters of cosmic-ray propagation models that correspond, respectively, to the best fit of cosmic-ray B/C data (MED) as well as the minimal (MIN) or maximal (MAX) antideuteron flux compatible with cosmic-ray B/C data. Figures taken from~\cite{Ibarra:2009tn,Donato:2003xg}.}
 \label{antideuteronparameters}
\end{table}

As for the case of antiprotons, no primary antideuterons are expected from astrophysical objects and the dominant background for dark matter searches are secondary antideuterons created in spallation processes of cosmic-ray protons and helium nuclei impinging on interstellar hydrogen and helium gas. In order to estimate the background for the signal from gravitino decay we employ here the astrophysical secondary antideuteron flux calculated in~\cite{Donato:2008yx}. We parametrize their results by assuming a similar functional dependence as for the other cosmic-ray backgrounds and find
\begin{equation}
  \frac{d\Phi_{\bar{d}}^{\text{sec}}}{d(T/\text{n})}\approx\frac{1.08\times10^{-7}\,\tau^{0.88}}{1+0.041\,\tau^{2.2}+2.6\times10^{-4}\,\tau ^4}\,(\text{GeV}\,\text{m}^2\,\text{s}\,\text{sr})^{-1},
\end{equation}
where $\tau=(T/\text{n})/$(GeV/n). This spectrum corresponds to the interstellar antideuteron flux where effects of solar modulation are not taken into account.

\begin{figure}[t]
 \centering
 \includegraphics[scale=0.8,bb=0 2 500 317]{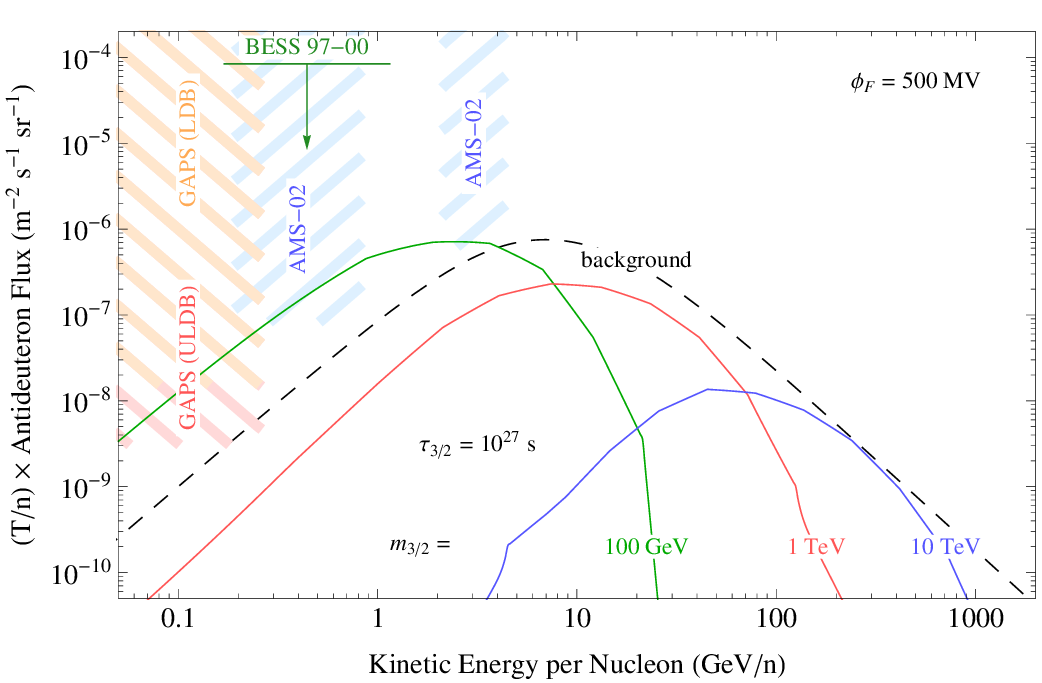} 
 \caption[Cosmic-ray antideuteron flux from gravitino decays compared to the expectation from astrophysical secondary production and the sensitivities of forthcoming experiments.]{Cosmic-ray antideuteron flux expected from the decay of gravitino dark matter in the MED propagation model compared to the expectation from astrophysical secondary production and the sensitivities of forthcoming experiments. The flux from gravitino decay is shown for a lifetime of $10^{27}\,$s and gravitino masses of 100\,GeV, 1\,TeV and 10\,TeV. The effect of solar modulation was taken into account assuming a Fisk potential of $\phi_F=500\,$MV.}
 \label{antidflux}
\end{figure}
In Figure~\ref{antidflux} we present the antideuteron spectrum from gravitino dark matter decays and compare it to the expected astrophysical background and the flux limit obtained by the BESS experiment~\cite{Fuke:2005it}. In addition, we present the projected sensitivity regions of the GAPS and AMS-02 experiments. In order to be able to compare the expected fluxes to the projected sensitivities we take into account the effect of solar modulation according to equation~(\ref{solarmod}) assuming a period at minimal solar activity with $\phi_F=500\,$MV.

In particular for rather low gravitino masses on the order of 100\,GeV the signal from gravitino decays can dominate the background even for lifetimes as large as $10^{27}\,$s that are not yet excluded by gamma-ray and antiproton observations. A signal in that range should be observed in forthcoming antideuteron searches practically without a background of antideuterons from astrophysical secondary production. In this respect it will also be interesting to see what antideuteron flux could be expected for even lower gravitino masses. We plan to study this region in future work using the spectra obtained from gravitino three-body decays~\cite{Covi:2011a}.

But also for larger gravitino masses the gravitino signal is at least at the same order as the background. This was not observed in earlier studies as the signal for large dark matter masses is artificially suppressed in the spherical approximation of the coalescence prescription. Therefore, one could argue that there are in principle also good prospects for the observation of an exotic component in the higher-energetic part of the spectrum, where currently no experiments are planned.

\subsection[Bounds on the Gravitino Lifetime from Charged Cosmic Rays]{Bounds on the Gravitino Lifetime from Observations of Charged Cosmic Rays}
\label{CRlifetime}

\begin{figure}[t]
 \centering
 \includegraphics[scale=0.9,bb=0 11 500 300]{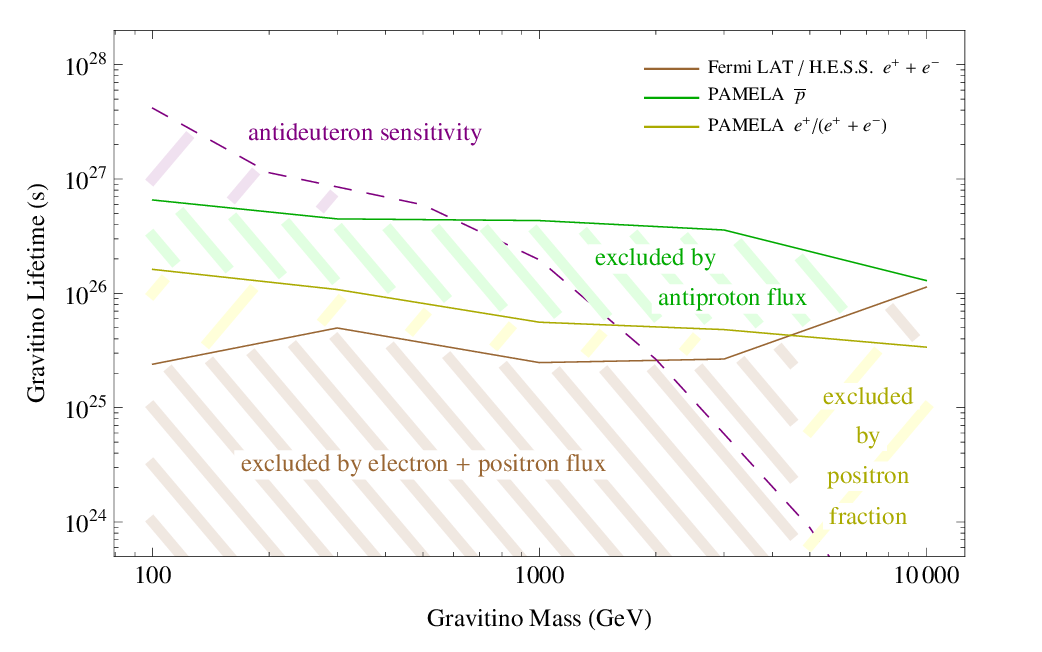}
 \caption[Bounds on the gravitino lifetime from observations of charged cosmic rays and sensitivity of forthcoming antideuteron experiments.]{Bounds on the gravitino lifetime from observations of charged cosmic rays and sensitivity of forthcoming antideuteron experiments.}
 \label{CRbound}
\end{figure}

Similar to the bounds derived from the diffuse gamma-ray flux, we can estimate conservative lower limits on the gravitino lifetime from the requirement that the gravitino signal in charged cosmic rays does not overshoot the error bars of the measurements. For this task we will assume a MED propagation model and consider the PAMELA measurement of the positron fraction, the measurements of the total electron + positron flux by Fermi LAT and H.E.S.S., and the PAMELA measurement of the antiproton flux. The resulting exclusion regions are presented in Figure~\ref{CRbound}.

As expected, observations of cosmic-ray antiprotons in general lead to the strongest constraints. The exclusion limit from observations of the positron fraction is weaker due to the observed excess above the expected astrophysical background. One can see from the boundary of the exclusion region that a lifetime of typically $10^{26}\,$s is required in order to explain the rise of the positron fraction in terms of a signal from dark matter decays. On the other hand, we clearly see that this possibility is ruled out for the case of gravitino dark matter by the strong constraints in the antiproton channel.

In addition, we present an estimate of the sensitivity of forthcoming antideuteron experiments to the gravitino parameter space. For this task we also employ the MED propagation model and require that the expected antideuteron flux from gravitino decays is at the lower edge of the sensitivity region of antideuteron experiments. In most cases the AMS-02 experiment provides the best sensitivity. Only for the lowest gravitino masses considered, the sensitivity of the GAPS experiment is better. This is due to the different energy ranges in which the experiments are sensitive to antideuterons. AMS-02 has a sensitivity region at larger antideuteron energies and is thus more sensitive to signals from heavier dark matter particles. On the other hand, in particular in the low mass range the antideuteron channel has the potential to be more sensitive than other charged cosmic ray channels and this is the region where the GAPS balloon can compete. For this reason it would also be very interesting to study the antideuteron signals from gravitino three-body decays at low gravitino masses~\cite{Covi:2011a}. It might well be that the sensitivity of the antideuteron channel is even higher than that of the gamma-ray channel. Clearly this also depends on the strength of the gamma-ray line from gravitino decays.

One comment is in order: Of course one should remember that these exclusion and sensitivity regions are only rough estimates as there is quite some uncertainty in the expected fluxes, for instance due to the choice of the propagation model. In addition, taking into account the expected flux contributions from astrophysical charged particle production can lead to much stricter limits on the gravitino parameters. In particular in the case of antiprotons the expected astrophysical background practically perfectly matches the observations, leaving only little space for an exotic contribution.

\section{Probing Gravitino Dark Matter with Neutrinos}
\label{neutrinos}

In this section we present a study on the neutrino signals from gravitino dark matter decays that was performed in analogy to our published phenomenological study of neutrino signals from generic decay channels of fermionic and scalar dark matter particles~\cite{Covi:2009xn}. This study is complementary to our previous study on tau neutrino signals from gravitino decays in~\cite{Covi:2008jy,Grefe:2008zz}.

\subsection{Neutrino Fluxes}
\label{neutrinofluxes}

In this section we want to present the neutrino flux expected from the decay of gravitino dark matter according to the discussion in the very beginning of this chapter.

\subsubsection{Propagation of Neutrinos}

\begin{table}
 \centering 
 \begin{tabular}{cccccc}
  \toprule
  mass hierarchy & $\sin^2\theta_{12}$ & $\sin^2\theta_{23}$ & $\sin^2\theta_{13}$ & $\Delta m_{21}^2\,(\text{eV}^2)$ & $\Delta m_{31}^2\,(\text{eV}^2)$ \\
  \midrule
  normal & 0.316 & 0.51 & 0.017 & $7.64\times10^{-5}$ & $2.45\times10^{-3}$ \\
  inverted & 0.316 & 0.52 & 0.020 & $7.64\times10^{-5}$ & $-2.34\times10^{-3}$ \\
  \bottomrule
 \end{tabular}
 \caption[Neutrino mixing parameters.]{Best-fit neutrino mixing parameters for the cases of normal and inverted neutrino mass hierarchies. Figures taken from~\cite{Schwetz:2011qt}.}
 \label{neutrinomixing}
\end{table}
As mentioned before, after the neutrinos are produced in the decay of gravitinos in the Milky Way halo or in extragalactic locations, they travel in straight lines to the Earth, essentially without any interactions. The only modifications to the fluxes during this time are due to flavor oscillations~\cite{Strumia:2006db}. In fact, using the experimental best-fit values for the neutrino mixing angles assuming normal (inverted) neutrino mass hierarchy (see Table~\ref{neutrinomixing}) and neglecting possible $CP$-violating effects, the neutrino oscillation probabilities in vacuum are given by
\begin{alignat}{2}
 P(\nu_e\leftrightarrow\nu_e) &=0.55\,, &\qquad\qquad P(\nu_e\leftrightarrow\nu_{\mu}) &=0.23\,, \nonumber\\
 P(\nu_e\leftrightarrow\nu_{\tau}) &=0.24\,, &\qquad\qquad P(\nu_{\mu}\leftrightarrow\nu_{\mu}) &=0.40\,, \\
 P(\nu_{\mu}\leftrightarrow\nu_{\tau}) &=0.37\,, &\qquad\qquad P(\nu_{\tau}\leftrightarrow\nu_{\tau}) &=0.39\:(0.38)\,.\nonumber
\end{alignat}
Thus a primary neutrino flux in a specific flavor is redistributed almost equally into all neutrino flavors during propagation and virtually any flavor information is lost. On the other hand, this means that nearly the same signal is present in any flavor and may allow to choose the best channel for discovery according to the background and efficiency of the detector.

\subsubsection{Neutrino Backgrounds}

Let us now discuss the background for the neutrino signal. In the GeV to TeV range the main background for the observation of neutrinos are neutrinos produced in cosmic-ray interactions with the Earth's atmosphere. Here we use the atmospheric neutrino fluxes calculated by Honda \textit{et al.}~\cite{Honda:2006qj}. The theoretical uncertainty of these fluxes is estimated to be better than 25\,\% in the GeV to TeV range, while the uncertainty in the ratio of the different flavors is significantly smaller. We extend the atmospheric neutrino fluxes to energies above 10\,TeV using the slopes given by Volkova \textit{et al.}~\cite{Volkova:2001th}. 

Conventional atmospheric electron and muon neutrinos are directly produced from pion and kaon decays. While electron neutrinos are practically unaffected by neutrino oscillations due to their large oscillation length, muon neutrinos, particularly at low energies, can be converted into tau neutrinos and provide the dominant tau neutrino background at energies below 1\,TeV. The conversion probability of muon neutrinos into tau neutrinos is given by~\cite{Strumia:2006db}
\begin{equation}
 \begin{split}
  P(\nu_{\mu}\rightarrow\nu_{\tau}) &=\sin^22\,\theta_{23}\,\sin^2\left( 1.27\,\frac{(\Delta m_{13}^2/\text{eV}^2)\,(L/\text{km})}{(E_\nu/\text{GeV})}\right) \\
  &\simeq\sin^2\left( 3\times 10^{-3}\,\frac{(L/\text{km})}{(E_\nu/\text{GeV})}\right) ,
 \end{split}
 \label{tauosc}
\end{equation}
where we employed the best-fit neutrino mixing parameters in the second step. In this expression, $E_\nu$ is the neutrino energy and $L$ is their propagation length after being produced in the atmosphere, which is given by
\begin{equation}
 L(\theta)=\sqrt{(R_{\oplus}\cos\theta)^2+2\,R_{\oplus}\,h+h^2}-R_{\oplus}\cos\theta
\end{equation}
as a function of the zenith angle $\theta$, with $R_{\oplus}\simeq 6.4\times 10^3\,$km being the Earth's radius and $h\simeq 15\,$km the mean altitude at which atmospheric neutrinos are produced~\cite{Strumia:2006db}.

In addition to this conventional atmospheric neutrino flux from pion and kaon decays there is a prompt neutrino flux from the decay of charmed particles that are also produced in cosmic-ray collisions with the atmosphere. The prompt neutrinos have a harder spectrum than the conventional ones and therefore dominate the background at higher energies (roughly above 10\,TeV for electron neutrinos and above 100\,TeV for muon neutrinos). Since these contributions are not well understood and in any case subdominant in the GeV to TeV energy range that is of interest for dark matter searches, we neglect them in the present study. On the other hand, the prompt flux of tau neutrinos starts to dominate around 1\,TeV (and at even smaller energies for down-going neutrinos). Thus we include this contribution using the parametrization~\cite{Pasquali:1998xf}
\begin{equation}
 \log_{10}\left[ E^3\,\frac{d\Phi_{\nu_{\tau}}}{dE}\left/ \left( \frac{\text{GeV}^2}{\text{cm}^2\,\text{s}\,\text{sr}}\right) \right. \right] =-A+Bx-Cx^2-Dx^3,
\end{equation}
where $x=\log_{10}\left( E/\text{GeV}\right) $, $A=6.69$, $B=1.05$, $C=0.150$ and $D=-0.00820$. This parametrization is valid in the energy range of 100\,GeV up to 1\,PeV. However, we point out that compared to the conventional atmospheric neutrino flux the prompt flux suffers from larger uncertainties.

Other neutrino backgrounds in the considered energy range are neutrinos produced in cosmic-ray interactions with the solar corona~\cite{Ingelman:1996mj} and those produced in cosmic-ray interactions with the interstellar medium of the Milky Way~\cite{Athar:2004um,Ingelman:1996md}. While the former is subdominant in diffuse searches for all flavors~\cite{Covi:2008jy,Grefe:2008zz} and can be excluded from the analysis by excluding neutrinos from the direction of the Sun, the latter represents an irreducible, ill-understood neutrino background for searches in the galactic disc direction. In fact, the flux of galactic neutrinos is expected to become comparable to the atmospheric electron neutrino background for the galactic disc direction and energies in the TeV range.

\subsubsection{General Detection Strategy and Use of Directionality}
\label{direction}

As it turns out that the neutrino signals from gravitino decays are in general dominated by the background it is important to devise strategies that minimize the background. In~\cite{Covi:2008jy,Grefe:2008zz} we proposed to use directionality in order to reduce the background in the tau neutrino channel. This is possible since the tau neutrino background at low energies comes mainly from the muon neutrino oscillation and is therefore strongly suppressed in the zenith direction. One could also consider to search for an enhanced muon neutrino signal only in the direction of the galactic center since this is where the maximal dark matter density is expected (\textit{cf.} Figure~\ref{profiles}). However, taking into account the typically low neutrino event rates, it is not always the best strategy to optimize the signal-to-background ratio. Instead, the statistical significance $\sigma=S/\sqrt{B}$ (number of signal events divided by the square root of the number of background events) is a better measure for comparing different detection strategies.

\begin{figure}[t]
 \centering
 \includegraphics[scale=0.44]{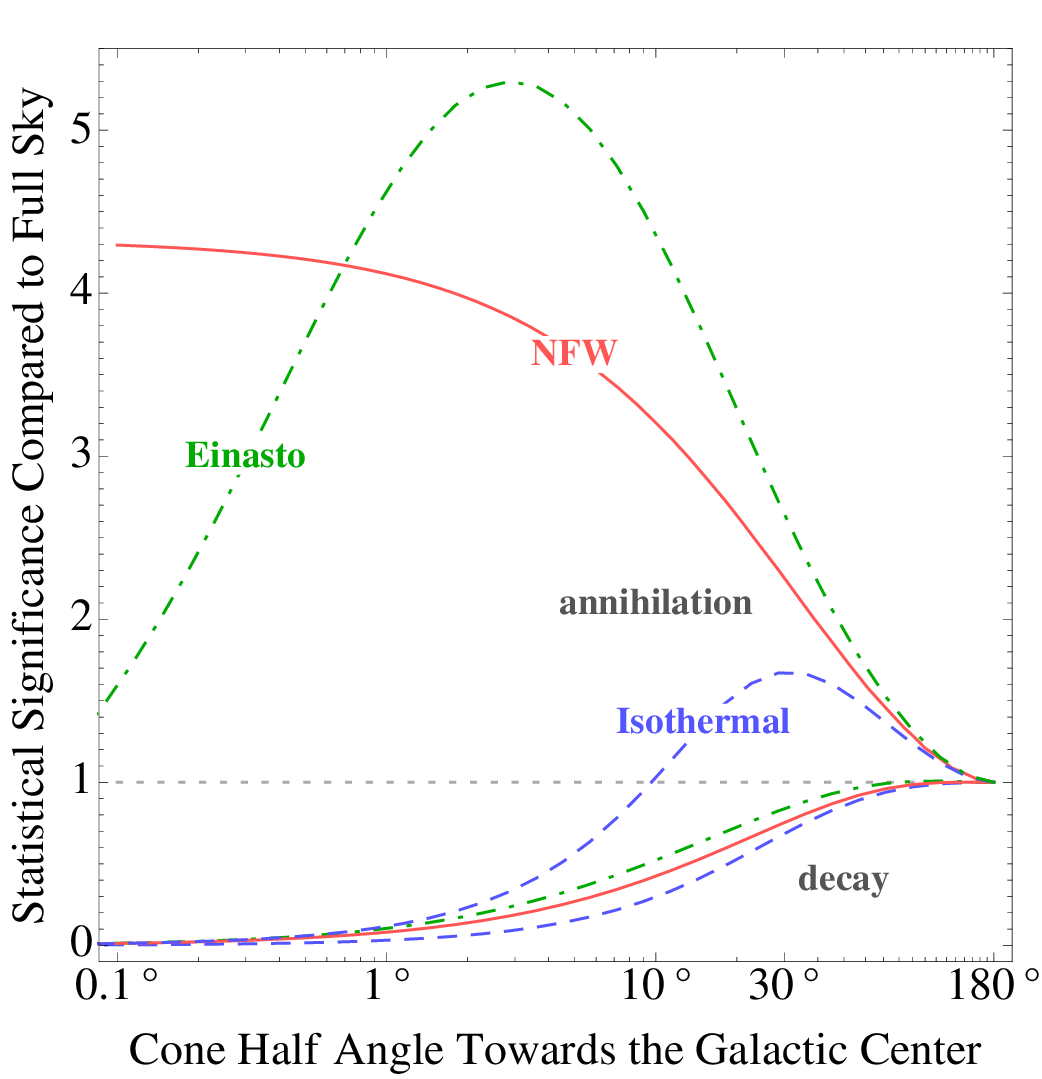}
 \includegraphics[scale=0.43]{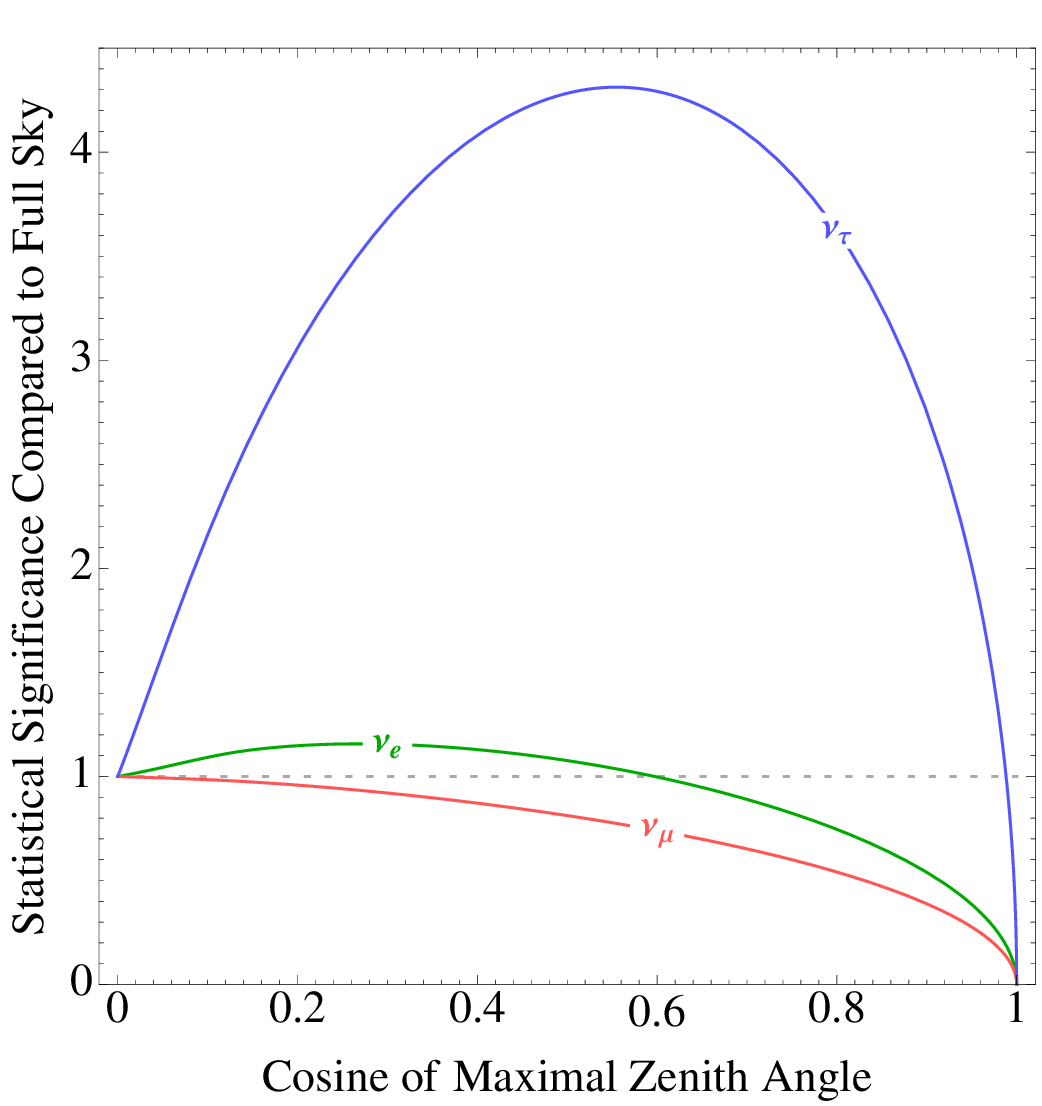}
 \caption[Dependence of the statistical significance of a neutrino signal from dark matter decays or annihilations on the size of the observed cone around the galactic center and dependence of the statistical significance of a neutrino signal from dark matter on the observed cone size around the zenith direction.]{\textit{Left:} Statistical significance of the neutrino signal as a function of the cone half angle towards the galactic center normalized to the significance of the full-sky signal for annihilating/decaying dark matter depending on the different density profiles. \textit{Right:} Statistical significance of the neutrino signal as a function of the maximal angle observed around the zenith direction normalized to the significance for the observation of the upper hemisphere. The curves are shown for a neutrino energy of 100\,GeV.}
\label{directionfigure}
\end{figure}

In the left panel of Figure~\ref{directionfigure} we show the significance of the signal as a function of the cone half angle around the galactic center normalized to the significance of the full-sky observation. This quantity is calculated as
\begin{equation}
 \frac{S}{\sqrt{B}}(\alpha)\Bigg/\frac{S}{\sqrt{B}}(180^\circ)=\frac{\int_0^\infty ds\int_0^\alpha\sin\theta\,d\theta\,\rho_{\text{halo}}^{(2)}(r(s,\,\theta))}{\int_0^\infty ds\int_0^\pi\sin\theta\,d\theta\,\rho_{\text{halo}}^{(2)}(r(s,\,\theta))}\times\sqrt{\frac{4\,\pi}{2\,\pi\left( 1-\cos\alpha\right) }}\,,
\end{equation}
where we changed the coordinate system to make use of the symmetry in the halo integration (see \textit{e.g.}~\cite{Grefe:2008zz}). The radius from the galactic center is then given by
\begin{equation}
 r(s,\,\theta)=\sqrt{s^2+R_{\odot}^2-2\,s\,R_{\odot}\cos{\theta}}
\end{equation}
and we integrate over the halo density (squared) for the case of dark matter decays (annihilations). Here we assume a background that does not depend on galactic coordinates like the dominating background of atmospheric neutrinos and neglect the contribution of galactic neutrinos, which is subdominant in the case of muon neutrinos. The background is therefore simply proportional to the observed solid angle. We see clearly that for annihilating dark matter the best way to detect the signal is indeed looking towards the galactic center. The cone half angle offering the best signal-to-square root of background ratio varies depending on the cuspiness of the profile: for an NFW profile we find a value of $\sim0^\circ$, for an Einasto profile a value around $3^\circ$ and for an isothermal profile a value around $30^\circ$. Note that in any case the gain of looking at the galactic center is not very large for a cored profile like the isothermal one.

In contrast, for the case of a decaying dark matter particle like the unstable gravitino, the best strategy is to measure the full-sky signal and not to concentrate on the region around the galactic center. In fact the gain coming from the enhanced dark matter density is counteracted by the smallness of the collecting area and so the significance of the signal goes quickly to zero as a function of the angle for any profile, even for cuspy profiles like the NFW profile. The observation of only a fraction of the sky around the galactic center direction leads to an increase in the signal-to-background ratio, but not of the significance. We therefore conclude that for decaying dark matter there is no advantage in looking only at the galactic center. The full-sky signal offers not only better statistics, but also a higher significance.

Considering the directionality of the atmospheric background instead of the signal, another good strategy might be to exploit the fact that the flux from the zenith and nadir directions is (depending on the neutrino flavor and the energy) a few times smaller than from the horizontal direction. Assuming a signal that does not depend on the zenith angle, which is practically true for the signal from dark matter decays, the observation of only a fraction of the sky around the zenith (and nadir) direction is again clearly leading to an increase in the signal-to-background ratio. In the case of muon neutrinos it turns out that the best value for the significance is achieved for a full-sky observation. For electron neutrinos the difference in the background fluxes from the zenith and horizontal directions is larger and a slight increase of the statistical significance could be achieved. As mentioned before, the tau neutrino background is strongly suppressed in the zenith direction since it is only produced from oscillations (\textit{cf.} equation~(\ref{tauosc})). Thus a large increase of the statistical significance is achieved if only a fraction of the upper hemisphere is observed. We show the dependence on the maximal zenith angle for an exemplary neutrino energy of 100\,GeV in the right panel of Figure~\ref{directionfigure}.

We can therefore conclude that exploiting the directionality of the signal or background, apart from the case of specific flavors like tau neutrinos as discussed in~\cite{Covi:2008jy,Grefe:2008zz}, is not promising for the first detection of a decaying dark matter candidate like the unstable gravitino. The largest rate and significance is achieved for a full-sky search, and this is the option we will discuss in the following. On the other hand, as discussed in the beginning of this chapter, directionality offers a clear way to disentangle a signal of decaying dark matter from a signal of annihilating dark matter, where looking into the galactic center should give an increase in significance. In addition, directional observation allows to distinguish the dark matter signal from point sources like dwarf galaxies, pulsars and supernova remnants.

\subsection{Neutrino and Muon Spectra}
\label{muonpectra}

\begin{figure}[t]
 \centering
 \includegraphics[scale=0.9,bb=0 11 500 300]{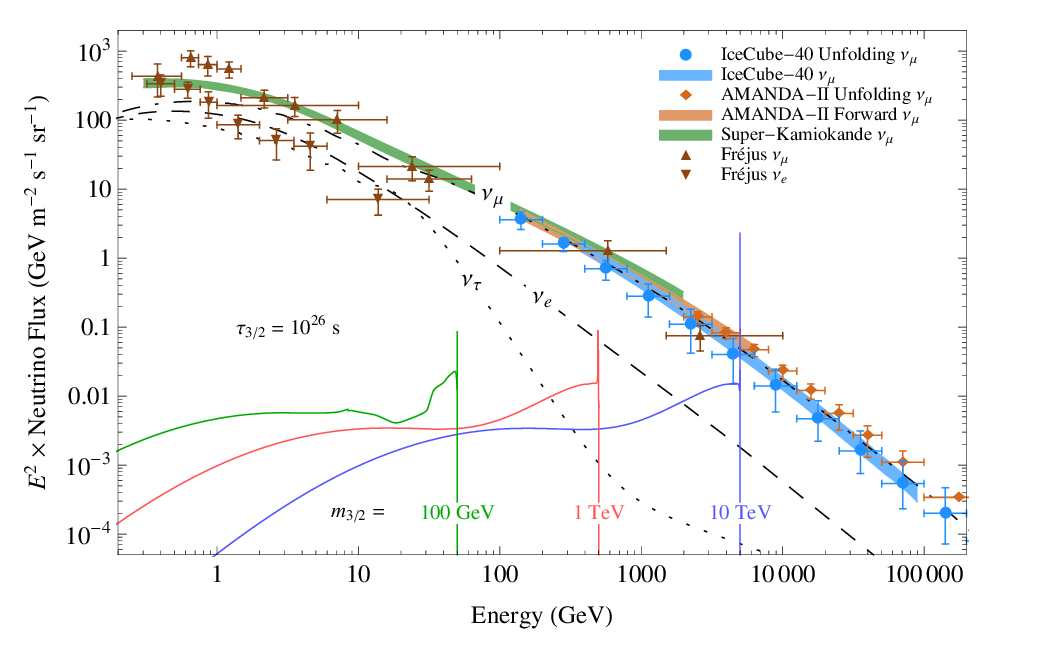}
 \caption[Expected neutrino spectrum from the decay of gravitino dark matter compared to the expected background of atmospheric neutrinos and data from neutrino experiments.]{Expected combined spectrum of neutrinos and antineutrinos from the decay of gravitino dark matter compared to the combined background of atmospheric neutrinos and antineutrinos and data of neutrino experiments. The expected flux is presented for a gravitino lifetime of $10^{26}\,$s and gravitino masses of 100\,GeV, 1\,TeV and 10\,TeV. Clearly visible are the monochromatic lines at the end of the spectra coming from the two-body decay channels with a neutrino in the final state. Due to the steeply falling atmospheric background the signal-to-background ratio at the endpoint of the decay spectrum increases significantly for larger gravitino masses if the lifetime is kept fixed.}
 \label{spectra-plot}
\end{figure}
The contributions to the neutrino flux from gravitino decays at extragalactic distances and in the Milky Way halo are calculated according to equations~(\ref{EGflux}) and~(\ref{haloflux}) using the neutrino spectra presented in Section~\ref{gravspectra}. The result after propagation to the Earth is shown in Figure~\ref{spectra-plot} together with the expected atmospheric background as simulated by Honda \textit{et al.}~\cite{Honda:2006qj} and the data measured by the Fr\'ejus~\cite{Daum:1994bf}, Super-Kamiokande~\cite{GonzalezGarcia:2006ay}, AMANDA-II~\cite{Collaboration:2009nf,Abbasi:2010qv} and IceCube~\cite{Abbasi:2010ie,Abbasi:2011jx} experiments. We see that for a lifetime of the order of $10^{26}$\,s the signal always lies below the measured background of muon neutrinos. The best signal-to-background ratio in the neutrino channel is achieved for the high-energy end of the spectrum, which contains information about the gravitino mass.

The neutrino spectra shown in Figure~\ref{spectra-plot} look rather distinctive, and an interesting question is whether they can be observed in a neutrino detector. In particular it is interesting to see if the neutrino line leads to an observable signal. However, we have to consider that neutrino detectors do not really measure neutrinos directly, but only the corresponding charged leptons or showers produced in interactions of neutrinos with the matter inside and around the detector.

\subsubsection{Neutrino Interactions}

Since we are mainly interested in neutrino energies much larger than nucleon masses we only take into account deep inelastic neutrino--nucleon scattering. Neutrino--electron elastic scattering is subdominant in this energy range and will be neglected~\cite{Strumia:2006db}.

The cross sections for deep inelastic scattering of (anti)neutrinos off nucleons at rest are given by~\cite{Strumia:2006db,Barger:2007xf}
\begin{equation}
 \begin{split}
  \frac{d\sigma_{\text{CC/NC}}^{\nu\, p,n}(E_{\nu}, y)}{dy} &\simeq\frac{2\,m_{p,n}\,G_F^2}{\pi}\,E_{\nu}\left( a_{\text{CC/NC}}^{\nu\,p,n}+b_{\text{CC/NC}}^{\nu\,p,n}\,(1-y)^2\right) \\
  &\simeq 3.2\times 10^{-38}\,\frac{\text{cm}^2}{\text{GeV}}\,E_{\nu}\left( a_{\text{CC/NC}}^{\nu\,p,n}+b_{\text{CC/NC}}^{\nu\,p,n}\,(1-y)^2\right) \\
 \end{split}
 \label{crosssection}
\end{equation}
with $a_{\text{CC}}^{\nu\,p,n}=0.15,\,0.25$, $b_{\text{CC}}^{\nu\,p,n}=0.04,\,0.06$ and $a_{\text{CC}}^{\bar{\nu}\,p,n}=b_{\text{CC}}^{\nu\,n,p}$, $b_{\text{CC}}^{\bar{\nu}\,p,n}=a_{\text{CC}}^{\nu\,n,p}$ for charged-current interactions, and $a_{\text{NC}}^{\nu\,p,n}=0.058,\,0.064$, $b_{\text{NC}}^{\nu\,p,n}=0.022,\,0.019$ and $a_{\text{NC}}^{\bar{\nu}\,p,n}=b_{\text{NC}}^{\nu\,p,n}$, $b_{\text{NC}}^{\bar{\nu}\,p,n}=a_{\text{NC}}^{\nu\,p,n}$ for neutral-current interactions. The inelasticity $y$ is given by
\begin{equation}
 y=1-\frac{E_\ell}{E_{\nu}}\qquad\text{or}\qquad y\simeq \frac{E_{\text{had}}}{E_{\nu}}\,,
\end{equation}
where $E_\ell$ is the energy of the generated lepton and $E_{\text{had}}$ is the energy of the generated hadronic shower. For the total neutrino--nucleon cross sections one obtains
\begin{equation}
 \begin{split}
  \sigma_{\text{CC/NC}}^{\nu\, p,n}(E_{\nu}) &\simeq\frac{2\,m_{p,n}G_F^2}{\pi}\,E_{\nu}\left( a_{\text{CC/NC}}^{\nu\,p,n}+\frac{1}{3}\,b_{\text{CC/NC}}^{\nu\,p,n}\right) .
 \end{split}
 \label{totalcross}
\end{equation}
As we can see, the total cross section is proportional to the energy of the incoming neutrino in the considered energy range. However, equation~(\ref{crosssection}) in principle holds only for neutrino energies up to the TeV region where the effect of the massive gauge boson propagators becomes relevant. Nevertheless, we will employ this parametrization, keeping in mind that at higher energies the cross-sections might be overestimated~\cite{Strumia:2006db}.

\subsubsection{Muon Neutrinos}

The charged-current deep inelastic scattering of a muon neutrino off a nucleus produces a hadronic shower and a muon. These track-like events can be clearly identified in Cherenkov detectors via the Cherenkov light cone of the relativistic muon.

\subsubsection*{Through-Going Muons}

\begin{figure}[t]
 \centering
 \includegraphics[scale=0.26]{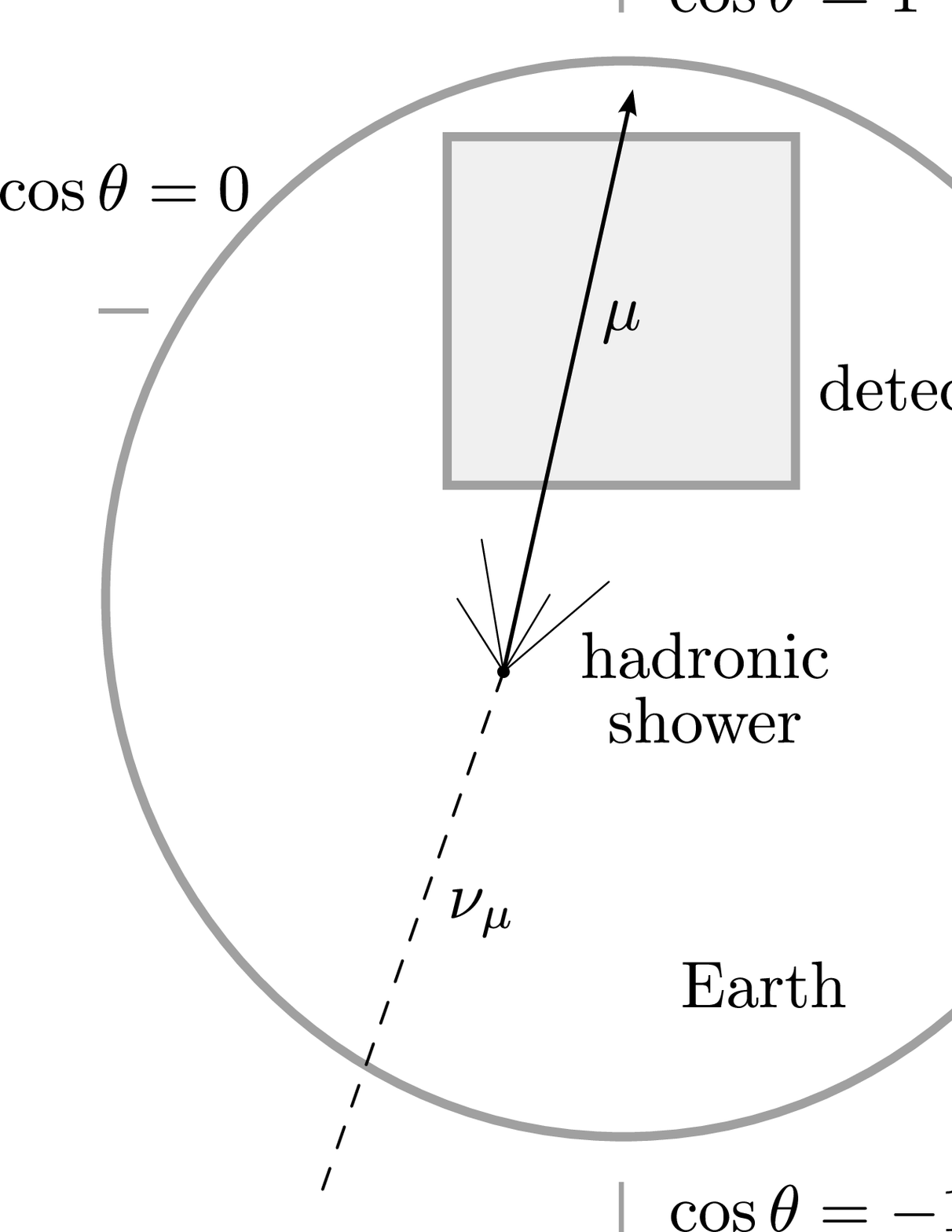}
 \hfill
 \includegraphics[scale=0.44]{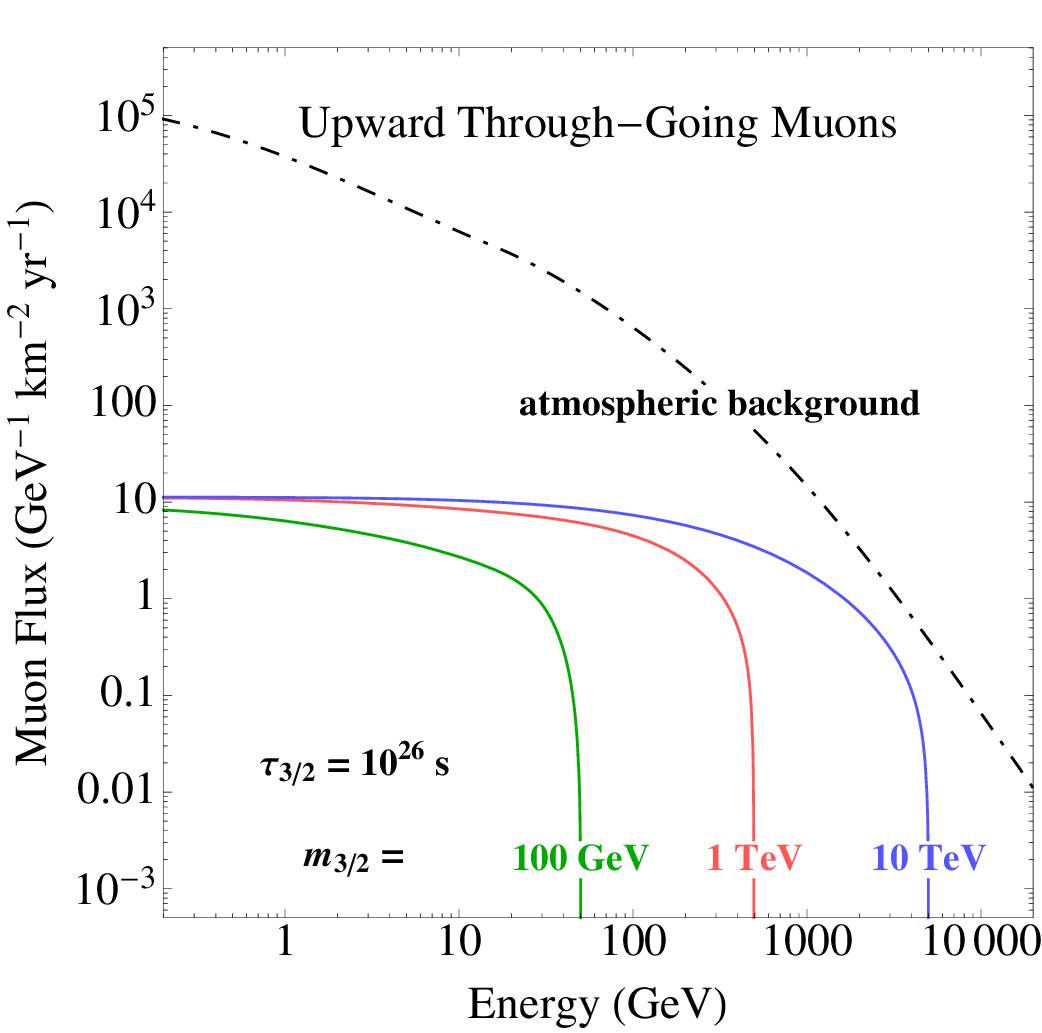}
 \caption[Event topology of muon neutrino-induced upward through-going muons and the expected flux of upward through-going muons from gravitino dark matter decays compared to the atmospheric background.]{\textit{Left:} Event topology of muon neutrino-induced upward through-going muons. A muon neutrino that traversed the Earth scatters in a charged-current interaction on matter surrounding the detector, thus producing a muon that penetrates the detector. \textit{Right:} Muon fluxes induced by muon neutrinos from gravitino dark matter decays compared to the atmospheric background for upward through-going muons in a neutrino detector surrounded by standard rock. The flux is computed for gravitino masses of 100\,GeV, 1\,TeV and 10\,TeV, and a lifetime of $10^{26}$\,s, corresponding to the neutrino spectra shown in Figure~\ref{spectra-plot}.}
 \label{SK-Mu-spectra}
\end{figure}

Since muons are rather long-lived ($c\tau_\mu=658.650\,$m), their range is only limited by energy loss during their passage through matter and not by their lifetime. Therefore Cherenkov detectors can also observe muons that are generated in the surrounding material of the detector (see left panel of Figure~\ref{SK-Mu-spectra}). This effect enhances the effective detector area for high-energetic muon neutrinos.

The average rate of muon energy loss can be written as~\cite{Nakamura:2010zzi}
\begin{equation}
 -\frac{dE_{\mu}}{dx}=\alpha(E_{\mu})+\beta(E_{\mu})\,E_{\mu}\,,
 \label{energyloss}
\end{equation}
where $\alpha(E_{\mu})$ describes the ionization energy loss and $\beta(E_{\mu})$ takes into account the energy loss due to radiative processes: $e^+e^-$ pair production, bremsstrahlung and photonuclear contributions. Both $\alpha(E_{\mu})$ and $\beta(E_{\mu})$ are slowly varying functions of the muon energy. As long as we can approximate $\alpha$ and $\beta$ as energy-independent, the range after which the muon energy drops below a threshold energy $E_{\mu}^{th}$ is given by
\begin{equation}
 R_{\mu}(E_{\mu},E_{\mu}^{\text{th}})=\frac{1}{\rho\,\beta}\,\ln\left[ \frac{\alpha+\beta E_{\mu}}{\alpha+\beta E_{\mu}^{\text{th}}}\right] ,
 \label{muonrange}
\end{equation}
where $\rho$ is the density of the medium. Important materials in connection with Cherenkov neutrino detectors are standard rock, water and ice. The relevant parameters for these materials are given in Table~\ref{materials}.
\begin{table}
 \centering
 \begin{tabular}{lcccc}
  \hline
  material & $\rho$ (g/cm$^3$) & $\left\langle Z/A\right\rangle $ & $\alpha$ (GeV\,cm$^2$/g) & $\beta$ (cm$^2$/g) \\
  \hline
  standard rock & 2.650 & 0.5 & $2.3\times 10^{-3}$ & $4.4\times 10^{-6}$ \\
  water & 1.000 & 0.55509 & $2.7\times 10^{-3}$ & $3.3\times 10^{-6}$ \\
  ice & 0.918 & 0.55509 & $2.7\times 10^{-3}$ & $3.3\times 10^{-6}$ \\
  \hline
 \end{tabular}
 \caption[Density, proton-number-to-mass-number ratio and approximate muon energy loss parameters for materials of interest in Cherenkov neutrino detectors.]{Density, proton-number-to-mass-number ratio and approximate muon energy loss parameters for materials of interest in Cherenkov neutrino detectors. The values of the density and the average proton-number-to-mass-number ratio are taken from~\cite{Lohmann:1985qg,Groom:2001kq}. The muon energy loss parameters $\alpha$ and $\beta$ are best-fit values from a fit of equation~(\ref{muonrange}) to the tabulated data in~\cite{Lohmann:1985qg,Groom:2001kq}.}
 \label{materials}
\end{table}
In fact, equation~(\ref{energyloss}) only describes the average muon energy loss and does not account for the stochastic nature of radiative muon energy loss processes which start to dominate at TeV energies ($E_\mu>\alpha/\beta$). Nonetheless, we will employ equation~(\ref{muonrange}) as a useful approximation for the muon range that allows us to determine the initial muon energy as an explicit function of the final muon energy and the muon range:
\begin{equation}
 E_{\mu}^0(E_{\mu},R_\mu)=E_{\mu}\e^{\rho\,\beta\,R_\mu}+\frac{\alpha}{\beta}\left( \e^{\rho\,\beta\,R_\mu}-1\right) .
 \label{muonenergy}
\end{equation}

The rate of muon neutrino-induced through-going muon events is given by
\begin{equation}
 \begin{split}
  \frac{dN}{dt}= &\int d\Omega\int_0^{\infty} dE_{\nu_{\mu}}\, \frac{d\Phi_{\nu_{\mu}}(E_{\nu_{\mu}},\theta,\phi)}{dE_{\nu_{\mu}}}\,A^{\text{eff}}_{\nu_{\mu}}(E_{\nu_{\mu}},\theta,\phi) \\
  = &\int d\Omega\int_{E_{\mu}^{\text{th}}}^{\infty} dE_{\nu_{\mu}}\int_{E_{\mu}^{\text{th}}}^{E_{\nu}}dE_{\mu}\, \frac{d\Phi_{\nu_{\mu}}(E_{\nu_{\mu}},\theta,\phi)}{dE_{\nu_{\mu}}}\left[ \frac{d\sigma_{\text{CC}}^{\nu p}(E_{\nu_{\mu}},E_{\mu})}{dE_{\mu}}\,n_p+(p\rightarrow n)\right] \\
  &\qquad\qquad\times R_{\mu}(E_{\mu},E_{\mu}^{\text{th}})\,A^{\text{eff}}_{\mu}(E_{\mu},\theta,\phi)\,\e^{-\sigma^{\nu N}(E_{\nu_{\mu}})\,n_N\,L(\theta)}+(\nu_\mu\rightarrow\bar{\nu}_\mu)\,,
 \end{split}
 \label{through-rate}
\end{equation}
where the number density of protons is given by $n_p=\rho\,M_p^{-1}\,N_A\left\langle Z/A\right\rangle $ and the density of neutrons by $n_n=\rho\,M_n^{-1}\,N_A(1-\left\langle Z/A\right\rangle )$. $N_A=6.022\times 10^{23}\,$mol$^{-1}$ is the Avogadro constant, $M_p\simeq M_n\simeq1\,$g\,mol$^{-1}$ is the molar mass of protons and neutrons, $\rho$ is the density of the material, and $\left\langle Z/A\right\rangle $ is the average ratio of the proton number and the mass number of the material as given in Table~\ref{materials}. Due to the small neutrino--nucleon cross-section the effect of the attenuation term that accounts for the absorption of part of the signal and background neutrino fluxes during the passage of the Earth is negligible in the considered energy range. However, since the neutrino--nucleon cross section rises with increasing neutrino energy, this effect becomes non-negligible at neutrino energies above 10\,TeV.

The muon neutrino effective area $A^{\text{eff}}_{\nu_{\mu}}$ is defined as the ratio of the rate of reconstructed events and the incident neutrino flux. It is calculated using Monte Carlo methods and incorporates the attenuation of the neutrino flux during the passage of the Earth, the neutrino--nucleon cross section, the range of the generated muon and the reconstruction and selection efficiencies. This effective area is usually provided by the experimental collaborations. The energy dependence of the neutrino effective area comes mainly from the energy dependence of the cross section (roughly $\propto E_{\nu}$) and the increase of the muon range with rising energy (roughly $\propto E_{\nu}$ for $E_{\nu}\ll\alpha/\beta\sim\mathcal{O}(1)\,$TeV and $\propto \ln E_{\nu}$ for $E_{\nu}\gg\mathcal{O}(1)\,$TeV). Notice that the muon effective area $A^{\text{eff}}_{\mu}$, on the other hand, is defined as the ratio of the rate of reconstructed events and the incident muon flux. This area incorporates only the geometry of the detector and the detection efficiency. It is roughly equal to the geometrical area but might have a slight energy dependence.

For the calculation of the spectrum of muon neutrino induced muons at the detector position we have to take into account the shift to lower energies due to the energy loss during muon propagation through matter~\cite{Erkoca:2009by}:
\begin{equation}
 \begin{split}
  \frac{d\phi_{\mu}}{dE_{\mu}}= &\int d\Omega\int_0^{\infty}dR_\mu\int_{E_{\mu}^0}^{\infty} dE_{\nu_{\mu}}\,\e^{\rho\,\beta\,R_\mu}\, \frac{d\Phi(E_{\nu_{\mu}},\theta,\phi)}{dE_{\nu_{\mu}}} \\
  &\qquad\times\left[ \frac{d\sigma_{\text{CC}}^{\nu p}(E_{\nu_{\mu}},E_{\mu}^0)}{dE_{\mu}^0}\,n_p+(p\rightarrow n)\right] _{E_{\mu}^0=E_{\mu}^0(E_{\mu},R_\mu)}+(\nu_\mu\rightarrow\bar{\nu}_\mu)\,,
 \end{split}
 \label{through-spec}
\end{equation}
where we neglected the attenuation term. In this expression the initial muon energy enters as an explicit function of the final muon energy as given by equation~(\ref{muonenergy}).

Using equation~(\ref{through-spec}) we calculate the flux of through-going muons induced by neutrinos from gravitino dark matter decay and show the results in the right panel of Figure~\ref{SK-Mu-spectra} for the case of a detector surrounded by standard rock like Super-Kamiokande. However, the result is also applicable for the case of detectors surrounded by water or ice since the dependence on the density of the material cancels in equation~(\ref{through-spec}) and the muon energy loss parameters are roughly similar for the different materials (\textit{cf.} Table~\ref{materials}). Since there is no possibility to veto for the overwhelming background of atmospheric muons, only up-going events and therefore a solid angle of $2\pi$ can be used for the analysis. We observe that the deep inelastic scattering transforms the incident neutrino spectrum into a softer muon spectrum. In addition, the energy loss in the muon propagation smooths out the spectrum making the edge corresponding to half the gravitino mass less clear.

\subsubsection*{Contained Muons}

\begin{figure}[t]
 \centering
 \includegraphics[scale=0.26]{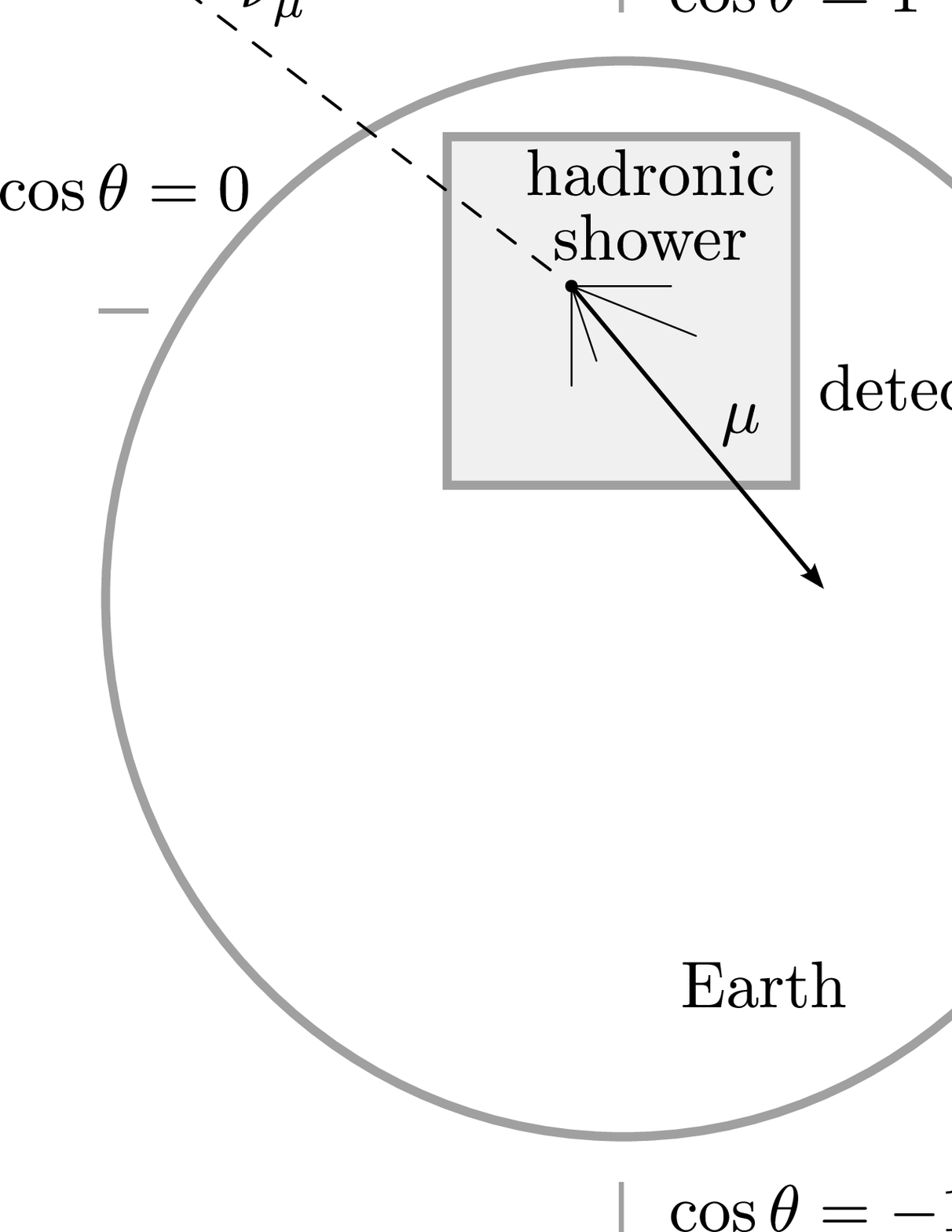}
 \hfill
 \includegraphics[scale=0.44]{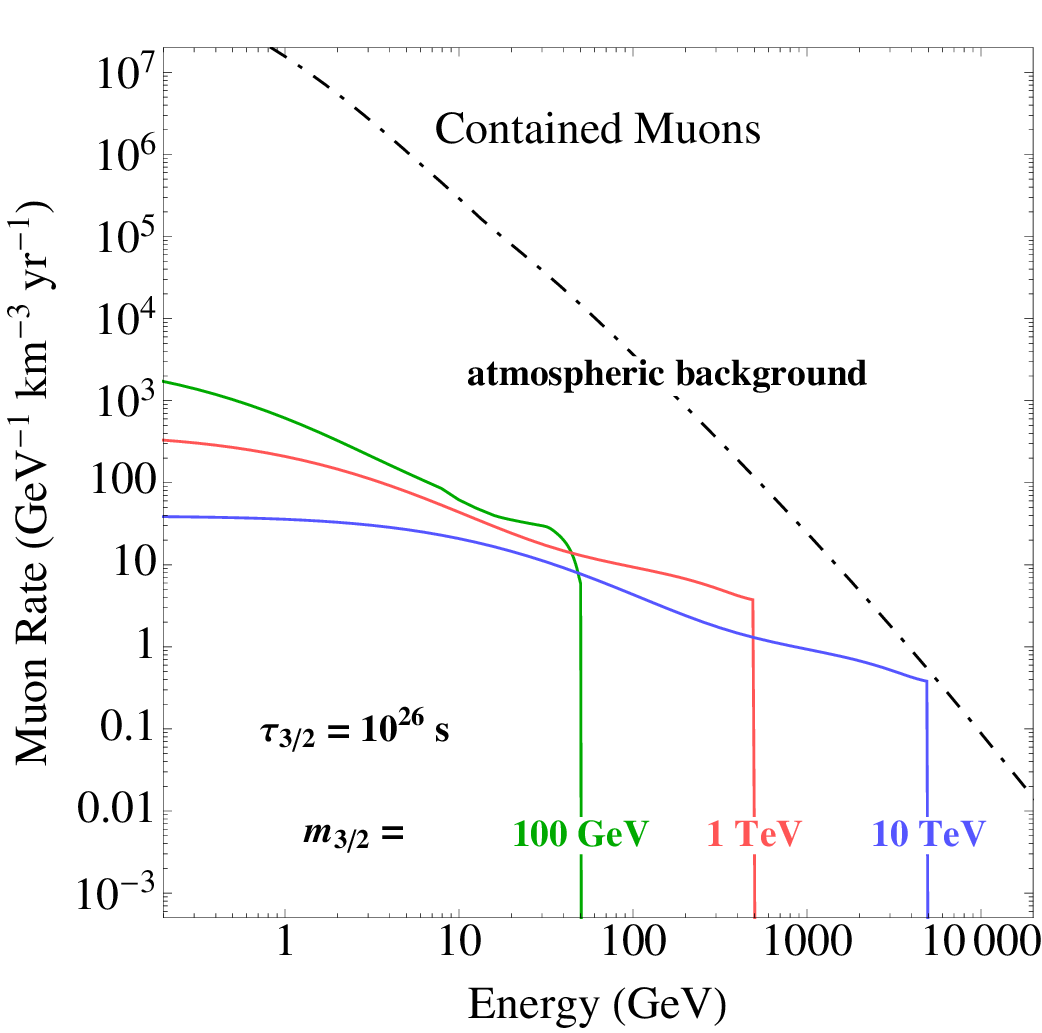}
 \caption[Event topology of muon neutrino-induced contained muons and the expected spectrum of contained muons from gravitino dark matter decays compared to the atmospheric background.]{\textit{Left:} Event topology of muon neutrino-induced contained muons. A muon neutrino scatters in a charged-current interaction on matter inside the detector, thus producing a muon that traverses the detector. \textit{Right:} Spectrum of contained muons induced by muon neutrinos from gravitino dark matter decays compared to the atmospheric background. The event rate per cubic kilometer of detector volume (filled with ice) is computed for gravitino masses of 100\,GeV, 1\,TeV and 10\,TeV, and a lifetime of $10^{26}$\,s, corresponding to the neutrino spectra shown in Figure~\ref{spectra-plot}.}
 \label{Mu-cont-spectra}
\end{figure}

These events are similar to through-going muons but in this case the neutrino--nucleon interaction takes place inside the instrumented volume (see left panel of Figure~\ref{Mu-cont-spectra}). If the muon track ends inside the detector the events are called contained. If the muon track leaves the detector one speaks of a partially contained event. The rate of muon neutrino induced contained track-like events per unit detector volume is given by\footnote{We do not discriminate between partially and completely contained events in this analysis.}
\begin{equation}
 \begin{split}
  \frac{dN}{dE_{\mu}\,dVdt} =&\int d\Omega\int_{E_{\mu}}^{\infty} dE_{\nu_{\mu}}\, \frac{d\Phi_{\nu_{\mu}}(E_{\nu_{\mu}},\theta,\phi)}{dE_{\nu_{\mu}}}\left[ \frac{d\sigma_{\text{CC}}^{\nu p}(E_{\nu_{\mu}},E_{\mu})}{dE_{\mu}}\,n_p+(p\rightarrow n)\right] \\
  &\qquad\quad\qquad+(\nu_\mu\rightarrow\bar{\nu}_\mu)\,,
 \end{split}
 \label{contained-rate}
\end{equation}
where we also neglected the attenuation term. In this case also the hadronic cascade is contained in the detector volume and therefore, by measuring the energy of the muon as well as of the hadronic cascade, it is in principle possible to reconstruct the total energy of the incident muon neutrino. In this case, however, one has to rely on the detection also of the hadronic cascade which, as we will discuss later, seems to be challenging. 
The effective volume of the detector for contained events corresponds roughly to the geometrical volume (apart from boundary effects and reconstruction efficiency) and it is not enhanced by the muon range, which as we have seen, grows as $E_\nu$. Therefore, the statistics for contained events is lower than for through-going events at large energies. For instance in the case of Super-Kamiokande the event rate above roughly 10\,GeV is dominated by through-going muons. On the other hand, in the energy range of interest for dark matter searches the muon range is of the order of one kilometer and therefore the expected rate of contained muons is comparable to the rate of through-going muons in detectors of cubic kilometer size. Thus these contained events might be equally important for dark matter searches at the new generation of neutrino telescopes. 

In addition, for downgoing contained muon events there is the interesting possibility to reduce the background of atmospheric muon neutrinos by the detection of a coincident muon that was produced in the same parent meson decay~\cite{Schonert:2008is}. This strategy could be used to increase the signal-to-background ratio for this channel, especially at large energies. However, we will not discuss this strategy quantitatively in this work. 

In the right panel of Figure~\ref{Mu-cont-spectra} we show the muon spectra for contained events calculated using equation~(\ref{contained-rate}) for the case of a detector volume filled with ice as in the case of the IceCube experiment. The result for a volume of water can easily be obtained by rescaling the event rate with the slightly higher density of water. In contrast to through-going muons there is no smoothing due to muon energy loss and the edge of the spectrum is clearer.

Here we only discussed the case where only the muon is measured since this is what can be done by the experiments at the moment. If the hadronic shower is also measured the total energy of the incident muon neutrino could be reconstructed and therefore in principle the complete spectral information of the neutrino flux as shown in Figure~\ref{spectra-plot} would be available for analyses. This is similar to the case of electron and tau neutrinos that is discussed in the next section. However, as will be discussed there, that channel offers a better signal-to-background ratio and a better energy resolution and will therefore be of more interest once the showers can be measured and used for analyses.

\subsubsection{Electron and Tau Neutrinos}
\label{showers}

\begin{figure}[t]
 \centering
 \includegraphics[scale=0.26]{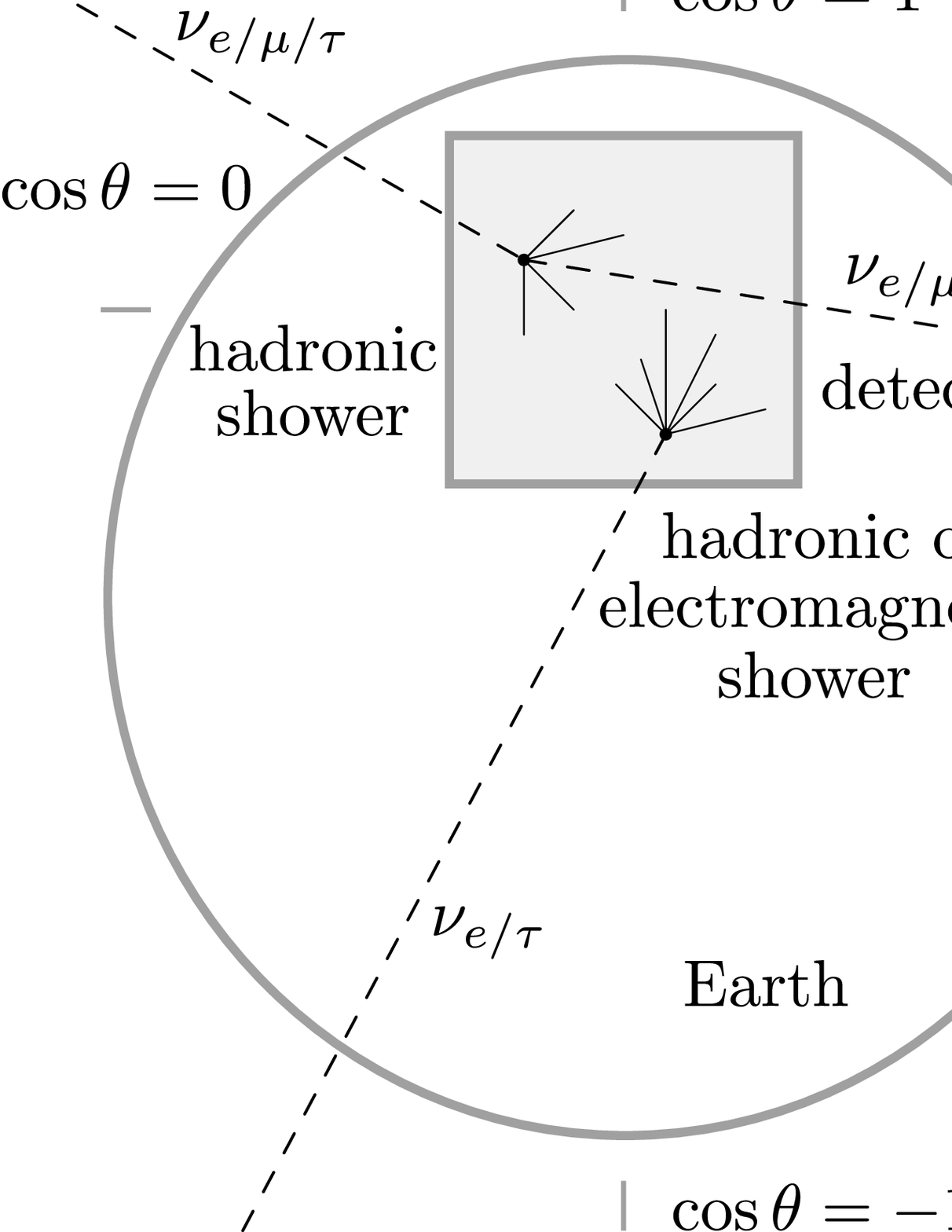}
 \hfill
 \includegraphics[scale=0.44]{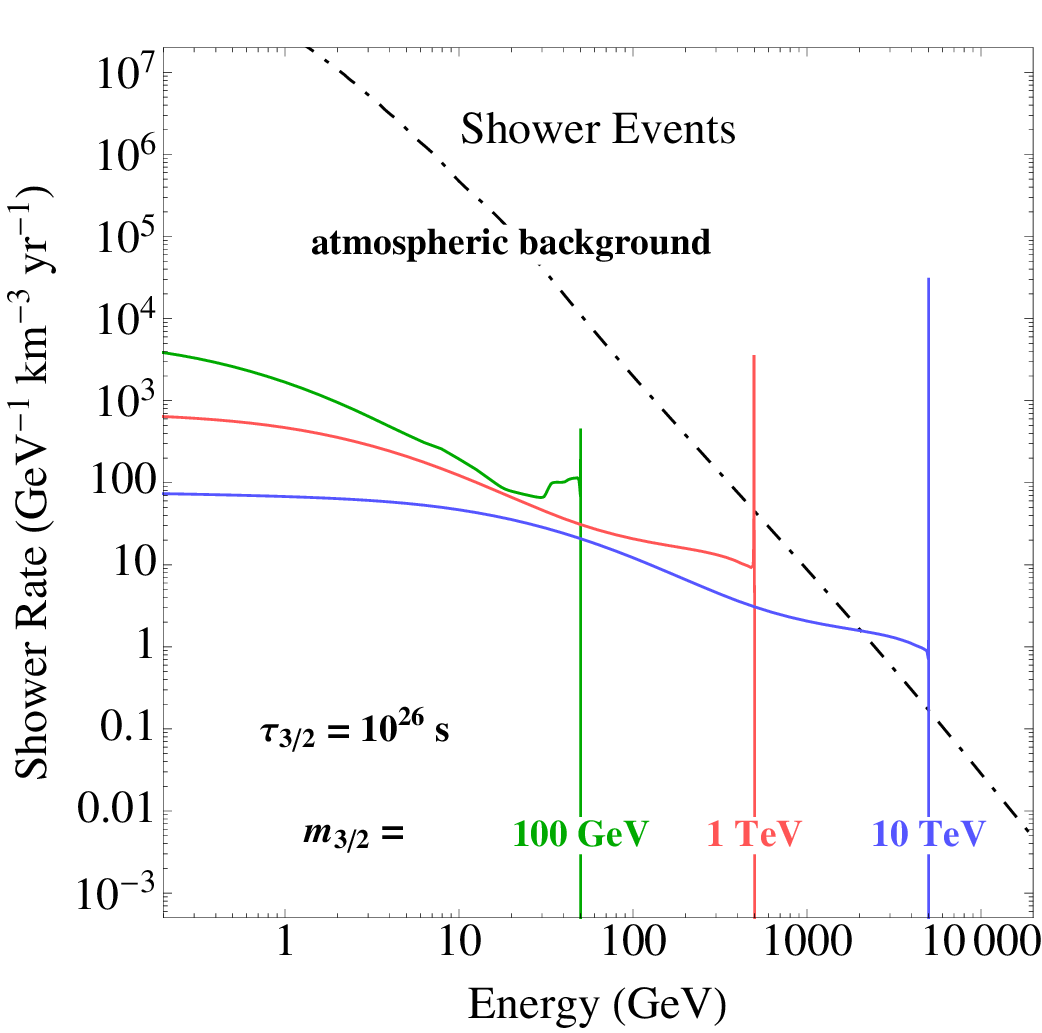}
 \caption[Event topology of neutrino-induced cascades and the expected spectrum of cascades from gravitino dark matter decays compared to the atmospheric background.]{\textit{Left:} Event topology of neutrino-induced cascades. A neutrino of arbitrary flavor scatters in a neutral-current interaction on matter inside the detector, thus producing a hadronic shower (upper track). In addition, electron and tau neutrinos scattering in charged-current interactions on matter inside the detector produce hadronic and electromagnetic showers (lower track). \textit{Right:} Spectrum of shower events for neutrinos from gravitino dark matter decays compared to the atmospheric background. The event rate per cubic kilometer of detector volume (filled with ice) is computed for gravitino masses of 100\,GeV, 1\,TeV and 10\,TeV, and a lifetime of $10^{26}$\,s, corresponding to the neutrino spectra shown in Figure~\ref{spectra-plot}.}
 \label{IC-Shower-spectra}
\end{figure}

The charged-current deep inelastic scattering of an electron neutrino off a nucleus produces a hadronic shower and an electron that immediately causes an electromagnetic shower. The charged-current deep inelastic scattering of a tau neutrino off a nucleus produces a hadronic shower and a tau lepton. Due to the short lifetime of the tau lepton ($c\tau_\tau=87.11\,\mu$m), at GeV to TeV energies it decays almost instantly and produces another shower at the interaction point. Thus, at energies below many TeV, neutrino detectors like IceCube cannot distinguish electron neutrino from tau neutrino events since both types produce similar showers in the detector~\cite{Cowen:2007ny} (see left panel of Figure~\ref{IC-Shower-spectra}). In these cases, however, the whole neutrino energy is deposited in the detector and therefore it may be possible in principle to reconstruct better the initial neutrino spectrum. On the other hand, the analysis for cascade-like events is much more difficult than the analysis for muon tracks~\cite{D'Agostino:2009sj}. Recently there has been a first study by the IceCube collaboration~\cite{Abbasi:2011ui} searching for atmospheric and extraterrestrial neutrino-induced cascades. Only few candidate events at energies above several TeV have been found so far, but the situation will improve in the future with refined analysis methods and higher statistics of recorded events. However, at this moment we cannot realistically estimate the sensitivity in shower events that finally will be reached and therefore we study this channel assuming an ideal detector.

Shower-like events are also characteristic of the neutrino--nucleon neutral-current interaction and for this reason probably only a combined analysis of neutral-current interactions for all neutrino flavors and charged-current interactions for tau and electron neutrinos will be feasible (see left panel of Figure~\ref{IC-Shower-spectra}). The total rate of neutrino-induced shower-like events per unit detector volume is given by
\begin{align}
 \frac{dN}{dE_{\text{shower}}\,dVdt}= &\int d\Omega\,\Bigg\lbrace \sum_{\ell\,=\,e,\tau}\left( \frac{d\Phi_{\nu_\ell}(E_{\nu_\ell},\theta,\phi)}{dE_{\nu_\ell}}\left[ \sigma_{\text{CC}}^{\nu p}(E_{\nu_\ell})\,n_p+(p\rightarrow n)\right] \right) _{E_{\nu_\ell}=E_{\text{shower}}} \nonumber\\
 &\qquad+\sum_{\ell\,=\,e,\mu,\tau}\,\int_{E_{\text{shower}}}^{\infty} \!\!\!\!dE_{\nu_\ell}\, \frac{d\Phi_{\nu_\ell}(E_{\nu_\ell},\theta,\phi)}{dE_{\nu_\ell}} \label{cascade-rate}\\
 &\qquad\qquad\times\left[ \frac{d\sigma_{\text{NC}}^{\nu p}(E_{\nu_\ell},E_{\text{shower}})}{dE_{\text{shower}}}\,n_p+(p\rightarrow n)\right] \Bigg\rbrace +(\nu_\ell\rightarrow\bar{\nu}_\ell)\,.\nonumber
\end{align}

We give in the right panel of Figure~\ref{IC-Shower-spectra} the signal and atmospheric background spectra calculated from equation~(\ref{cascade-rate}) for a detector volume filled with ice as in the case of the IceCube experiment. Note that in this case the muon neutrinos contribute only via their neutral-current interactions which are weaker by a factor of about three compared to charged-current interactions (\textit{cf.} equation~(\ref{totalcross})). Still, since the atmospheric muon neutrino flux is a factor of 20 larger than the electron neutrino flux at TeV energies, the atmospheric muon neutrinos provide the dominant background. At the same time the signal rate is increased by roughly a factor of three. This is because, due to neutrino oscillations, the signal is roughly equal in all neutrino flavors and, therefore, the signal rate from the charged-current interactions of electron and tau neutrinos is the same as for the muon neutrinos. In addition, the combined neutral-current signal of all flavors contributes at the same level as the charged-current signal of one flavor. In summary, cascade-like events will offer a signal-to-background ratio that is roughly one order of magnitude larger than in the case of through-going muons and, therefore, they appear to be a very promising channel, once they can be measured.

\subsection{Rates and Bounds}
\label{Bounds}

In this section we want to discuss the rate of events from gravitino dark matter decay expected in neutrino detectors and what bounds on the gravitino lifetime can be expected from forthcoming neutrino experiments.

\subsubsection*{Current Bounds from Super-Kamiokande and IceCube}

The collaborations of various neutrino experiments have performed searches for dark matter signals using neutrino data~\cite{Ackermann:2005fr,Achterberg:2006jf,Landsman:2006mc,Lim:2009jy,Braun:2009fr,Danninger:2009uf,Rott:2009hr,Heros:2010ss}. However, all of the current studies focus on annihilation signals of WIMP dark matter, in particular neutralinos and the lightest Kaluza--Klein particle. In general, the expected neutrino spectrum from annihilations as well as the angular distribution of the signal enter explicitly in the derived exclusion limits. Therefore, it is difficult to apply these limits to the case of unstable gravitino dark matter.

Nevertheless, there are two exclusion limits from Super-Kamiokande and IceCube that we will employ to derive a bound on the gravitino lifetime.

\paragraph{Super-Kamiokande}

Super-Kamiokande is a 50\,kt water Cherenkov detector with a fiducial mass of 22.5\,kt. The muon effective area is 1200\,m$^2$ (with a slight zenith angle dependence due to the cylindrical shape of the detector) and is identical to the geometrical area since the reconstruction and selection efficiencies are virtually 100\,\%~\cite{Fukuda:1998ah}. Super-Kamiokande has been looking for a neutrino signal from WIMP annihilations in the center of the Sun, the center of the Earth and in the galactic center using 1679.6 days of data~\cite{Desai:2004pq}. No excess has been found so far, and this can also be used to obtain a constraint in the case of decaying dark matter.

\begin{figure}[t]
 \centering
 \includegraphics[scale=.44]{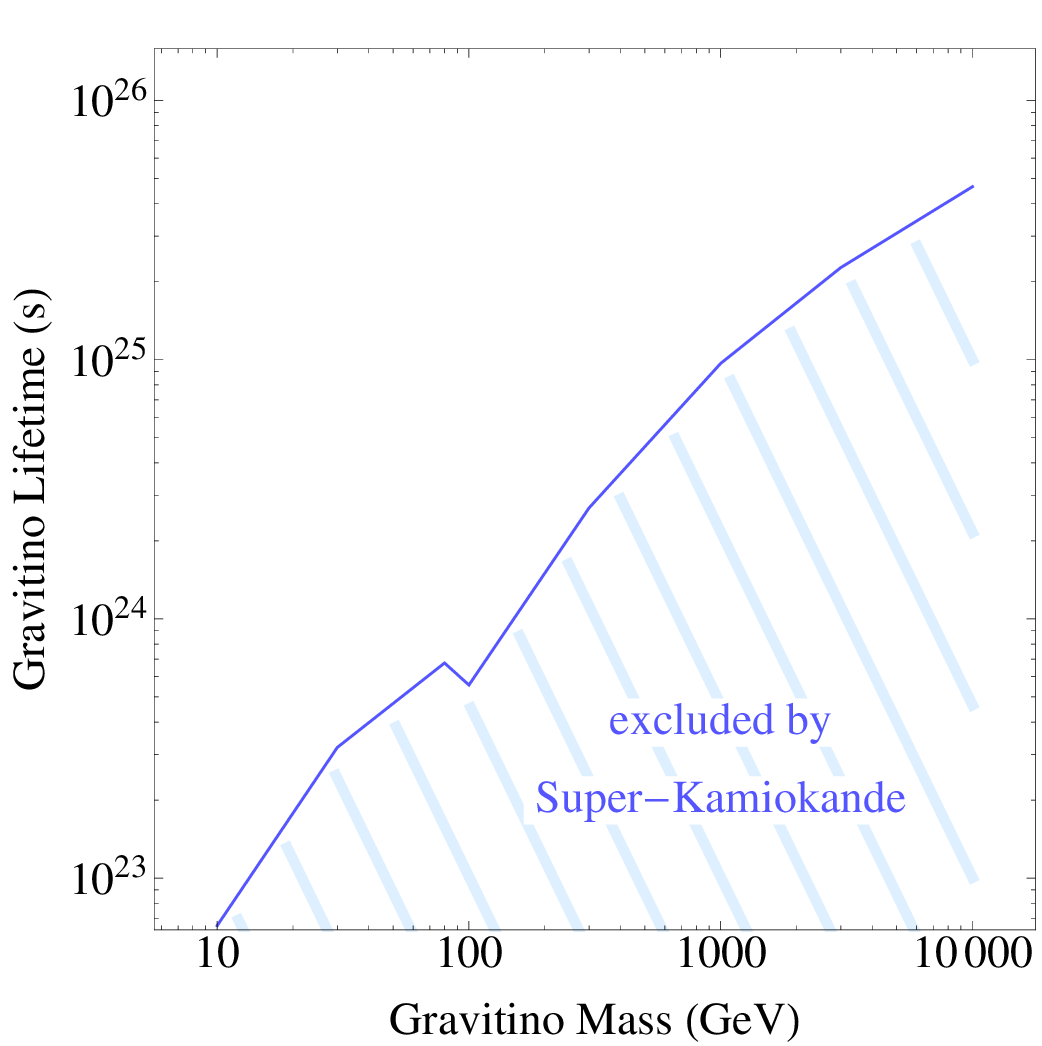}
 \includegraphics[scale=.44]{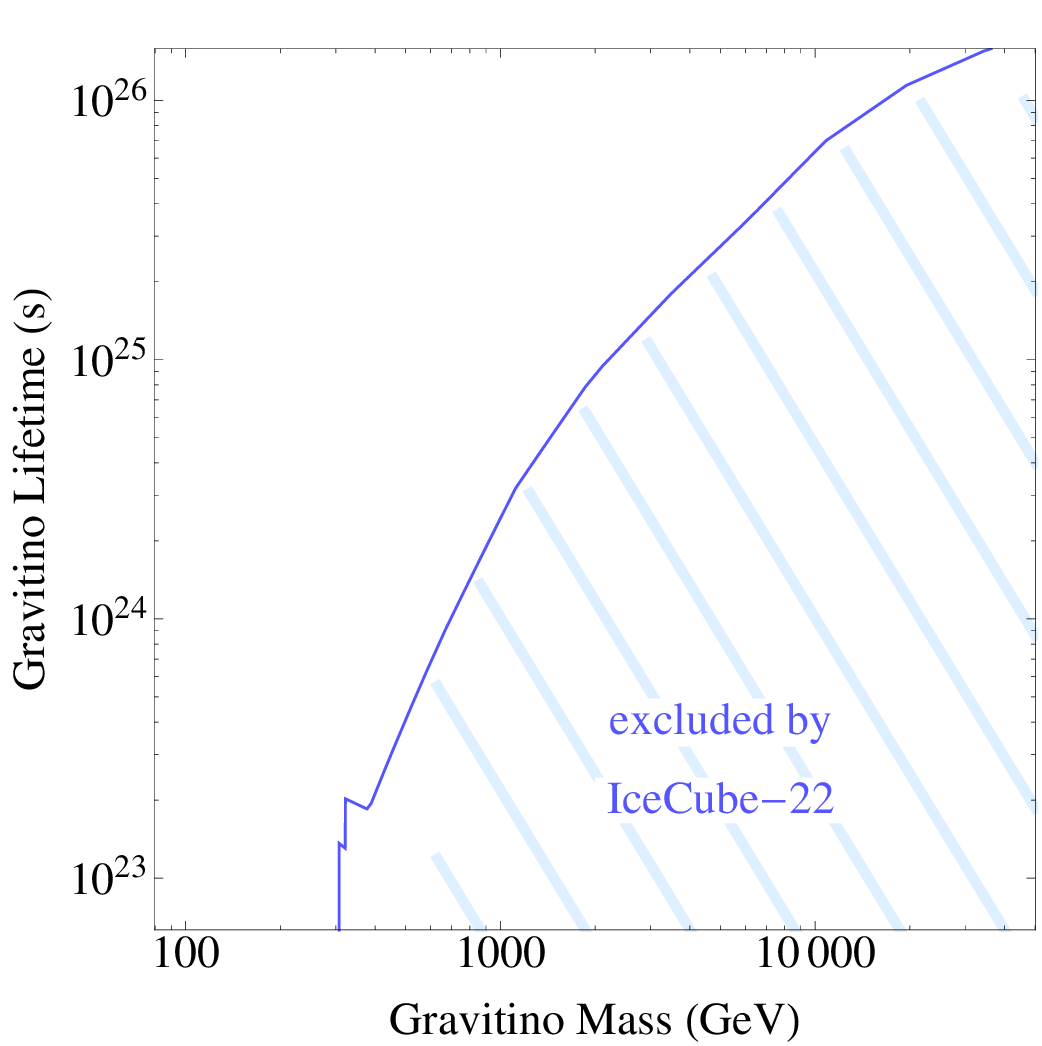}
 \caption[Present bounds on the gravitino lifetime from Super-Kamiokande and IceCube-22.]{\textit{Left:} 90\,\% C.L.\ exclusion region in the lifetime vs.\ mass plane for decaying gravitino dark matter from the non-observation of an excess in the Super-Kamiokande data (\textit{cf.} text). \textit{Right:} 90\,\% C.L.\ exclusion region in the lifetime vs.\ mass plane for decaying gravitino dark matter from a search for monochromatic neutrinos from the galactic halo with the IceCube-22 detector (\textit{cf.} text). Both bounds are stronger for larger gravitino masses since the integrated muon flux increases due to the increasing neutrino--nucleon cross section and the increasing muon range in spite of the decreasing neutrino flux.}
\label{SK-limit}
\end{figure}

We calculate the flux of upward through-going muons induced by muon neutrinos from gravitino dark matter decays according to a variation of equation~(\ref{through-rate}):\footnote{Since we are not interested in the spectrum of muons in this case, we do not need to perform an integration over the muon range as in equation~(\ref{through-spec}). Thus we can save one step of numerical integration while still obtaining the correct total muon flux.}
\begin{equation}
 \begin{split}
  \phi_{\mu}= &\int d\Omega\int_{E_{\mu}^{\text{th}}}^{\infty}dE_\mu\,R_\mu(E_\mu,E_{\mu}^{\text{th}})\int_{E_{\mu}}^{\infty}  dE_{\nu_{\mu}}\,\frac{d\Phi(E_{\nu_{\mu}},\theta,\phi)}{dE_{\nu_{\mu}}} \\ &\qquad\qquad\qquad\qquad\times\left[ \frac{d\sigma_{\text{CC}}^{\nu p}(E_{\nu_{\mu}},E_{\mu})}{dE_{\mu}}\,n_p+(p\rightarrow n)\right] +(\nu_\mu\rightarrow\bar{\nu}_\mu)\,.
 \end{split}
\end{equation}
The Super-Kamiokande detector is surrounded by rock and we employ a detection threshold of $E_\mu^{\text{th}}=1.6\,$GeV for upgoing muons~\cite{Desai:2004pq}. We compare the resulting flux with the 90\,\% C.L.\ flux limit of excess neutrino-induced upward through-going muons provided by the Super-Kamiokande collaboration for the galactic center~\cite{Desai:2004pq}.\footnote{The flux limits from the analyses of the directions towards the Sun and the Earth result in weaker limits for the case of decaying dark matter.} As discussed in Section~\ref{direction}, the strongest bounds are obtained for the largest field of view. The exclusion region in the gravitino parameter space derived from the limit in the $30^\circ$ half-angle cone around the galactic center is given in the left panel of Figure~\ref{SK-limit}.

\paragraph{IceCube}

The IceCube neutrino telescope is a Cherenkov detector consisting of 86 strings with optical sensors deployed at depths between 1.5\,km and 2.5\,km in the ice at the South Pole~\cite{Karle:2010xx}. With a surface area of one square kilometer the instrumented volume amounts to one cubic kilometer, a factor of 20\,000 larger than Super-Kamiokande. On the other hand, IceCube is much more loosely instrumented compared to Super-Kamiokande and thus has a higher detection threshold of $\mathcal{O}(100)\,$GeV.

A recent search for neutrino signals from dark matter in the galactic halo has been performed using data taken with the IceCube detector on 276 days during a stage of construction when it consisted of 22 strings~\cite{Abbasi:2011eq}. In this study the expected angular distribution of the dark matter signal from the galactic halo was used to constrain the dark matter contribution by comparing the observed event rates from two sky regions: one directed towards the galactic center and one directed away from the galactic center. An exclusion limit was also presented for the lifetime of a dark matter particle decaying into two monochromatic neutrinos.

We can use this limit to derive a conservative lower limit on the gravitino lifetime by taking into account only the neutrino-line contributions from gravitino decays:
\begin{equation}
 \begin{split}
  \tau_{3/2}(m_{3/2}) &\gtrsim\frac{1}{2}\Bigg( \BR\left( \gamma\,\nu_i\right) \times\tau_{\nu\nu}(m_{3/2})+\BR\left( Z\nu_i\right) \times\tau_{\nu\nu}\left( m_{3/2}\left( 1-\frac{m_Z^2}{m_{3/2}^2}\right) \right) \\
  &\qquad\qquad+\BR\left( h\,\nu_i\right) \times\tau_{\nu\nu}\left( m_{3/2}\left( 1-\frac{m_h^2}{m_{3/2}^2}\right) \right) \Bigg)\,.
 \end{split}
\end{equation}
Our result is presented in the right panel of Figure~\ref{SK-limit}.

The bounds obtained from the Super-Kamiokande and IceCube analyses become stronger for larger masses although the neutrino flux is inversely proportional to the gravitino mass for a fixed gravitino lifetime. This is due to the increasing neutrino--nucleon cross section and the increasing muon range. Note, however, that these present bounds do not have sensitivity to the parameter region preferred by the PAMELA excess yet, which corresponds to a gravitino lifetime of the order of $10^{26}$\,s and masses larger than 200\,GeV.

We observe that the current limit from Super-Kamiokande is still stronger than the IceCube limit. This is probably due to the longer data-taking period of Super-Kamiokande, the reduced size of the IceCube 22 string configuration compared to the full IceCube array and the higher detection threshold of IceCube. As we will discuss in the following, we expect the full IceCube detector to improve these results significantly.

\subsubsection{Rates and Bounds for Present and Future Experiments}

Assuming decaying gravitino dark matter with a lifetime of $10^{26}$\,s, we can now compute the expected signal rates for present and future experiments. These results can be easily generalized to arbitrary lifetimes, by recalling that the flux is proportional to $1/\tau_{3/2}$. We give the rates for some typical detectors of different sizes, \textit{i.e.} Super-Kamiokande~\cite{Fukuda:2002uc}, AMANDA~\cite{DeYoung:2008nc}, ANTARES~\cite{Collaboration:2011ns} and IceCube~\cite{Karle:2010xx}. The results for Super-Kamiokande can be easily scaled up to the Hyper-Kamiokande~\cite{Nakamura:2003hk} or UNO~\cite{Wilkes:2005rg} size by multiplying by a factor 10 or 20 (for a Hyper-Kamiokande mass of 500\,kt and a Hyper-Kamiokande/UNO mass of 1\,Mt, respectively). The result for KM3NeT~\cite{Katz:2011zz} will be very similar to that expected for IceCube.

We would like to stress here that Super-Kamiokande is still taking data, and that the full ANTARES detector was completed in summer 2008 and is also operational. The AMANDA-II detector was decommissioned in summer 2009, but has since been substituted by the IceCube detector, which has already taken data during various stages of construction and was completed in late 2010. The other experiments are still in the planning phase: KM3NeT is a proposed cubic-kilometer sized underwater neutrino telescope in the Mediterranean Sea, which will probably have an effective volume comparable to IceCube, but will be able to look at the galactic center, while the proposed Hyper-Kamiokande detector and the Underwater Neutrino Observatory (UNO) are water Cherenkov detectors similar to Super-Kamiokande but of megaton scale.

For the case of IceCube we also take into account the DeepCore subdetector~\cite{Wiebusch:2009jf}. It is designed to lower the energy threshold of the experiment to roughly 10\,GeV and to increase the sensitivity at low energies. This detector consists of six additional strings with less spacing between the digital optical modules compared to IceCube. The combination of IceCube and DeepCore can use the outer layers of IceCube as a veto to atmospheric muons and therefore has a $4\,\pi$ sensitivity for fully and partially contained events, but a considerably smaller effective volume.

\begin{figure}[t]
 \centering
 \includegraphics[scale=.44]{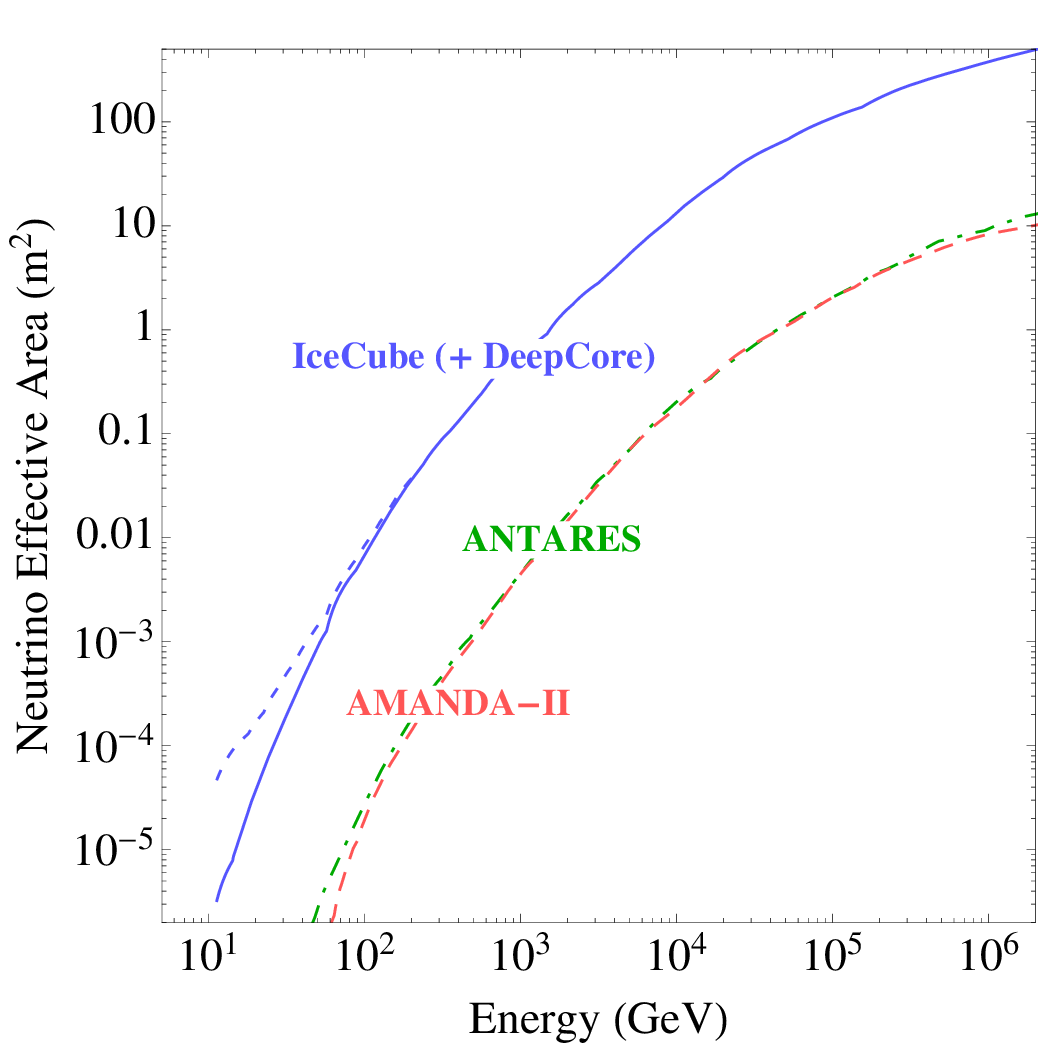}
 \includegraphics[scale=.44,bb=0 -3 500 500]{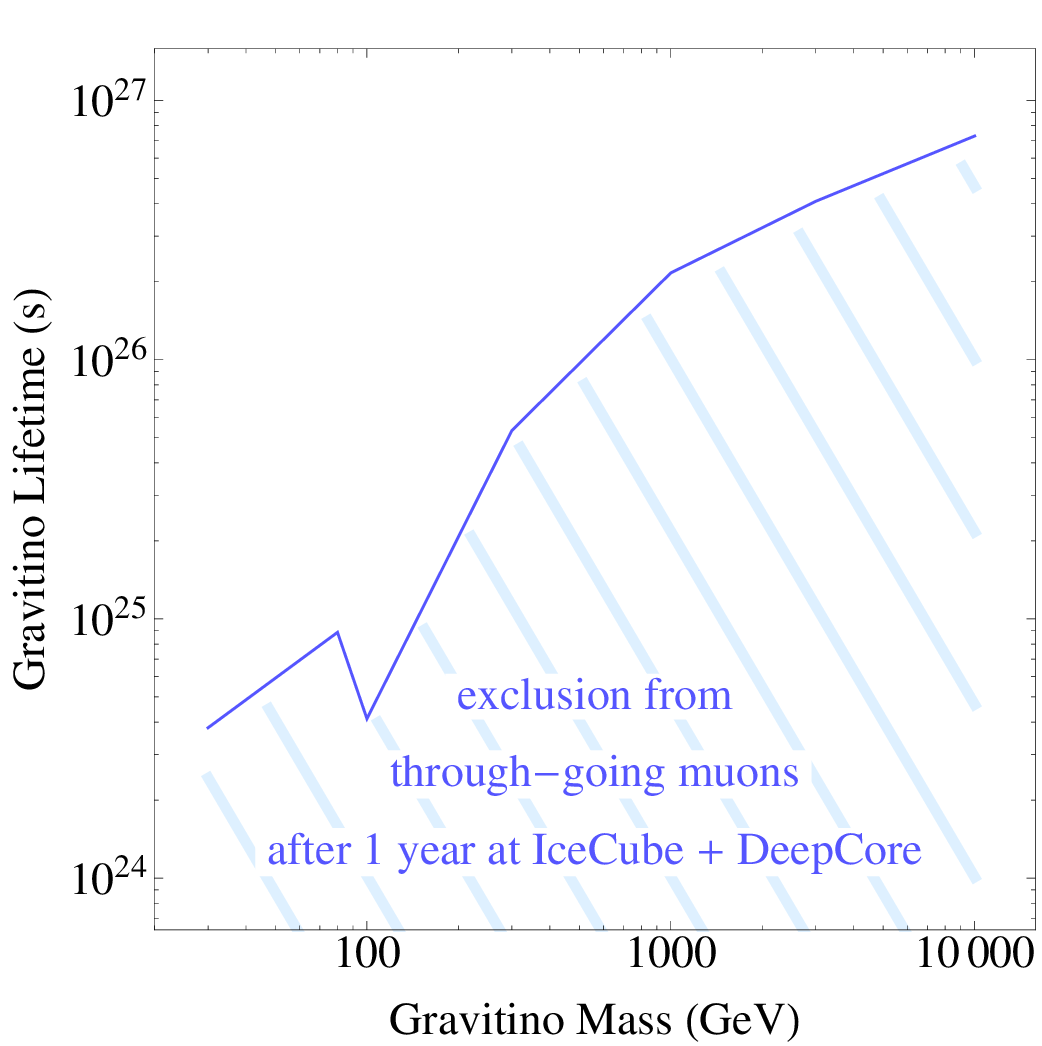}
 \caption[Neutrino effective areas and prospects for bounds on the gravitino dark matter lifetime from the IceCube experiment.]{\textit{Left:} Neutrino effective areas of the AMANDA-II, ANTARES and IceCube-80 experiments. The increased effective area due to the DeepCore extension of the IceCube detector is shown as a dashed line (see text for references). \textit{Right:} 90\,\% C.L.\ exclusion prospects in the lifetime vs.\ mass plane for decaying gravitino dark matter from the non-observation of a statistically significant excess in the total rate of neutrino-induced upward through-going muons observed at IceCube + DeepCore in one year.}
 \label{ICprospects}
\end{figure}

For the calculation of rates of upward through-going muon events we use the neutrino effective areas for AMANDA, ANTARES and the IceCube 80 strings configuration from~\cite{Montaruli:2009kv} and integrate over the muon spectrum. For the combined IceCube + DeepCore detector we amend the effective area in the low-energy range using the neutrino effective area given in~\cite{Wiebusch:2009jf}. The effective areas are reproduced in the left panel of Figure~\ref{ICprospects}. In the case of Super-Kamiokande we calculate the rate using equation~(\ref{through-rate}) with standard rock as the surrounding material, a muon effective area of 1200\,m$^2$ and a threshold muon energy of 1.6\,GeV.

\begin{table}
 \centering
 \begin{tabular}{cccccc}
  \hline
  neutrino source & Super-K & AMANDA & ANTARES & IceCube & IC+DeepCore \\
  \hline
  atmospheric $\nu_\mu$ & $4.3\times10^2$ & $1.5\times10^3$ & $1.8\times10^3$ & $3.0\times10^5$ & $3.5\times10^5$ \\
  $\psi_{3/2}$ (100\,GeV) & $8.6\times10^{-2}$ & --- & $5.1\times10^{-2}$ & $1.8\times10^1$ & $3.1\times10^1$ \\
  $\psi_{3/2}$ (300\,GeV) & $4.2\times10^{-1}$ & $1.1\times10^{0}$ & $1.4\times10^{0}$ & $3.4\times10^2$ & $4.1\times10^2$ \\
  $\psi_{3/2}$ (1\,TeV) & $1.5\times10^0$ & $8.0\times10^0$ & $9.6\times10^0$ & $1.6\times10^3$ & $1.6\times10^3$ \\
  $\psi_{3/2}$ (3\,TeV) & $3.5\times10^0$ & $2.4\times10^1$ & $2.8\times10^1$ & $3.1\times10^3$ & $3.1\times10^3$ \\
  $\psi_{3/2}$ (10\,TeV) & $7.2\times10^0$ & $6.3\times10^1$ & $6.5\times10^1$ & $5.5\times10^3$ & $5.5\times10^3$ \\
  \hline
 \end{tabular}
 \caption[Rates of upward through-going muon events from the atmospheric neutrino background and gravitino dark matter decays at contemporary neutrino experiments.]{Number of upward through-going muon events per year from the atmospheric neutrino background and gravitino dark matter decays at contemporary neutrino experiments. The rates are given for a gravitino lifetime of $10^{26}$\,s and gravitino masses of 100\,GeV, 300\,GeV, 1\,TeV, 3\,TeV and 10\,TeV. Note that Super-Kamiokande expects at most a few muons per year from gravitino decays at this lifetime while the combination of IceCube and DeepCore expects tens up to thousands of muons.}
 \label{rates}
\end{table}
We see from Table~\ref{rates} that a sizable number of muon events is expected for a gravitino lifetime of $10^{26}$\,s, especially for experiments of cubic kilometer scale, such as to become significant above the atmospheric background even for a gravitino mass of 300\,GeV. Of course for larger masses the significance becomes greater due to the increasing signal rate. Note that here we did not make use of any spectral information which would lead to an even greater statistical significance of the gravitino decay signal since in that case larger gravitino masses would also benefit from the falling background of atmospheric neutrinos.

Requiring the combined number of signal and background events not to exceed the background above the 90\,\% C.L.,\footnote{In the Gaussian approximation a 90\,\% C.L. upper limit corresponds to $S/\sqrt{B}<1.28$.} similar to the case of Super-Kamiokande in Figure~\ref{SK-limit}, we can then give in the right panel of Figure~\ref{ICprospects} a forecast of the exclusion region which may be obtained from kilometer-cubed experiments using one year of data. The larger statistics of the future experiments will improve the Super-Kamiokande and IceCube-22 bounds by more than an order of magnitude and explore the region of gravitino lifetimes above $10^{25}$\,s, for masses larger than 200\,GeV. Note that for ten years of data the lifetime limit will become stronger approximately by a factor of three. For lower gravitino masses a very important role will be played by DeepCore, which will considerably improve the IceCube performance between 10--100\,GeV masses, and also by the megaton water detectors which are expected to strengthen the Super-Kamiokande bounds by an order of magnitude down to masses of a few GeV. This low-mass region is plagued by a stronger atmospheric background, but it is still remarkable that even there gravitino lifetimes larger than $10^{24}$\,s will be probed in future neutrino experiments.

\subsubsection{Energy Resolution and Reconstructed Spectra}

Once a signal has been detected, the question arises if it will also be possible to reconstruct the neutrino spectra and extract some information on the nature of the dark matter particle. For this purpose one important factor is the energy resolution of the neutrino detectors. We will take here for reference the IceCube detector, for which the energy resolution is $\sigma=\log_{10}(E_{\text{max}}/E_{\text{min}})=0.3$--0.4 for track-like events and $\sigma=\log_{10}(E_{\text{max}}/E_{\text{min}})=0.18$ for cascade-like events~\cite{Resconi:2008fe}.

\begin{figure}[t]
 \centering
 \includegraphics[scale=.44]{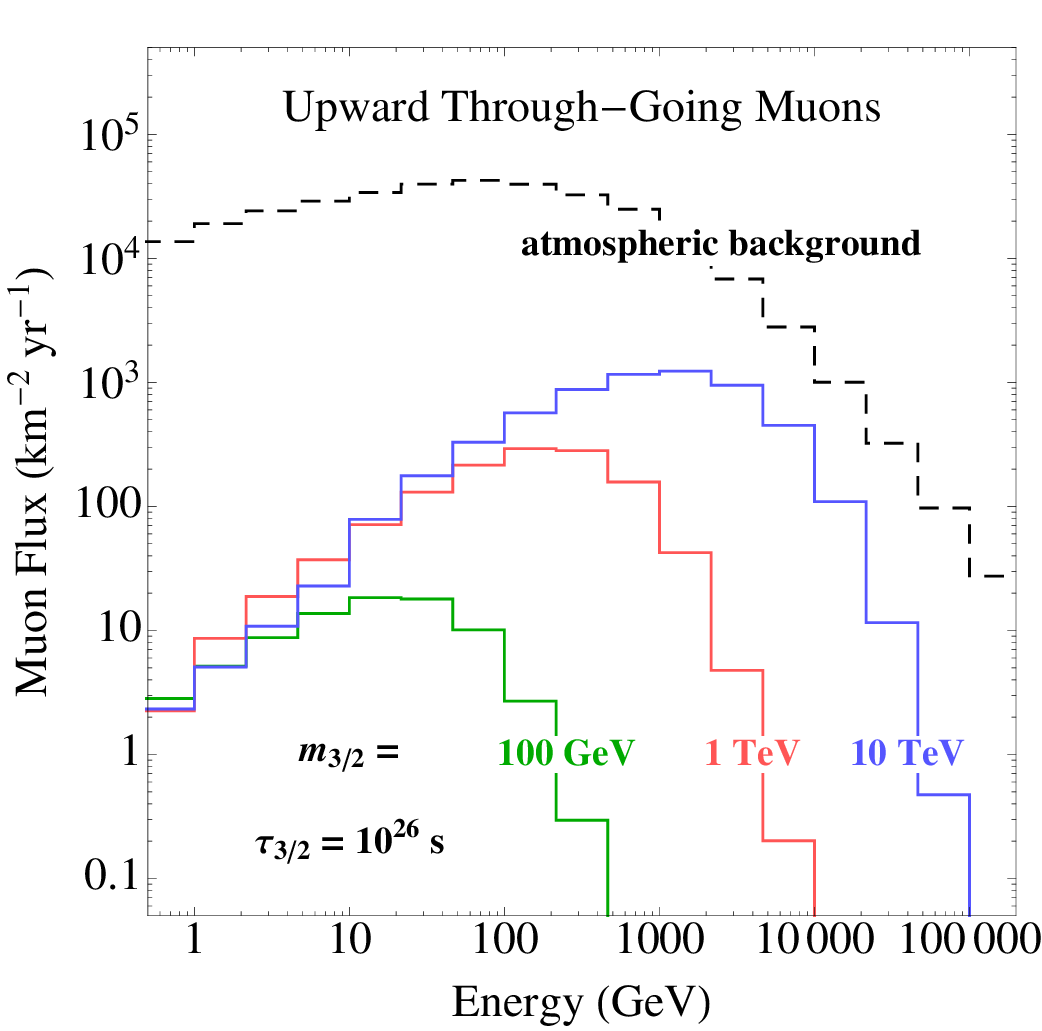}
 \includegraphics[scale=.44]{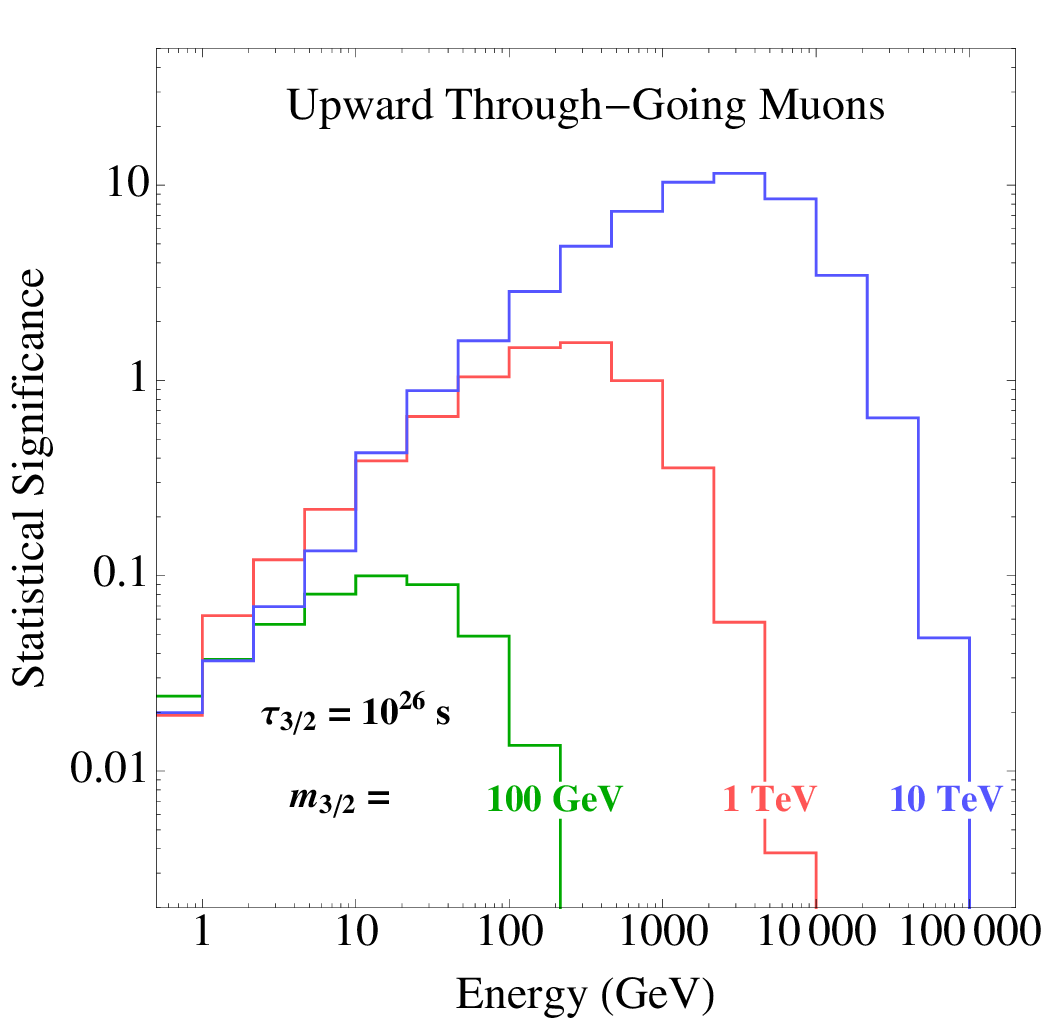}
 \caption[Flux of upward through-going muons expected from decaying gravitino dark matter compared to the atmospheric background and the statistical significance of the expected signal.]{\textit{Left:} Flux of upward through-going muons expected from decaying gravitino dark matter compared to the atmospheric background. The flux is computed for gravitino masses of 100\,GeV, 1\,TeV and 10\,TeV, and a lifetime of $10^{26}$\,s using an energy resolution of 0.3 in $\log_{10} E$ and three bins per decade. \textit{Right:} Statistical significance ($S/\sqrt{B}$) of the signal of through-going muons as shown in the left panel calculated for every single energy bin using one year of data assuming an ideal detector with an effective area of 1\,km$^2$.}
 \label{significance1}
\end{figure}
\begin{figure}[t]
 \centering
 \includegraphics[scale=.44]{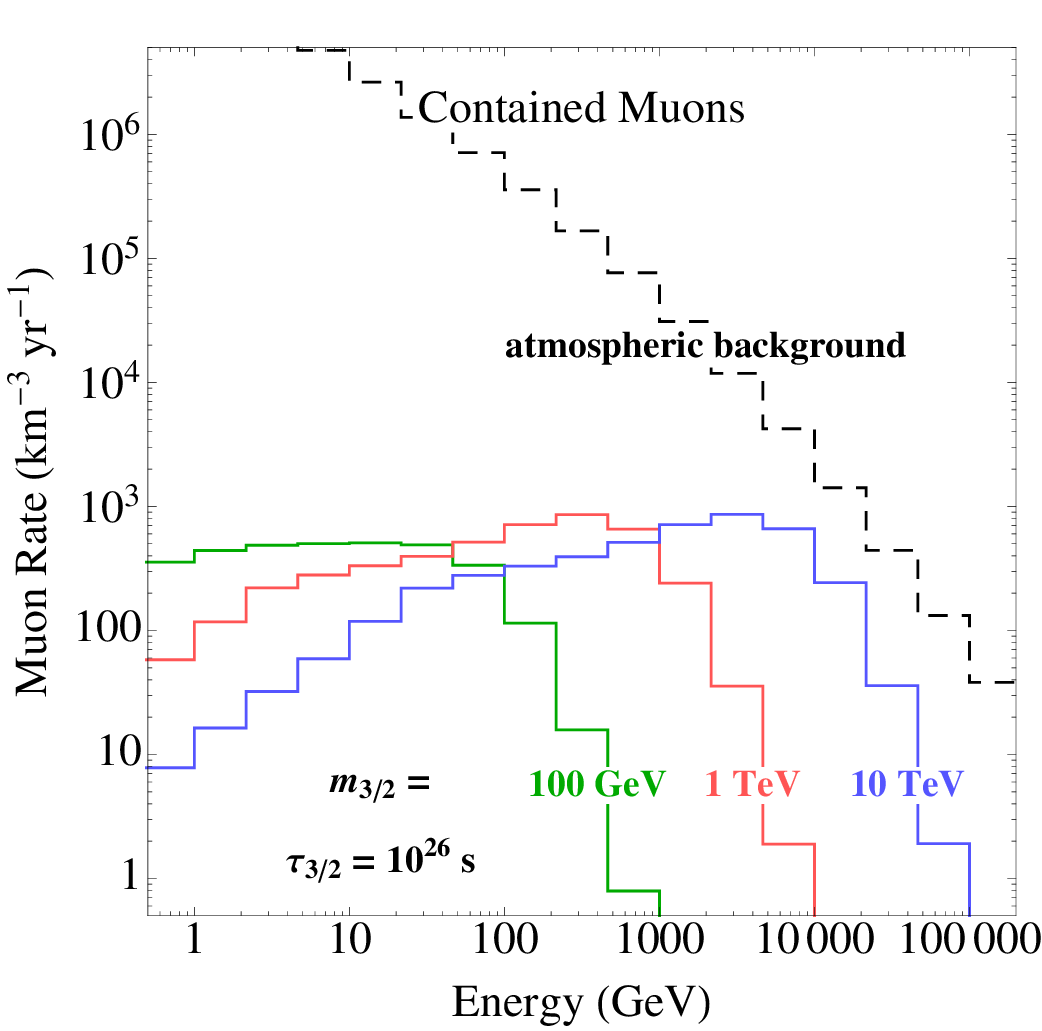}
 \includegraphics[scale=.44]{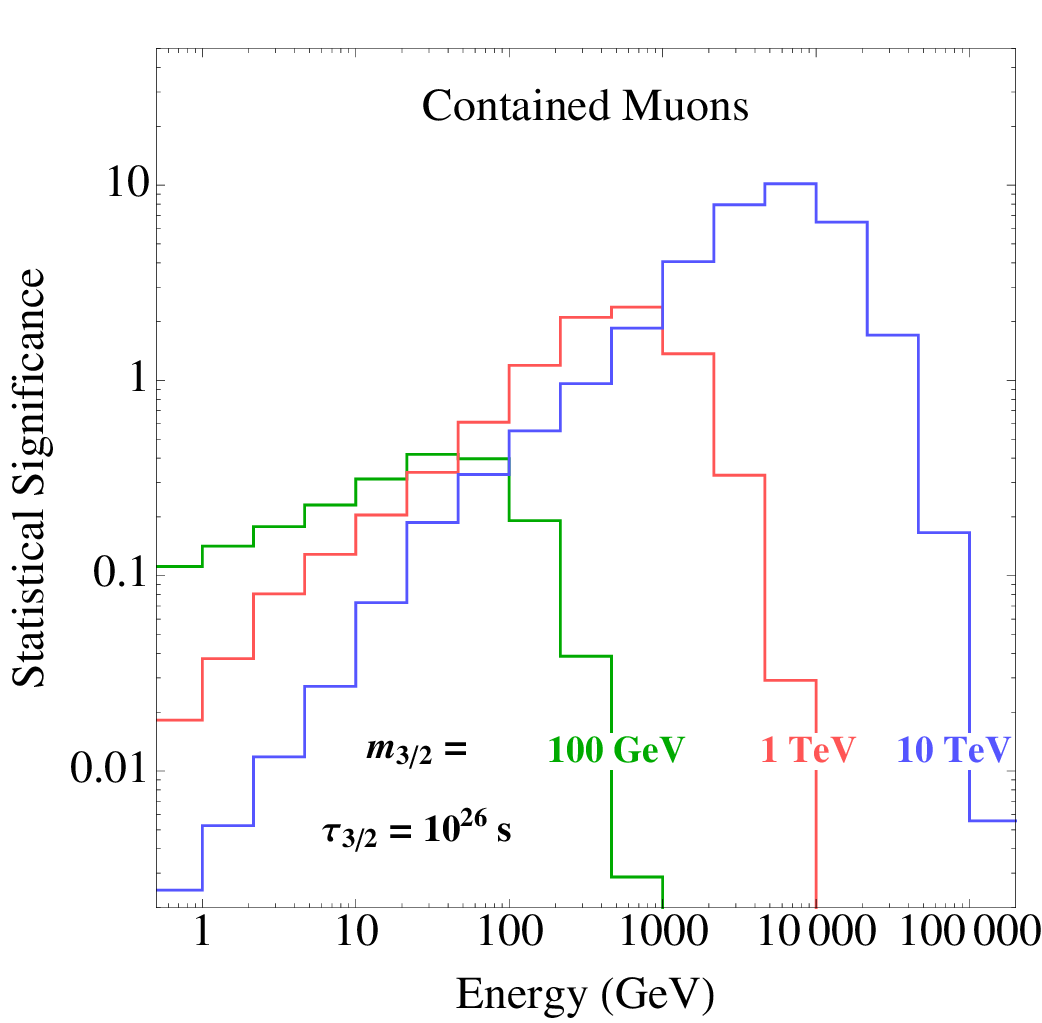}
 \caption[Rate of contained muons expected from decaying gravitino dark matter compared to the atmospheric background and the statistical significance of the expected signal.]{\textit{Left:} Rate of contained muons per cubic kilometer of detector volume from decaying gravitino dark matter compared to the atmospheric background. The rates are computed for gravitino masses of 100\,GeV, 1\,TeV and 10\,TeV, and a lifetime of $10^{26}$\,s using an energy resolution of 0.3 in $\log_{10} E$ and three bins per decade. \textit{Right:} Statistical significance ($S/\sqrt{B}$) of the contained muon signal as shown in the left panel calculated for every single energy bin using one year of data assuming an ideal detector with an effective volume of 1\,km$^3$.}
 \label{significance2}
\end{figure}
\begin{figure}[p]
 \centering
 \includegraphics[scale=.44]{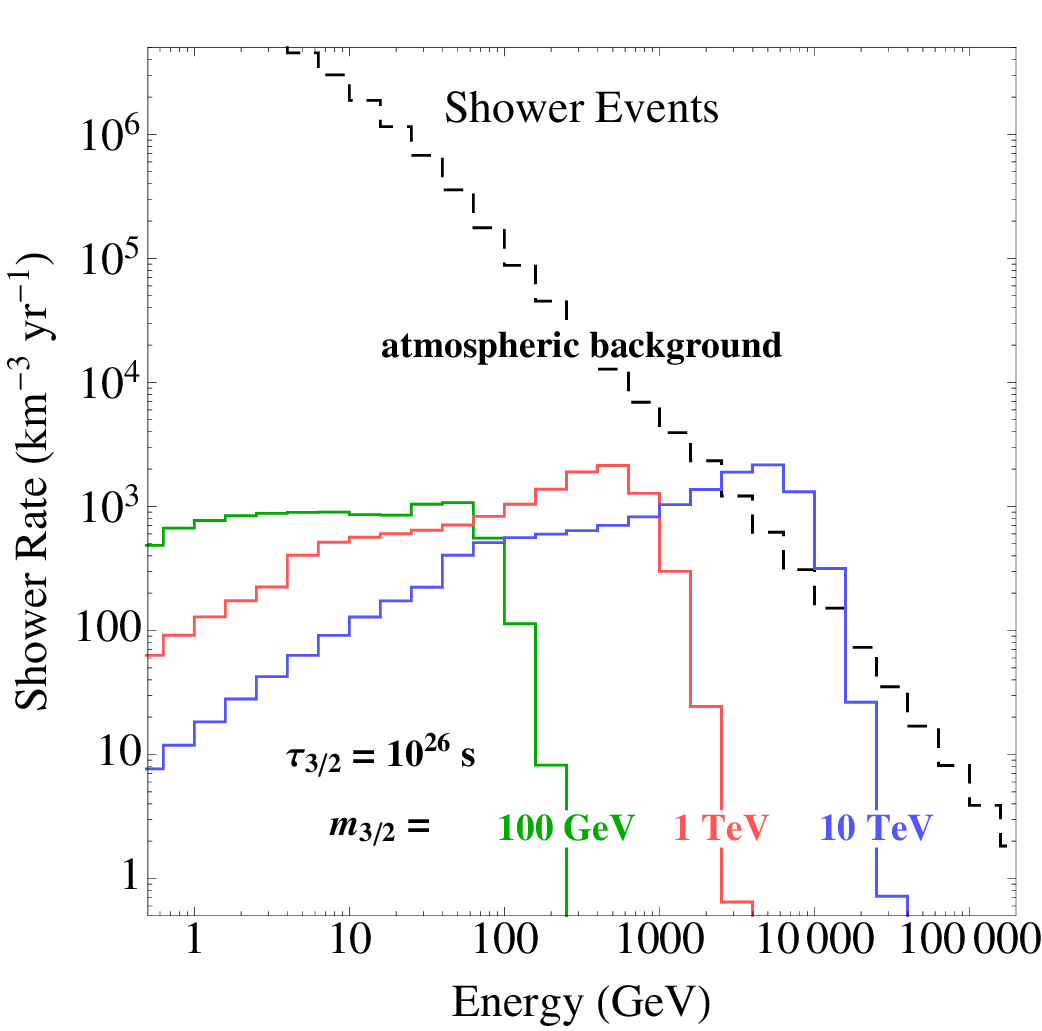}
 \includegraphics[scale=.44]{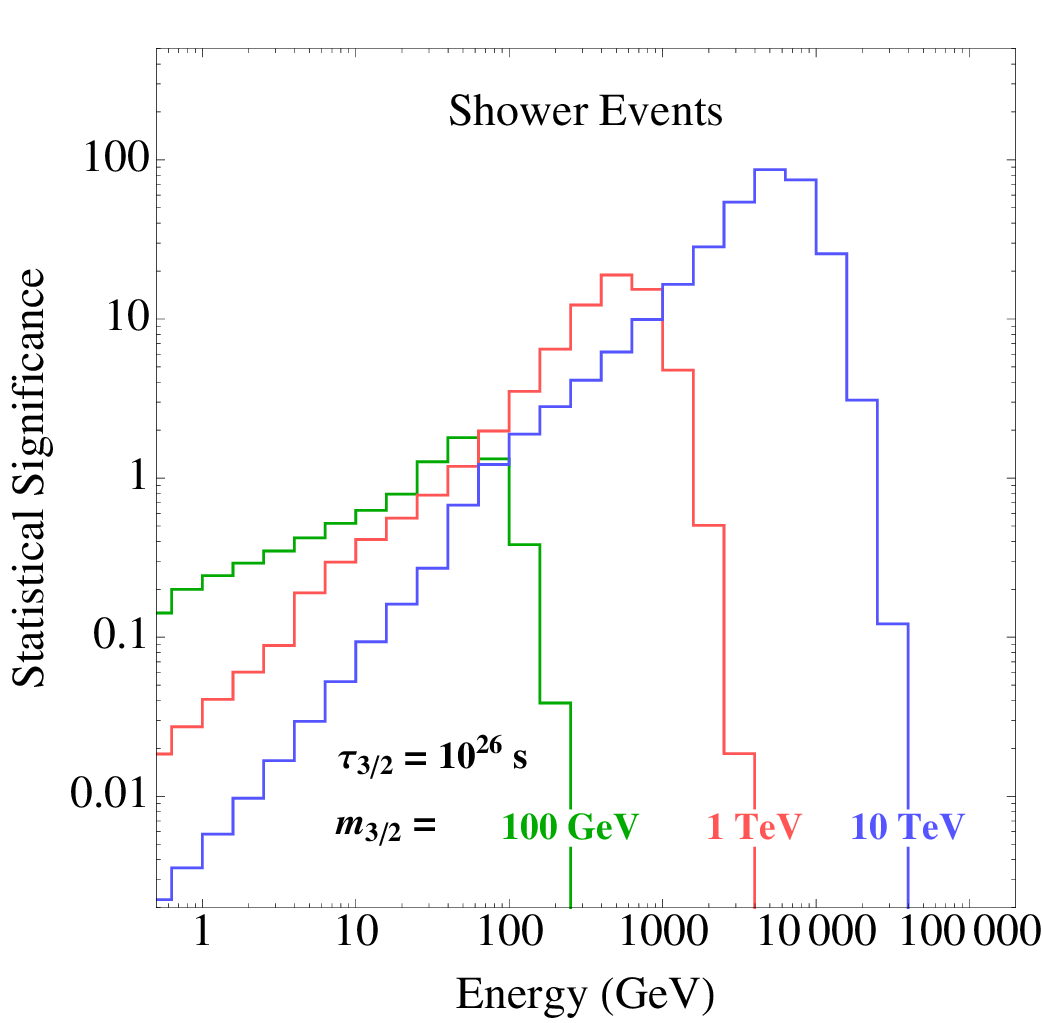}
 \caption[Rate of cascade events expected from decaying gravitino dark matter compared to the atmospheric background and the statistical significance of the expected signal.]{\textit{Left:} Rate of electromagnetic and hadronic showers per cubic kilometer of detector volume from decaying gravitino dark matter compared to the atmospheric background. The rates are computed for gravitino masses of 100\,GeV, 1\,TeV and 10\,TeV, and a lifetime of $10^{26}$\,s using an energy resolution of 0.18 in $\log_{10} E$ and five bins per decade. \textit{Right:} Statistical significance ($S/\sqrt{B}$) of the shower signal as shown in the left panel calculated for every single energy bin using one year of data assuming an ideal detector with an effective volume of 1\,km$^3$.}
 \label{significance3}
\end{figure}
Technically we apply the energy resolution by a convolution of the signal spectrum with a Gaussian distribution in the logarithm of the energy:
\begin{equation}
 \frac{dN_{\text{Gauss}}}{dE}=\frac{\exp\left( -(\sigma\ln10)^2/2\right) }{\sqrt{2\,\pi}\,\sigma\ln10}\int_0^{\infty}dE'\,\frac{1}{E'}\exp\left( -\frac{1}{2}\left( \frac{\log_{10}E/E'}{\sigma}\right) ^2\right) \frac{dN}{dE'}\,.
\end{equation}
In Figures~\ref{significance1}--\ref{significance3} we show the resulting histograms for the signal from gravitino decays and the atmospheric background using an energy resolution of 0.3 in $\log_{10} E$ and three bins per decade for upward through-going and contained muons, and an energy resolution of 0.18 in $\log_{10} E$ and five bins per decade for shower events. These figures can be compared to Figures~\ref{SK-Mu-spectra}--\ref{IC-Shower-spectra} which show the spectra unbinned and without finite energy resolution. Also shown is the significance of the signal over the background in different bins for a lifetime of $10^{26}$\,s using one year of data assuming an ideal detector with an effective area of 1\,km$^2$ for upward through-going muons and an effective volume of 1\,km$^3$ for contained muons and cascades.

\begin{table}[p]
 \centering
 \begin{tabular}{cccc}
  \hline
  gravitino mass & through-going muons & contained muons & shower events \\
  \hline
  100\,GeV & $2.0\times10^{24}\,$s & $8.4\times10^{24}\,$s & $3.6\times10^{25}\,$s \\
  300\,GeV & $8.5\times10^{24}\,$s & $2.3\times10^{25}\,$s & $1.3\times10^{26}\,$s \\
  1\,TeV & $3.1\times10^{25}\,$s & $4.7\times10^{25}\,$s & $3.8\times10^{26}\,$s \\
  3\,TeV & $8.3\times10^{25}\,$s & $9.1\times10^{25}\,$s & $7.3\times10^{26}\,$s \\
  10\,TeV & $2.3\times10^{26}\,$s & $2.0\times10^{26}\,$s & $1.7\times10^{27}\,$s \\
  \hline
 \end{tabular}
 \caption[Gravitino lifetimes for various gravitino masses corresponding to a five sigma statistical significance in the most significant energy bin after one year of observation in an idealized neutrino detector.]{Gravitino lifetimes for various gravitino masses corresponding to a $S/\sqrt{B}=5$ significance in the most significant energy bin after one year of observation in an idealized detector with an effective muon area of 1\,km$^2$ and an effective volume for contained muons and showers of 1\,km$^3$. Notice that the sensitivity obtained with through-going and contained muons is quite similar. At larger masses the bound from through-going muons is stronger since the statistics increases due to the longer muon range at higher energies. However, neglecting reconstruction efficiencies the strongest constraint is obtained from shower events since that channel offers the best signal-to-background ratio (see discussion in Section~\ref{showers}).}
 \label{sensitivity}
\end{table}
We see that for nearly all gravitino masses the signal will appear with a large statistical significance in more than a single bin and it is obvious that the neutrino signal from gravitino decays is not following a power law like the atmospheric one. Thus it is clear that using spectral information it will be possible to set much stricter limits on the gravitino parameter space than shown in Figure~\ref{ICprospects}. In order to give an idea of the sensitivities that can be obtained using spectral information, we show in Table~\ref{sensitivity} the values of the gravitino lifetime for several gravitino masses that correspond to a 5 $\sigma$ signal in the most significant energy bin after one year of observation for an idealized detector with an effective muon area of 1\,km$^2$ and an effective volume of 1\,km$^3$ for contained muons and shower events. We observe that the limits from through-going and contained muons are better but not far from those shown in Figure~\ref{ICprospects}, while the shower events in principle allow to constrain even one order of magnitude larger lifetimes. Using not only the dominant energy bin from Figures~\ref{significance1} to \ref{significance3} but a combination of several energy bins optimized for the expected signal from gravitino decays it will be possible to set even stronger constraints on the gravitino lifetime.

On the other hand, discriminating the spectra expected from gravitino decay from those expected for other dark matter candidates will not be that straightforward, especially if the mass of the decaying particle is unknown (\textit{cf.} the discussion in~\cite{Covi:2009xn}). After convolution with the energy resolution, the spectrum expected for gravitino dark matter and those for generic channels of scalar or fermionic dark matter particles appear quite similar, especially within their statistical error.

\section{Constraints on the Gravitino Dark Matter Parameter Space}

In this final section we want to summarize the limits on the parameter space of decaying gravitino dark matter coming from indirect searches in the gamma-ray, positron, antiproton, antideuteron and neutrino channels. While we have presented bounds on the gravitino lifetime from the different channels in the previous sections, here we want to use these limits to constrain the amount of bilinear $R$-parity violation.

The upper limits on the parameter $\xi$ can be calculated from the decay widths presented in Section~\ref{gravdecay} and the lower limits on the gravitino lifetime in the following way:
\begin{equation}
  \xi\leq\left( \frac{\Gamma_{\text{tot}}(m_{3/2})}{\xi^2}\times\tau_{3/2}(m_{3/2})\right) ^{-1/2}.
\end{equation}
Since the gravitino decay width scales like $m_{3/2}^3$ one would generically expect a bound on $\xi$ that scales like $m_{3/2}^{-3/2}$ for a lifetime bound that is independent of the gravitino mass. Rather flat lifetime bounds are obtained from practically all cosmic ray observations except for neutrinos, where the limits become stronger for larger gravitino masses. However, we also need to take into account the dependence of the mixing parameters on the gravitino mass. As we chose the gaugino masses to be proportional to the gravitino mass, the gaugino--neutrino and gaugino--charged lepton mixing parameters scale like $1/m_{3/2}$ (\textit{cf.} the discussion in Section~\ref{branchingratios}). Therefore, we find that the decay width for the two-body decay into photon and neutrino actually scales like $m_{3/2}$. For the other decay channels the decay width is dominated by diagrams not involving any mixing parameters for larger gravitino masses and it therefore really scales like $m_{3/2}^3$.

\begin{figure}[t]
 \centering
 \includegraphics[scale=0.8]{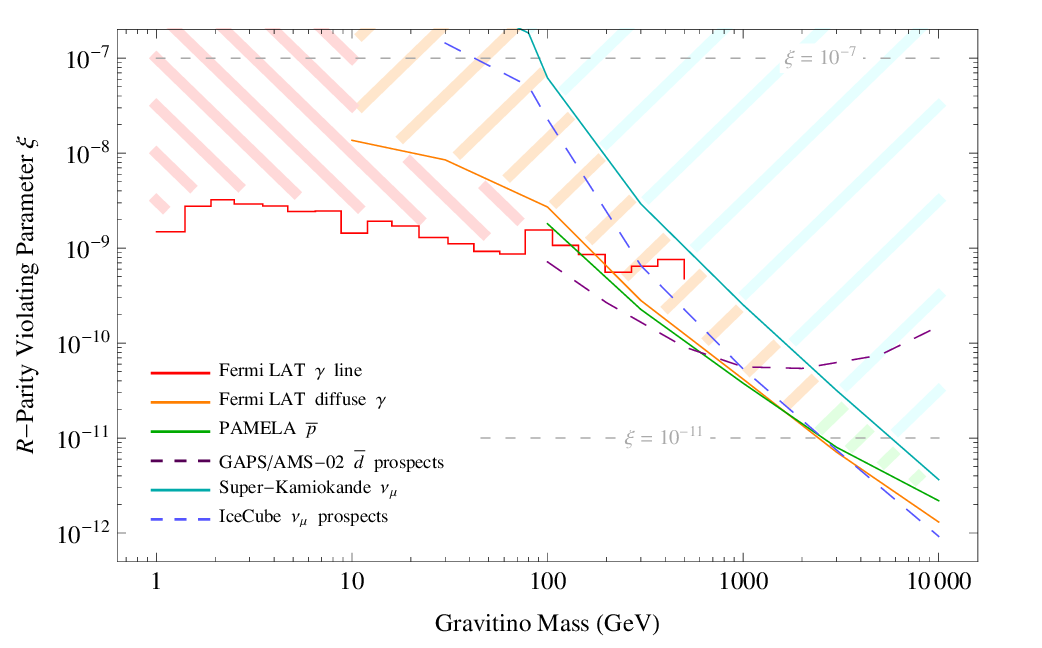} 
 \caption[Summarized bounds from indirect dark matter searches on the gravitino dark matter parameter space.]{Summarized bounds from indirect dark matter searches on the gravitino dark matter parameter space. We present the current bounds from cosmic-ray observations and future prospects for antideuteron and neutrino searches. In addition, we mark the boundaries of the cosmologically preferred range of the $R$-parity breaking parameter.}
 \label{xibound}
\end{figure}
In Figure~\ref{xibound} we present the exclusion regions in the $\xi-m_{3/2}$ parameter space obtained via the method described above. At low gravitino masses the dominant bound comes from photon-line searches. Above $m_{3/2}=100\,$GeV the bounds from the continuum contributions to the gamma-ray and antiproton fluxes provide the most stringent constraints. The current limits of Super-Kamiokande from the neutrino flux are weaker by almost one order of magnitude. However, future bounds from IceCube are expected to be competitive to those of gamma rays and antiprotons. Beyond that, using the more sensitive channel of neutrino-induced showers and employing spectral information the neutrino bounds could even dominate the large gravitino mass region.

In scenarios where the photon line is absent an important role is expected to be played by observations of charged cosmic rays for low gravitino masses. Since bounds from antiprotons rely on background subtractions in an energy range where the effect of solar modulation introduces some uncertainty, the almost background-free antideuteron channel is expected to provide the best exclusion limit in this case.

We also show the boundaries of the cosmologically preferred range of the $R$-parity breaking parameter $\xi$. It is quite remarkable that indirect dark matter searches are probing exactly the allowed region derived from completely different arguments. In particular, for masses above $\mathcal{O}(1)\,$TeV practically the complete favored range of $\xi$ is already excluded. It will be interesting to see how future observations or complementary limits from collider searches will further constrain this parameter space.

\chapter{Direct Detection of Gravitino Dark Matter}
\label{gravitinoDD}

The method of direct dark matter detection is based on the observation of nuclear recoils in underground detectors due to scattering processes of nuclei with dark matter particles traversing the Earth. In particular for WIMP dark matter candidates a signal is expected since their interactions with matter should be of the order of weak scale interactions.

However, as the interactions of gravitinos are strongly suppressed compared to the standard WIMP case, typically no observable signal is expected in direct detection experiments from gravitino--nucleon scattering. In this chapter we want to discuss quantitatively how the situation changes for gravitino dark matter in models with broken $R$ parity. Although the rate for elastic scatterings is not changed by additional $R$-parity violating interactions, the slight violation of $R$ parity allows for inelastic scatterings off nucleons that could be significantly enhanced in strength compared to the elastic scattering processes.

Let us start the discussion with a very brief overview of the method of direct dark matter detection. A more extensive review on this topic and further references can be found, for instance, in~\cite{Jungman:1995df}.

\section{Direct Searches for Dark Matter}

As the solar system orbits around the galactic center, it is expected that dark matter particles from the galactic halo cross the Earth in large numbers. In~\cite{Goodman:1984dc} it was first proposed to search for signals from scatterings of these dark matter particles in underground detectors. The general idea is to observe the recoil of target nuclei induced by elastic scatterings of dark matter particles off those nuclei. Usually, one distinguishes between spin-independent and spin-dependent interactions: Spin-dependent interactions are determined by the coupling of the dark matter particle's spin and the spin of the target nucleus. In many cases this type of interaction dominates in the scattering off a single proton or light nuclei. Spin-independent interactions, on the other hand, have the advantage that they coherently add up and thus can be significantly enhanced when scattering off heavy nuclei is considered. In the following we will only consider the latter case.

The differential recoil rate for spin-independent dark matter--nucleus scattering processes is given by~\cite{Jungman:1995df}
\begin{equation}
  \frac{dR}{dE_R}=\frac{2\,\rho_{\text{loc}}}{m_{\text{DM}}}\,\frac{d\sigma}{d|\vec{k}|^2}\,v\,f(v)\,dv\,,
\end{equation}
where $\vec{k}$ is the momentum of the recoiling nucleus and the recoil energy is given by
\begin{equation}
  E_R\simeq\frac{|\vec{k}|^2}{2\,m_N}
\end{equation}
in the non-relativistic case. The elastic scattering of dark matter particles with masses of the order of the electroweak scale typically leads to recoil energies in the range from 10 to 100\,keV. The four-momentum transfer $t$ of the scattering process is given by
\begin{equation}
  t=\left( k'-k\right) ^2
  =2\,m_N^2-2\,m_N\sqrt{m_N^2+|\vec{k}|^2}\simeq-|\vec{k}|^2,
\end{equation}
where $k$ and $k'$ are the four-momenta of the nucleus before and after the scattering, and $\vec{k}$ is the non-relativistic recoil momentum. Thus the differential scattering cross section is typically described in terms of $|\vec{k}|^2$. The differential scattering cross section of an elastic scattering process can also be written as the product of the cross section at zero momentum transfer $\sigma_0$ and a nuclear form factor
\begin{equation}
  \frac{d\sigma}{d|\vec{k}|^2}=\frac{\sigma_0}{4\,\mu^2\,v^2}\,F^2(|\vec{k}|)\,,
\end{equation}
where the dark matter--nucleus reduced mass is defined as
\begin{equation}
  \mu=\frac{m_N\,m_{\text{DM}}}{m_N+m_{\text{DM}}}\,.
\end{equation}
In elastic processes there is a relation between the momentum transfer, the relative velocity and the scattering angle $\theta^*$ in the center-of-momentum frame:
\begin{equation}
  |\vec{k}|^2=2\,\mu^2\,v^2\left( 1-\cos\theta^*\right) \qquad\text{so that}\qquad E_R\simeq\frac{\mu^2\,v^2}{m_N}\left( 1-\cos\theta^*\right) .
\end{equation}
The expected signal is determined assuming that the dark matter particles are distributed in the halo according to an isothermal profile with a local dark matter density $\rho_{\text{loc}}\simeq0.3\,\text{GeV}\,\text{cm}^{-3}$. In addition, usually a Maxwell--Boltzmann velocity distribution $f(v)$ with a characteristic velocity of $v_0\sim 220\,\text{km}\,\text{s}^{-1}$ is assumed for the dark matter particles~\cite{Jungman:1995df}. The total rate is then determined by integrating over the dark matter velocity distribution:
\begin{equation}
  \frac{dR}{dE_R}=\frac{\rho_{\sigma_0\,\text{loc}}}{2\,m_{\text{DM}}\,\mu}\,F^2(|\vec{k}|)\int_{v_\text{min}}^{\infty}\frac{f(v)}{v}\,dv\,,
\end{equation}
where the minimal velocity is given by the condition $v_\text{min}=\sqrt{E_R\,m_N/(2\,\mu^2)}$. In the popular case of neutralino WIMP dark matter one expects elastic scattering cross sections of the order of $10^{-10}$ to $10^{-6}\,$pb corresponding to a maximal event rate of about one event per kilogram detector material and day~\cite{Chung:2003fi}. Most recent direct detection experiments have excluded spin-independent WIMP elastic scattering cross sections down to a level of $10^{-7}$ to $10^{-8}\,$pb (see for instance~\cite{Ahmed:2009zw,Armengaud:2011cy,Aprile:2011hi}).

On the other hand, it is claimed that the DAMA experiment observes an annual modulation signal from dark matter with a significance of more than $8\,\sigma$~\cite{Bernabei:2000qi,Bernabei:2008yi}. An annual modulation in the rate of dark matter scattering events is a generic signal from dark matter particles in the galactic halo. It is expected due to the variation of the Earth's velocity with respect to the halo because of the Earth's orbit around the Sun. The DAMA observation is generally considered as being in conflict with the null results of other experiments, although various possibilities to achieve compatibility are discussed in the literature (see \textit{e.g.}~\cite{Savage:2008er}).

One proposal to reconcile the observations of DAMA with the majority of other direct detection experiments is inelastic dark matter~\cite{TuckerSmith:2001hy,TuckerSmith:2004jv}. In this framework the dark matter particle is accompanied by an almost degenerate state that is heavier by $\mathcal{O}(100)\,$keV. Assuming a highly suppressed elastic scattering cross section, an incident dark matter particle can only inelastically scatter off nuclei, leading to a transition to the heavier state. This means, however, that not all dark matter particles crossing the Earth can scatter off nuclei but only those with a particular minimal velocity. This effect leads to very a strong suppression of the signal in detectors like CDMS and EDELWEISS using rather light target nuclei, thereby giving a possible explanation for the compatibility of the DAMA observation using rather heavy iodine nuclei. However, very recently this model has been almost completely ruled out by new data of the XENON100 experiment~\cite{Aprile:2011ts}.

Let us now turn to the case of gravitino dark matter. As the elastic scattering cross section of gravitinos on nuclei is proportional to $1/\MP^4$ there is no chance to observe these processes in underground detectors. In the case of additional $R$-parity violating interactions, however, inelastic scattering processes whose cross sections are only suppressed by $1/\MP^2$ and the amount of $R$-parity breaking become possible. In this case the inelasticity is in the opposite direction as for the framework mentioned above. A gravitino can scatter off nuclei, thereby being transformed into a neutrino. In this process the mass of the gravitino is released as energy and passed over to the neutrino and the target nucleus. In this respect, one could refer to the gravitino with broken $R$ parity as anti-inelastic dark matter.

\section[Gravitino Dark Matter with Bilinear \texorpdfstring{$R$}{R}-Parity Violation]{Gravitino Dark Matter with Bilinear \boldmath$R$-Parity Violation}

Here we want to discuss the inelastic scattering of gravitino dark matter off target nucleons in the framework of bilinear $R$-parity breaking. We separately consider three cases: the exchange of a lightest Higgs boson, the exchange of a $Z$ boson and the exchange of a photon. In all cases the gravitino is destroyed in the scattering process while a neutrino is generated in the final state. In particular the photon exchange channel might be interesting since it is not available for practically all other dark matter candidates and potentially leads to an enhanced cross section because of the massless propagator.

\subsection*{Inelastic Gravitino--Nucleon Scattering via Higgs Exchange}

At tree level there are two Feynman diagrams contributing to the inelastic gravitino scattering off a nucleon via the exchange of the lightest Higgs boson:
\begin{equation*}
 \parbox{6.3cm}{
  \begin{picture}(183,173) (107,-22)
    \SetWidth{0.5}
    \SetColor{Black}
    \Line[dash,dashsize=4,arrow,arrowpos=0.5,arrowlength=3.75,arrowwidth=1.5,arrowinset=0.2](198,62)(198,96)
    \Line[dash,dashsize=4](198,62)(198,29)
    \Line[arrow,arrowpos=0.5,arrowlength=3.75,arrowwidth=1.5,arrowinset=0.2](198,96)(256,130)
    \Line[arrow,arrowpos=0.5,arrowlength=3.75,arrowwidth=1.5,arrowinset=0.2](198,29)(256,-5)
    \Line[arrow,arrowpos=0.5,arrowlength=3.75,arrowwidth=1.5,arrowinset=0.2](140,-5)(198,29)
    \Line[double,sep=4](140,130)(198,96)
    \Vertex(198,96){2.5}
    \Vertex(198,29){2.5}
    \Line(193.5,67)(202.5,58)\Line(202.5,67)(193.5,58)
    \Text(138,131)[rb]{\normalsize{\Black{$\psi_{3/2}$}}}
    \Text(257,131)[lb]{\normalsize{\Black{$\nu_i$}}}
    \Line(198,96)(227,113)
    \Text(206,46)[l]{\normalsize{\Black{$h$}}}
    \Text(206,79)[l]{\normalsize{\Black{$\tilde{\nu}_i$}}}
    \Text(138,-6)[rt]{\normalsize{\Black{$N$}}}
    \Text(257,-6)[lt]{\normalsize{\Black{$N$}}}
  \end{picture}
 }
 \quad+\quad
 \parbox{6.3cm}{
  \begin{picture}(183,173) (107,-22)
    \SetWidth{0.5}
    \SetColor{Black}
    \Line[dash,dashsize=4](198,96)(198,29)
    \Line[arrow,arrowpos=0.5,arrowlength=3.75,arrowwidth=1.5,arrowinset=0.2](227,113)(256,130)
    \Line(198,96)(227,113)
    \Line[arrow,arrowpos=0.5,arrowlength=3.75,arrowwidth=1.5,arrowinset=0.2](198,29)(256,-5)
    \Line[arrow,arrowpos=0.5,arrowlength=3.75,arrowwidth=1.5,arrowinset=0.2](140,-5)(198,29)
    \Line[double,sep=4](140,130)(198,96)
    \Vertex(198,96){2.5}
    \Vertex(198,29){2.5}
    \Line(225,107.001)(229,118.999)\Line(221.001,115)(232.999,111)
    \Text(214,102)[lt]{\normalsize{\Black{$\tilde{h}$}}}
    \Text(138,131)[rb]{\normalsize{\Black{$\psi_{3/2}$}}}
    \Text(257,131)[lb]{\normalsize{\Black{$\nu_i$}}}
    \Text(206,62)[l]{\normalsize{\Black{$h$}}}
    \Text(138,-6)[rt]{\normalsize{\Black{$N$}}}
    \Text(257,-6)[lt]{\normalsize{\Black{$N$}}}
  \end{picture}
 }.
\end{equation*}
For the coupling of the Higgs boson to the nucleons we need to know the effective mass of the quarks inside the nucleon. Typically, this is expressed in the form~\cite{Jungman:1995df}
\begin{equation}
 \left\langle N|m_q\,\bar{q}\,q|N\right\rangle =m_N\,f_{T_q}^N\,.
\end{equation}
For the cases of protons and neutrons the coefficients for the light quarks are given by~\cite{Ellis:2000ds}
\begin{alignat}{2}
  f_{T_u}^p &=0.020\pm0.004\,, &\qquad\qquad f_{T_u}^n &=0.014\pm0.003\,, \nonumber\\
  f_{T_d}^p &=0.026\pm0.005\,, & f_{T_d}^n &=0.036\pm0.008\,, \\
  f_{T_s}^p &=0.118\pm0.062\,, & f_{T_s}^n &=0.118\pm0.062\,. \nonumber
\end{alignat}
In addition, there is a coupling to the gluon content of the nucleons via heavy quark loops. This leads to coefficients for the heavy quarks of the form~\cite{Shifman:1978zn}
\begin{equation}
 f_{T_{c,\,b\,,t}}^N=\frac{2}{27}\,f_{T_G}^N\qquad\text{with}\qquad f_{T_G}^N=1-f_{T_u}^N-f_{T_d}^N-f_{T_s}^N\,.
\end{equation}
For the cases of protons and neutrons this leads to practically equal coefficients for the heavy quarks:
\begin{equation}
 f_{T_{c,\,b\,,t}}^p\simeq0.062\qquad\text{and}\qquad f_{T_{c,\,b\,,t}}^n\simeq0.062\,.
\end{equation}
The calculation of the differential scattering cross section is presented in Appendix~\ref{gravitinoscattering}. Here we directly present the final result:
\begin{align}
  \frac{d\sigma_N}{dt} &=\frac{\xi_i^2\,m_{3/2}^2\,m_N^2\left( \sum_qf_{T_q}^N\right) ^2}{1536\,\pi\,v^2\,m_h^4\,\MP^2\abs{\vec{v}}^2}\abs{1+\frac{1}{2}\,\frac{m_Z^2}{m_{\tilde{\nu}_i}^2}\cos2\,\beta-2\sin\beta\,U_{\tilde{H}_u^0\tilde{Z}}-2\cos\beta\,U_{\tilde{H}_d^0\tilde{Z}}}^2 \nonumber\\
  &\qquad\qquad\times\left( 1-\frac{t}{m_{3/2}^2}\right) ^3\left( 1-\frac{t}{2\,m_N^2}\right) .
\end{align}
The four-momentum transfer $t$ can be expressed in terms of the scattering angle $\theta$ between the direction of the incoming gravitino and the scattered neutrino:
\begin{equation}
  t=(p'-p)^2=m_{3/2}^2-2\left( E\abs{\vec{p}\,'}-\abs{\vec{p}\,}\abs{\vec{p}\,'}\cos\theta\right)\,.
\end{equation}
Energy conservation requires that the momentum of the final state neutrino is given by
\begin{equation}
  |\vec{p}\,'|=\frac{m_{3/2}^2+2\,m_NE}{2\left( E+m_N-\abs{\vec{p}\,}\cos\theta\right) }\,.
\end{equation}
This leads finally to the relation
\begin{equation}
  t=m_{3/2}^2-\frac{\left( E-\abs{\vec{p}\,}\cos\theta\right) \left( m_{3/2}^2+2\,m_NE\right) }{m_N+E-\abs{\vec{p}\,}\cos\theta}\,.
\end{equation}
The limiting values of the four-momentum transfer are then given by
\begin{equation}
  t_0=t(\theta=0^\circ)=m_{3/2}^2-\frac{\left( E-\abs{\vec{p}\,}\right) \left( m_{3/2}^2+2\,m_NE\right) }{m_N+E-\abs{\vec{p}\,}}
\end{equation}
and
\begin{equation}
  t_1=t(\theta=180^\circ)=m_{3/2}^2-\frac{\left( E+\abs{\vec{p}\,}\right) \left( m_{3/2}^2+2\,m_NE\right) }{m_N+E+\abs{\vec{p}\,}}\,,
\end{equation}
and the size of the kinematically allowed range is given by
\begin{equation}
  t_0-t_1=\frac{2\abs{\vec{p}\,}m_N\left( m_{3/2}^2+2\,m_NE\right) }{\left( m_N+E-\abs{\vec{p}\,}\right)\left( m_N+E+\abs{\vec{p}\,}\right)}\,.
  \label{trange}
\end{equation}
The kinematics in this situation strongly differs from the case of elastic WIMP scattering, where the momentum transfer is determined by the velocity of the dark matter particles. Here the momentum transfer is mainly given by the gravitino mass with only a slight dependence on the velocity. Therefore, one needs much lighter gravitino masses to get a recoil signal in the 10--100\,keV range probed by current experiments. We then find that we will need gravitinos with a mass of $\mathcal{O}$(1--100)\,MeV.\footnote{A consistent cosmological scenario with a gravitino mass in the MeV range can be constructed in specific models of gauge mediation~\cite{Fujii:2002yx,Fujii:2002fv,Lemoine:2005hu,Jedamzik:2005ir}.} In this case the gravitino is much lighter than the target nucleon, $m_{3/2}\ll m_N$, and the results simplify to
\begin{equation}
  t\simeq -m_{3/2}^2\left( 1-2\abs{\vec{v}}\cos\theta\right) \qquad\text{and}\qquad t_0-t_1\simeq4\abs{\vec{v}}m_{3/2}^2\,.
  \label{trangeapx}
\end{equation}
Since the differential cross section does not change significantly over the kinematically allowed range, the total cross section is approximately given by
\begin{equation}
  \sigma_N\simeq\frac{\xi_i^2\,m_{3/2}^4\,m_N^2\left( \sum_qf_{T_q}^N\right) ^2}{48\,\pi\,v^2\,m_h^4\,\MP^2\abs{\vec{v}}}\,,
\end{equation}
where we neglected a possible suppression by a nuclear form factor. Numerically we find
\begin{equation}
  \sigma_p\simeq1.2\times10^{-63}\,\text{pb}\left( \frac{\xi_i}{10^{-7}}\right) ^2\left( \frac{m_{3/2}}{10\,\text{MeV}}\right) ^4\left( \frac{115\,\text{GeV}}{m_h}\right) ^4\left( \frac{220\,\text{km}\,\text{s}^{-1}}{\abs{\vec{v}}}\right) 
\end{equation}
and
\begin{equation}
  \sigma_n\simeq1.3\times10^{-63}\,\text{pb}\left( \frac{\xi_i}{10^{-7}}\right) ^2\left( \frac{m_{3/2}}{10\,\text{MeV}}\right) ^4\left( \frac{115\,\text{GeV}}{m_h}\right) ^4\left( \frac{220\,\text{km}\,\text{s}^{-1}}{\abs{\vec{v}}}\right) 
\end{equation}
for the cases of gravitino--proton and gravitino--neutron scattering, respectively. These values are extremely far below the current experimental limits on elastic scattering cross sections which are on the order of $10^{-8}\,$pb. Therefore, it seems hopeless to detect a signal from gravitino dark matter in an underground laboratory, even in the case of larger amounts of $R$-parity violation. Nevertheless, we also want to consider the remaining two cases.
\newpage

\subsection*{Inelastic Gravitino--Nucleon Scattering via \boldmath$Z$ Exchange}

At tree level there are four diagrams contributing to the inelastic gravitino--nucleon scattering via the exchange of a $Z$ boson:
\begin{equation*}
 \parbox{6.3cm}{
  \begin{picture}(183,173) (107,-22)
    \SetWidth{0.5}
    \SetColor{Black}
    \Photon(198,96)(198,29){-6}{3}
    \Line[arrow,arrowpos=0.5,arrowlength=3.75,arrowwidth=1.5,arrowinset=0.2](227,113)(256,130)
    \Line[arrow,arrowpos=0.5,arrowlength=3.75,arrowwidth=1.5,arrowinset=0.2](198,29)(256,-5)
    \Line[arrow,arrowpos=0.5,arrowlength=3.75,arrowwidth=1.5,arrowinset=0.2](140,-5)(198,29)
    \Line[double,sep=4](140,130)(198,96)
    \Vertex(198,96){2.5}
    \Vertex(198,29){2.5}
    \Text(138,131)[rb]{\normalsize{\Black{$\psi_{3/2}$}}}
    \Text(257,131)[lb]{\normalsize{\Black{$\nu_i$}}}
    \Line(198,96)(227,113)
    \Photon(198,96)(227,113){-6}{2}
    \Line(225,107.001)(229,118.999)\Line(221.001,115)(232.999,111)
    \Text(216,101)[lt]{\normalsize{\Black{$\tilde{Z}$}}}
    \Text(206,62)[l]{\normalsize{\Black{$Z$}}}
    \Text(138,-6)[rt]{\normalsize{\Black{$N$}}}
    \Text(257,-6)[lt]{\normalsize{\Black{$N$}}}
  \end{picture}
 }
 \quad+\quad
 \parbox{6.3cm}{
  \begin{picture}(183,173) (107,-22)
    \SetWidth{0.5}
    \SetColor{Black}
    \Photon(198,96)(198,29){-6}{3}
    \Line[arrow,arrowpos=0.5,arrowlength=3.75,arrowwidth=1.5,arrowinset=0.2](198,96)(256,130)
    \Line[arrow,arrowpos=0.5,arrowlength=3.75,arrowwidth=1.5,arrowinset=0.2](198,29)(256,-5)
    \Line[arrow,arrowpos=0.5,arrowlength=3.75,arrowwidth=1.5,arrowinset=0.2](140,-5)(198,29)
    \Line[double,sep=4](140,130)(198,96)
    \Vertex(198,96){2.5}
    \Vertex(198,29){2.5}
    \Text(138,131)[rb]{\normalsize{\Black{$\psi_{3/2}$}}}
    \Text(257,131)[lb]{\normalsize{\Black{$\nu_i$}}}
    \Text(206,62)[l]{\normalsize{\Black{$Z$}}}
    \Text(138,-6)[rt]{\normalsize{\Black{$N$}}}
    \Text(257,-6)[lt]{\normalsize{\Black{$N$}}}
    \Line[dash,dashsize=4,arrow,arrowpos=0.5,arrowlength=3.75,arrowwidth=1.5,arrowinset=0.2](179,85)(198,96)
    \Text(178,86)[rt]{\normalsize{\Black{$v_i$}}}
  \end{picture}
 }
\end{equation*}
\begin{equation*}
 +\quad
 \parbox{6.3cm}{
  \begin{picture}(183,173) (107,-22)
    \SetWidth{0.5}
    \SetColor{Black}
    \Photon(198,96)(198,29){-6}{3}
    \Line[arrow,arrowpos=0.5,arrowlength=3.75,arrowwidth=1.5,arrowinset=0.2](227,113)(256,130)
    \Line(198,96)(227,113)
    \Line[arrow,arrowpos=0.5,arrowlength=3.75,arrowwidth=1.5,arrowinset=0.2](198,29)(256,-5)
    \Line[arrow,arrowpos=0.5,arrowlength=3.75,arrowwidth=1.5,arrowinset=0.2](140,-5)(198,29)
    \Line[double,sep=4](140,130)(198,96)
    \Vertex(198,96){2.5}
    \Vertex(198,29){2.5}
    \Line(225,107.001)(229,118.999)\Line(221.001,115)(232.999,111)
    \Text(214,101)[lt]{\normalsize{\Black{$\tilde{H}_{u,\,d}^0$}}}
    \Text(138,131)[rb]{\normalsize{\Black{$\psi_{3/2}$}}}
    \Text(257,131)[lb]{\normalsize{\Black{$\nu_i$}}}
    \Text(206,62)[l]{\normalsize{\Black{$Z$}}}
    \Text(138,-6)[rt]{\normalsize{\Black{$N$}}}
    \Text(257,-6)[lt]{\normalsize{\Black{$N$}}}
    \Line[dash,dashsize=4](179,85)(198,96)
    \Text(178,86)[rt]{\normalsize{\Black{$v_{u,\,d}$}}}
  \end{picture}
 }.
\end{equation*}
For the coupling of the $Z$ boson to the nucleons we only consider the vector part of the interaction. In this case only the charges of the valence quarks contribute. The coupling to protons and neutrons is then calculated as
\begin{equation}
 \begin{split}
 \text{proton:}\qquad\quad\sum_qC_V &=2\,C_V(u)+C_V(d)=\frac{1}{4}-\sin^2\theta_W\,, \\
 \text{neutron:}\qquad\quad\sum_qC_V &=C_V(u)+2\,C_V(d)=-\frac{1}{4}\,.
 \end{split}
\end{equation}
Since $\sin^2\theta_W\approx1/4$, the $Z$ coupling to protons is suppressed compared to the coupling to neutrons. We present the calculation of the differential cross section in Appendix~\ref{gravitinoscattering}. The full result is rather lengthy so we directly show the approximate result for the total cross section. Using the same approximations as for the Higgs channel we find
\begin{equation}
 \begin{split}
  \sigma_n &\simeq\frac{\xi_i^2\,g_Z^2\,m_{3/2}^2}{128\,\pi\,m_Z^2\,\MP^2\abs{\vec{v}}}\bigg( U_{\tilde{Z}\tilde{Z}}^2\frac{m_{3/2}^2}{m_Z^2}+\abs{1+s_\beta\,U_{\tilde{H}_u^0\tilde{Z}}-c_\beta\,U_{\tilde{H}_d^0\tilde{Z}}}^2 \\
  &\qquad\qquad\qquad\qquad-\frac{2}{3}\,\frac{m_{3/2}}{m_Z}\,U_{\tilde{Z}\tilde{Z}}\left( 1+s_\beta \RE U_{\tilde{H}_u^0\tilde{Z}}-c_\beta \RE U_{\tilde{H}_d^0\tilde{Z}}\right) \!\bigg)\,.
 \end{split}
\end{equation}
We observe that the contribution from the 4-vertex dominates the cross section. Neglecting the small higgsino--zino mixing parameters, this leads to the following numerical result for the gravitino--neutron cross section via $Z$ exchange:
\begin{equation}
  \sigma_n\simeq4.7\times10^{-50}\,\text{pb}\left( \frac{\xi_i}{10^{-7}}\right) ^2\left( \frac{m_{3/2}}{10\,\text{MeV}}\right) ^2\left( \frac{220\,\text{km}\,\text{s}^{-1}}{\abs{\vec{v}}}\right) .
\end{equation}
This value is still far below the reach of detectors but it is thirteen orders of magnitude larger than the Higgs exchange contribution. Let us now turn to the case of photon exchange.

\subsection*{Inelastic Gravitino--Nucleon Scattering via Photon Exchange}

At tree level there is only one diagram contributing to the inelastic gravitino--nucleon scattering via the exchange of a photon:
\begin{equation*}
 \parbox{6.3cm}{
  \begin{picture}(183,173) (107,-22)
    \SetWidth{0.5}
    \SetColor{Black}
    \Photon(198,96)(198,29){-6}{3}
    \Line[arrow,arrowpos=0.5,arrowlength=3.75,arrowwidth=1.5,arrowinset=0.2](227,113)(256,130)
    \Line[arrow,arrowpos=0.5,arrowlength=3.75,arrowwidth=1.5,arrowinset=0.2](198,29)(256,-5)
    \Line[arrow,arrowpos=0.5,arrowlength=3.75,arrowwidth=1.5,arrowinset=0.2](140,-5)(198,29)
    \Line[double,sep=4](140,130)(198,96)
    \Vertex(198,96){2.5}
    \Vertex(198,29){2.5}
    \Text(138,131)[rb]{\normalsize{\Black{$\psi_{3/2}$}}}
    \Text(257,131)[lb]{\normalsize{\Black{$\nu_i$}}}
    \Line(198,96)(227,113)
    \Photon(198,96)(227,113){-6}{2}
    \Line(225,107.001)(229,118.999)\Line(221.001,115)(232.999,111)
    \Text(216,101)[lt]{\normalsize{\Black{$\tilde{\gamma}$}}}
    \Text(206,62)[l]{\normalsize{\Black{$\gamma$}}}
    \Text(138,-6)[rt]{\normalsize{\Black{$N$}}}
    \Text(257,-6)[lt]{\normalsize{\Black{$N$}}}
  \end{picture}
 }.
\end{equation*}
The photon coupling to the nucleons is also a vector interaction. From the calculation in Appendix~\ref{gravitinoscattering} we find the differential cross section
\begin{align}
  \frac{d\sigma_N}{dt}
  &=\frac{\xi_i^2\left( \sum_qQ\right) ^2e^2\abs{U_{\tilde{\gamma}\tilde{Z}}}^2}{192\,\pi\,t\,\MP^2\abs{\vec{v}}^2} \Bigg( 1-\frac{3}{4}\,\frac{m_{3/2}^2}{m_N^2}-5\,\frac{E}{m_N}-\frac{3}{4}\,\frac{t}{m_N^2}-\frac{t}{m_{3/2}^2}-6\,\frac{E^2}{m_{3/2}^2} \\
  &\qquad\quad-\frac{t\,E}{m_N\,m_{3/2}^2}+\frac{1}{4}\,\frac{t^2}{m_N^2\,m_{3/2}^2}+\frac{t}{m_{3/2}^2}\left( 2\,\left( \frac{E^2}{m_{3/2}^2}+\frac{t\,E}{m_N\,m_{3/2}^2}\right) +\frac{1}{4}\,\frac{t^2}{m_N^2\,m_{3/2}^2}\right) \!\Bigg)\,. \nonumber
\end{align}
In this case the coupling to neutrons vanishes as they are electromagnetically neutral. The coupling to protons is simply given by the elementary charge. We find the approximate total cross section
\begin{equation}
  \sigma_p\simeq\frac{\xi_i^2\,e^2\abs{U_{\tilde{\gamma}\tilde{Z}}}^2}{8\,\pi\,\MP^2\abs{\vec{v}}}=\frac{\xi_i^2\,\alpha\abs{U_{\tilde{\gamma}\tilde{Z}}}^2}{2\,\MP^2\abs{\vec{v}}}\,.
\end{equation}
It is important to note that this result is independent of the masses of the gravitino and the nucleon. We find the following numerical result for the gravitino--proton cross section via photon exchange:
\begin{equation}
  \sigma_p\simeq3.4\times10^{-43}\,\text{pb}\left( \frac{\xi_i}{10^{-7}}\right) ^2\left( \frac{220\,\text{km}\,\text{s}^{-1}}{\abs{\vec{v}}}\right) \left( \frac{\abs{U_{\tilde{\gamma}\tilde{Z}}}^2}{0.1}\right) .
\end{equation}
Although this number is still way to small to hope for a detection in this channel, it is very interesting that the cross section for the photon exchange channel is roughly seven orders of magnitude larger than for the $Z$ boson exchange channel and even twenty orders of magnitude larger than for the Higgs exchange channel. This shows that the massless propagator really leads to a significant enhancement of the cross section, even if -- unfortunately -- still far below the reach of any proposed underground detector.

\chapter{Conclusions and Outlook}
\label{conclusions}

After several decades of thorough theoretical studies and a multitude of ambitious experiments the nature of dark matter still remains one of the greatest unresolved mysteries in cosmology. A particularly well-motivated particle candidate for the dark matter is the gravitino in theories with a slight violation of $R$ parity as it naturally leads to a consistent cosmological scenario avoiding all cosmological gravitino problems. After inflation the universe is reheated to a very high temperature as required in order to generate the observed baryon asymmetry via the mechanism of thermal leptogenesis. In this early hot phase gravitinos are produced in scattering processes of the thermal plasma, thus leading to a relic density that is consistent with the observed dark matter density for generic values of the particle physics parameters. A tiny amount of $R$-parity breaking induces a sufficiently fast decay of the other supersymmetric particles so that the particle content of the thermal plasma reduces to that expected from the standard model of particle physics well before the onset of primordial nucleosynthesis. Together with the upper limit on the amount of $R$-parity breaking from the condition that a once generated baryon asymmetry is not washed out again by lepton number-violating processes, this leads to an allowed range for the gravitino lifetime that is several orders of magnitude above the current age of the universe.

In this thesis we have studied in detail the phenomenological consequences of unstable gravitino dark matter for indirect and direct detection experiments. We started with a discussion of the effect of $R$-parity violating mass mixing in the fermionic sector, providing new analytical approximate formulae for the mixing parameters. Equipped with these tools we presented a detailed calculation of gravitino three-body decay widths, thereby extending and correcting results previously obtained in the literature. Moreover, we extended the results for the gravitino two-body decay widths by adding a set of Feynman diagrams that were neglected in earlier calculations.

Using the results for the three-body decay widths we determined the gravitino branching ratios in the decoupling limit for an exemplary choice of parameters and discussed the differences to the two-body results and the implications for indirect searches for gravitino dark matter. We also found that there are regions in the parameter space where the two-body decay into a photon and a neutrino is suppressed. This could lead to interesting differences in the phenomenology for light gravitinos. We then discussed the spectra of particles produced in gravitino decays. For the case of gravitino two-body decays we presented the spectra of all possible stable final state particles as obtained with the help of PYTHIA including also the spectra for deuterons and antideuterons.\footnote{The deuteron spectra were generated by my collaborator Gilles Vertongen employing a Monte Carlo approach instead of the commonly used spherical coalescence model.} In the case of gravitino three-body decays we only presented the spectra of particles directly produced in the three-body decay as we were lacking a numerical tool to treat the subsequent fragmentation processes. This is a point that definitely should be worked out in future studies.

Moreover, we presented an extensive study of the cosmic-ray signals expected from gravitino dark matter decays, finding that gravitino decays cannot explain the observed rise in the positron fraction as this would lead to gamma-ray and and antiproton signals overshooting the observed fluxes. Actually, requiring that the gravitino signal does not exceed the observed fluxes, we estimated conservative lower limits on the gravitino lifetime at the level of $10^{26}$--$10^{27}\,$s. Searches for photon lines lead to significantly stronger lifetime limits reaching almost $10^{29}\,$s for gravitino masses below 100\,GeV if the photon line from gravitino decays is not suppressed.  We also elaborated on the prospects of future antideuteron searches, finding that the antideuteron channel potentially allows to put even stronger limits on the gravitino lifetime, especially at low masses. This is due to the fact that forthcoming experiments are planned to search for antideuterons at low energies where the astrophysical background is small compared to the signal expectation. But also for larger gravitino masses the antideuteron signal from gravitino decays might be observable on top of the astrophysical background. This is in contrast to findings in earlier analyses where the spectrum of high-energetic antideuterons was artificially suppressed by the spherical approximation of the antideuteron coalescence prescription. As we were lacking the spectra of final state particles for gravitino three-body decays, we concentrated in these discussions on the case of gravitinos heavier than the $W$ boson mass. However, since the phenomenology of gravitino three-body decays promises to lead to interesting predictions we plan to perform a study for lighter gravitinos in future work.

In addition, we presented a detailed analysis of the expected neutrino signals from gravitino decays, discussing also the different event topologies and the respective expected rates in neutrino detectors. We found that the neutrino channel is particularly useful to constrain rather heavy gravitinos since the observed event rates get enhanced at high energies. The most promising detection channel are neutrino-induced shower events as they provide the best energy resolution and suffer the least background of atmospheric neutrinos. Finally, we employed the combination of limits on the gravitino lifetime from various cosmic-ray channels to place constraints on the underlying $R$-parity violating parameter. Interestingly, indirect dark matter searches exactly probe values for the strength of $R$-parity violation that are favored by arguments from cosmology. Actually, we find that about one half of the cosmologically favored parameter range is already excluded by indirect searches over the range of gravitino masses considered in this work.

In the last part of this thesis we discussed the prospects for a direct detection of gravitino dark matter. It is interesting that the violation of $R$ parity allows for inelastic scatterings off nucleons with a significantly enhanced cross section. The kinematics of these processes is very different from elastic scatterings and also from the proposed inelastic dark matter models. We find that the momentum transferred to the nucleon is practically independent of the dark matter velocity distribution in the galactic halo and simply given by the gravitino mass. This could lead to a spectacular new kind of signals at dark matter detectors. Additionally, in contrast to other dark matter candidates the gravitino can scatter via photon exchange, leading to an enhancement of the cross section due to the massless propagator. Unfortunately, the suppression by the Planck scale and the small $R$-parity breaking parameters is still too strong and there seems to be no hope to ever detect a gravitino scattering signal in an underground detector.\smallskip

Now is a very exciting time for dark matter studies. Direct detection experiments will soon completely probe the parameter space for neutralino dark matter and other generic weakly interacting massive particles. At the same time the LHC will be able to test a large portion of the parameter space of softly broken supersymmetric theories and might soon deliver a first observation of supersymmetric particles. In particular, the LHC will be able to test scenarios of gravitino dark matter with broken $R$-parity by searching for effects of long-lived massive particles. As the decay length of the next-to-lightest supersymmetric particle depends on the strength of bilinear $R$-parity breaking, the LHC will allow to extend the astrophysical bounds by roughly one order of magnitude for gravitino masses at the order of the electroweak scale.

Finally, forthcoming indirect detection experiments will allow to search for dark matter signals with much higher precision. In particular the AMS-02 experiment that was successfully launched into space on the last flight of space shuttle Endeavor and subsequently installed at the International Space Station just a few weeks ago will greatly improve on the observations of charged cosmic rays in the coming years. It is capable of measuring positrons and electrons up to energies of almost a TeV. Due to its spectrometric measurement it is able to distinguish positrons from electrons up to energies of several hundred GeV thus allowing to observe a possible cutoff in the positron spectrum due to an endpoint in the spectrum from dark matter annihilations or decays. In addition, the observation of the spectrum of antiprotons up to almost a TeV will allow to decide if a deviation from the astrophysical background as predicted by heavy dark matter explanations of the PAMELA and Fermi LAT data really exists. Last but not least AMS-02 will be the first experiment to significantly improve on the sensitivity for antideuterons, thereby possibly allowing for the detection of a low-energetic dark matter signal practically free from astrophysical backgrounds.

\newpage

\subsection*{Note Added}

Last week an analysis appeared on the arXiv that also presents a calculation of the gravitino three-body decay widths~\cite{Diaz:2011pc}. Their analysis confirms that the results for the decay widths presented in~\cite{Choi:2010jt} were incorrect. In addition, their formulae seem to be in accord with our results. The Higgs exchange channel was not considered in this work, but the interference between the photon, $Z$ and $W$ channels for the $\ell_i^+\ell_i^-\nu_i$ final state was taken into account. This part was neglected in our calculation.

\phantomsection
\addcontentsline{toc}{chapter}{Acknowledgments}
\chapter*{Acknowledgments}

First and foremost I would like to thank my fianc\'ee Melanie who supported me during all the time I was working on this thesis.

Many thanks go also to my supervisor Laura Covi for many helpful discussions and valuable advice, and to Jan Louis for acting as second examiner of this thesis.

I am also grateful to Alejandro Ibarra, David Tran and Gilles Vertongen for a fruitful collaboration on parts of the topics discussed in this thesis.

In addition, I would like to thank my office colleagues Vladimir Mitev, Jasper Hasenkamp, Kai Schmitz and Sebastian Schmidt for many helpful discussions and an enjoyable atmosphere in the office.

I would also like to thank Jan Hajer for helpful discussions about bilinear $R$-parity breaking, and Kai Schmitz, Gilles Vertongen, my brother Christian and in particular Jasper Hasenkamp for helpful suggestions on the manuscript of this thesis.

Finally, many thanks go to the entire DESY theory group for an enjoyable time during the years of preparing this thesis.\smallskip

In addition, I acknowledge the support of the ``Impuls- und Vernetzungsfonds'' of the Helmholtz Association under contract number VH-NG-006 and of the Deutsche Forschungsgemeinschaft within the Collaborative Research Centre 676.

{\appendix

\chapter{Units and Physical Constants}
\label{constants}
In this appendix we summarize the values of all the physical and astrophysical constants that are used throughout this thesis. At times we work in units where the reduced Planck constant, the speed of light and the Boltzmann constant obey $\hbar=c=k=1$. In this case, we have the conversion factors
\begin{equation*}
 \begin{split}
  1\,\text{eV} &=1.160\,4505(20)\times 10^4\,\text{K}\,, \\
  1\,\text{eV} &=1.602\,176\,487(40)\times 10^{-19}\,\text{J}\,, \\
  1\,\text{GeV} &=1.782\,661\,758(44)\times 10^{-27}\,\text{kg}\,, \\
  1\,\text{GeV}^{-1} &=1.973\,269\,631(49)\times 10^{-16}\,\text{m}\,, \\
  1\,\text{GeV}^{-1} &=6.582\,118\,99(16)\times 10^{-25}\,\text{s}\,, \\
  1\,\text{GeV}^{-2} &=0.389\,379\,304(19)\,\text{mb}\,.
 \end{split}
\end{equation*}
All figures are taken from the \textit{The Review of Particle Physics}~\cite{Nakamura:2010zzi}. Numbers in parentheses represent the one standard deviation uncertainty in the last digits. \medskip

\noindent
\begin{tabular*}{\linewidth}{lllll}
 \toprule
 Quantity & $\;$ & Symbol & $\;$ & Value \\
 \midrule
 parsec & & pc & & $ 3.085\,6776\times 10^{16}\,\text{m} $ \\
 Solar distance from galactic center & & $ r_{\odot} $ & & $ 8.4(4)\,\text{kpc} $ \\
 Earth mean equatorial radius & & $ R_{\oplus} $ & & $ 6.378\,137\times 10^6\,\text{m} $ \\
 local halo density & & $ \rho_{\text{loc}} $ & & 0.3\,GeV\,cm$^{-3}$ (within factor 2--3) \\
 present-day CMB temperature & & $ T_0 $ & & $ 2.725(1)\,\text{K} $ \\
 normalized Hubble constant & & $ h $ & & $ 0.72(3) $ \\
 critical density & & $ \rho_c $ & & $ 1.053\,68(11)\times 10^{-5}\, h^2\,\text{GeV}\,\text{cm}^{-3} $ \\
 matter density & & $ \Omega_mh^2 $ & & $ 0.133(6) $ \\
 baryon density & & $ \Omega_bh^2 $ & & $ 0.0227(6) $ \\
 dark matter density & & $ \Omega_{\text{DM}}h^2 $ & & $ 0.110(6) $ \\
 dark energy density & & $ \Omega_{\Lambda} $ & & $ 0.74(3) $ \\
 total energy density & & $ \Omega_{\text{tot}} $ & & $ 1.006(6) $ \\
 baryon-to-photon ratio & & $ \eta $ & & $ 6.23(17)\times 10^{-10} $ \\
 \bottomrule
\end{tabular*}

\noindent
\begin{tabular*}{\linewidth}{lllll}
 \toprule
 Quantity & $\qquad$ & Symbol & $\qquad$ & Value \\
 \midrule
 barn & & b & & $10^{-28}\,\text{m}^2$ \\
 electron mass & & $ m_e $ & & $ 510.998\,910(13)\,\text{keV} $ \\
 muon mass & & $ m_{\mu} $ & & $ 105.658\,367(4)\,\text{MeV} $ \\
 muon mean life & & $ \tau_{\mu} $ & & $ 2.197\,034(21)\times 10^{-6}\,\text{s} $ \\
 tau mass & & $ m_{\tau} $ & & $ 1.776\,82(16)\,\text{GeV} $ \\
 tau mean life & & $ \tau_{\tau} $ & & $ 290.6(1.0)\times 10^{-15}\,\text{s} $ \\
 up quark mass & & $m_u$ & & 1.7--3.3\,MeV \\
 down quark mass & & $m_d$ & & 4.1--5.8\,MeV \\
 strange quark mass & & $m_s$ & & $101_{-21}^{+29}\,$MeV \\
 charm quark mass & & $m_c$ & & $1.27_{-0.09}^{+0.07}\,$GeV \\
 bottom quark mass & & $m_b$ & & $4.19_{-0.06}^{+0.18}\,$GeV \\
 top quark mass & & $m_t$ & & $172.0(0.9)(1.3)\,$GeV \\
 $W$ boson mass & & $ m_W $ & & $ 80.399(23)\,\text{GeV} $ \\
 $W$ boson decay width & & $ \Gamma_W $ & & $ 2.085(42)\,\text{GeV} $ \\
 $Z$ boson mass & & $ m_Z $ & & $ 91.1876(21)\,\text{GeV} $ \\
 $Z$ boson decay width & & $ \Gamma_Z $ & & $ 2.4952(23)\,\text{GeV} $ \\
 neutron mass & & $ m_n $ & & $ 939.565\,346(23)\,\text{MeV} $ \\
 neutron mean life & & $ \tau_n $ & & $ 885.7(8)\,\text{s} $ \\
 proton mass & & $ m_p $ & & $ 938.272\,013(23)\,\text{MeV} $ \\
 deuteron mass & & $m_d$ & & 1875.612\,793(47)\,MeV \\
 fine-structure constant & & $\alpha$ & & $1/137.035\,999\,679(94)$ \\
 Fermi coupling constant & & $G_F$ & & $1.116\,37(1)\times 10^{-5}\,$GeV$^{-2}$ \\
 weak mixing angle & & $\sin^2\theta_W(m_Z)$ & & $ 0.231\,16(13) $ \\
 strong coupling constant & & $\alpha_s(m_Z)$ & & 0.1184(7) \\
 gravitational constant & & $ G_N $ & & $ 6.708\,81(67)\times 10^{-39}\,\text{GeV}^{-2} $ \\
 Avogadro constant & & $N_A$ & & $6.022\,141\,79(30)\times10^{23}\,\text{mol}^{-1}$ \\
 \bottomrule
\end{tabular*}
\bigskip

\noindent
The values for the neutrino mixing parameters are taken from the recent analysis in~\cite{Schwetz:2011qt}. If two values are given, the first is the value assuming normal neutrino mass hierarchy and the second is the value for inverted hierarchy.
\bigskip

\noindent
\begin{tabular*}{\linewidth}{lllll}
 \toprule
 Quantity & Symbol & Value \\
 \midrule
 solar neutrino mixing angle & $\sin^2{\theta_{12}}$ & $0.316(16)$ \\
 atmospheric mixing angle & $\sin^2{\theta_{23}}$ & $0.51(6)$,\, $0.52(6)$ \\
 third neutrino mixing angle & $\sin^2{\theta_{13}}$ & $0.017(_{-9}^{+7})$,\, $0.020(_{-9}^{+8})$ \\
 solar neutrino mass difference & $\Delta m_{21}^2$ & $ 7.64(_{-18}^{+19})\times 10^{-5}\,\text{eV}^2$ \\
 atmospheric mass difference & $\Delta m_{31}^2$ & $2.45(9)\times 10^{-3}\,\text{eV}^2$,\, $-2.34(_{-9}^{+10})\times 10^{-3}\,\text{eV}^2$ \\
 \bottomrule
\end{tabular*}

\chapter{Notation, Conventions and Formulae}
\label{notation}

\paragraph{Four-Vectors and Tensors}
Lorentz indices are depicted by small Greek letters, \textit{e.g.} $ \mu=0,\,1,\,2,\,3 $. The metric of Minkowski space is chosen to be
\begin{equation}
 g_{\mu\nu}=g^{\mu\nu}=
 \begin{pmatrix}
  +1 & 0 & 0 & 0 \\
  0 & -1 & 0 & 0 \\
  0 & 0 & -1 & 0 \\
  0 & 0 & 0 & -1
 \end{pmatrix}.
 \label{metric}
\end{equation}
Lorentz indices of four-vectors and higher rank tensors are raised and lowered using the space-time metric: 
\begin{equation}
 a_{\mu}=g_{\mu\nu}\,a^{\nu}\qquad\text{and}\qquad b^{\mu}=g^{\mu\nu}\,b_{\nu}\,. 
\end{equation}
We use in all cases a sum convention for Lorentz indices: 
\begin{equation}
 a^{\mu}b_{\mu}\equiv\sum_{\mu=0}^3a^{\mu}b_{\mu}\,. 
\end{equation}

\paragraph{Antisymmetric Symbols}
The totally antisymmetric tensor $\varepsilon^{\mu\nu\rho\sigma}$ is defined such that
\begin{equation}
 \varepsilon^{0123}=+1\,.
\end{equation}
The indices of the totally antisymmetric tensor can be lowered using the space-time metric:
\begin{equation}
 \varepsilon_{\mu\nu\rho\sigma}=g_{\mu\alpha}\,g_{\nu\beta}\,g_{\rho\gamma}\,g_{\sigma\delta}\,\varepsilon^{\alpha\beta\gamma\delta}.
\end{equation}
The totally antisymmetric tensor in three dimensions, $\varepsilon_{ijk}$, is chosen such that
\begin{equation}
 \varepsilon_{123}=+1\,.
\end{equation}

\paragraph{Gamma Matrices}
The Dirac gamma matrices are defined to form a Clifford algebra with the anticommutation relation
\begin{equation}
 \left\lbrace \gamma^{\mu},\,\gamma^{\nu}\right\rbrace =2\,g^{\mu\nu}. 
 \label{clifford}
\end{equation}
In most cases it is convenient to suppress the spinor indices of the gamma matrices, so we do not write them here. We also want to introduce an explicit representation of the gamma matrices. In the Weyl basis (also called chiral basis) the Dirac gamma matrices read
\begin{equation}
  \gamma^\mu=
  \begin{pmatrix}
  0 & \sigma^\mu \\
  \bar{\sigma}^\mu & 0
  \end{pmatrix}.
\end{equation}
In this expression the sigma matrices are given by $\sigma^\mu=\left( \mathbf{1}, \sigma_i\right) $ and $\bar{\sigma}^\mu=\left( \mathbf{1}, -\sigma_i\right) $, where the Pauli matrices are given by
\begin{equation}
 \sigma_1=
 \begin{pmatrix}
  0 & 1 \\
  1 & 0
 \end{pmatrix},\quad 
 \sigma_2=
 \begin{pmatrix}
  0 & -i \\
  i & 0
 \end{pmatrix},\quad 
 \sigma_3=
 \begin{pmatrix}
  1 & 0 \\
  0 & -1
 \end{pmatrix},
 \label{PauliMatrix}
\end{equation}
satisfying the identity $\sigma_i\,\sigma_j=\delta_{ij}+i\,\varepsilon_{ijk}\,\sigma_k\,$. Using the algebra of the gamma matrices and the explicit form of the space-time metric (\ref{metric}) it can be easily seen that
\begin{equation}
 \left( \gamma^0\right) ^2=\mathbf{1}\qquad\text{and}\qquad\left( \gamma^i\right) ^2=-\mathbf{1}\,.
\end{equation}
Hermitian conjugation of the gamma matrices yields 
\begin{equation}
 \left( \gamma^{\mu}\right) ^{\dagger} =\gamma^0\gamma^{\mu}\gamma^0. 
\end{equation}
The Lorentz indices of gamma matrices are raised and lowered by the metric, just as in the case of four-vectors. That implies 
\begin{equation}
 \gamma_0=\gamma^0\qquad\text{and}\qquad\gamma_i=-\gamma^i. 
\end{equation}
Thus we see that raising and lowering of Lorentz indices is equivalent to hermitian conjugation. In addition to $\gamma^0,\,\gamma^1,\,\gamma^2,\,\gamma^3,$ a fifth gamma matrix can be defined:
\begin{equation}
 \gamma^5\equiv i\gamma^0\gamma^1\gamma^2\gamma^3=-\frac{i}{4!}\,\varepsilon_{\mu\nu\rho\sigma} \gamma^{\mu}\gamma^{\nu}\gamma^{\rho}\gamma^\sigma. 
\end{equation}
Using the algebra (\ref{clifford}) it can be easily shown that $ \gamma^5 $ has the following properties: 
\begin{equation}
  \left\lbrace \gamma^5,\,\gamma^{\mu}\right\rbrace =0\,,\qquad\left( \gamma^5\right) ^2= 1\,,\qquad\left( \gamma^5\right) ^{\dagger}=\gamma^5.
\end{equation}
In the Weyl basis $\gamma^5$ has the following explicit form:
\begin{equation}
  \gamma^5=
  \begin{pmatrix}
  -\mathbf{1} & 0 \\
  0 & \mathbf{1}
  \end{pmatrix}.
\end{equation}
Using this expression we can define the chirality projection operators
\begin{equation}
 P_L\equiv\frac{1}{2}\left( 1-\gamma^5\right) =
 \begin{pmatrix}
 \mathbf{1} & 0 \\
 0 & 0
 \end{pmatrix}\qquad\text{and}\qquad P_R\equiv\frac{1}{2}\left( 1+\gamma^5\right) =
 \begin{pmatrix}
 0 & 0 \\
 0 & \mathbf{1}
 \end{pmatrix}.
 \label{projectors}
\end{equation}
These operators possess the projector properties
\begin{equation}
 P_L^2=P_L\,,\qquad P_R^2=P_R\qquad\text{and}\qquad P_LP_R=P_RP_L=0\,.
 \label{chirality}
\end{equation}

\paragraph{Identities for Gamma Matrices}
In the calculation of squared matrix elements we encounter traces of gamma matrices. For these the following identities can be derived from the algebra~(\ref{clifford}):
\begin{equation}
 \begin{split}
  \Tr\left( \mathbf{1}\right) &=4\,, \\
  \Tr\left( \text{odd number of $ \gamma $s}\right) &=0\,, \\
  \Tr\left( \gamma^{\mu}\gamma^{\nu}\right) &=4\,g^{\mu\nu}, \\
  \Tr\left( \gamma^{\mu}\gamma^{\nu}\gamma^{\rho}\gamma^{\sigma}\right) &=4\left( g^{\mu\nu}g^{\rho\sigma}-g^{\mu\rho}g^{\nu\sigma}+g^{\mu\sigma}g^{\nu\rho}\right) , \\
  \Tr\left( \gamma^5\right) &=0\,, \\
  \Tr\left( \gamma^{\mu}\gamma^{\nu}\gamma^5\right) &=0\,, \\
  \Tr\left( \gamma^{\mu}\gamma^{\nu}\gamma^{\rho}\gamma^{\sigma}\gamma^5\right) &=-4\,i\,\varepsilon^{\mu\nu\rho\sigma}. 
 \end{split}
 \label{GammaTrace}
\end{equation}
We also encounter contractions of gamma matrices. Using the algebra (\ref{clifford}), we obtain the following identities:
\begin{equation}
 \begin{split}
  \gamma^{\mu}\gamma_{\mu} &=4\,, \\
  \gamma^{\mu}\gamma^{\nu}\gamma_{\mu} &=-2\,\gamma^{\nu}, \\
  \gamma^{\mu}\gamma^{\nu}\gamma^{\rho}\gamma_{\mu} &=4\,g^{\nu\rho}, \\
  \gamma^{\mu}\gamma^{\nu}\gamma^{\rho}\gamma^{\sigma}\gamma_{\mu} &=-2\,\gamma^{\sigma}\gamma^{\rho}\gamma^{\nu}. \\
 \end{split}
 \label{GammaContraction}
\end{equation}

\paragraph{Two-Component and Four-Component Spinors}
A four-component Dirac spinor $\psi_D$ is made up of two two-component spinors, $\chi_\alpha$ and $\eta^{\dagger\dot{\alpha}}$, in the following way:
\begin{equation}
  \psi_D=
  \begin{pmatrix}
  \chi_\alpha \\
  \eta^{\dagger\dot{\alpha}}
  \end{pmatrix}.
\end{equation}
The two-component spinors $\chi_\alpha$ and $\eta^{\dagger\dot{\alpha}}$ are complex, anticommuting objects with two distinct types of spinor indices, $\alpha=1,\,2$ and $\dot{\alpha}=1,\,2$. The spinor indices are depicted by small Greek letters from the beginning of the alphabet. They can be raised and lowered using the antisymmetric symbols
\begin{equation}
  \varepsilon^{\alpha\beta},\quad\varepsilon_{\alpha\beta}\,,\quad\varepsilon^{\dot{\alpha}\dot{\beta}}\quad\text{and}\quad\varepsilon_{\dot{\alpha}\dot{\beta}}\,,\qquad\text{where}\quad\varepsilon^{12}=1\quad\text{and}\quad\varepsilon_{12}=-1.
\end{equation}
In many places the spinor indices are not written explicitly. Then they can be reconstructed according to the contraction conventions
\begin{equation}
  \chi\,\eta=\chi^\alpha\eta_\alpha\qquad\text{and}\qquad\chi^\dagger\eta^\dagger=\chi_{\dot{\alpha}}^\dagger\eta^{\dagger\dot{\alpha}}.
\end{equation}
The Dirac adjoint and the charge conjugate of a Dirac spinor are, respectively, given by
\begin{align}
  \bar{\psi}_D &\equiv\psi_D^\dagger\gamma^0=\left( \eta^\alpha\;\;\chi_{\dot{\alpha}}^\dagger\right) , \\
  \psi_D^c &\equiv C\bar{\psi}_D^T=
  \begin{pmatrix}
  \eta_\alpha \\
  \chi^{\dagger\dot{\alpha}}
  \end{pmatrix},
\end{align}
where we introduced the charge-conjugation matrix $C$ with the properties
\begin{equation}
  C^{\dagger}=C^{-1},\qquad C^T=-C\qquad\text{and}\qquad C\Gamma^{i\,T}C^{\dagger}=\eta_i\Gamma^i,
\end{equation}
and
\begin{equation}
 \eta_i =\left\lbrace \begin{array}{rl}
 1\quad\text{for}\;\Gamma^i & =1, \gamma^{\mu}\gamma^5, \gamma^5 \\
 -1\quad\text{for}\;\Gamma^i & =\gamma^{\mu}, \sigma^{\mu\nu}
 \end{array}.\right.
\end{equation}
In the above expression we use the definition $ \sigma^{\mu\nu}\equiv\frac{i}{2}\left[ \gamma^{\mu},\,\gamma^{\nu}\right] ,\quad\mu<\nu\,$.
Using the chirality projectors from equation~(\ref{projectors}) we define the left- and right-handed Dirac spinors
\begin{equation}
  P_L\psi_D=
  \begin{pmatrix}
  \chi_\alpha \\
  0
  \end{pmatrix},\qquad 
  P_R\psi_D=
  \begin{pmatrix}
  0 \\
  \eta^{\dagger\dot{\alpha}}
  \end{pmatrix}.
\end{equation}
A four-component Majorana spinor $\psi_M$ is defined to be invariant under charge conjugation. Thus a Majorana spinor is a Dirac spinor with the condition that $\eta=\chi$, \textit{i.e.}
\begin{equation}
  \psi_M=
  \begin{pmatrix}
  \chi_\alpha \\
  \chi^{\dagger\dot{\alpha}}
  \end{pmatrix}
  =\psi_M^c\equiv C\bar{\psi}_M^T\,.
\end{equation}
A very common object in theories of particles with half-integral spin is the contraction of gamma matrices with the particle momentum. Therefore, we use the Feynman slash notation 
\begin{equation}
 \slashed{p}\equiv\gamma^{\mu}p_{\mu}\,. 
\end{equation}
In order to obtain Lorentz scalars, in the formulation of theories of particles with half-integral spin there occur products of adjoint spinors and spinors. The adjoint spinor and adjoint vector-spinor are, respectively, defined as 
\begin{equation}
 \bar{\psi}\equiv\psi^{\dagger}\gamma^0\qquad\text{and}\qquad\bar{\psi}_{\mu}\equiv\psi_{\mu}^{\dagger}\gamma^0.
\end{equation}

\paragraph{Dirac Spinors}
Free spin-1/2 fermions are described by the Dirac equation,
\begin{equation}
 \left( i\slashed{\partial}-m\right) \psi(x)=0\,,
\end{equation}
that has four linearly independent plane-wave solutions:
\begin{equation}
 \psi(x)=u^s(p)\,e^{-ip\cdot x}\qquad\text{and}\qquad\psi(x)=v^s(p)\,e^{ip\cdot x},\quad s=\pm\frac{1}{2}\,, 
\end{equation}
where the Dirac spinors $u$ and $v$ have to obey the following constraints:
\begin{equation}
 \left( \slashed{p}-m\right) u^s(p)=0\qquad\text{and}\qquad\left( \slashed{p}+m\right) v^s(p)=0\,. 
\end{equation}
The polarization sums for spin-1/2 fermions are then given by~\cite{Peskin:1995ev}
\begin{equation}
 \begin{split} 
  \sum_s u^s(p)\,\bar{u}^s(p) &=\slashed{p}+m\,, \\
  \sum_s v^s(p)\,\bar{v}^s(p) &=\slashed{p}-m\,,
 \end{split}
 \label{fermionPolarization}
\end{equation}
where the sum is performed over the fermion helicity states $s=\pm\frac{1}{2}$\,.

\paragraph{Polarization Vectors}
Free massless spin-1 particles in the Lorentz gauge ($ \partial^{\mu}A_{\mu}=0 $) and free massive spin-1 particles obey the Proca equation~\cite{Collins:1989kn}
\begin{equation}
 \left( \partial^{\nu}\partial_{\nu}+M^2\right) A_{\mu}=0\,.
\end{equation}
This equation has the plane-wave solutions
\begin{equation}
 A_{\mu}=\epsilon_{\mu}(p)\,e^{-ip\cdot x},
\end{equation}
where $ \epsilon_{\mu} $ is a polarization vector. The condition $ \partial^{\mu}A_{\mu}=0 $ demands that
\begin{equation}
 p^{\mu}\epsilon_{\mu}=0\,,
\end{equation}
reducing the number of independent polarization vectors to three. For massless spin-1 particles we have the freedom to make the additional gauge transformation
\begin{equation}
 A_{\mu}\rightarrow A'_{\mu}=A_{\mu}+\partial^{\mu}\Lambda\qquad\text{with}\qquad\partial^{\mu}\partial_{\mu}\Lambda=0\,. 
\end{equation}
The condition for $ \Lambda $ is required by the Lorentz gauge condition. This additional gauge freedom reduces the number of independent polarization vectors for massless spin-1 particles to two.
The polarization sum for massless spin-1 particles is then given by~\cite{Peskin:1995ev}
\begin{equation}
 \sum_{\lambda}\epsilon_{\mu}^{\lambda\,*}(p)\,\epsilon_{\nu}^{\lambda}(p)=-g_{\mu\nu}\,,
 \label{vectorPolarization}
\end{equation}
where the sum is performed over the two polarization states of the massless spin-1 particle $\lambda=\pm1$, and the polarization sum for massive spin-1 particles is given by
\begin{equation}
 \sum_{\lambda}\epsilon_{\mu}^{\lambda\,*}(p)\,\epsilon_{\nu}^{\lambda}(p)=-\left( g_{\mu\nu}-\frac{p_{\mu}p_{\nu}}{m_A^2}\right) , 
\end{equation}
where the sum is performed over the three polarization states of the massive spin-1 particle $\lambda=\pm1,\,0$.

\chapter{Feynman Rules}
\label{feynmanrules}

In this appendix we provide the necessary Feynman rules for the calculation of the gravitino cross sections and decay widths. This set is taken from~\cite{Pradler:2007ne} and amended by the addition of the rules for the negative frequency solution of the gravitino. The Feynman rules are given in the unitary gauge in which Goldstone fields are absent from the theory. This is convenient since we are only performing tree-level calculations in a theory with broken electroweak symmetry.

Gravitinos are depicted as double solid lines, chiral fermions are drawn as single solid lines and scalars are drawn as dashed lines. Gauge bosons are represented by wiggled lines and gauginos by solid lines with additional wiggled lines.

The continuous fermion flow is independent of the fermion number flow carried by fermions and sfermions and also of the momentum direction.

\subsubsection*{External Lines}
The momentum $ p $ flows from the left side to the right side for the external lines shown below. \smallskip

\noindent
Scalar particles:
\begin{equation*}
 \parbox{2.0cm}{
 \begin{picture}(52,3) (-2,-1)
    \SetWidth{0.5}
    \SetColor{Black}
    \Vertex(50,1){2.5}
    \DashArrowLine(0,1)(50,1){4}
 \end{picture}
 }\quad\parbox{2.0cm}{
 \begin{picture}(52,3) (-2,-1)
    \SetWidth{0.5}
    \SetColor{Black}
    \Vertex(0,1){2.5}
    \DashArrowLine(0,1)(50,1){4}
 \end{picture}
 }\quad\parbox{2.0cm}{
 \begin{picture}(52,3) (-2,-1)
    \SetWidth{0.5}
    \SetColor{Black}
    \Vertex(50,1){2.5}
    \DashArrowLine(50,1)(0,1){4}
 \end{picture}
 }\quad\parbox{2.0cm}{
 \begin{picture}(52,3) (-2,-1)
    \SetWidth{0.5}
    \SetColor{Black}
    \Vertex(0,1){2.5}
    \DashArrowLine(50,1)(0,1){4}
 \end{picture}
 }\quad =\quad 1\,.
\end{equation*}
Gauginos and matter fermions:
\begin{align*}
 \parbox{2.0cm}{
 \begin{picture}(52,10) (-2,-3)
    \SetWidth{0.5}
    \SetColor{Black}
    \Photon(0,1)(50,1){3}{3}
    \Line(0,1)(50,1)
    \Vertex(50,1){2.5}
    \ArrowLine(5,8)(45,8)
 \end{picture}
 }\qquad\parbox{2.0cm}{
 \begin{picture}(52,10) (-2,-3)
    \SetWidth{0.5}
    \SetColor{Black}
    \Vertex(50,1){2.5}
    \ArrowLine(0,1)(50,1)
    \ArrowLine(5,8)(45,8)
 \end{picture}
 }\qquad\parbox{2.0cm}{
 \begin{picture}(52,10) (-2,-3)
    \SetWidth{0.5}
    \SetColor{Black}
    \Vertex(50,1){2.5}
    \ArrowLine(50,1)(0,1)
    \ArrowLine(5,8)(45,8)
 \end{picture}
 }\quad & =\quad u^s(p)\,, \\
 \parbox{2.0cm}{
 \begin{picture}(52,10) (-2,-3)
    \SetWidth{0.5}
    \SetColor{Black}
    \Photon(0,1)(50,1){3}{3}
    \Line(0,1)(50,1)
    \Vertex(0,1){2.5}
    \ArrowLine(5,8)(45,8)
 \end{picture}
 }\qquad\parbox{2.0cm}{
 \begin{picture}(50,12) (-2,-3)
    \SetWidth{0.5}
    \SetColor{Black}
    \Vertex(0,1){2.5}
    \ArrowLine(0,1)(50,1)
    \ArrowLine(5,8)(45,8)
 \end{picture}
 }\qquad\parbox{2.0cm}{
 \begin{picture}(52,12) (-2,-3)
    \SetWidth{0.5}
    \SetColor{Black}
    \Vertex(0,1){2.5}
    \ArrowLine(50,1)(0,1)
    \ArrowLine(5,8)(45,8)
 \end{picture}
 }\quad & =\quad\bar{u}^s(p)\,, \\
 \parbox{2.0cm}{
 \begin{picture}(52,10) (-2,-3)
    \SetWidth{0.5}
    \SetColor{Black}
    \Photon(0,1)(50,1){3}{3}
    \Line(0,1)(50,1)
    \Vertex(0,1){2.5}
    \ArrowLine(45,8)(5,8)
 \end{picture}
 }\qquad\parbox{2.0cm}{
 \begin{picture}(52,12) (-2,-3)
    \SetWidth{0.5}
    \SetColor{Black}
    \Vertex(0,1){2.5}
    \ArrowLine(0,1)(50,1)
    \ArrowLine(45,8)(5,8)
 \end{picture}
 }\qquad\parbox{2.0cm}{
 \begin{picture}(52,12) (-2,-3)
    \SetWidth{0.5}
    \SetColor{Black}
    \Vertex(0,1){2.5}
    \ArrowLine(50,1)(0,1)
    \ArrowLine(45,8)(5,8)
 \end{picture}
 }\quad & =\quad v^s(p)\,, \\
 \parbox{2.0cm}{
 \begin{picture}(52,10) (-2,-3)
    \SetWidth{0.5}
    \SetColor{Black}
    \Photon(0,1)(50,1){3}{3}
    \Line(0,1)(50,1)
    \Vertex(50,1){2.5}
    \ArrowLine(45,8)(5,8)
 \end{picture}
 }\qquad\parbox{2.0cm}{
 \begin{picture}(52,12) (-2,-3)
    \SetWidth{0.5}
    \SetColor{Black}
    \Vertex(50,1){2.5}
    \ArrowLine(0,1)(50,1)
    \ArrowLine(45,8)(5,8)
 \end{picture}
 }\qquad\parbox{2.0cm}{
 \begin{picture}(52,12) (-2,-3)
    \SetWidth{0.5}
    \SetColor{Black}
    \Vertex(50,1){2.5}
    \ArrowLine(50,1)(0,1)
    \ArrowLine(45,8)(5,8)
 \end{picture}
 }\quad & =\quad\bar{v}^s(p)\,.
\end{align*}
Gauge bosons:
\begin{align*}
 \parbox{2.8cm}{
 \begin{picture}(77,16) (-2,0)
    \SetWidth{0.5}
    \SetColor{Black}
    \Photon(25,8)(75,8){3}{3}
    \Text(10,8)[]{\normalsize{\Black{$\mu,a$}}}
    \Vertex(75,8){2.5}
 \end{picture}
 }\quad =\quad\epsilon_{\mu}^a(p)\,,\quad & \quad\parbox{2.8cm}{
 \begin{picture}(77,16) (-2,0)
    \SetWidth{0.5}
    \SetColor{Black}
    \Photon(0,8)(50,8){3}{3}
    \Text(65,8)[]{\normalsize{\Black{$\mu,a$}}}
    \Vertex(0,8){2.5}
 \end{picture}
 }\quad =\quad\epsilon_{\mu}^{*a}(p)\,.
\end{align*}
Gravitinos:
\begin{align*}
 \parbox{2.4cm}{
 \begin{picture}(67,16) (-2,0)
    \SetWidth{0.5}
    \SetColor{Black}
    \Line(15,9)(65,9)\Line(15,5)(65,5)%%JaxoDrawID:DoubleLine(2)
    \Vertex(65,7){2.5}
    \ArrowLine(20,14)(60,14)
    \Text(5,7)[]{\normalsize{\Black{$\mu$}}}
 \end{picture}
 }\quad =\quad\psi_{\mu}^{+\,s}(p)\,,\quad & \quad\parbox{2.4cm}{
 \begin{picture}(67,16) (-2,0)
    \SetWidth{0.5}
    \SetColor{Black}
    \Line(0,9)(50,9)\Line(0,5)(50,5)%%JaxoDrawID:DoubleLine(2)
    \Vertex(0,7){2.5}
    \ArrowLine(5,14)(45,14)
    \Text(60,7)[]{\normalsize{\Black{$\mu$}}}
 \end{picture}
 }\quad =\quad\bar{\psi}_{\mu}^{+\,s}(p)\,, \\ 
 \parbox{2.4cm}{
 \begin{picture}(67,16) (-2,0)
    \SetWidth{0.5}
    \SetColor{Black}
    \Line(0,9)(50,9)\Line(0,5)(50,5)%%JaxoDrawID:DoubleLine(2)
    \Vertex(0,7){2.5}
    \ArrowLine(45,14)(5,14)
    \Text(60,7)[]{\normalsize{\Black{$\mu$}}}
 \end{picture}
 }\quad =\quad\psi_{\mu}^{-\,s}(p)\,,\quad & \quad\parbox{2.4cm}{
 \begin{picture}(67,16) (-2,0)
    \SetWidth{0.5}
    \SetColor{Black}
    \Line(15,9)(65,9)\Line(15,5)(65,5)%%JaxoDrawID:DoubleLine(2)
    \Vertex(65,7){2.5}
    \ArrowLine(60,14)(20,14)
    \Text(5,7)[]{\normalsize{\Black{$\mu$}}}
 \end{picture}
 }\quad =\quad\bar{\psi}_{\mu}^{-\,s}(p)\,.
\end{align*}

\subsubsection*{Propagators}
The momentum $ p $ flows from the left side to the right side for the propagators shown below. \medskip
\\
Scalar particles:
\begin{align*}
 \parbox{2.8cm}{
 \begin{picture}(82,5) (-2,-1)
    \SetWidth{0.5}
    \SetColor{Black}
    \DashArrowLine(15,1)(65,1){4}
    \Vertex(15,1){2.5}
    \Vertex(65,1){2.5}
    \Text(5,1)[]{\normalsize{\Black{$i$}}}
    \Text(75,1)[]{\normalsize{\Black{$j$}}}
 \end{picture}
 }\quad &=\quad\frac{i}{p^2-m_\phi^2+im_\phi\Gamma_\phi}\,\delta^{ij}\,.
 \intertext{Matter fermions: }
 \parbox{2.8cm}{
 \begin{picture}(82,10) (-2,-3)
    \SetWidth{0.5}
    \SetColor{Black}
    \ArrowLine(15,1)(65,1)
    \Vertex(15,1){2.5}
    \Vertex(65,1){2.5}
    \ArrowLine(20,8)(60,8)
    \Text(5,1)[]{\normalsize{\Black{$i$}}}
    \Text(75,1)[]{\normalsize{\Black{$j$}}}
 \end{picture}
 }\quad &=\quad\frac{i\left( \slashed{p}+m_\chi\right) }{p^2-m_\chi^2+im_\chi\Gamma_\chi}\,\delta^{ij}\,, \\
 \parbox{2.8cm}{
 \begin{picture}(82,10) (-2,-3)
    \SetWidth{0.5}
    \SetColor{Black}
    \ArrowLine(15,1)(65,1)
    \Vertex(15,1){2.5}
    \Vertex(65,1){2.5}
    \ArrowLine(60,8)(20,8)
    \Text(5,1)[]{\normalsize{\Black{$i$}}}
    \Text(75,1)[]{\normalsize{\Black{$j$}}}
 \end{picture}
 }\quad &=\quad\frac{i\left( -\slashed{p}+m_{\chi}\right) }{p^2-m_{\chi}^2+im_\chi\Gamma_\chi}\,\delta^{ij}\,.
 \intertext{Gauginos: }
 \parbox{2.8cm}{
 \begin{picture}(82,10) (-2,-3)
    \SetWidth{0.5}
    \SetColor{Black}
    \Photon(15,1)(65,1){3}{3}
    \Line(15,1)(65,1)
    \Vertex(15,1){2.5}
    \Vertex(65,1){2.5}
    \ArrowLine(20,8)(60,8)
    \Text(5,1)[]{\normalsize{\Black{$a$}}}
    \Text(75,1)[]{\normalsize{\Black{$b$}}}
 \end{picture}
 }\quad &=\quad\frac{i\left( \slashed{p}+m_{\lambda}\right) }{p^2-m_{\lambda}^2+im_\lambda\Gamma_\lambda}\,\delta^{ab}\,.
 \intertext{Massless and massive gauge bosons: }
 \parbox{2.8cm}{
 \begin{picture}(82,10) (-2,-3)
    \SetWidth{0.5}
    \SetColor{Black}
    \Photon(15,1)(65,1){3}{3}
    \Vertex(15,1){2.5}
    \Vertex(65,1){2.5}
    \Text(0,1)[]{\normalsize{\Black{$a,\mu$}}}
    \Text(80,1)[]{\normalsize{\Black{$b,\nu$}}}
 \end{picture}
 }\quad &=\quad-\frac{ig_{\mu\nu}}{p^2}\,\delta^{ab}\,, \\
 \parbox{2.8cm}{
 \begin{picture}(82,10) (-2,-3)
    \SetWidth{0.5}
    \SetColor{Black}
    \Photon(15,1)(65,1){3}{3}
    \Vertex(15,1){2.5}
    \Vertex(65,1){2.5}
    \Text(0,1)[]{\normalsize{\Black{$a,\mu$}}}
    \Text(80,1)[]{\normalsize{\Black{$b,\nu$}}}
 \end{picture}
 }\quad &=\quad-\frac{i\left( g_{\mu\nu}-p_{\mu}p_{\nu}/m_A^2\right) }{p^2-m_A^2+im_A\Gamma_A}\,\delta^{ab}\,.
\end{align*}

\subsubsection*{Gravitino Vertices}

The momentum $ p $ flows into the vertex for the vertices shown below.
\begin{align*}
 \parbox{2.3cm}{
 \begin{picture}(90,85) (0,0)
    \SetWidth{0.5}
    \SetColor{Black}
    \ArrowLine(70,17)(55,43)
    \Vertex(55,43){2.5}
    \Line(25,45)(55,45)\Line(25,41)(55,41)%%JaxoDrawID:DoubleLine(2)
    \DashArrowLine(55,43)(70,69){4}
    \Text(15,43)[]{\normalsize{\Black{$\mu$}}}
    \Text(75,77)[]{\normalsize{\Black{$i,p$}}}
    \Text(75,8)[]{\normalsize{\Black{$j$}}}
    \ArrowArc(29.96,-2.04)(35.04,30.99,89.94)
 \end{picture}
 } &=-\frac{i}{\sqrt{2}\,\MP}\,\delta_{ij}\slashed{p}\gamma^{\mu}P_L\,, & \parbox{2.3cm}{
 \begin{picture}(90,85) (0,0)
    \SetWidth{0.5}
    \SetColor{Black}
    \ArrowLine(70,17)(55,43)
    \Vertex(55,43){2.5}
    \Line(25,45)(55,45)\Line(25,41)(55,41)%%JaxoDrawID:DoubleLine(2)
    \DashArrowLine(55,43)(70,69){4}
    \Text(15,43)[]{\normalsize{\Black{$\mu$}}}
    \Text(75,77)[]{\normalsize{\Black{$i,p$}}}
    \Text(75,8)[]{\normalsize{\Black{$j$}}}
    \ArrowArcn(29.96,-2.04)(35.04,89.94,30.99)
 \end{picture}
 } &=-\frac{i}{\sqrt{2}\,\MP}\,\delta_{ij}P_L\gamma^{\mu}\slashed{p}\,, \\
 \parbox{2.3cm}{
 \begin{picture}(90,85) (0,0)
    \SetWidth{0.5}
    \SetColor{Black}
    \ArrowLine(55,43)(70,17)
    \Vertex(55,43){2.5}
    \Line(25,45)(55,45)\Line(25,41)(55,41)%%JaxoDrawID:DoubleLine(2)
    \DashArrowLine(70,69)(55,43){4}
    \Text(15,43)[]{\normalsize{\Black{$\mu$}}}
    \Text(75,77)[]{\normalsize{\Black{$i,p$}}}
    \Text(75,8)[]{\normalsize{\Black{$j$}}}
    \ArrowArc(29.96,-2.04)(35.04,30.99,89.94)
 \end{picture}
 } &=-\frac{i}{\sqrt{2}\,\MP}\,\delta_{ij}\slashed{p}\gamma^{\mu}P_R\,, & \parbox{2.3cm}{
 \begin{picture}(90,85) (0,0)
    \SetWidth{0.5}
    \SetColor{Black}
    \ArrowLine(55,43)(70,17)
    \Vertex(55,43){2.5}
    \Line(25,45)(55,45)\Line(25,41)(55,41)%%JaxoDrawID:DoubleLine(2)
    \DashArrowLine(70,69)(55,43){4}
    \Text(15,43)[]{\normalsize{\Black{$\mu$}}}
    \Text(75,77)[]{\normalsize{\Black{$i,p$}}}
    \Text(75,8)[]{\normalsize{\Black{$j$}}}
    \ArrowArcn(29.96,-2.04)(35.04,89.94,30.99)
 \end{picture}
 } &=-\frac{i}{\sqrt{2}\,\MP}\,\delta_{ij}P_R\gamma^{\mu}\slashed{p}\,, \\
 \parbox{2.3cm}{
 \begin{picture}(100,86) (5,-5)
    \SetWidth{0.5}
    \SetColor{Black}
    \Vertex(55,38){2.5}
    \Photon(30,63)(55,38){3}{2.5}
    \DashArrowLine(55,38)(80,63){4}
    \ArrowLine(80,13)(55,38)
    \Line(28.59,14.41)(53.59,39.41)\Line(31.41,11.59)(56.41,36.59)%%JaxoDrawID:DoubleLine(2)
    \ArrowArc(55.5,-2.12)(23.13,40.84,139.16)
    \Text(23,70)[]{\normalsize{\Black{$a,\rho$}}}
    \Text(87,70)[]{\normalsize{\Black{$i$}}}
    \Text(87,6)[]{\normalsize{\Black{$j$}}}
    \Text(23,6)[]{\normalsize{\Black{$\mu$}}}
 \end{picture}
 } &=-\frac{ig_{\alpha}}{\sqrt{2}\,\MP}T_{a,\,ij}^{(\alpha)}P_L\gamma^{\rho}\gamma^{\mu}, & \parbox{2.3cm}{
 \begin{picture}(100,86) (5,-5)
    \SetWidth{0.5}
    \SetColor{Black}
    \Vertex(55,38){2.5}
    \Photon(30,63)(55,38){3}{2.5}
    \DashArrowLine(55,38)(80,63){4}
    \ArrowLine(80,13)(55,38)
    \Line(28.59,14.41)(53.59,39.41)\Line(31.41,11.59)(56.41,36.59)%%JaxoDrawID:DoubleLine(2)
    \ArrowArcn(55.5,-2.12)(23.13,139.16,40.84)
    \Text(23,70)[]{\normalsize{\Black{$a,\rho$}}}
    \Text(87,70)[]{\normalsize{\Black{$i$}}}
    \Text(87,6)[]{\normalsize{\Black{$j$}}}
    \Text(23,6)[]{\normalsize{\Black{$\mu$}}}
 \end{picture}
 } &=-\frac{ig_{\alpha}}{\sqrt{2}\,\MP}T_{a,\,ij}^{(\alpha)}P_L\gamma^{\mu}\gamma^{\rho}, \\
 \parbox{2.3cm}{
 \begin{picture}(100,86) (5,-5)
    \SetWidth{0.5}
    \SetColor{Black}
    \Vertex(55,38){2.5}
    \Photon(30,63)(55,38){3}{2.5}
    \DashArrowLine(80,63)(55,38){4}
    \ArrowLine(55,38)(80,13)
    \Line(28.59,14.41)(53.59,39.41)\Line(31.41,11.59)(56.41,36.59)%%JaxoDrawID:DoubleLine(2)
    \ArrowArc(55.5,-2.12)(23.13,40.84,139.16)
    \Text(23,70)[]{\normalsize{\Black{$a,\rho$}}}
    \Text(87,70)[]{\normalsize{\Black{$i$}}}
    \Text(87,6)[]{\normalsize{\Black{$j$}}}
    \Text(23,6)[]{\normalsize{\Black{$\mu$}}}
 \end{picture}
 } &=-\frac{ig_{\alpha}}{\sqrt{2}\,\MP}T_{a,\,ji}^{(\alpha)}P_R\gamma^{\rho}\gamma^{\mu}, & \parbox{2.3cm}{
 \begin{picture}(100,86) (5,-5)
    \SetWidth{0.5}
    \SetColor{Black}
    \Vertex(55,38){2.5}
    \Photon(30,63)(55,38){3}{2.5}
    \DashArrowLine(80,63)(55,38){4}
    \ArrowLine(55,38)(80,13)
    \Line(28.59,14.41)(53.59,39.41)\Line(31.41,11.59)(56.41,36.59)%%JaxoDrawID:DoubleLine(2)
    \ArrowArcn(55.5,-2.12)(23.13,139.16,40.84)
    \Text(23,70)[]{\normalsize{\Black{$a,\rho$}}}
    \Text(87,70)[]{\normalsize{\Black{$i$}}}
    \Text(87,6)[]{\normalsize{\Black{$j$}}}
    \Text(23,6)[]{\normalsize{\Black{$\mu$}}}
 \end{picture}
 } &=-\frac{ig_{\alpha}}{\sqrt{2}\,\MP}T_{a,\,ji}^{(\alpha)}P_R\gamma^{\mu}\gamma^{\rho}, \\
 \parbox{2.3cm}{
 \begin{picture}(90,85) (0,0)
    \SetWidth{0.5}
    \SetColor{Black}
    \Line(55,43)(70,17)
    \Vertex(55,43){2.5}
    \Line(25,45)(55,45)\Line(25,41)(55,41)%%JaxoDrawID:DoubleLine(2)
    \Photon(55,43)(70,69){4}{2}
    \Photon(55,43)(70,17){4}{2}
    \Text(15,43)[]{\normalsize{\Black{$\mu$}}}
    \Text(75,77)[]{\normalsize{\Black{$b,\rho,p$}}}
    \Text(75,8)[]{\normalsize{\Black{$a$}}}
    \ArrowArc(29.96,-2.04)(35.04,30.99,89.94)
 \end{picture}
 } &=-\frac{i}{4\,\MP}\,\delta_{ab}\left[ \slashed{p}, \gamma^{\rho}\right] \gamma^{\mu}, & \parbox{2.3cm}{
 \begin{picture}(90,85) (0,0)
    \SetWidth{0.5}
    \SetColor{Black}
    \Line(55,43)(70,17)
    \Vertex(55,43){2.5}
    \Line(25,45)(55,45)\Line(25,41)(55,41)%%JaxoDrawID:DoubleLine(2)
    \Photon(55,43)(70,69){4}{2}
    \Photon(55,43)(70,17){4}{2}
    \Text(15,43)[]{\normalsize{\Black{$\mu$}}}
    \Text(75,77)[]{\normalsize{\Black{$b,\rho,p$}}}
    \Text(75,8)[]{\normalsize{\Black{$a$}}}
    \ArrowArcn(29.96,-2.04)(35.04,89.94,30.99)
 \end{picture}
 } &=-\frac{i}{4\,\MP}\,\delta_{ab}\gamma^{\mu}\left[ \slashed{p}, \gamma^{\rho}\right] , \\
 \parbox{2.3cm}{
 \begin{picture}(100,86) (5,-5)
    \SetWidth{0.5}
    \SetColor{Black}
    \Vertex(55,38){2.5}
    \Photon(30,63)(55,38){3}{2.5}
    \Photon(80,63)(55,38){3}{2.5}
    \Line(80,13)(55,38)
    \Photon(80,13)(55,38){3}{2.5}
    \Line(28.59,14.41)(53.59,39.41)\Line(31.41,11.59)(56.41,36.59)%%JaxoDrawID:DoubleLine(2)
    \ArrowArc(55.5,-2.12)(23.13,40.84,139.16)
    \Text(23,70)[]{\normalsize{\Black{$b,\nu$}}}
    \Text(87,70)[]{\normalsize{\Black{$c,\rho$}}}
    \Text(87,6)[]{\normalsize{\Black{$a$}}}
    \Text(23,6)[]{\normalsize{\Black{$\mu$}}}
 \end{picture}
 } &=-\frac{g_{\alpha}}{4\,\MP}f^{(\alpha)\,abc}\left[ \gamma^{\nu},\,\gamma^{\rho}\right] \gamma^{\mu},\hspace*{-10mm} & \parbox{2.3cm}{
 \begin{picture}(100,86) (5,-5)
    \SetWidth{0.5}
    \SetColor{Black}
    \Vertex(55,38){2.5}
    \Photon(30,63)(55,38){3}{2.5}
    \Photon(80,63)(55,38){3}{2.5}
    \Line(80,13)(55,38)
    \Photon(80,13)(55,38){3}{2.5}
    \Line(28.59,14.41)(53.59,39.41)\Line(31.41,11.59)(56.41,36.59)%%JaxoDrawID:DoubleLine(2)
    \ArrowArcn(55.5,-2.12)(23.13,139.16,40.84)
    \Text(23,70)[]{\normalsize{\Black{$b,\nu$}}}
    \Text(87,70)[]{\normalsize{\Black{$c,\rho$}}}
    \Text(87,6)[]{\normalsize{\Black{$a$}}}
    \Text(23,6)[]{\normalsize{\Black{$\mu$}}}
 \end{picture}
 } &=-\frac{g_{\alpha}}{4\,\MP}f^{(\alpha)\,abc}\gamma^{\mu}\left[ \gamma^{\nu},\,\gamma^{\rho}\right] .
\end{align*}

\subsubsection*{Gauge Vertices}

\begin{align*}
 \parbox{2.3cm}{
 \begin{picture}(90,85) (0,0)
    \SetWidth{0.5}
    \SetColor{Black}
    \ArrowLine(55,43)(70,17)
    \Vertex(55,43){2.5}
    \ArrowLine(25,43)(55,43)
    \Photon(55,43)(70,69){4}{2}
    \Text(15,43)[]{\normalsize{\Black{$i$}}}
    \Text(75,77)[]{\normalsize{\Black{$a,\,\mu$}}}
    \Text(75,8)[]{\normalsize{\Black{$j$}}}
   % \ArrowArcn(29.96,-2.04)(35.04,89.94,30.99)
 \end{picture}
 } &=-ig_\alpha T_{a,\,ij}^{(\alpha)}\gamma^\mu P_L\,,
\end{align*}
\begin{align*}
 \parbox{2.3cm}{
 \begin{picture}(90,85) (0,0)
    \SetWidth{0.5}
    \SetColor{Black}
    \ArrowLine(70,17)(55,43)
    \Vertex(55,43){2.5}
    \Line(25,43)(55,43)
    \Photon(25,43)(55,43){4}{2}
    \DashArrowLine(55,43)(70,69){4}
    \Text(15,43)[]{\normalsize{\Black{$a$}}}
    \Text(75,77)[]{\normalsize{\Black{$i$}}}
    \Text(75,8)[]{\normalsize{\Black{$j$}}}
    \ArrowArcn(32.46,1.96)(35.04,89.94,30.99)
    \ArrowArc(29.96,-2.04)(35.04,30.99,89.94)
 \end{picture}
 } &=-i\sqrt{2}g_{\alpha}T_{a,\,ij}^{(\alpha)}P_L\,, & \parbox{2.3cm}{
 \begin{picture}(90,85) (0,0)
    \SetWidth{0.5}
    \SetColor{Black}
    \ArrowLine(55,43)(70,17)
    \Vertex(55,43){2.5}
    \Line(25,43)(55,43)
    \Photon(25,43)(55,43){4}{2}
    \DashArrowLine(70,69)(55,43){4}
    \Text(15,43)[]{\normalsize{\Black{$a$}}}
    \Text(75,77)[]{\normalsize{\Black{$i$}}}
    \Text(75,8)[]{\normalsize{\Black{$j$}}}
    \ArrowArcn(32.46,1.96)(35.04,89.94,30.99)
    \ArrowArc(29.96,-2.04)(35.04,30.99,89.94)
 \end{picture}
 } &=-i\sqrt{2}g_{\alpha}T_{a,\,ji}^{(\alpha)}P_R\,.
\end{align*}

\subsubsection*{Yukawa Vertices}

\begin{align*}
 \parbox{2.3cm}{
 \begin{picture}(90,85) (0,0)
    \SetWidth{0.5}
    \SetColor{Black}
    \ArrowLine(55,43)(70,17)
    \Vertex(55,43){2.5}
    \ArrowLine(25,43)(55,43)
    \DashLine(70,69)(55,43){4}
    \Text(15,43)[]{\normalsize{\Black{$f_i$}}}
    \Text(75,77)[]{\normalsize{\Black{$h$}}}
    \Text(75,8)[]{\normalsize{\Black{$f_j$}}}
 \end{picture}
 } &=\frac{i\,m_f}{\sqrt{2}\,v}\,\delta_{ij}\,.
\end{align*}

\chapter{Kinematics of Scattering and Decay Processes}
\label{kinematics}

In this appendix we collect formulae about kinematics that are relevant for the calculation of scattering cross sections and decay widths in this work. Most of the formulae are taken from~\cite{Nakamura:2010zzi}.

\subsection*{Two-Body Decays}
\begin{equation*}
  \begin{picture}(169,116) (95,-57)
    \SetWidth{0.5}
    \SetColor{Black}
    \Line[arrow,arrowpos=0.5,arrowlength=3.75,arrowwidth=1.5,arrowinset=0.2](96,-1)(163,-1)
    \Line[arrow,arrowpos=0.5,arrowlength=3.75,arrowwidth=1.5,arrowinset=0.2](191,9)(258,48)
    \Line[arrow,arrowpos=0.5,arrowlength=3.75,arrowwidth=1.5,arrowinset=0.2](191,-11)(258,-50)
    \GOval(180,-1)(18,18)(0){0.8}
    \Line[arrow,arrowpos=0.5,arrowlength=3.75,arrowwidth=1.5,arrowinset=0.2](112,9)(146,9)
    \Line[arrow,arrowpos=0.5,arrowlength=3.75,arrowwidth=1.5,arrowinset=0.2](202,28)(236,48)
    \Line[arrow,arrowpos=0.5,arrowlength=3.75,arrowwidth=1.5,arrowinset=0.2](202,-30)(236,-50)
    \Text(129,16)[]{\normalsize{\Black{$p$}}}
    \Text(221,41)[rb]{\normalsize{\Black{$p_1$}}}
    \Text(221,-42)[rt]{\normalsize{\Black{$p_2$}}}
  \end{picture}
\end{equation*}
In this case the scalar products of the four-momenta can be easily obtained using four-momentum conservation. Starting from
\begin{equation}
  p=p_1+p_2\,,
\end{equation}
we multiply by the three different four-momenta to obtain
\begin{equation}
 \begin{split}
  \left( p\cdot p\right) &=\left( p\cdot p_1\right) +\left( p\cdot p_2\right) , \\
  \left( p\cdot p_1\right) &=\left( p_1\cdot p_1\right) +\left( p_1\cdot p_2\right) , \\
  \left( p\cdot p_2\right) &=\left( p_1\cdot p_2\right) +\left( p_2\cdot p_2\right) .
 \end{split}
\end{equation}
Using the fact that $ \left( p\cdot p\right) =M^2 $ and $ \left( p_i\cdot p_i\right) =m_i^2 $ then yields the relations
\begin{equation}
 \begin{split}
  \left( p\cdot p_1\right) &=\frac{M^2+m_1^2-m_2^2}{2}\,, \\
  \left( p\cdot p_2\right) &=\frac{M^2-m_1^2+m_2^2}{2}\,, \\
  \left( p_1\cdot p_2\right) &=\frac{M^2-m_1^2-m_2^2}{2}\,.
  \label{twobodyscalar}
 \end{split}
\end{equation}
The total decay width in a two-body decay process is given by
\begin{equation}
  \Gamma=\frac{\bar{\abs{\mathcal{M}}^2}}{8\,\pi}\,\frac{\abs{\vec{p}_1}}{M^2}
  \label{twobodywidth}
\end{equation}
if we average over the spin states of the decaying particle. Four-momentum conservation requires that the final state momenta are identical in the center-of-mass frame and given by
\begin{equation}
  \abs{\vec{p}_1}=\abs{\vec{p}_2}=\frac{1}{2\,M}\sqrt{\left( M^2-\left( m_1+m_2\right) ^2\right) \left( M^2-\left( m_1-m_2\right) ^2\right) }\,,
  \label{twobodymomentum}
\end{equation}
while the energies in the center-of-mass frame are given by
\begin{equation}
  E_1=\frac{M^2+m_1^2-m_2^2}{2\,M}\qquad\text{and}\qquad E_2=\frac{M^2-m_1^2+m_2^2}{2\,M}\,.
\end{equation}

\subsection*{Three-Body Decays}
\begin{equation*}
  \begin{picture}(169,116) (95,-57)
    \SetWidth{0.5}
    \SetColor{Black}
    \Line[arrow,arrowpos=0.5,arrowlength=3.75,arrowwidth=1.5,arrowinset=0.2](96,-1)(163,-1)
    \Line[arrow,arrowpos=0.5,arrowlength=3.75,arrowwidth=1.5,arrowinset=0.2](191,9)(258,48)
    \Line[arrow,arrowpos=0.5,arrowlength=3.75,arrowwidth=1.5,arrowinset=0.2](191,-11)(258,-50)
    \Line[arrow,arrowpos=0.5,arrowlength=3.75,arrowwidth=1.5,arrowinset=0.2](196,-1)(263,-1)
    \GOval(180,-1)(18,18)(0){0.8}
    \Line[arrow,arrowpos=0.5,arrowlength=3.75,arrowwidth=1.5,arrowinset=0.2](112,9)(146,9)
    \Line[arrow,arrowpos=0.5,arrowlength=3.75,arrowwidth=1.5,arrowinset=0.2](202,28)(236,48)
    \Line[arrow,arrowpos=0.5,arrowlength=3.75,arrowwidth=1.5,arrowinset=0.2](213,9)(247,9)
    \Line[arrow,arrowpos=0.5,arrowlength=3.75,arrowwidth=1.5,arrowinset=0.2](202,-30)(236,-50)
    \Text(129,16)[]{\normalsize{\Black{$p$}}}
    \Text(221,41)[rb]{\normalsize{\Black{$p_1$}}}
    \Text(221,-42)[rt]{\normalsize{\Black{$p_3$}}}
    \Text(230,16)[]{\normalsize{\Black{$p_2$}}}
  \end{picture}
\end{equation*}
Starting from four-momentum conservation,
\begin{equation}
 p=p_1+p_2+p_3\,,
\end{equation}
we can express all scalar products of the four-momenta in terms of the particle masses, $p^2=M^2$ and $p_i^2=m_i^2$, and two additional parameters, \textit{e.g.} the energies $E_2$ and $E_3$ in the rest frame of the decaying particle or the invariant masses $m_{12}^2$ and $m_{13}^2$:
\begin{equation}
  \begin{split}
   m_{12}^2 &=\left( p_1+p_2\right) ^2=\left( p-p_3\right) ^2=M^2+m_3^2-2\,M E_3\,, \\
   m_{13}^2 &=\left( p_1+p_3\right) ^2=\left( p-p_2\right) ^2=M^2+m_2^2-2\,M E_2\,.
  \end{split}
\end{equation}
From these expressions we can directly derive all scalar products of the four-momenta:
\begin{alignat}{2}
  \left( p\cdot p_3\right) &=\frac{M^2+m_3^2-m_{12}^2}{2}\,, &\qquad\qquad \left( p_1\cdot p_2\right) &=\frac{m_{12}^2-m_1^2-m_2^2}{2}\,, \nonumber\\
  \left( p\cdot p_2\right) &=\frac{M^2+m_2^2-m_{13}^2}{2}\,, &\qquad\qquad \left( p_1\cdot p_3\right) &=\frac{m_{13}^2-m_1^2-m_3^2}{2}\,, \label{threebodyscalar}\\
  \left( p\cdot p_1\right) &=\frac{m_{12}^2+m_{13}^2-m_2^2-m_3^2}{2}\,, & \left( p_2\cdot p_3\right) &=\frac{M^2+m_1^2-m_{12}^2-m_{13}^2}{2}\,. \nonumber
\end{alignat}
Alternatively, one can use the following set of scalar products:
\begin{equation}
 \left( p\cdot p_1\right) =M \left( M-E_2-E_3\right) ,\quad\left( p\cdot p_2\right) =M E_2,\quad\left( p\cdot p_3\right) =M E_3\,,
\end{equation}
and
\begin{equation}
 \begin{split}
  \left( p_1\cdot p_2\right) &=\frac{M^2-m_1^2-m_2^2+m_3^2}{2}-M E_3\,, \\
  \left( p_1\cdot p_3\right) &=\frac{M^2-m_1^2+m_2^2-m_3^2}{2}-M E_2\,, \\
  \left( p_2\cdot p_3\right) &=M\left( E_2+E_3\right) -\frac{M^2-m_1^2+m_2^2+m_3^2}{2}\,.
 \end{split}
\end{equation}
When we average over the spin states of the decaying particle, its partial decay width turns out to be
\begin{equation}
  \frac{d\Gamma}{dm_{12}^2\,dm_{13}^2}=\frac{1}{(2\,\pi)^3}\,\frac{\bar{\abs{\mathcal{M}}^2}}{32\,M^3}\,,
  \label{threebodywidth}
\end{equation}
where the limiting values of the invariant masses are given by
\begin{equation}
  (m_1+m_2)^2\leq m_{12}^2\leq(M-m_3)^2
  \label{m12range}
\end{equation}
and
\begin{equation}
  \begin{split}
    m_{13,\,min/max}^2 &=m_1^2+m_3^2+\frac{1}{2\,m_{12}^2}\,\bigg(\left( m_{12}^2+m_1^2-m_2^2\right) \left( M^2-m_{12}^2-m_3^2\right) \\
    &\qquad\qquad\qquad\qquad\qquad\mp\sqrt{\lambda(m_{12}^2, m_1^2, m_2^2)}\sqrt{\lambda(M^2, m_{12}^2, m_3^2}\bigg)\,.
  \end{split}
  \label{m13range}
\end{equation}
In the last expression we used the abbreviation
\begin{equation}
  \lambda(x, y, z)=x^2+y^2+z^2-2\,x\,y-2\,x\,z-2\,y\,z\,.
\end{equation}
Alternatively, one can write the partial decay width in the form
\begin{equation}
  \frac{d\Gamma}{dE_2\,dE_3}=\frac{1}{(2\,\pi)^3}\,\frac{\bar{\abs{\mathcal{M}}^2}}{8\,M}\,.
\end{equation}

\subsection*{Two-Body Reactions}
\begin{equation*}
  \begin{picture}(158,116) (103,-57)
    \SetWidth{0.5}
    \SetColor{Black}
    \Line[arrow,arrowpos=0.5,arrowlength=3.75,arrowwidth=1.5,arrowinset=0.2](104,48)(171,9)
    \Line[arrow,arrowpos=0.5,arrowlength=3.75,arrowwidth=1.5,arrowinset=0.2](193,9)(260,48)
    \Line[arrow,arrowpos=0.5,arrowlength=3.75,arrowwidth=1.5,arrowinset=0.2](193,-11)(260,-50)
    \Line[arrow,arrowpos=0.5,arrowlength=3.75,arrowwidth=1.5,arrowinset=0.2](104,-50)(171,-11)
    \GOval(182,-1)(18,18)(0){0.8}
    \Line[arrow,arrowpos=0.5,arrowlength=3.75,arrowwidth=1.5,arrowinset=0.2](126,48)(160,28)
    \Line[arrow,arrowpos=0.5,arrowlength=3.75,arrowwidth=1.5,arrowinset=0.2](204,28)(238,48)
    \Line[arrow,arrowpos=0.5,arrowlength=3.75,arrowwidth=1.5,arrowinset=0.2](126,-50)(160,-30)
    \Line[arrow,arrowpos=0.5,arrowlength=3.75,arrowwidth=1.5,arrowinset=0.2](204,-30)(238,-50)
    \Text(144,41)[lb]{\normalsize{\Black{$p$}}}
    \Text(222,40)[rb]{\normalsize{\Black{$p'$}}}
    \Text(222,-43)[rt]{\normalsize{\Black{$k'$}}}
    \Text(143,-43)[lt]{\normalsize{\Black{$k$}}}
  \end{picture}
\end{equation*}
For the discussion of two-to-two scattering processes we want to introduce the Lorentz-invariant Mandelstam variables:
\begin{equation}
 \begin{split}
  s&=\left( p+k\right) ^2=\left( p'+k'\right) ^2=M^2+m^2+2\left( p\cdot k\right) =M'^2+m'^2+2\left( p'\cdot k'\right) , \\
  t&=\left( p-p'\right) ^2=\left( k-k'\right) ^2=M^2+M'^2-2\left( p\cdot p'\right) =m^2+m'^2-2\left( k\cdot k'\right) , \\
  u&=\left( p-k'\right) ^2=\left( k-p'\right) ^2=M^2+m'^2-2\left( p\cdot k'\right) =m^2+M'^2-2\left( k\cdot p'\right) ,
 \end{split}
\end{equation}
where the squared four-momenta are replaced by the particle masses squared ($p^2=M^2$, $p'^2=M'^2$, $k^2=m^2$ and $k'^2=m'^2$). The Mandelstam variables satisfy the relation
\begin{equation}
  s+t+u=M^2+m^2+M'^2+m'^2.
 \label{MandelstamSum}
\end{equation}
We can now write down the six scalar products of the four-momenta in terms of the Mandelstam variables and the particle masses:
\begin{alignat}{2}
  \left( p\cdot k\right) &=\frac{s-M^2-m^2}{2}\,, &\qquad\qquad \left( p'\cdot k'\right) &=\frac{s-M'^2-m'^2}{2}\,, \nonumber\\
  \left( p\cdot p'\right) &=\frac{M^2+M'^2-t}{2}\,, &\qquad\qquad \left( k\cdot k'\right) &=\frac{m^2+m'^2-t}{2}\,, \label{scatteringscalarproduct}\\
  \left( p\cdot k'\right) &=\frac{s+t-M'^2-m^2}{2}\,, & \left( k\cdot p'\right) &=\frac{s+t-M^2-m'^2}{2}\,. \nonumber
\end{alignat}
In the last two expressions we replaced $u$ using equation~(\ref{MandelstamSum}).
The differential cross section of a two-body reaction is given by
\begin{equation}
  \frac{d\sigma}{dt}=\frac{\abs{\mathcal{M}}^2}{64\,\pi\,s\abs{\vec{p}_{cm}}^2}\,,
  \label{crosssec}
\end{equation}
where the limiting values of $t$ are given by
\begin{equation}
  t_{0/1}=\frac{\left( M^2-M'^2-m^2+m'^2\right) ^2}{4\,s}-\left( p_{cm}\mp p'_{cm}\right) ^2.
\end{equation}
The particle momenta in the center-of-mass frame are given by
\begin{equation}
  p_{cm}=\sqrt{E_{cm}^2-M^2}\quad\text{and}\quad p'_{cm}=\sqrt{E_{cm}'^2-M'^2}\,,
\end{equation}
while the center-of-mass energies read
\begin{equation}
  E_{cm}=\frac{s+M^2-m^2}{2\,\sqrt{s}}\quad\text{and}\quad E'_{cm}=\frac{s+M'^2-m'^2}{2\,\sqrt{s}}\,.
\end{equation}
In the laboratory frame, where the particle with four-momentum $k$ is at rest, the scattering cross section can be rewritten using the relation
\begin{equation}
  \abs{\vec{p}_{cm}}^2=\frac{\abs{\vec{p}_{lab}}^2m^2}{s}\,.
\end{equation}

\chapter{Calculation of Gravitino Decay Widths}
\label{gravitinodecay}

In this appendix we compute the decay channels of the LSP gravitino in the framework of bilinear $R$-parity breaking. The relative magnitude of the decay widths of the different decay channels determines the branching ratios for these channels. In addition, the exact shape of the differential decay widths determines the spectra of the final state particles. Therefore, this is a crucial input for the prediction of spectra of gravitino decay products. For a discussion of these topics see Sections~\ref{gravdecay} and~\ref{gravspectra}.

\section{Two-Body Decays}

Here we want to calculate the decay widths of the two additional effective 3-vertex diagrams coming from 4-vertex diagrams with a Higgs VEV at one of the legs:
\begin{equation*}
 \parbox{5.5cm}{
  \begin{picture}(172,162) (92,-96)
    \SetWidth{0.5}
    \SetColor{Black}
    \Line(211,-55.999)(215,-44.001)\Line(207.001,-48)(218.999,-52)
    \Line[arrow,arrowpos=0.5,arrowlength=3.75,arrowwidth=1.5,arrowinset=0.2](213,-51)(232,-84)
    \Line(194,-17)(213,-51)
    \Photon(194,-17)(232,51){-5}{4}
    \Line[double,sep=4](117,-17)(194,-17)
    \Vertex(194,-17){2.5}
    \Line[arrow,arrowpos=1,arrowlength=3.75,arrowwidth=1.5,arrowinset=0.2](197,11)(210,33)
    \Line[arrow,arrowpos=1,arrowlength=3.75,arrowwidth=1.5,arrowinset=0.2](226,-50)(239,-72)
    \Line[arrow,arrowpos=1,arrowlength=3.75,arrowwidth=1.5,arrowinset=0.2](143,-6)(169,-6)
    \Text(104,-17)[]{\normalsize{\Black{$\psi_{3/2}$}}}
    \Text(156,-4)[b]{\normalsize{\Black{$p$}}}
    \Text(203,22)[rb]{\normalsize{\Black{$k$}}}
    \Text(235,-62)[lb]{\normalsize{\Black{$q$}}}
    \Text(237,58)[]{\normalsize{\Black{$Z$}}}
    \Text(219,-28)[]{\normalsize{\Black{$\tilde{H}_{u,\,d}^0$}}}
    \Text(182,-40)[]{\normalsize{\Black{$v_{u,\,d}$}}}
    \Text(237,-91)[]{\normalsize{\Black{$\nu_i$}}}
    \Arc[arrow,arrowpos=0.5,arrowlength=3.75,arrowwidth=1.5,arrowinset=0.2,clock](133.852,-120.786)(81.877,92.697,26.698)
    \Line[dash,dashsize=4](185,-34)(195,-17)
  \end{picture}
 }
 \text{and}\qquad
 \parbox{5.5cm}{
  \begin{picture}(172,162) (92,-96)
    \SetWidth{0.5}
    \SetColor{Black}
    \Line(211,-55.999)(215,-44.001)\Line(207.001,-48)(218.999,-52)
    \Line[arrow,arrowpos=0.5,arrowlength=3.75,arrowwidth=1.5,arrowinset=0.2](213,-51)(232,-84)
    \Line(194,-17)(213,-51)
    \Photon(194,-17)(232,51){-5}{4}
    \Line[double,sep=4](117,-17)(194,-17)
    \Vertex(194,-17){2.5}
    \Line[arrow,arrowpos=1,arrowlength=3.75,arrowwidth=1.5,arrowinset=0.2](197,11)(210,33)
    \Line[arrow,arrowpos=1,arrowlength=3.75,arrowwidth=1.5,arrowinset=0.2](226,-50)(239,-72)
    \Line[arrow,arrowpos=1,arrowlength=3.75,arrowwidth=1.5,arrowinset=0.2](143,-6)(169,-6)
    \Text(104,-17)[]{\normalsize{\Black{$\psi_{3/2}$}}}
    \Text(156,-4)[b]{\normalsize{\Black{$p$}}}
    \Text(203,22)[rb]{\normalsize{\Black{$k$}}}
    \Text(235,-62)[lb]{\normalsize{\Black{$q$}}}
    \Text(237,58)[]{\normalsize{\Black{$W^+$}}}
    \Text(219,-28)[]{\normalsize{\Black{$\tilde{H}_{d}^-$}}}
    \Text(182,-40)[]{\normalsize{\Black{$v_d$}}}
    \Text(237,-91)[]{\normalsize{\Black{$\ell_i^-$}}}
    \Arc[arrow,arrowpos=0.5,arrowlength=3.75,arrowwidth=1.5,arrowinset=0.2,clock](133.852,-120.786)(81.877,92.697,26.698)
    \Line[dash,dashsize=4](185,-34)(195,-17)
  \end{picture}
 }.
\end{equation*}
Using the rotation of the neutral higgsino fields into the neutrino as discussed in Section~\ref{Rbreaking} we find the following decay amplitude for the first Feynman diagram:
\begin{align}
  i\mathcal{M}_Z &=-\bar{u}^r(q)\,P_R\,\Big\lbrace\!N_{\nu_i\,\tilde{H}_u^0}^{7*}v_u\left( g\,\frac{\sigma_{3,\,22}}{2}\cos{\theta_W}-g'Y_{H_u}\sin{\theta_W}\right) \nonumber\\
  &\qquad\quad+N_{\nu_i\,\tilde{H}_d^0}^{7*}v_d\left( g\,\frac{\sigma_{3,\,11}}{2}\cos{\theta_W}-g'Y_{H_d}\sin{\theta_W}\right) \!\Big\rbrace\,\frac{i}{\sqrt{2}\,\MP}\,\gamma^\mu\gamma^\rho\,\psi_{\mu}^{+\,s}(p)\,\epsilon_{\rho}^{\lambda\,*}(k) \nonumber\\
  &\simeq-\frac{i\,g_Z\,\xi_i}{2\,\sqrt{2}\,\MP}\left\lbrace v_u\,U^*_{\tilde{H}_u^0\tilde{Z}}-v_d\,U^*_{\tilde{H}_d^0\tilde{Z}}\right\rbrace \bar{u}^r(q)\,P_R\,\gamma^\mu\gamma^\rho\,\psi_{\mu}^{+\,s}(p)\,\epsilon_{\rho}^{\lambda\,*}(k) \\
  &=-\frac{i\,m_Z\,\xi_i}{2\,\MP}\left\lbrace \sin\beta\,U^*_{\tilde{H}_u^0\tilde{Z}}-\cos\beta\,U^*_{\tilde{H}_d^0\tilde{Z}}\right\rbrace \bar{u}^r(q)\,P_R\,\gamma^\mu\gamma^\rho\,\psi_{\mu}^{+\,s}(p)\,\epsilon_{\rho}^{\lambda\,*}(k)\,. \nonumber
\end{align}
This additional diagram is exactly of the same form as the 4-vertex diagram that was already taken into account in previous calculations. Therefore, we can simply alter the previously obtained result into the following expression:
\begin{equation}
  \begin{split}
    \Gamma\left( \psi_{3/2}\rightarrow Z\nu_i\right) &\simeq\frac{\xi_i^2\,m_{3/2}^3\,\beta_Z^2}{64\,\pi\,\MP^2}\left\lbrace \,U_{\tilde{Z}\tilde{Z}}^2f_Z+\frac{1}{6}\abs{1+s_\beta\,U_{\tilde{H}_u^0\tilde{Z}}-c_\beta\,U_{\tilde{H}_d^0\tilde{Z}}}^2h_Z\right. \\ 
    &\left. \quad\qquad-\frac{8}{3}\frac{m_Z}{m_{3/2}}\,U_{\tilde{Z}\tilde{Z}}\left( 1+s_\beta \RE U_{\tilde{H}_u^0\tilde{Z}}-c_\beta \RE U_{\tilde{H}_d^0\tilde{Z}}\right) j_Z \right\rbrace .
  \end{split}
\end{equation}
In order to write down the decay amplitude of the second diagram we need the rotation of the charged down-type higgsino into the left-handed charged leptons. We find
\begin{equation}
 \begin{split}
  i\mathcal{M}_W &=-\bar{u}^r(q)\,P_R\,U_{\ell_i\,\tilde{H}_d^-}^{5*}v_d\,\frac{g}{\sqrt{2}}\left( \frac{\sigma_{1,\,21}}{2}-i\,\frac{\sigma_{2,\,21}}{2}\right) \frac{i}{\sqrt{2}\,\MP}\,\gamma^\mu\gamma^\rho\,\psi_{\mu}^{+\,s}(p)\,\epsilon_{\rho}^{\lambda\,*}(k) \\
  &\simeq\frac{i\,g\,\xi_i\,v_d}{\sqrt{2}\,\MP}\,U^*_{\tilde{H}_d^-\tilde{W}}\,\bar{u}^r(q)\,P_R\,\gamma^\mu\gamma^\rho\,\psi_{\mu}^{+\,s}(p)\,\epsilon_{\rho}^{\lambda\,*}(k) \\
  &=\frac{i\sqrt{2}\,m_W\,\xi_i}{\sqrt{2}\,\MP}\,\cos\beta\,U^*_{\tilde{H}_d^-\tilde{W}}\,\bar{u}^r(q)\,P_R\,\gamma^\mu\gamma^\rho\,\psi_{\mu}^{+\,s}(p)\,\epsilon_{\rho}^{\lambda\,*}(k)\,.
 \end{split}
\end{equation}
This additional diagram is also exactly of the same form as the 4-vertex diagram that was already taken into account. Therefore, we can simply alter the previously obtained result into the following expression:
\begin{equation}
  \begin{split}
    \Gamma\left( \psi_{3/2}\rightarrow W^+\ell_i^-\right) &\simeq\frac{\xi_i^2\,m_{3/2}^3\,\beta_W^2}{32\,\pi\,\MP^2}\left\lbrace \,U_{\tilde{W}\tilde{W}}^2f_W+\frac{1}{6}\abs{1-\sqrt{2}\,c_\beta\,U_{\tilde{H}_d^-\tilde{W}}}^2h_W\right. \\ 
    &\left. \quad\qquad-\frac{8}{3}\frac{m_W}{m_{3/2}}\,U_{\tilde{W}\tilde{W}}\left( 1-\sqrt{2}\,c_\beta \RE U_{\tilde{H}_d^-\tilde{W}}\right) j_W \right\rbrace .
  \end{split}
\end{equation}
We also want to include a small correction for the decay into the lightest Higgs particle and a neutrino. In the previous calculation in~\cite{Grefe:2008zz} we neglected the $D$-term contribution to the sneutrino VEV. This affected also the mixing that was assumed between the lightest Higgs particle and the sneutrinos. We simply replace this part in the final result with the complete expression found in Section~\ref{Rbreaking}:
\begin{equation}
  \Gamma\left( \psi_{3/2}\rightarrow h\,\nu_i\right) \simeq\frac{\xi_i^2\,m_{3/2}^3\,\beta_h^4}{384\,\pi\,\MP^2}\abs{\frac{m_{\tilde{\nu}_i}^2+\frac{1}{2}\,m_Z^2\cos2\,\beta}{m_h^2-m_{\tilde{\nu}_i}^2}+2\sin{\beta}\,U_{\tilde{H}_u^0\tilde{Z}}+2\cos{\beta}\,U_{\tilde{H}_d^0\tilde{Z}}}^2.\!\!\!
\end{equation}

\section{Three-Body Decays}

In this section we calculate in detail the decay widths of the three-body gravitino decay processes $\psi_{3/2}\rightarrow \gamma^*\,\nu\rightarrow f\,\bar{f}\,\nu$, $\psi_{3/2}\rightarrow Z^*\,\nu\rightarrow f\,\bar{f}\,\nu$, $\psi_{3/2}\rightarrow {W^+}^*\,\ell^-\rightarrow f\,\bar{f'}\,\ell^-$ and $\psi_{3/2}\rightarrow h^*\,\nu\rightarrow f\,\bar{f}\,\nu$. The related discussion in the text can be found in Section~\ref{threebody}.

\subsection[\texorpdfstring{$\psi_{3/2}\rightarrow \gamma^*/Z^*\,\nu\rightarrow f\,\bar{f}\,\nu$}{\textpsi\ --> \textgamma*/Z* \textnu --> f f \textnu}]{\boldmath$\psi_{3/2}\rightarrow \gamma^*/Z^*\,\nu\rightarrow f\,\bar{f}\,\nu$}

At tree level there are five diagrams contributing to the decay of a gravitino into a fermion-antifermion pair and a neutrino:
\begin{equation*}
 \parbox{5.5cm}{
  \begin{picture}(172,162) (92,-96)
    \SetWidth{0.5}
    \SetColor{Black}
    \Line(211,-55.999)(215,-44.001)\Line(207.001,-48)(218.999,-52)
    \Line[arrow,arrowpos=0.5,arrowlength=3.75,arrowwidth=1.5,arrowinset=0.2](213,-50)(232,-84)
    \Line(194,-17)(213,-51)
    \Photon(194,-17)(213,17){-5}{2}
    \Photon(194,-17)(213,-51){-5}{2}
    \Line[arrow,arrowpos=0.5,arrowlength=3.75,arrowwidth=1.5,arrowinset=0.2](213,17)(232,51)
    \Line[arrow,arrowpos=0.5,arrowlength=3.75,arrowwidth=1.5,arrowinset=0.2](233,-17)(214,17)
    \Line[double,sep=4](117,-17)(194,-17)
    \Vertex(194,-17){2.5}
    \Vertex(213,17){2.5}
    \Line[arrow,arrowpos=1,arrowlength=3.75,arrowwidth=1.5,arrowinset=0.2](207,28)(220,50)
    \Line[arrow,arrowpos=1,arrowlength=3.75,arrowwidth=1.5,arrowinset=0.2](226,17)(239,-5)
    \Line[arrow,arrowpos=1,arrowlength=3.75,arrowwidth=1.5,arrowinset=0.2](226,-50)(239,-72)
    \Line[arrow,arrowpos=1,arrowlength=3.75,arrowwidth=1.5,arrowinset=0.2](143,-6)(169,-6)
    \Text(104,-17)[]{\normalsize{\Black{$\psi_{3/2}$}}}
    \Text(156,-4)[b]{\normalsize{\Black{$p$}}}
    \Text(234,6)[lb]{\normalsize{\Black{$k_2$}}}
    \Text(213,39)[rb]{\normalsize{\Black{$k_1$}}}
    \Text(235,-62)[lb]{\normalsize{\Black{$q$}}}
    \Text(237,58)[]{\normalsize{\Black{$f$}}}
    \Text(237,-24)[]{\normalsize{\Black{$\bar{f}$}}}
    \Text(188,6)[]{\normalsize{\Black{$\gamma, Z$}}}
    \Text(188,-39)[]{\normalsize{\Black{$\tilde{\gamma}, \tilde{Z}$}}}
    \Text(237,-91)[]{\normalsize{\Black{$\nu_i$}}}
    \Arc[arrow,arrowpos=0.5,arrowlength=3.75,arrowwidth=1.5,arrowinset=0.2,clock](133.852,-120.786)(81.877,92.697,26.698)
  \end{picture}
 }
 \qquad+\qquad
 \parbox{5.5cm}{
  \begin{picture}(172,162) (92,-96)
    \SetWidth{0.5}
    \SetColor{Black}
    \Line[arrow,arrowpos=0.5,arrowlength=3.75,arrowwidth=1.5,arrowinset=0.2](194,-17)(232,-84)
    \Photon(194,-17)(213,17){-5}{2}
    \Line[arrow,arrowpos=0.5,arrowlength=3.75,arrowwidth=1.5,arrowinset=0.2](213,17)(232,51)
    \Line[arrow,arrowpos=0.5,arrowlength=3.75,arrowwidth=1.5,arrowinset=0.2](233,-17)(214,17)
    \Line[double,sep=4](117,-17)(194,-17)
    \Vertex(194,-17){2.5}
    \Vertex(213,17){2.5}
    \Line[arrow,arrowpos=1,arrowlength=3.75,arrowwidth=1.5,arrowinset=0.2](207,28)(220,50)
    \Line[arrow,arrowpos=1,arrowlength=3.75,arrowwidth=1.5,arrowinset=0.2](226,17)(239,-5)
    \Line[arrow,arrowpos=1,arrowlength=3.75,arrowwidth=1.5,arrowinset=0.2](217,-34)(230,-56)
    \Line[arrow,arrowpos=1,arrowlength=3.75,arrowwidth=1.5,arrowinset=0.2](143,-6)(169,-6)
    \Text(104,-17)[]{\normalsize{\Black{$\psi_{3/2}$}}}
    \Text(156,-4)[b]{\normalsize{\Black{$p$}}}
    \Text(234,6)[lb]{\normalsize{\Black{$k_2$}}}
    \Text(213,39)[rb]{\normalsize{\Black{$k_1$}}}
    \Text(226,-46)[lb]{\normalsize{\Black{$q$}}}
    \Text(237,58)[]{\normalsize{\Black{$f$}}}
    \Text(237,-24)[]{\normalsize{\Black{$\bar{f}$}}}
    \Text(193,6)[]{\normalsize{\Black{$Z$}}}
    \Text(182,-40)[]{\normalsize{\Black{$v_i$}}}
    \Text(237,-91)[]{\normalsize{\Black{$\nu_i$}}}
    \Arc[arrow,arrowpos=0.5,arrowlength=3.75,arrowwidth=1.5,arrowinset=0.2,clock](133.852,-120.786)(81.877,92.697,26.698)
    \Line[dash,dashsize=4,arrow,arrowpos=0.5,arrowlength=3.75,arrowwidth=1.5,arrowinset=0.2](185,-34)(195,-17)
  \end{picture}
 }
\end{equation*}
and in addition the diagrams coming from the 4-vertex with a Higgs VEV at one of the legs that were also discussed for the two-body decay:
\begin{equation*}
 \parbox{5.5cm}{
  \begin{picture}(172,162) (92,-96)
    \SetWidth{0.5}
    \SetColor{Black}
    \Line(211,-55.999)(215,-44.001)\Line(207.001,-48)(218.999,-52)
    \Line[arrow,arrowpos=0.5,arrowlength=3.75,arrowwidth=1.5,arrowinset=0.2](213,-51)(232,-84)
    \Photon(194,-17)(213,17){-5}{2}
    \Line[arrow,arrowpos=0.5,arrowlength=3.75,arrowwidth=1.5,arrowinset=0.2](213,17)(232,51)
    \Line[arrow,arrowpos=0.5,arrowlength=3.75,arrowwidth=1.5,arrowinset=0.2](233,-17)(214,17)
    \Line(194,-17)(213,-51)
    \Line[double,sep=4](117,-17)(194,-17)
    \Vertex(194,-17){2.5}
    \Vertex(213,17){2.5}
    \Line[arrow,arrowpos=1,arrowlength=3.75,arrowwidth=1.5,arrowinset=0.2](207,28)(220,50)
    \Line[arrow,arrowpos=1,arrowlength=3.75,arrowwidth=1.5,arrowinset=0.2](226,17)(239,-5)
    \Line[arrow,arrowpos=1,arrowlength=3.75,arrowwidth=1.5,arrowinset=0.2](226,-50)(239,-72)
    \Line[arrow,arrowpos=1,arrowlength=3.75,arrowwidth=1.5,arrowinset=0.2](143,-6)(169,-6)
    \Text(104,-17)[]{\normalsize{\Black{$\psi_{3/2}$}}}
    \Text(156,-4)[b]{\normalsize{\Black{$p$}}}
    \Text(234,6)[lb]{\normalsize{\Black{$k_2$}}}
    \Text(213,39)[rb]{\normalsize{\Black{$k_1$}}}
    \Text(235,-62)[lb]{\normalsize{\Black{$q$}}}
    \Text(237,58)[]{\normalsize{\Black{$f$}}}
    \Text(237,-24)[]{\normalsize{\Black{$\bar{f}$}}}
    \Text(193,6)[]{\normalsize{\Black{$Z$}}}
    \Text(219,-28)[]{\normalsize{\Black{$\tilde{H}_{u,\,d}^0$}}}
    \Text(182,-40)[]{\normalsize{\Black{$v_{u,\,d}$}}}
    \Text(237,-91)[]{\normalsize{\Black{$\nu_i$}}}
    \Arc[arrow,arrowpos=0.5,arrowlength=3.75,arrowwidth=1.5,arrowinset=0.2,clock](133.852,-120.786)(81.877,92.697,26.698)
    \Line[dash,dashsize=4](185,-34)(195,-17)
  \end{picture}
 }.
\end{equation*}
We find for the diagram with photon exchange
\begin{equation}
 \begin{split}
  i\mathcal{M}_\gamma &=\bar{u}^t(q)\,P_R\,N_{\nu_i\,\tilde{\gamma}}^{7*}\,\frac{i}{4\,\MP}\,\gamma^{\mu}\left[ \slashed{p}-\slashed{q},\,\gamma^{\rho}\right] \psi_{\mu}^{+\,s}(p)\,\frac{ig_{\rho\nu}}{(p-q)^2}\,\bar{u}^q(k_1)\,i\,Q\,e\,\gamma^\nu v^r(k_2) \\
  &\simeq\frac{i\,Q\,e\,\xi_i}{4\,\MP}\,\frac{g_{\rho\nu}}{(p-q)^2}\,U^*_{\tilde{\gamma}\tilde{Z}}\,\bar{u}^t(q)\,P_R\,\gamma^\mu\left[ \slashed{p}-\slashed{q},\,\gamma^\rho\right] \psi_{\mu}^{+\,s}(p)\,\bar{u}^q(k_1)\,\gamma^\nu v^r(k_2) \\
  &=-\frac{i\,Q\,e\,\xi_i\,U^*_{\tilde{\gamma}\tilde{Z}}}{\MP\,(p-q)^2}\,\bar{u}^t(q)\,P_R\left( q^\mu\gamma^\rho-g^{\mu\rho}(\slashed{q}-\slashed{p})\right) \psi_{\mu}^{+\,s}(p)\,\bar{u}^q(k_1)\,\gamma_\rho\,v^r(k_2)\,.
 \end{split}
\end{equation}
In the last step we used the fact that $p^\mu\psi_{\mu}^{+\,s}(p)=0$ (see equation~(\ref{modeRarita})) and thus
\begin{equation}
 \begin{split}
  \gamma^\mu\left[ \slashed{p}-\slashed{q},\,\gamma^\rho\right] \psi_{\mu}^{+\,s}(p) &=4\left( (p^\mu-q^\mu)\gamma^\rho-g^{\mu\rho}(\slashed{p}-\slashed{q})\right) \psi_{\mu}^{+\,s}(p) \\
  &=-4\left( q^\mu\gamma^\rho-g^{\mu\rho}(\slashed{q}-\slashed{p})\right) \psi_{\mu}^{+\,s}(p)\,.
 \end{split}
\end{equation}
The photino--zino mixing parameter $U_{\tilde{\gamma}\tilde{Z}}$ has already been introduced in equation~(\ref{UgammaZ}). The amplitude for the first $Z$ boson exchange diagram reads
\begin{align}
  i\mathcal{M}_{Z_1} &=\bar{u}^t(q)\,P_R\,N_{\nu_i\,\tilde{Z}}^{7*}\,\frac{i}{4\,\MP}\,\gamma^{\mu}\left[ \slashed{p}-\slashed{q},\,\gamma^{\rho}\right] \psi_{\mu}^{+\,s}(p) \nonumber\\
  &\qquad\quad\times\frac{i\left( g_{\rho\nu}-\frac{(p_\rho-q_\rho)(p_\nu-q_\nu)}{m_Z^2}\right) }{(p-q)^2-m_Z^2+im_Z\Gamma_Z}\,\bar{u}^q(k_1)\,\frac{ig}{\cos\theta_W}\,\gamma^\nu\left( C_V+C_A\gamma^5\right) v^r(k_2) \nonumber\\
  &\simeq\frac{ig_Z\,\xi_i}{4\,\MP}\,\frac{g_{\rho\nu}-\frac{(p_\rho-q_\rho)(p_\nu-q_\nu)}{m_Z^2}}{(p-q)^2-m_Z^2+im_Z\Gamma_Z}\,U^*_{\tilde{Z}\tilde{Z}} \\
  &\qquad\quad\times\bar{u}^t(q)\,P_R\,\gamma^{\mu}\left[ \slashed{p}-\slashed{q},\,\gamma^{\rho}\right] \psi_{\mu}^{+\,s}(p)\,\bar{u}^q(k_1)\,\gamma^\nu\left( C_V+C_A\gamma^5\right) v^r(k_2) \nonumber\\
  &=-\frac{ig_Z\,\xi_i}{\MP}\,\frac{U^*_{\tilde{Z}\tilde{Z}}}{(p-q)^2-m_Z^2+im_Z\Gamma_Z} \nonumber\\
  &\qquad\quad\times\bar{u}^t(q)\,P_R\left( q^\mu\gamma^\rho-g^{\mu\rho}(\slashed{q}-\slashed{p})\right) \psi_{\mu}^{+\,s}(p)\,\bar{u}^q(k_1)\,\gamma_\rho\left( C_V+C_A\gamma^5\right) v^r(k_2)\,. \nonumber
\end{align}
Here we used the $Z$ boson coupling constant $g_Z=g/\cos\theta_W$. The zino--zino mixing parameter $U_{\tilde{Z}\tilde{Z}}$ has already been introduced in equation~(\ref{UZZ}). The coefficients of the $V-A$ structure of the vertex of the $Z$ boson with two fermions were introduced in equation~(\ref{CVCA}). The amplitude for the second $Z$ boson exchange diagram is given by
\begin{equation}
 \begin{split}
  i\mathcal{M}_{Z_2} &=-\bar{u}^t(q)\,P_R\,\frac{i\,v_i}{\sqrt{2}\,\MP}\left( g\,\frac{\sigma_{3,\,11}}{2}\cos{\theta_W}-g'Y_{\nu_L}\sin{\theta_W}\right) \gamma^{\mu}\gamma^{\rho}\psi_{\mu}^{+\,s}(p) \\
  &\qquad\quad\times\frac{i\left( g_{\rho\nu}-\frac{(p_\rho-q_\rho)(p_\nu-q_\nu)}{m_Z^2}\right) }{(p-q)^2-m_Z^2+im_Z\Gamma_Z}\,\bar{u}^q(k_1)\,\frac{ig}{\cos\theta_W}\,\gamma^\nu\left( C_V+C_A\gamma^5\right) v^r(k_2) \\
  &=\frac{ig_Z\,m_Z\,\xi_i}{2\,\MP}\,\frac{g_{\rho\nu}-\frac{(p_\rho-q_\rho)(p_\nu-q_\nu)}{m_Z^2}}{(p-q)^2-m_Z^2+im_Z\Gamma_Z} \\
  &\qquad\quad\times\bar{u}^t(q)\,P_R\,\gamma^{\mu}\gamma^{\rho}\psi_{\mu}^{+\,s}(p)\,\bar{u}^q(k_1)\,\gamma^\nu\left( C_V+C_A\gamma^5\right) v^r(k_2) \\
  &=\frac{ig_Z\,m_Z\,\xi_i}{\MP}\,\frac{g^{\mu\rho}+\frac{q^\mu(p^\rho-q^\rho)}{m_Z^2}}{(p-q)^2-m_Z^2+im_Z\Gamma_Z} \\
  &\quad\qquad\times\bar{u}^t(q)\,P_R\,\psi_\mu^{+\,s}(p)\,\bar{u}^q(k_1)\,\gamma_\rho\left( C_V+C_A\gamma^5\right) v^r(k_2)\,.
 \end{split}
\end{equation}
Here we used the fact that $\gamma^\mu\psi_{\mu}^{+\,s}(p)=0$ (see equation~(\ref{modeRarita})) and thus
\begin{equation}
  \gamma^{\mu}\gamma^{\rho}\psi_{\mu}^{+\,s}(p)=\left( 2\,g^{\mu\rho}-\gamma^{\rho}\gamma^{\mu}\right) \psi_{\mu}^{+\,s}(p)=2\,\psi^{+\,s\,\rho}(p)\,.
\end{equation}
The last two diagrams have the same vertex structure as the previous one. In analogy to the two-body decay amplitude we find
\begin{equation}
 \begin{split}
  i\mathcal{M}_{Z_3} &=\frac{ig_Z\,m_Z\,\xi_i}{\MP}\left\lbrace \sin\beta\,U^*_{\tilde{H}_u^0\tilde{Z}}-\cos\beta\,U^*_{\tilde{H}_d^0\tilde{Z}}\right\rbrace \,\frac{g^{\mu\rho}+\frac{q^\mu(p^\rho-q^\rho)}{m_Z^2}}{(p-q)^2-m_Z^2+im_Z\Gamma_Z} \\
  &\quad\qquad\times\bar{u}^t(q)P_R\,\psi_\mu^{+\,s}(p)\,\bar{u}^q(k_1)\,\gamma_\rho\left( C_V+C_A\gamma^5\right) v^r(k_2) \\
  &=i\mathcal{M}_{Z_2}\times\left\lbrace \sin\beta\,U^*_{\tilde{H}_u^0\tilde{Z}}-\cos\beta\,U^*_{\tilde{H}_d^0\tilde{Z}}\right\rbrace .
 \end{split}
\end{equation}
With these expressions for the individual amplitudes we can now write down the squared matrix element:
\begin{align}
  \bar{\abs{\mathcal{M}}^2}= &\;\frac{1}{4}\sum_{s,\,t,\,q,\,r}\left( i\mathcal{M}_\gamma+i\mathcal{M}_{Z_1}+i\mathcal{M}_{Z_2}+i\mathcal{M}_{Z_3}\right) \left( i\mathcal{M}_\gamma+i\mathcal{M}_{Z_1}+i\mathcal{M}_{Z_2}+i\mathcal{M}_{Z_3}\right) ^* \nonumber\\
  = &-\frac{\xi_i^2}{4\,\MP^2}\Bigg[ \frac{Q^2e^2\abs{U_{\tilde{\gamma}\tilde{Z}}}^2}{(p-q)^4}\Tr\left[ \left( \slashed{k}_1+m_f\right) \gamma_\rho\left( \slashed{k}_2-m_f\right) \gamma_\sigma\right] \nonumber\\
  &\qquad\times\Tr\left[ \slashed{q}P_R\left( q^\mu\gamma^\rho-g^{\mu\rho}(\slashed{q}-\slashed{p})\right) \slashed{p}\,\Phi_{\mu\nu}(p)\left( q^\nu\gamma^\sigma-g^{\nu\sigma}(\slashed{q}-\slashed{p})\right) \right] \nonumber\\
  &\quad-2\,Q\,e\,g_Z\,U_{\tilde{Z}\tilde{Z}}\,\frac{\RE U_{\tilde{\gamma}\tilde{Z}}\left( m_Z^2-s\right) +\IM U_{\tilde{\gamma}\tilde{Z}}\,m_Z\,\Gamma_Z}{\left( (p-q)^2-m_Z^2\right) ^2+m_Z^2\Gamma_Z^2} \nonumber\\
  &\qquad\times\Tr\left[ \left( \slashed{k}_1+m_f\right) \gamma_\rho\left( C_V+C_A\gamma^5\right) \left( \slashed{k}_2-m_f\right) \gamma_\sigma\right] \nonumber\\
  &\qquad\times\Tr\left[ \slashed{q}P_R\left( q^\mu\gamma^\rho-g^{\mu\rho}(\slashed{q}-\slashed{p})\right) \slashed{p}\,\Phi_{\mu\nu}(p)\left( q^\nu\gamma^\sigma-g^{\nu\sigma}(\slashed{q}-\slashed{p})\right) \right] \nonumber\\
  &\quad+\frac{g_Z^2\,U_{\tilde{Z}\tilde{Z}}^2}{\left( (p-q)^2-m_Z^2\right) ^2+m_Z^2\Gamma_Z^2} \nonumber\\
  &\qquad\times\Tr\left[ \left( \slashed{k}_1+m_f\right) \gamma_\rho\left( C_V+C_A\gamma^5\right) \left( \slashed{k}_2-m_f\right) \left( C_V-C_A\gamma^5\right) \gamma_\sigma\right] \nonumber\\
  &\qquad\times\Tr\left[ \slashed{q}P_R\left( q^\mu\gamma^\rho-g^{\mu\rho}(\slashed{q}-\slashed{p})\right) \slashed{p}\,\Phi_{\mu\nu}(p)\left( q^\nu\gamma^\sigma-g^{\nu\sigma}(\slashed{q}-\slashed{p})\right) \right] \nonumber\\
  &\quad-\frac{2\,g_Z^2\,m_{3/2}\,m_Z\,U_{\tilde{Z}\tilde{Z}}\left( 1+s_\beta \RE U_{\tilde{H}_u^0\tilde{Z}}-c_\beta \RE U_{\tilde{H}_d^0\tilde{Z}}\right) }{\left( (p-q)^2-m_Z^2\right) ^2+m_Z^2\Gamma_Z^2}\left( g^{\nu\sigma}+\frac{q^\nu(p^\sigma-q^\sigma)}{m_Z^2}\right) \nonumber\\
  &\qquad\times\Tr\left[ \left( \slashed{k}_1+m_f\right) \gamma_\rho\left( C_V+C_A\gamma^5\right) \left( \slashed{k}_2-m_f\right) \left( C_V-C_A\gamma^5\right) \gamma_\sigma\right] \\
  &\qquad\times\Tr\left[ \slashed{q}P_R\left( q^\mu\gamma^\rho-g^{\mu\rho}(\slashed{q}-\slashed{p})\right) \Phi_{\mu\nu}(p)\right] \nonumber\\
  &\quad+g_Z^2\,m_Z^2\abs{1+s_\beta\,U_{\tilde{H}_u^0\tilde{Z}}-c_\beta\,U_{\tilde{H}_d^0\tilde{Z}}}^2\frac{\left( g^{\mu\rho}+\frac{q^\mu(p^\rho-q^\rho)}{m_Z^2}\right) \left( g^{\nu\sigma}+\frac{q^\nu(p^\sigma-q^\sigma)}{m_Z^2}\right) }{\left( (p-q)^2-m_Z^2\right) ^2+m_Z^2\Gamma_Z^2} \nonumber\\
  &\qquad\times\Tr\left[ \left( \slashed{k}_1+m_f\right) \gamma_\rho\left( C_V+C_A\gamma^5\right) \left( \slashed{k}_2-m_f\right) \left( C_V-C_A\gamma^5\right) \gamma_\sigma\right] \nonumber\\
  &\qquad\times\Tr\left[ \slashed{q}P_R\,\slashed{p}\,\Phi_{\mu\nu}(p)\right] \nonumber\\
  &\quad+2\,Q\,e\,g_Z\,m_{3/2}\,m_Z\left( 1+s_\beta \RE U_{\tilde{H}_u^0\tilde{Z}}-c_\beta \RE U_{\tilde{H}_d^0\tilde{Z}}\right) \nonumber\\
  &\qquad\times\frac{\RE U_{\tilde{\gamma}\tilde{Z}}\left( m_Z^2-s\right) +\IM U_{\tilde{\gamma}\tilde{Z}}\,m_Z\,\Gamma_Z}{\left( (p-q)^2-m_Z^2\right) ^2+m_Z^2\Gamma_Z^2}\left( g^{\nu\sigma}+\frac{q^\nu(p^\sigma-q^\sigma)}{m_Z^2}\right) \nonumber\\
  &\qquad\times\Tr\left[ \slashed{q}P_R\left( q^\mu\gamma^\rho-g^{\mu\rho}(\slashed{q}-\slashed{p})\right) \Phi_{\mu\nu}(p)\right] \nonumber\\
  &\qquad\times\Tr\left[ \left( \slashed{k}_1+m_f\right) \gamma_\rho\left( C_V+C_A\gamma^5\right) \left( \slashed{k}_2-m_f\right) \gamma_\sigma\right] \Bigg]\,.\nonumber
\end{align}
A number of traces of gamma matrices appear in this expression. Using the Mathematica package Feyn~Calc~\cite{Mertig:1990an} we find that the traces originating from the fermion current part of the amplitude are given by
\begin{equation}
 \begin{split}
  &\Tr\left[ \left( \slashed{k}_1+m_f\right) \gamma_\rho\left( \slashed{k}_2-m_f\right) \gamma_\sigma\right] \\
  &\qquad=4\left[ k_{1\rho} k_{2\sigma}+k_{1\sigma} k_{2\rho}-g_{\rho\sigma}\left( m_f^2+(k_1\cdot k_2)\right) \right] , \\
  &\Tr\left[ \left( \slashed{k}_1+m_f\right) \gamma_\rho\left( C_V+C_A\gamma^5\right) \left( \slashed{k}_2-m_f\right) \gamma_\sigma\right] \\
  &\qquad=4\,C_V\left[ k_{1\rho} k_{2\sigma}+k_{1\sigma} k_{2\rho}-g_{\rho\sigma}\left( m_f^2+(k_1\cdot k_2)\right) \right] +4\,i\,C_A\,\varepsilon_{\rho\sigma\delta\lambda}k_1^\delta k_2^\lambda\,, \\
  &\Tr\left[ \left( \slashed{k}_1+m_f\right) \gamma_\rho\left( C_V+C_A\gamma^5\right) \left( \slashed{k}_2-m_f\right) \left( C_V-C_A\gamma^5\right) \gamma_\sigma\right] \\
  &\qquad=4\left( C_V^2+C_A^2\right) \left[ k_{1\rho} k_{2\sigma}+k_{1\sigma} k_{2\rho}-g_{\rho\sigma}\left( m_f^2+(k_1\cdot k_2)\right) \right] +8\,C_A^2\,g_{\rho\sigma}\,m_f^2 \\
  &\qquad\quad\quad+8\,i\,C_V\,C_A\,\varepsilon_{\rho\sigma\delta\lambda}k_1^\delta k_2^\lambda\,,
 \end{split}
\end{equation}
while the traces including the gravitino field turn out to be
\begin{equation}
 \begin{split}
  &\Tr\left[ \slashed{q}\,P_R\left( q^\mu\gamma^\rho-g^{\mu\rho}(\slashed{q}-\slashed{p})\right) \slashed{p}\,\Phi_{\mu\nu}(p)\left( q^\nu\gamma^\sigma-g^{\nu\sigma}(\slashed{q}-\slashed{p})\right) \right] \\
  &\qquad=\frac{4}{3\,m_{3/2}^2}\left[ \left( p\cdot q\right) \left( g^{\rho\sigma}\left( m_{3/2}^4-m_{3/2}^2\left( p\cdot q\right) +\left( p\cdot q\right) ^2\right) \right. \right. \\
  &\qquad\quad\quad-p^\rho\left. \left( m_{3/2}^2\,p^\sigma+q^\sigma\left( p\cdot q\right) \right) \Big) +q^\rho\left( m_{3/2}^4\,q^\sigma-p^\sigma\left( p\cdot q\right) ^2\right) \right] \\
  &\qquad\quad+\frac{2\,i}{3\,m_{3/2}^2}\left( m_{3/2}^4+2\,m_{3/2}^2\left( p\cdot q\right) -2\left( p\cdot q\right) ^2\right) \varepsilon^{\rho\sigma\delta\lambda}p_\delta\,q_\lambda\,, \\
  &\Tr\left[ \slashed{q}P_R\left( q^\mu\gamma^\rho-g^{\mu\rho}(\slashed{q}-\slashed{p})\right) \Phi_{\mu\nu}(p)\right] \\
  &\qquad=\frac{2}{3\,m_{3/2}^2}\left[ \left( p\cdot q\right) \left( \delta_\nu^\rho\left( 2\,m_{3/2}^2-\left( p\cdot q\right) \right) +p_\nu\left( 2\,p^\rho+q^\rho\right) \right. \right. \\
  &\left. \qquad\quad\quad+\left. q_\nu\left( m_{3/2}^2\,q^\rho+p^\rho\left( p\cdot q\right) \right) \right) -i\left( m_{3/2}^2+\left( p\cdot q\right) \right) g_{\nu\sigma}\,\varepsilon^{\sigma\rho\delta\lambda}p_\delta\,q_\lambda\right] , \\
  &\Tr\left[ \slashed{q}P_R\,\slashed{p}\,\Phi_{\mu\nu}(p)\right] \\
  &\qquad=\frac{2}{3\,m_{3/2}^2}\left[ 2\left( m_{3/2}^2\,g_{\mu\nu}-p_\mu p_\nu\right) \left( p\cdot q\right) +i\,m_{3/2}^2\,\varepsilon_{\mu\nu\rho\sigma}p^\rho\,q^\sigma\right] .
 \end{split}
 \label{gravitinotrace}
\end{equation}
In these expressions we already replaced the squared four-momenta with the squared particle masses:
\begin{equation}
  p^2=m_{3/2}^2\,,\quad k_1^2=k_2^2=m_f^2\simeq 0\,,\quad q^2=m_\nu^2\simeq 0\,.
\end{equation}
In most cases it is convenient to neglect the masses of the final state particles. Now we want to use the complete kinematics of this decay process in order to calculate the corresponding decay width. Using equation~(\ref{threebodyscalar}) we find
\begin{alignat}{2}
  \left( p\cdot q\right) &\simeq\frac{m_{3/2}^2-s}{2}\,, &\qquad\qquad \left( k_1\cdot k_2\right) &\simeq\frac{s}{2}\,, \nonumber\\
  \left( p\cdot k_2\right) &\simeq\frac{m_{3/2}^2-t}{2}\,, &\qquad\qquad \left( k_1\cdot q\right) &\simeq\frac{t}{2}\,, \\
  \left( p\cdot k_1\right) &\simeq\frac{s+t}{2}\,, & \left( k_2\cdot q\right) &\simeq\frac{m_{3/2}^2-s-t}{2}\,, \nonumber
\end{alignat}
where we renamed the invariant masses used to parametrize the kinematics:
\begin{equation}
  s\equiv m_{12}^2=(p-q)^2=(k_1+k_2)^2,\qquad t\equiv m_{13}^2=(p-k_2)^2=(k_1+q)^2.
\end{equation}
From equations~(\ref{m12range}) and~(\ref{m13range}) the ranges of the invariant masses are found to be
\begin{equation}
  0\lesssim t\lesssim m_{3/2}^2-s \qquad\text{and}\qquad 0\lesssim s\lesssim m_{3/2}^2\,.
\end{equation}
Using equation~(\ref{threebodywidth}) we can now write down the differential decay width for a gravitino decaying into a fermion-antifermion pair and a neutrino:
\begin{align}
  \frac{d\Gamma}{ds\,dt}\simeq &\;\frac{\xi_i^2}{1536\,\pi^3\,m_{3/2}^5\,\MP^2}\,\Bigg[\frac{e^2\,Q^2}{s}\abs{U_{\tilde{\gamma}\tilde{Z}}}^2 \nonumber\\
  &\qquad\times\left( 3\,m_{3/2}^6-3\,m_{3/2}^4\,(s+2\,t)+m_{3/2}^2\left( s^2+8\,s\,t+6\,t^2\right) -s\left( s^2+2\,s\,t+2\,t^2\right) \right) \nonumber\\
  &\quad+\frac{g_Z}{\left( s-m_Z^2\right) ^2+m_Z^2\,\Gamma_Z^2}\,\bigg\lbrace g_Z\,s\,U_{\tilde{Z}\tilde{Z}}^2\Big((C_V-C_A)^2\left( 3\,m_{3/2}^4-s^2\right) \nonumber\\
  &\qquad\quad\times\left( m_{3/2}^2-s-2\,t\right) +\left( C_V^2+C_A^2\right) \left( m_{3/2}^2\left( 2\,s^2+8\,s\,t+6\,t^2\right) -2\,s\,(s+t)^2 \right) \!\Big) \nonumber\\
  &\qquad-2\,g_Z\,s\,m_{3/2}\,m_Z\,U_{\tilde{Z}\tilde{Z}}\left( 1+s_\beta \RE U_{\tilde{H}_u^0\tilde{Z}}-c_\beta \RE U_{\tilde{H}_d^0\tilde{Z}}\right) \nonumber\\
  &\qquad\quad\times\Big(-2\,C_V\,C_A\left( 3\,m_{3/2}^2-s\right) \left( m_{3/2}^2-s-2\,t\right) \nonumber\\
  &\qquad\qquad+\left( C_V^2+C_A^2\right) \left( 3\,m_{3/2}^4-2\,m_{3/2}^2\,(s+t)-s^2+2\,s\,t+2\,t^2\right) \!\Big) \\
  &\qquad+2\,g_Z\,m_Z^2\abs{1+s_\beta\,U_{\tilde{H}_u^0\tilde{Z}}-c_\beta\,U_{\tilde{H}_d^0\tilde{Z}}}^2\Big(-2\,m_{3/2}^2\,s\,C_V\,C_A\left( m_{3/2}^2-s-2\,t\right) \nonumber\\
  &\qquad\quad+\left( C_V^2+C_A^2\right) \left( m_{3/2}^4\,(2\,s+t)-m_{3/2}^2\left( 2\,s^2+2\,s\,t+t^2\right) +s\,t\,(s+t)\right) \Big) \nonumber\\
  &\qquad-2\,e\,Q\left( \RE U_{\tilde{\gamma}\tilde{Z}}\left( m_Z^2-s\right) +\IM U_{\tilde{\gamma}\tilde{Z}}\,m_Z\,\Gamma_Z\right) \nonumber\\
  &\qquad\quad\times\bigg(U_{\tilde{Z}\tilde{Z}}\Big(C_V\big( 3\,m_{3/2}^6-3\,m_{3/2}^4\,(s+2\,t)+m_{3/2}^2\left( s^2+8\,s\,t+6\,t^2\right) \nonumber\\
  &\qquad\qquad\quad-s\left( s^2+2\,s\,t+2\,t^2\right) \big)-C_A\left( 3\,m_{3/2}^4-s^2\right) \left( m_{3/2}^2-s-2\,t\right) \Big) \nonumber\\
  &\qquad\qquad-m_{3/2}\,m_Z\left( 1+s_\beta \RE U_{\tilde{H}_u^0\tilde{Z}}-c_\beta \RE U_{\tilde{H}_d^0\tilde{Z}}\right) \nonumber\\
  &\qquad\qquad\quad\times\Big(C_V\left( 3\,m_{3/2}^4-2\,m_{3/2}^2\,(s+t)-s^2+2\,s\,t+2\,t^2\right) \nonumber\\
  &\qquad\qquad\qquad-C_A\left( 3\,m_{3/2}^2-s\right) \left( m_{3/2}^2-s-2\,t\right) \Big) \bigg) \bigg\rbrace \Bigg]\,. \nonumber
\end{align}
After integrating over the invariant mass $t$ we find as our final result the following differential decay width:
\begin{align}
  \frac{d\Gamma}{ds} &\simeq\frac{\xi_i^2\,m_{3/2}^3\,\beta_s^2}{768\,\pi^3\MP^2}\Bigg[ \frac{e^2\,Q^2}{s}\abs{U_{\tilde{\gamma}\tilde{Z}}}^2f_s+\frac{g_Z}{\left( s-m_Z^2\right) ^2+m_Z^2\,\Gamma_Z^2}\,\bigg\lbrace\,g_Z\,U_{\tilde{Z}\tilde{Z}}^2\,s\left( C_V^2+C_A^2\right) f_s \nonumber\\
  &\qquad\quad\qquad-\frac{8}{3}\,\frac{m_Z}{m_{3/2}}\,g_Z\,U_{\tilde{Z}\tilde{Z}}\left( 1+s_\beta \RE U_{\tilde{H}_u^0\tilde{Z}}-c_\beta \RE U_{\tilde{H}_d^0\tilde{Z}}\right) s\left( C_V^2+C_A^2\right) j_s \nonumber\\
  &\qquad\quad\qquad+\frac{1}{6}\,g_Z\,m_Z^2\abs{1+s_\beta\,U_{\tilde{H}_u^0\tilde{Z}}-c_\beta\,U_{\tilde{H}_d^0\tilde{Z}}}^2\left( C_V^2+C_A^2\right) h_s \\
  &\qquad\quad\qquad+e\,Q\left( \RE U_{\tilde{\gamma}\tilde{Z}}\left( m_Z^2-s\right) +\IM U_{\tilde{\gamma}\tilde{Z}}\,m_Z\,\Gamma_Z\right) C_V \nonumber\\
  &\qquad\qquad\qquad\times\left( 2\,U_{\tilde{Z}\tilde{Z}}\,f_s +\frac{8}{3}\frac{m_Z}{m_{3/2}}\left( 1+s_\beta \RE U_{\tilde{H}_u^0\tilde{Z}}-c_\beta \RE U_{\tilde{H}_d^0\tilde{Z}}\right) j_s\right) \bigg\rbrace \Bigg]\,. \nonumber
\end{align}
In this case the kinematic functions $\beta_s, f_s, j_s$ and $h_s$ are defined corresponding to the kinematic functions in the two-body decays (see equation(\ref{kinematicfunctions})):
\begin{alignat}{2}
  \beta_s &=1-\frac{s}{m_{3/2}^2}\,, &\qquad\qquad f_s &=1+\frac{2}{3}\,\frac{s}{m_{3/2}^2}+\frac{1}{3}\,\frac{s^2}{m_{3/2}^4}\,, \nonumber\\
  j_s &=1+\frac{1}{2}\,\frac{s}{m_{3/2}^2}\,, &\qquad\qquad h_s &=1+10\,\frac{s}{m_{3/2}^2}+\frac{s^2}{m_{3/2}^4}\,.
\end{alignat}
We can now verify that above the threshold for $Z$ boson production and using the narrow-width approximation (NWA) the part of this result coming from $Z$ boson exchange coincides with the two-body decay result as given in equation~(\ref{gravitinowidths}), \textit{i.e.}:
\begin{equation}
  \Gamma_{\text{NWA}}\left( \psi_{3/2}\rightarrow Z^*\nu_i\rightarrow f\,\bar{f}\,\nu_i\right) =\Gamma\left( \psi_{3/2}\rightarrow Z\,\nu_i\right) \times\BR\left( Z\rightarrow f\,\bar{f}\right) .
\end{equation}
The narrow-width approximation states that in the limit of a vanishing width-to-mass ratio the propagator can be replaced by a $\delta$-function:
\begin{equation}
  \lim_{\Gamma/m\rightarrow 0}\frac{1}{\left( s-m^2\right) ^2+m^2\Gamma^2}=\frac{\pi}{m\,\Gamma}\,\delta(s-m^2)\,.
\end{equation}
Using this replacement we find for the three-body decay width
\begin{equation}
  \begin{split}
   &\Gamma_{\text{NWA}}\left( \psi_{3/2}\rightarrow Z^*\nu_i\rightarrow f\,\bar{f}\,\nu_i\right) =\int_0^{m_{3/2}^2}\!\!ds\,\frac{d\Gamma\left( \psi_{3/2}\rightarrow Z^*\nu_i\rightarrow f\,\bar{f}\,\nu_i\right)}{ds}\bigg|_{\text{NWA}} \\
   &\qquad=\frac{\xi_i^2\,m_{3/2}^3\,\beta_Z^2}{64\,\pi\,\MP^2}\left( U_{\tilde{Z}\tilde{Z}}^2\,f_Z-\frac{8}{3}\,\frac{m_Z}{m_{3/2}}\,U_{\tilde{Z}\tilde{Z}}\left( 1+s_\beta \RE U_{\tilde{H}_u^0\tilde{Z}}-c_\beta \RE U_{\tilde{H}_d^0\tilde{Z}}\right) j_Z\right. \\
   &\qquad\left.\qquad+\frac{1}{6}\abs{1+s_\beta\,U_{\tilde{H}_u^0\tilde{Z}}-c_\beta\,U_{\tilde{H}_d^0\tilde{Z}}}^2h_Z\right) \times\frac{\Gamma\left( Z\rightarrow f\,\bar{f}\right) }{\Gamma_Z}\,,
  \end{split}
\end{equation}
which is exactly the equivalence we wanted to verify. In the last expression we used the partial width of the $Z$ boson decay into a fermion pair (see for instance~\cite{Nakamura:2010zzi}):
\begin{equation}
  \Gamma\left( Z\rightarrow f\,\bar{f}\right) =\frac{\sqrt{2}\,G_F\,m_Z^3}{3\,\pi}\left( C_V^2+C_A^2\right) =\frac{g_Z^2\,m_Z}{12\,\pi}\left( C_V^2+C_A^2\right) .
\end{equation}

\subsection[\texorpdfstring{$\psi_{3/2}\rightarrow {W^+}^*\,\ell^-\rightarrow f\,\bar{f'}\,\ell^-$}{\textpsi\ --> W* l --> f f' l}]{\boldmath$\psi_{3/2}\rightarrow {W^+}^*\,\ell^-\rightarrow f\,\bar{f'}\,\ell^-$}

At tree level there are three diagrams contributing to the decay of a gravitino into two fermions and a charged lepton:
\begin{equation*}
 \parbox{5.5cm}{
  \begin{picture}(172,162) (92,-96)
    \SetWidth{0.5}
    \SetColor{Black}
    \Line(211,-55.999)(215,-44.001)\Line(207.001,-48)(218.999,-52)
    \Line[arrow,arrowpos=0.5,arrowlength=3.75,arrowwidth=1.5,arrowinset=0.2](213,-50)(232,-84)
    \Line(194,-17)(213,-51)
    \Photon(194,-17)(213,17){-5}{2}
    \Photon(194,-17)(213,-51){-5}{2}
    \Line[arrow,arrowpos=0.5,arrowlength=3.75,arrowwidth=1.5,arrowinset=0.2](213,17)(232,51)
    \Line[arrow,arrowpos=0.5,arrowlength=3.75,arrowwidth=1.5,arrowinset=0.2](233,-17)(214,17)
    \Line[double,sep=4](117,-17)(194,-17)
    \Vertex(194,-17){2.5}
    \Vertex(213,17){2.5}
    \Line[arrow,arrowpos=1,arrowlength=3.75,arrowwidth=1.5,arrowinset=0.2](207,28)(220,50)
    \Line[arrow,arrowpos=1,arrowlength=3.75,arrowwidth=1.5,arrowinset=0.2](226,17)(239,-5)
    \Line[arrow,arrowpos=1,arrowlength=3.75,arrowwidth=1.5,arrowinset=0.2](226,-50)(239,-72)
    \Line[arrow,arrowpos=1,arrowlength=3.75,arrowwidth=1.5,arrowinset=0.2](143,-6)(169,-6)
    \Text(104,-17)[]{\normalsize{\Black{$\psi_{3/2}$}}}
    \Text(156,-4)[b]{\normalsize{\Black{$p$}}}
    \Text(234,6)[lb]{\normalsize{\Black{$k_2$}}}
    \Text(213,39)[rb]{\normalsize{\Black{$k_1$}}}
    \Text(235,-62)[lb]{\normalsize{\Black{$q$}}}
    \Text(237,58)[]{\normalsize{\Black{$f$}}}
    \Text(237,-24)[]{\normalsize{\Black{$\bar{f}'$}}}
    \Text(191,6)[]{\normalsize{\Black{$W^+$}}}
    \Text(191,-39)[]{\normalsize{\Black{$\tilde{W}^-$}}}
    \Text(237,-91)[]{\normalsize{\Black{$\ell_i^-$}}}
    \Arc[arrow,arrowpos=0.5,arrowlength=3.75,arrowwidth=1.5,arrowinset=0.2,clock](133.852,-120.786)(81.877,92.697,26.698)
  \end{picture}
 }
 \qquad+\qquad
 \parbox{5.5cm}{
  \begin{picture}(172,162) (92,-96)
    \SetWidth{0.5}
    \SetColor{Black}
    \Line[arrow,arrowpos=0.5,arrowlength=3.75,arrowwidth=1.5,arrowinset=0.2](194,-17)(232,-84)
    \Photon(194,-17)(213,17){-5}{2}
    \Line[arrow,arrowpos=0.5,arrowlength=3.75,arrowwidth=1.5,arrowinset=0.2](213,17)(232,51)
    \Line[arrow,arrowpos=0.5,arrowlength=3.75,arrowwidth=1.5,arrowinset=0.2](233,-17)(214,17)
    \Line[double,sep=4](117,-17)(194,-17)
    \Vertex(194,-17){2.5}
    \Vertex(213,17){2.5}
    \Line[arrow,arrowpos=1,arrowlength=3.75,arrowwidth=1.5,arrowinset=0.2](207,28)(220,50)
    \Line[arrow,arrowpos=1,arrowlength=3.75,arrowwidth=1.5,arrowinset=0.2](226,17)(239,-5)
    \Line[arrow,arrowpos=1,arrowlength=3.75,arrowwidth=1.5,arrowinset=0.2](217,-34)(230,-56)
    \Line[arrow,arrowpos=1,arrowlength=3.75,arrowwidth=1.5,arrowinset=0.2](143,-6)(169,-6)
    \Text(104,-17)[]{\normalsize{\Black{$\psi_{3/2}$}}}
    \Text(156,-4)[b]{\normalsize{\Black{$p$}}}
    \Text(234,6)[lb]{\normalsize{\Black{$k_2$}}}
    \Text(213,39)[rb]{\normalsize{\Black{$k_1$}}}
    \Text(226,-46)[lb]{\normalsize{\Black{$q$}}}
    \Text(237,58)[]{\normalsize{\Black{$f$}}}
    \Text(237,-24)[]{\normalsize{\Black{$\bar{f}'$}}}
    \Text(191,6)[]{\normalsize{\Black{$W^+$}}}
    \Text(182,-40)[]{\normalsize{\Black{$v_i$}}}
    \Text(237,-91)[]{\normalsize{\Black{$\ell_i^-$}}}
    \Arc[arrow,arrowpos=0.5,arrowlength=3.75,arrowwidth=1.5,arrowinset=0.2,clock](133.852,-120.786)(81.877,92.697,26.698)
    \Line[dash,dashsize=4,arrow,arrowpos=0.5,arrowlength=3.75,arrowwidth=1.5,arrowinset=0.2](185,-34)(195,-17)
  \end{picture}
 }
\end{equation*}
and in addition the diagram coming from the 4-vertex with a down-type Higgs VEV at one of the legs that was also discussed for the two-body decay:
\begin{equation*}
 \parbox{5.5cm}{
  \begin{picture}(172,162) (92,-96)
    \SetWidth{0.5}
    \SetColor{Black}
    \Line(211,-55.999)(215,-44.001)\Line(207.001,-48)(218.999,-52)
    \Line[arrow,arrowpos=0.5,arrowlength=3.75,arrowwidth=1.5,arrowinset=0.2](213,-51)(232,-84)
    \Photon(194,-17)(213,17){-5}{2}
    \Line[arrow,arrowpos=0.5,arrowlength=3.75,arrowwidth=1.5,arrowinset=0.2](213,17)(232,51)
    \Line[arrow,arrowpos=0.5,arrowlength=3.75,arrowwidth=1.5,arrowinset=0.2](233,-17)(214,17)
    \Line(194,-17)(213,-51)
    \Line[double,sep=4](117,-17)(194,-17)
    \Vertex(194,-17){2.5}
    \Vertex(213,17){2.5}
    \Line[arrow,arrowpos=1,arrowlength=3.75,arrowwidth=1.5,arrowinset=0.2](207,28)(220,50)
    \Line[arrow,arrowpos=1,arrowlength=3.75,arrowwidth=1.5,arrowinset=0.2](226,17)(239,-5)
    \Line[arrow,arrowpos=1,arrowlength=3.75,arrowwidth=1.5,arrowinset=0.2](226,-50)(239,-72)
    \Line[arrow,arrowpos=1,arrowlength=3.75,arrowwidth=1.5,arrowinset=0.2](143,-6)(169,-6)
    \Text(104,-17)[]{\normalsize{\Black{$\psi_{3/2}$}}}
    \Text(156,-4)[b]{\normalsize{\Black{$p$}}}
    \Text(234,6)[lb]{\normalsize{\Black{$k_2$}}}
    \Text(213,39)[rb]{\normalsize{\Black{$k_1$}}}
    \Text(235,-62)[lb]{\normalsize{\Black{$q$}}}
    \Text(237,58)[]{\normalsize{\Black{$f$}}}
    \Text(237,-24)[]{\normalsize{\Black{$\bar{f}'$}}}
    \Text(193,6)[]{\normalsize{\Black{$W^+$}}}
    \Text(219,-28)[]{\normalsize{\Black{$\tilde{H}_{d}^-$}}}
    \Text(182,-40)[]{\normalsize{\Black{$v_{d}$}}}
    \Text(237,-91)[]{\normalsize{\Black{$\ell_i^-$}}}
    \Arc[arrow,arrowpos=0.5,arrowlength=3.75,arrowwidth=1.5,arrowinset=0.2,clock](133.852,-120.786)(81.877,92.697,26.698)
    \Line[dash,dashsize=4](185,-34)(195,-17)
  \end{picture}
 }.
\end{equation*}
The amplitude for the first diagram is given by
\begin{align}
  i\mathcal{M}_{W_1} &=\bar{u}^t(q)\,P_R\,U_{\ell_i\,\tilde{W}}^{5*}\,\frac{i}{4\,\MP}\,\gamma^{\mu}\left[ \slashed{p}-\slashed{q},\,\gamma^{\rho}\right] \psi_{\mu}^{+\,s}(p)\,\frac{i\left( g_{\rho\nu}-\frac{(p_\rho-q_\rho)(p_\nu-q_\nu)}{m_W^2}\right) }{(p-q)^2-m_W^2+im_W\Gamma_W} \nonumber\\
  &\qquad\quad\times\bar{u}^q(k_1)\,\frac{ig}{\sqrt{2}}\,\gamma^\nu\left( \frac{\sigma_{1,\,21}}{2}-i\,\frac{\sigma_{2,\,21}}{2}\right) P_L\,v^r(k_2) \nonumber\\
  &\simeq\frac{ig\,\xi_i}{4\,\MP}\,U^*_{\tilde{W}\tilde{W}}\,\frac{g_{\rho\nu}-\frac{(p_\rho-q_\rho)(p_\nu-q_\nu)}{m_W^2}}{(p-q)^2-m_W^2+im_W\Gamma_W} \\
  &\qquad\quad\times\bar{u}^t(q)\,P_R\,\gamma^{\mu}\left[ \slashed{p}-\slashed{q},\,\gamma^{\rho}\right] \psi_{\mu}^{+\,s}(p)\,\bar{u}^q(k_1)\,\gamma^\nu P_L\,v^r(k_2) \nonumber\\
  &=-\frac{ig\,\xi_i}{\MP}\,\frac{U^*_{\tilde{W}\tilde{W}}}{(p-q)^2-m_W^2+im_W\Gamma_W} \nonumber\\
  &\qquad\quad\times\bar{u}^t(q)\,P_R\left( q^\mu\gamma^\rho-g^{\mu\rho}(\slashed{q}-\slashed{p})\right) \psi_{\mu}^{+\,s}(p)\,\bar{u}^q(k_1)\,\gamma_\rho P_L\,v^r(k_2)\,, \nonumber
\end{align}
while the amplitude for the second diagram reads
\begin{align}
  i\mathcal{M}_{W_2} &=-\bar{u}^t(q)\,P_R\,\frac{i\,v_i}{\sqrt{2}\,\MP}\,\frac{g}{\sqrt{2}}\left( \frac{\sigma_{1,\,21}}{2}-i\,\frac{\sigma_{2,\,21}}{2}\right) \gamma^{\mu}\gamma^{\rho}\,\psi_{\mu}^{+\,s}(p) \\
  &\qquad\quad\times\frac{i\left( g_{\rho\nu}-\frac{(p_\rho-q_\rho)(p_\nu-q_\nu)}{m_W^2}\right) }{(p-q)^2-m_W^2+im_W\Gamma_W}\,\bar{u}^q(k_1)\,\frac{ig}{\sqrt{2}}\,\gamma^\nu\left( \frac{\sigma_{1,\,21}}{2}-i\,\frac{\sigma_{2,\,21}}{2}\right) P_L\,v^r(k_2) \nonumber\\
  &\simeq\frac{ig\,m_W\,\xi_i}{2\,\MP}\,\frac{g_{\rho\nu}-\frac{(p_\rho-q_\rho)(p_\nu-q_\nu)}{m_W^2}}{(p-q)^2-m_W^2+im_W\Gamma_W}\,\bar{u}^t(q)\,P_R\,\gamma^{\mu}\gamma^{\rho}\,\psi_{\mu}^{+\,s}(p)\,\bar{u}^q(k_1)\,\gamma^\nu P_L\,v^r(k_2) \nonumber\\
  &=\frac{ig\,m_W\,\xi_i}{\MP}\,\frac{g_{\rho\nu}+\frac{q_\rho(p_\nu-q_\nu)}{m_W^2}}{(p-q)^2-m_W^2+im_W\Gamma_W}\,\bar{u}^t(q)\,P_R\,\psi^{+\,s\,\rho}(p)\,\bar{u}^q(k_1)\,\gamma^\nu P_L\,v^r(k_2)\,. \nonumber
\end{align}
The last diagram has the same vertex structure as the previous one. In analogy to the two-body decay amplitude we find
\begin{equation}
 \begin{split}
  i\mathcal{M}_{W_3} &=-\frac{ig\,m_W\,\xi_i}{\MP}\,\sqrt{2}\cos\beta\,U^*_{\tilde{H}_d^-\tilde{W}}\,\frac{g_{\rho\nu}+\frac{q_\rho(p_\nu-q_\nu)}{m_W^2}}{(p-q)^2-m_W^2+im_W\Gamma_W} \\
  &\qquad\quad\times\bar{u}^t(q)\,P_R\,\psi^{+\,s\,\rho}(p)\,\bar{u}^q(k_1)\,\gamma^\nu P_L\,v^r(k_2) \\
  &=-i\mathcal{M}_{W_2}\times\sqrt{2}\cos\beta\,U^*_{\tilde{H}_d^-\tilde{W}}\,.
 \end{split}
\end{equation}
Using these individual amplitudes we find the squared matrix element of this process to be given by
\begin{align}
  \bar{\abs{\mathcal{M}}^2}= &\;\frac{1}{4}\sum_{s,t,q,r}\left( i\mathcal{M}_{W_1}+i\mathcal{M}_{W_2}+i\mathcal{M}_{W_3}\right) \left( i\mathcal{M}_{W_1}+i\mathcal{M}_{W_2}+i\mathcal{M}_{W_3}\right) ^* \nonumber\\
  \simeq &\;\frac{g^2\xi_i^2}{4\,\MP^2}\,\frac{1}{\left( (p-q)^2-m_W^2\right) ^2+m_W^2\Gamma_W^2}\Big[ U_{\tilde{W}\tilde{W}}^2\Tr\left[ \left( \slashed{k}_1+m_f\right) \gamma_\rho P_L \left( \slashed{k}_2-m_{f'}\right) P_R\,\gamma_\sigma\right] \nonumber\\
  &\qquad\times\Tr\left[ \left( \slashed{q}+m_\ell\right) P_R\left( q^\mu\gamma^\rho-g^{\mu\rho}(\slashed{q}-\slashed{p})\right) P_{\mu\nu}^+(p)\left( q^\nu\gamma^\sigma-g^{\nu\sigma}(\slashed{q}-\slashed{p})\right) P_L\right] \nonumber\\
  &\quad-2\,m_W\,U_{\tilde{W}\tilde{W}}\left( 1-\sqrt{2}\,c_\beta\RE U_{\tilde{H}_d^-\tilde{W}}\right) \left( g^{\nu\sigma}+\frac{q^\nu(p^\sigma-q^\sigma)}{m_W^2}\right) \nonumber\\
  &\qquad\times\Tr\left[ \left( \slashed{k}_1+m_f\right) \gamma_\rho P_L \left( \slashed{k}_2-m_{f'}\right) P_R\,\gamma_\sigma\right] \nonumber\\
  &\qquad\times\Tr\left[ \left( \slashed{q}+m_\ell\right) P_R\left( q^\mu\gamma^\rho-g^{\mu\rho}(\slashed{q}-\slashed{p})\right) P_{\mu\nu}^+(p)\,P_L\right] \nonumber\\
  &\quad+m_W^2\abs{1-\sqrt{2}\,c_\beta\,U_{\tilde{H}_d^-\tilde{W}}}^2\left( g_{\mu\rho}+\frac{q_\mu(p_\rho-q_\rho)}{m_W^2}\right) \left( g_{\nu\sigma}+\frac{q_\nu(p_\sigma-q_\sigma)}{m_W^2}\right) \\
  &\qquad\times\Tr\left[ \left( \slashed{k}_1+m_f\right) \gamma^\rho P_L\,\left( \slashed{k}_2-m_{f'}\right) P_R\,\gamma^\sigma\right] \Tr\left[ \left( \slashed{q}+m_\ell\right) P_R\,P^{+\,\mu\nu}(p)\,P_L\right] \Big] \nonumber\\
  = &-\frac{g^2\xi_i^2}{4\,\MP^2}\,\frac{1}{\left( (p-q)^2-m_W^2\right) ^2+m_W^2\Gamma_W^2}\Big[ U_{\tilde{W}\tilde{W}}^2\Tr\left[ \slashed{k}_1\gamma_\rho P_L\slashed{k}_2\gamma_\sigma\right] \nonumber\\
  &\qquad\times\Tr\left[ \slashed{q}P_R\left( q^\mu\gamma^\rho-g^{\mu\rho}(\slashed{q}-\slashed{p})\right) \slashed{p}\,\Phi_{\mu\nu}(p)\left( q^\nu\gamma^\sigma-g^{\nu\sigma}(\slashed{q}-\slashed{p})\right) \right] \nonumber\\
  &\quad-2\,m_{3/2}\,m_W\,U_{\tilde{W}\tilde{W}}\left( 1-\sqrt{2}\,c_\beta\RE U_{\tilde{H}_d^-\tilde{W}}\right) \left( g^{\nu\sigma}+\frac{q^\nu(p^\sigma-q^\sigma)}{m_W^2}\right) \nonumber\\
  &\qquad\times\Tr\left[ \slashed{k}_1\gamma_\rho P_L\slashed{k}_2\gamma_\sigma\right] \Tr\left[ \slashed{q}P_R\left( q^\mu\gamma^\rho-g^{\mu\rho}(\slashed{q}-\slashed{p})\right) \Phi_{\mu\nu}(p)\right] \nonumber\\
  &\quad+m_W^2\abs{1-\sqrt{2}\,c_\beta\,U_{\tilde{H}_d^-\tilde{W}}}^2\left( g_{\mu\rho}+\frac{q_\mu(p_\rho-q_\rho)}{m_W^2}\right) \left( g_{\nu\sigma}+\frac{q_\nu(p_\sigma-q_\sigma)}{m_W^2}\right) \nonumber\\
  &\qquad\times\Tr\left[ \slashed{k}_1\gamma^\rho P_L\,\slashed{k}_2\gamma^\sigma\right] \Tr\left[ \slashed{q}\,P_R\,\slashed{p}\,\Phi^{\mu\nu}(p)\right] \Big]\,. \nonumber
\end{align}
Also in this squared matrix element a number of traces of gamma matrices appear. The trace originating from the fermion current part of the amplitude is given by
\begin{equation}
  \Tr\left[ \slashed{k}_1\gamma^\rho P_L\,\slashed{k}_2\gamma^\sigma\right] =2\left( k_1^\rho k_2^\sigma+k_1^\sigma k_2^\rho-g^{\rho\sigma}\left( k_1\cdot k_2\right) -i\varepsilon^{\rho\sigma\delta\lambda}k_{1\,\delta}k_{2\,\lambda}\right) ,
\end{equation}
while the traces including the gravitino field are exactly the same as those given in equation~(\ref{gravitinotrace}). In these expressions we already replaced the squared four-momenta with the squared particle masses:
\begin{equation}
  p^2=m_{3/2}^2\,,\quad k_1^2=m_f^2\simeq 0\,,\quad k_2^2=m_{f'}^2\simeq 0\,,\quad q^2=m_\ell^2\simeq 0\,.
\end{equation}
Neglecting the masses of the final state particles the kinematics of this decay process is the same as in the previously discussed decay via photon/$Z$ boson exchange. Using equation~(\ref{threebodywidth}) we find the following differential decay width for a gravitino decaying into two fermions and a charged lepton:
\begin{equation}
 \begin{split}
  \frac{d\Gamma}{ds\,dt}\simeq &\;\frac{g^2\,\xi_i^2}{1536\,\pi^3\,m_{3/2}^5\,\MP^2\left( \left( s-m_W^2\right) ^2+m_W^2\,\Gamma_W^2\right) } \\
  &\quad\times\Big[\,s\,U_{\tilde{W}\tilde{W}}^2\left( m_{3/2}^2-t\right) \left( 3\,m_{3/2}^4-3\,m_{3/2}^2\left( s+t\right) +s\,t\right) \\
  &\qquad-2\,s\,m_{3/2}\,m_W\,U_{\tilde{W}\tilde{W}}\left( 1-\sqrt{2}\,c_\beta\RE U_{\tilde{H}_d^-\tilde{W}}\right) \\
  &\qquad\quad\times\left( 3\,m_{3/2}^4-m_{3/2}^2\left( 3\,s+4\,t\right) +t\left( 2\,s+t\right) \right) \\
  &\qquad+m_W^2\abs{1-\sqrt{2}\,c_\beta\,U_{\tilde{H}_d^-\tilde{W}}}^2 \\
  &\qquad\quad\times\left( m_{3/2}^4\,(3\,s+t)-m_{3/2}^2\left( 3\,s^2+4\,s\,t+t^2\right) +s\,t\,(s+t)\right) \Big]\,.
 \end{split}
\end{equation}
After integrating over the invariant mass $t$ we find as our final result the differential decay width
\begin{align}
  \frac{d\Gamma}{ds} &\simeq\frac{g^2\,\xi_i^2\,m_{3/2}^3\,\beta_s^2}{1536\,\pi^3\MP^2\left( \left( s-m_W^2\right) ^2+m_W^2\,\Gamma_W^2\right) }\bigg( s\,U_{\tilde{W}\tilde{W}}^2\,f_s \\
  &\qquad-\frac{8}{3}\,\frac{m_W}{m_{3/2}}\,s\,U_{\tilde{W}\tilde{W}}\left( 1-\sqrt{2}\,c_\beta\RE U_{\tilde{H}_d^-\tilde{W}}\right) j_s+\frac{1}{6}\,m_W^2\abs{1-\sqrt{2}\,c_\beta\,U_{\tilde{H}_d^-\tilde{W}}}^2h_s\bigg)\,, \nonumber
\end{align}
where the kinematic functions $\beta_s, f_s, j_s$ and $h_s$ are those given in equation~(\ref{kinematicfunc}). As in the case of gravitino decays via $Z$ propagator we can verify that above the threshold for $W^{\pm}$ production and using the narrow-width approximation this result coincides with the two-body decay result as given in equation~(\ref{gravitinowidths}), \textit{i.e.}:
\begin{equation}
  \Gamma_{\text{NWA}}\left( \psi_{3/2}\rightarrow {W^+}^*\ell_i^-\rightarrow f\,\bar{f}'\,\ell_i^-\right) =\Gamma\left( \psi_{3/2}\rightarrow W^+\ell_i^-\right) \times\BR\left( W^+\rightarrow f\,\bar{f}'\right) .
\end{equation}
Replacing the propagator according to equation~(\ref{narrowwidth}) we find
\begin{equation}
  \begin{split}
   &\Gamma_{\text{NWA}}\left( \psi_{3/2}\rightarrow {W^+}^*\ell_i^-\rightarrow f\,\bar{f}'\,\ell_i^-\right) =\int_0^{m_{3/2}^2}\!\!ds\,\frac{d\Gamma\left( \psi_{3/2}\rightarrow {W^+}^*\ell_i^-\rightarrow f\,\bar{f}'\,\ell_i^-\right)}{ds}\bigg|_{\text{NWA}} \\
   &\qquad=\frac{\xi_i^2\,m_{3/2}^3\,\beta_W^2}{32\,\pi\,\MP^2}\bigg( U_{\tilde{W}\tilde{W}}^2\,f_W-\frac{8}{3}\,\frac{m_W}{m_{3/2}}\,U_{\tilde{W}\tilde{W}}\left( 1-\sqrt{2}\,c_\beta\RE U_{\tilde{H}_d^-\tilde{W}}\right) j_W \\
   &\qquad\qquad\quad+\frac{1}{6}\abs{1-\sqrt{2}\,c_\beta\,U_{\tilde{H}_d^-\tilde{W}}}^2h_W\bigg) \times\frac{\Gamma\left( W^+\rightarrow f\,\bar{f}'\right) }{\Gamma_W}\,,
  \end{split}
\end{equation}
which is exactly the equivalence we wanted to verify. In the last expression we used the partial width of the $W^\pm$ boson decay into fermions~\cite{Nakamura:2010zzi}:
\begin{equation}
  \Gamma\left( W^+\rightarrow f\,\bar{f}'\right) =\frac{\sqrt{2}\,G_F\,m_W^3}{12\,\pi}=\frac{g^2\,m_W}{48\,\pi}\,.
\end{equation}

\subsection[\texorpdfstring{$\psi_{3/2}\rightarrow h^*\,\nu\rightarrow f\,\bar{f}\,\nu$}{\textpsi\ --> h* \textnu --> f f \textnu}]{\boldmath$\psi_{3/2}\rightarrow h^*\,\nu\rightarrow f\,\bar{f}\,\nu$}

At tree level there are two Feynman diagrams contributing to the decay of a gravitino into a fermion-antifermion pair and a neutrino via an intermediate lightest Higgs boson:
\begin{equation*}
 \parbox{5.5cm}{
  \begin{picture}(172,162) (92,-96)
    \SetWidth{0.5}
    \SetColor{Black}
    \Line(201,5)(205,-7)\Line(197,-3)(209,1)
    \Line[arrow,arrowpos=0.5,arrowlength=3.75,arrowwidth=1.5,arrowinset=0.2](194,-17)(232,-84)
    \Line[dash,dashsize=4,arrow,arrowpos=0.75,arrowlength=3.75,arrowwidth=1.5,arrowinset=0.2](213,17)(194,-17)
    \Line[arrow,arrowpos=0.5,arrowlength=3.75,arrowwidth=1.5,arrowinset=0.2](213,17)(232,51)
    \Line[arrow,arrowpos=0.5,arrowlength=3.75,arrowwidth=1.5,arrowinset=0.2](233,-17)(214,17)
    \Line[double,sep=4](117,-17)(194,-17)
    \Vertex(194,-17){2.5}
    \Vertex(213,17){2.5}
    \Line[arrow,arrowpos=1,arrowlength=3.75,arrowwidth=1.5,arrowinset=0.2](207,28)(220,50)
    \Line[arrow,arrowpos=1,arrowlength=3.75,arrowwidth=1.5,arrowinset=0.2](226,17)(239,-5)
    \Line[arrow,arrowpos=1,arrowlength=3.75,arrowwidth=1.5,arrowinset=0.2](217,-34)(230,-56)
    \Line[arrow,arrowpos=1,arrowlength=3.75,arrowwidth=1.5,arrowinset=0.2](143,-6)(169,-6)
    \Text(104,-17)[]{\normalsize{\Black{$\psi_{3/2}$}}}
    \Text(156,-4)[b]{\normalsize{\Black{$p$}}}
    \Text(234,6)[lb]{\normalsize{\Black{$k_2$}}}
    \Text(213,39)[rb]{\normalsize{\Black{$k_1$}}}
    \Text(226,-46)[lb]{\normalsize{\Black{$q$}}}
    \Text(237,58)[]{\normalsize{\Black{$f$}}}
    \Text(237,-24)[]{\normalsize{\Black{$\bar{f}$}}}
    \Text(198,15)[]{\normalsize{\Black{$h$}}}
    \Text(212,-15)[]{\normalsize{\Black{$\tilde{\nu}_i^*$}}}
    \Text(237,-91)[]{\normalsize{\Black{$\nu_i$}}}
    \Arc[arrow,arrowpos=0.5,arrowlength=3.75,arrowwidth=1.5,arrowinset=0.2,clock](133.852,-120.786)(81.877,92.697,26.698)
  \end{picture}
 }\qquad
 +\qquad
 \parbox{5.5cm}{
  \begin{picture}(172,162) (92,-96)
    \SetWidth{0.5}
    \SetColor{Black}
    \Line(211,-55.999)(215,-44.001)\Line(207.001,-48)(218.999,-52)
    \Line[arrow,arrowpos=0.5,arrowlength=3.75,arrowwidth=1.5,arrowinset=0.2](213,-51)(232,-84)
    \Line[dash,dashsize=4](194,-17)(213,17)
    \Line[arrow,arrowpos=0.5,arrowlength=3.75,arrowwidth=1.5,arrowinset=0.2](213,17)(232,51)
    \Line[arrow,arrowpos=0.5,arrowlength=3.75,arrowwidth=1.5,arrowinset=0.2](233,-17)(214,17)
    \Line[double,sep=4](117,-17)(194,-17)
    \Line(194,-17)(213,-51)
    \Vertex(194,-17){2.5}
    \Vertex(213,17){2.5}
    \Line[arrow,arrowpos=1,arrowlength=3.75,arrowwidth=1.5,arrowinset=0.2](207,28)(220,50)
    \Line[arrow,arrowpos=1,arrowlength=3.75,arrowwidth=1.5,arrowinset=0.2](226,17)(239,-5)
    \Line[arrow,arrowpos=1,arrowlength=3.75,arrowwidth=1.5,arrowinset=0.2](226,-50)(239,-72)
    \Line[arrow,arrowpos=1,arrowlength=3.75,arrowwidth=1.5,arrowinset=0.2](143,-6)(169,-6)
    \Text(104,-17)[]{\normalsize{\Black{$\psi_{3/2}$}}}
    \Text(156,-4)[b]{\normalsize{\Black{$p$}}}
    \Text(234,6)[lb]{\normalsize{\Black{$k_2$}}}
    \Text(213,39)[rb]{\normalsize{\Black{$k_1$}}}
    \Text(235,-62)[lb]{\normalsize{\Black{$q$}}}
    \Text(237,58)[]{\normalsize{\Black{$f$}}}
    \Text(237,-24)[]{\normalsize{\Black{$\bar{f}$}}}
    \Text(193,6)[]{\normalsize{\Black{$h$}}}
    \Text(193,-39)[]{\normalsize{\Black{$\tilde{h}$}}}
    \Text(237,-91)[]{\normalsize{\Black{$\nu_i$}}}
    \Arc[arrow,arrowpos=0.5,arrowlength=3.75,arrowwidth=1.5,arrowinset=0.2,clock](133.852,-120.786)(81.877,92.697,26.698)
  \end{picture}
 }.
\end{equation*}
The first diagram involves a mixing between the sneutrino and the lightest Higgs boson and the corresponding amplitude reads
\begin{align}
  i\mathcal{M}_{h_1} &=\bar{u}^t(q)\,P_R\,\frac{i}{\sqrt{2}\,\MP}\,\gamma^{\mu}\left( \slashed{p}-\slashed{q}\right) \psi_{\mu}^{+\,s}(p)\,\frac{i}{\left( p-q\right) ^2-m_{\tilde{\nu}_i}^2+i\,m_{\tilde{\nu}_i}\Gamma_{\tilde{\nu}_i}} \nonumber\\
  &\quad\qquad\times\frac{i\,\xi_i}{\sqrt{2}}\left( m_{\tilde{\nu}_i}^2+\frac{1}{2}\,m_Z^2\cos2\,\beta\right) \frac{i}{\left( p-q\right) ^2-m_h^2+i\,m_h\Gamma_h}\,\frac{i\,m_f}{\sqrt{2}\,v}\,\bar{u}^q(k_1)\,v^r(k_2) \nonumber\\
  &\simeq\frac{i\,m_f\,\xi_i}{v\sqrt{2}\,\MP\left( \left( p-q\right) ^2-m_h^2+i\,m_h\Gamma_h\right) }\,\frac{m_{\tilde{\nu}_i}^2+\frac{1}{2}\,m_Z^2\cos2\,\beta}{\left( p-q\right) ^2-m_{\tilde{\nu}_i}^2} \\
  &\quad\qquad\times\bar{u}^t(q)\,P_R\,q^{\mu}\,\psi_{\mu}^{+\,s}(p)\,\bar{u}^q(k_1)\,v^r(k_2)\,, \nonumber
\end{align}
where we used the fact that $p^\mu\psi_{\mu}^{+\,s}(p)=\gamma^\mu\psi_{\mu}^{+\,s}(p)=0$ (see equation~(\ref{modeRarita})) and thus
\begin{equation}
  \gamma^\mu\left( \slashed{p}-\slashed{q}\right) \psi_{\mu}^{+\,s}(p)=\left( 2\,(p^\mu-q^\mu)+(\slashed{p}-\slashed{q})\,\gamma^\mu\right) \psi_{\mu}^{+\,s}(p)=-2\,q^\mu\,\psi_{\mu}^{+\,s}(p)\,.
\end{equation}
In addition, we assumed that the width of the sneutrino is negligible compared to its mass. The second diagram involves a mixing between the higgsino partner of the lightest Higgs boson and the neutrino. The amplitude for this diagram is given by
\begin{align}
  i\mathcal{M}_{h_2} &=\bar{u}^t(q)\,P_R\,\sqrt{2}\left\lbrace \sin\beta\,N_{\nu_i\,\tilde{H}_u^0}^{7*}+\cos\beta\,N_{\nu_i\,\tilde{H}_d^0}^{7*}\right\rbrace \frac{i}{\sqrt{2}\,\MP}\,\gamma^{\mu}\left( \slashed{p}-\slashed{q}\right) \psi_{\mu}^{+\,s}(p)\,\nonumber\\
  &\quad\qquad\times\frac{i}{\left( p-q\right) ^2-m_h^2+i\,m_h\Gamma_h}\,\frac{i\,m_f}{\sqrt{2}\,v}\,\bar{u}^q(k_1)\,v^r(k_2) \nonumber\\
  &\simeq\frac{i\,m_f\,\xi_i}{v\sqrt{2}\,\MP\left( \left( p-q\right) ^2-m_h^2+i\,m_h\Gamma_h\right) }\,\left( 2\sin{\beta}\,U^*_{\tilde{H}_u^0\tilde{Z}}+2\cos{\beta}\,U^*_{\tilde{H}_d^0\tilde{Z}}\right) \\
  &\quad\qquad\times\bar{u}^t(q)\,P_R\,q^{\mu}\,\psi_{\mu}^{+\,s}(p)\,\bar{u}^q(k_1)\,v^r(k_2)\,. \nonumber
\end{align}
The squared matrix element of this process is then given by
\begin{equation}
 \begin{split}
  \bar{\abs{\mathcal{M}}^2} &=\frac{1}{4}\sum_{s,t,q,r}\left( i\mathcal{M}_{h_1}+i\mathcal{M}_{h_2}\right) \left( i\mathcal{M}_{h_1}+i\mathcal{M}_{h_2}\right) ^* \\
  &=-\frac{m_f^2\,\xi_i^2}{8\,v^2\,\MP^2}\,\frac{1}{\left( \left( p-q\right) ^2-m_h^2\right) ^2+m_h^2\Gamma_h^2} \\
  &\qquad\quad\times\abs{\frac{m_{\tilde{\nu}_i}^2+\frac{1}{2}\,m_Z^2\cos2\,\beta}{s-m_{\tilde{\nu}_i}^2}+2\sin\beta\,U_{\tilde{H}_u^0\tilde{Z}}+2\cos\beta\,U_{\tilde{H}_d^0\tilde{Z}}}^2 \\
  &\qquad\quad\times q^\mu\,q^\nu\Tr\left[ \slashed{q}\,P_R\,\slashed{p}\,\Phi_{\mu\nu}(p)\right] \Tr\left[ \left( \slashed{k}_1+m_f\right) \left( \slashed{k}_2-m_f\right) \right] .
 \end{split}
\end{equation}
The traces appearing in this expression are given by
\begin{equation}
 \begin{split}
  q^\mu\,q^\nu\Tr\left[ \slashed{q}\,P_R\,\slashed{p}\,\Phi_{\mu\nu}(p)\right] &=-\frac{4}{3\,m_{3/2}^2}\left( p\cdot q\right) ^3, \\
  \Tr\left[ \left( \slashed{k}_1+m_f\right) \left( \slashed{k}_2-m_f\right) \right] &=4\left( \left( k_1\cdot k_2\right) -m_f^2\right) ,
 \end{split}
\end{equation}
where we already replaced the squared four-momenta with the squared particle masses:
\begin{equation}
  p^2=m_{3/2}^2\,,\quad k_1^2=k_2^2=m_f^2\simeq 0\,,\quad q^2=m_\nu^2\simeq 0\,.
\end{equation}
The only scalar products appearing in the squared matrix element are given by
\begin{equation}
  \left( p\cdot q\right) =\frac{m_{3/2}^2-s}{2}\qquad\text{and}\qquad\left( k_1\cdot k_2\right) =\frac{s}{2}\,,
\end{equation}
and thus the differential decay width reads
\begin{equation}
  \frac{d\Gamma}{ds\,dt}\simeq\frac{\xi_i^2\,m_{3/2}\,m_f^2\,\beta_s^3\,s\abs{\frac{m_{\tilde{\nu}_i}^2+\frac{1}{2}\,m_Z^2\cos2\,\beta}{s-m_{\tilde{\nu}_i}^2}+2\sin{\beta}\,U_{\tilde{H}_u^0\tilde{Z}}+2\cos{\beta}\,U_{\tilde{H}_d^0\tilde{Z}}}^2}{6144\,\pi^3\MP^2\,v^2\left( \left( s-m_h^2\right) ^2+m_h^2\,\Gamma_h^2\right) }\,.
\end{equation}
The differential decay width is independent of the invariant mass $t$ and thus the integration over $t$ just results in an additional factor of $(m_{3/2}^2-s)=m_{3/2}^2\,\beta_s$ leading to
\begin{equation}
  \frac{d\Gamma}{ds}\simeq\frac{\xi_i^2\,m_{3/2}^3\,m_f^2\,\beta_s^4\,s\abs{\frac{m_{\tilde{\nu}_i}^2+\frac{1}{2}\,m_Z^2\cos2\,\beta}{s-m_{\tilde{\nu}_i}^2}+2\sin{\beta}\,U_{\tilde{H}_u^0\tilde{Z}}+2\cos{\beta}\,U_{\tilde{H}_d^0\tilde{Z}}}^2}{6144\,\pi^3\MP^2\,v^2\left( \left( s-m_h^2\right) ^2+m_h^2\,\Gamma_h^2\right) }\,.
\end{equation}
Similar to the other decay channels we can verify that above the threshold for $h$ production and using the narrow-width approximation this result coincides with the two-body decay result as given in equation~(\ref{gravitinowidths}), \textit{i.e.}:
\begin{equation}
  \Gamma_{\text{NWA}}\left( \psi_{3/2}\rightarrow h^*\nu_i\rightarrow f\,\bar{f}\,\nu_i\right) =\Gamma\left( \psi_{3/2}\rightarrow h\,\nu_i\right) \times\BR\left( h\rightarrow f\,\bar{f}\right) .
\end{equation}
Replacing the propagator according to equation~(\ref{narrowwidth}) we find
\begin{align}
  &\Gamma_{\text{NWA}}\left( \psi_{3/2}\rightarrow h^*\nu_i\rightarrow f\,\bar{f}\,\nu_i\right) =\int_0^{m_{3/2}^2}\!\!ds\,\frac{d\Gamma\left( \psi_{3/2}\rightarrow h^*\nu_i\rightarrow f\,\bar{f}\,\nu_i\right)}{ds}\bigg|_{\text{NWA}} \\
  &\qquad=\frac{\xi_i^2\,m_{3/2}^3\,\beta_h^4}{384\,\pi\,\MP^2}\abs{\frac{m_{\tilde{\nu}_i}^2+\frac{1}{2}\,m_Z^2\cos2\,\beta}{m_h^2-m_{\tilde{\nu}_i}^2}+2\sin{\beta}\,U_{\tilde{H}_u^0\tilde{Z}}+2\cos{\beta}\,U_{\tilde{H}_d^0\tilde{Z}}}^2\times\frac{\Gamma\left( h\rightarrow f\,\bar{f}\right) }{\Gamma_h}\,, \nonumber
\end{align}
which is exactly the equivalence we wanted to verify. In the last expression we used the partial width of the Standard Model Higgs boson decay into a pair of fermions~\cite{Nakamura:2010zzi}:
\begin{equation}
  \Gamma\left( h\rightarrow f\,\bar{f}\right) =\frac{G_F\,m_f^2\,m_h}{4\,\pi\sqrt{2}}=\frac{m_f^2\,m_h}{16\,\pi\,v^2}\,.
\end{equation}

\chapter{Calculation of Gravitino--Nucleon Cross Sections}
\label{gravitinoscattering}

In this appendix we compute the scattering cross sections of the LSP gravitino with target nucleons in the framework of bilinear $R$-parity breaking. We separately consider three cases: the exchange of a lightest Higgs boson, the exchange of a $Z$ boson and the exchange of a photon. For a discussion of this topic see Chapter~\ref{gravitinoDD}.

\section[Inelastic Gravitino--Nucleon Scattering via Higgs Exchange]{Inelastic Gravitino--Nucleon Scattering via Higgs Exchange}

At tree level there are two Feynman diagrams contributing to the inelastic gravitino scattering off a nucleon via the exchange of the lightest Higgs boson:
\begin{equation*}
 \parbox{6.3cm}{
  \begin{picture}(183,173) (107,-22)
    \SetWidth{0.5}
    \SetColor{Black}
    \Line[dash,dashsize=4,arrow,arrowpos=0.5,arrowlength=3.75,arrowwidth=1.5,arrowinset=0.2](198,62)(198,96)
    \Line[dash,dashsize=4](198,62)(198,29)
    \Line[arrow,arrowpos=0.5,arrowlength=3.75,arrowwidth=1.5,arrowinset=0.2](198,96)(256,130)
    \Line[arrow,arrowpos=0.5,arrowlength=3.75,arrowwidth=1.5,arrowinset=0.2](198,29)(256,-5)
    \Line[arrow,arrowpos=0.5,arrowlength=3.75,arrowwidth=1.5,arrowinset=0.2](140,-5)(198,29)
    \Line[double,sep=4](140,130)(198,96)
    \Line[arrow,arrowpos=1,arrowlength=3.75,arrowwidth=1.5,arrowinset=0.2](223,96)(242,107)
    \Line[arrow,arrowpos=1,arrowlength=3.75,arrowwidth=1.5,arrowinset=0.2](153,107)(172,96)
    \Line[arrow,arrowpos=1,arrowlength=3.75,arrowwidth=1.5,arrowinset=0.2](153,18)(172,29)
    \Line[arrow,arrowpos=1,arrowlength=3.75,arrowwidth=1.5,arrowinset=0.2](223,29)(242,18)
    \Vertex(198,96){2.5}
    \Vertex(198,29){2.5}
    \Line(193.5,67)(202.5,58)\Line(202.5,67)(193.5,58)
    \Text(138,131)[rb]{\normalsize{\Black{$\psi_{3/2}$}}}
    \Text(257,131)[lb]{\normalsize{\Black{$\nu_i$}}}
    \Line(198,96)(227,113)
    \Text(206,46)[l]{\normalsize{\Black{$h$}}}
    \Text(206,79)[l]{\normalsize{\Black{$\tilde{\nu}_i$}}}
    \Text(138,-6)[rt]{\normalsize{\Black{$N$}}}
    \Text(257,-6)[lt]{\normalsize{\Black{$N$}}}
    \Text(164,99)[rt]{\normalsize{\Black{$p$}}}
    \Text(233,102)[lt]{\normalsize{\Black{$p'$}}}
    \Text(164,25)[rb]{\normalsize{\Black{$k$}}}
    \Text(233,25)[lb]{\normalsize{\Black{$k'$}}}
    \Arc[arrow,arrowpos=0.5,arrowlength=3.75,arrowwidth=1.5,arrowinset=0.2](197.5,185.75)(67.752,-124.628,-55.372)
  \end{picture}
 }
 \quad+\quad
 \parbox{6.3cm}{
  \begin{picture}(183,173) (107,-22)
    \SetWidth{0.5}
    \SetColor{Black}
    \Line[dash,dashsize=4](198,96)(198,29)
    \Line[arrow,arrowpos=0.5,arrowlength=3.75,arrowwidth=1.5,arrowinset=0.2](227,113)(256,130)
    \Line(198,96)(227,113)
    \Line[arrow,arrowpos=0.5,arrowlength=3.75,arrowwidth=1.5,arrowinset=0.2](198,29)(256,-5)
    \Line[arrow,arrowpos=0.5,arrowlength=3.75,arrowwidth=1.5,arrowinset=0.2](140,-5)(198,29)
    \Line[double,sep=4](140,130)(198,96)
    \Line[arrow,arrowpos=1,arrowlength=3.75,arrowwidth=1.5,arrowinset=0.2](238,104)(257,115)
    \Line[arrow,arrowpos=1,arrowlength=3.75,arrowwidth=1.5,arrowinset=0.2](153,107)(172,96)
    \Line[arrow,arrowpos=1,arrowlength=3.75,arrowwidth=1.5,arrowinset=0.2](153,18)(172,29)
    \Line[arrow,arrowpos=1,arrowlength=3.75,arrowwidth=1.5,arrowinset=0.2](223,29)(242,18)
    \Vertex(198,96){2.5}
    \Vertex(198,29){2.5}
    \Line(225,107.001)(229,118.999)\Line(221.001,115)(232.999,111)
    \Text(214,102)[lt]{\normalsize{\Black{$\tilde{h}$}}}
    \Text(138,131)[rb]{\normalsize{\Black{$\psi_{3/2}$}}}
    \Text(257,131)[lb]{\normalsize{\Black{$\nu_i$}}}
    \Text(206,62)[l]{\normalsize{\Black{$h$}}}
    \Text(138,-6)[rt]{\normalsize{\Black{$N$}}}
    \Text(257,-6)[lt]{\normalsize{\Black{$N$}}}
    \Text(164,99)[rt]{\normalsize{\Black{$p$}}}
    \Text(247,110)[lt]{\normalsize{\Black{$p'$}}}
    \Text(164,25)[rb]{\normalsize{\Black{$k$}}}
    \Text(233,25)[lb]{\normalsize{\Black{$k'$}}}
    \Arc[arrow,arrowpos=0.5,arrowlength=3.75,arrowwidth=1.5,arrowinset=0.2](197.5,185.75)(67.752,-124.628,-55.372)
  \end{picture}
 }.
\end{equation*}
Similar to the case of gravitino three-body decay the amplitude is given by
\begin{align}
  i\mathcal{M}_{h_1} &=\sum_q\bar{u}^t(p')\,P_R\,\frac{i}{\sqrt{2}\,\MP}\,\gamma^{\mu}\left( \slashed{p}-\slashed{p}'\right) \psi_{\mu}^{+\,s}(p)\,\frac{i}{\left( p-p'\right) ^2-m_{\tilde{\nu}_i}^2+i\,m_{\tilde{\nu}_i}\Gamma_{\tilde{\nu}_i}} \\
  &\quad\qquad\times\frac{i\,\xi_i}{\sqrt{2}}\left( m_{\tilde{\nu}_i}^2+\frac{1}{2}\,m_Z^2\cos2\,\beta\right) \frac{i}{\left( p-p'\right) ^2-m_h^2+i\,m_h\Gamma_h}\,\frac{i\,m_q}{\sqrt{2}\,v}\,\bar{u}^q(k')\,u^r(k) \nonumber\\
  &\simeq-\sum_q\frac{i\,\xi_i\,m_N\,f_{T_q}^N}{v\sqrt{2}\,m_h^2\,\MP}\left( 1+\frac{1}{2}\,\frac{m_Z^2}{m_{\tilde{\nu}_i}^2}\cos2\,\beta\right) \bar{u}^t(p')\,P_R\,p'^{\mu}\,\psi_{\mu}^{+\,s}(p)\,\bar{u}^q(k')\,u^r(k)\,, \nonumber
\end{align}
where we need to sum over all quarks and their contribution to the nucleon mass. In addition, we only consider the case where $(p-p')^2\ll m_h^2,\,m_{\tilde{\nu}_i}^2$ so that the propagator reduces to a contact interaction. The amplitude for the second diagram is given by
\begin{align}
  i\mathcal{M}_{h_2} &=\sum_q\bar{u}^t(p')\,P_R\,\sqrt{2}\left\lbrace \sin\beta\,N_{\nu_i\,\tilde{H}_u^0}^{7*}+\cos\beta\,N_{\nu_i\,\tilde{H}_d^0}^{7*}\right\rbrace \frac{i}{\sqrt{2}\,\MP}\,\gamma^{\mu}\left( \slashed{p}-\slashed{p}'\right) \psi_{\mu}^{+\,s}(p)\,\nonumber\\
  &\quad\qquad\times\frac{i}{\left( p-p'\right) ^2-m_h^2+i\,m_h\Gamma_h}\,\frac{i\,m_q}{\sqrt{2}\,v}\,\bar{u}^q(k')\,u^r(k) \\
  &\simeq\sum_q\frac{i\,\xi_i\,m_N\,f_{T_q}^N}{v\sqrt{2}\,m_h^2\,\MP}\,\left( 2\sin{\beta}\,U^*_{\tilde{H}_u^0\tilde{Z}}+2\cos{\beta}\,U^*_{\tilde{H}_d^0\tilde{Z}}\right) \bar{u}^t(p')\,P_R\,p'^{\mu}\,\psi_{\mu}^{+\,s}(p)\,\bar{u}^q(k')\,u^r(k)\,. \nonumber
\end{align}
The squared matrix element of this process is then given by
\begin{align}
  \bar{\abs{\mathcal{M}}^2} &=\frac{1}{8}\sum_{s,t,q,r}\left( i\mathcal{M}_{h_1}+i\mathcal{M}_{h_2}\right) \left( i\mathcal{M}_{h_1}+i\mathcal{M}_{h_2}\right) ^* \\
  &=-\frac{\xi_i^2\,m_N^2\left( \sum_qf_{T_q}^N\right) ^2}{16\,v^2\,m_h^4\,\MP^2}\abs{1+\frac{1}{2}\,\frac{m_Z^2}{m_{\tilde{\nu}_i}^2}\cos2\,\beta-2\sin\beta\,U_{\tilde{H}_u^0\tilde{Z}}-2\cos\beta\,U_{\tilde{H}_d^0\tilde{Z}}}^2 \nonumber\\
  &\qquad\quad\times p'^\mu\,p'^\nu\Tr\left[ \slashed{p}'P_R\,\slashed{p}\,\Phi_{\mu\nu}(p)\right] \Tr\left[ \left( \slashed{k}'+m_q\right) \left( \slashed{k}+m_q\right) \right] . \nonumber
\end{align}
Neglecting the masses of the individual quarks, the traces appearing in this expression are given by
\begin{equation}
 \begin{split}
  p'^\mu\,p'^\nu\Tr\left[ \slashed{p}'P_R\,\slashed{p}\,\Phi_{\mu\nu}(p)\right] &=-\frac{4}{3\,m_{3/2}^2}\left( p\cdot p'\right) ^3, \\
  \Tr\left[ \left( \slashed{k}'_1+m_q\right) \left( \slashed{k}+m_q\right) \right] &\simeq4\left( k\cdot k'\right) ,
 \end{split}
\end{equation}
where we already replaced the squared four-momenta with the squared particle masses:
\begin{equation}
  p^2=m_{3/2}^2\,,\quad p'^2=m_\nu^2\simeq 0\,,\quad k^2=k'^2=m_N^2\,.
\end{equation}
The only scalar products appearing in the squared matrix element are given by (\textit{cf.} equation~(\ref{scatteringscalarproduct}))
\begin{equation}
  \left( p\cdot p'\right) =\frac{m_{3/2}^2-t}{2}\qquad\text{and}\qquad\left( k\cdot k'\right) =m_N^2-\frac{t}{2}\,.
\end{equation}
Using equation~(\ref{crosssec}) we can now write down the differential cross section:
\begin{align}
  \frac{d\sigma}{dt} &=\frac{\bar{\abs{\mathcal{M}}^2}}{64\,\pi\,s\abs{\vec{p}_{1cm}}^2}=\frac{\bar{\abs{\mathcal{M}}^2}}{64\,\pi\abs{\vec{p}_{1lab}}^2m_N^2}\simeq\frac{\bar{\abs{\mathcal{M}}^2}}{64\,\pi\,m_{3/2}^2\,m_N^2\abs{\vec{v}}^2} \nonumber\\
  &=\frac{\xi_i^2\,m_{3/2}^2\,m_N^2\left( \sum_qf_{T_q}^N\right) ^2}{1536\,\pi\,v^2\,m_h^4\,\MP^2\abs{\vec{v}}^2}\abs{1+\frac{1}{2}\,\frac{m_Z^2}{m_{\tilde{\nu}_i}^2}\cos2\,\beta-2\sin\beta\,U_{\tilde{H}_u^0\tilde{Z}}-2\cos\beta\,U_{\tilde{H}_d^0\tilde{Z}}}^2 \nonumber\\
  &\qquad\qquad\times\left( 1-\frac{t}{m_{3/2}^2}\right) ^3\left( 1-\frac{t}{2\,m_N^2}\right) .
\end{align}

\section[\texorpdfstring{Inelastic Gravitino--Nucleon Scattering via $Z$ Exchange}{Inelastic Gravitino-Nucleon Scattering via Z Exchange}]{Inelastic Gravitino--Nucleon Scattering via \boldmath{$Z$} Exchange}

At tree level there are four diagrams contributing to the inelastic gravitino--nucleon scattering via the exchange of a $Z$ boson:
\begin{equation*}
 \parbox{6.3cm}{
  \begin{picture}(183,173) (107,-22)
    \SetWidth{0.5}
    \SetColor{Black}
    \Photon(198,96)(198,29){-6}{3}
    \Line[arrow,arrowpos=0.5,arrowlength=3.75,arrowwidth=1.5,arrowinset=0.2](227,113)(256,130)
    \Line[arrow,arrowpos=0.5,arrowlength=3.75,arrowwidth=1.5,arrowinset=0.2](198,29)(256,-5)
    \Line[arrow,arrowpos=0.5,arrowlength=3.75,arrowwidth=1.5,arrowinset=0.2](140,-5)(198,29)
    \Line[double,sep=4](140,130)(198,96)
    \Line[arrow,arrowpos=1,arrowlength=3.75,arrowwidth=1.5,arrowinset=0.2](238,104)(257,115)
    \Line[arrow,arrowpos=1,arrowlength=3.75,arrowwidth=1.5,arrowinset=0.2](153,107)(172,96)
    \Line[arrow,arrowpos=1,arrowlength=3.75,arrowwidth=1.5,arrowinset=0.2](153,18)(172,29)
    \Line[arrow,arrowpos=1,arrowlength=3.75,arrowwidth=1.5,arrowinset=0.2](223,29)(242,18)
    \Vertex(198,96){2.5}
    \Vertex(198,29){2.5}
    \Text(138,131)[rb]{\normalsize{\Black{$\psi_{3/2}$}}}
    \Text(257,131)[lb]{\normalsize{\Black{$\nu_i$}}}
    \Line(198,96)(227,113)
    \Photon(198,96)(227,113){-6}{2}
    \Line(225,107.001)(229,118.999)\Line(221.001,115)(232.999,111)
    \Text(216,101)[lt]{\normalsize{\Black{$\tilde{Z}$}}}
    \Text(206,62)[l]{\normalsize{\Black{$Z$}}}
    \Text(138,-6)[rt]{\normalsize{\Black{$N$}}}
    \Text(257,-6)[lt]{\normalsize{\Black{$N$}}}
    \Text(164,99)[rt]{\normalsize{\Black{$p$}}}
    \Text(247,110)[lt]{\normalsize{\Black{$p'$}}}
    \Text(164,25)[rb]{\normalsize{\Black{$k$}}}
    \Text(233,25)[lb]{\normalsize{\Black{$k'$}}}
    \Arc[arrow,arrowpos=0.5,arrowlength=3.75,arrowwidth=1.5,arrowinset=0.2](197.5,185.75)(67.752,-124.628,-55.372)
  \end{picture}
 }
 \quad+\quad
 \parbox{6.3cm}{
  \begin{picture}(183,173) (107,-22)
    \SetWidth{0.5}
    \SetColor{Black}
    \Photon(198,96)(198,29){-6}{3}
    \Line[arrow,arrowpos=0.5,arrowlength=3.75,arrowwidth=1.5,arrowinset=0.2](198,96)(256,130)
    \Line[arrow,arrowpos=0.5,arrowlength=3.75,arrowwidth=1.5,arrowinset=0.2](198,29)(256,-5)
    \Line[arrow,arrowpos=0.5,arrowlength=3.75,arrowwidth=1.5,arrowinset=0.2](140,-5)(198,29)
    \Line[double,sep=4](140,130)(198,96)
    \Line[arrow,arrowpos=1,arrowlength=3.75,arrowwidth=1.5,arrowinset=0.2](223,96)(242,107)
    \Line[arrow,arrowpos=1,arrowlength=3.75,arrowwidth=1.5,arrowinset=0.2](153,107)(172,96)
    \Line[arrow,arrowpos=1,arrowlength=3.75,arrowwidth=1.5,arrowinset=0.2](153,18)(172,29)
    \Line[arrow,arrowpos=1,arrowlength=3.75,arrowwidth=1.5,arrowinset=0.2](223,29)(242,18)
    \Vertex(198,96){2.5}
    \Vertex(198,29){2.5}
    \Text(138,131)[rb]{\normalsize{\Black{$\psi_{3/2}$}}}
    \Text(257,131)[lb]{\normalsize{\Black{$\nu_i$}}}
    \Text(206,62)[l]{\normalsize{\Black{$Z$}}}
    \Text(138,-6)[rt]{\normalsize{\Black{$N$}}}
    \Text(257,-6)[lt]{\normalsize{\Black{$N$}}}
    \Text(164,99)[rt]{\normalsize{\Black{$p$}}}
    \Text(233,102)[lt]{\normalsize{\Black{$p'$}}}
    \Text(164,25)[rb]{\normalsize{\Black{$k$}}}
    \Text(233,25)[lb]{\normalsize{\Black{$k'$}}}
    \Arc[arrow,arrowpos=0.5,arrowlength=3.75,arrowwidth=1.5,arrowinset=0.2](197.5,185.75)(67.752,-124.628,-55.372)
    \Line[dash,dashsize=4,arrow,arrowpos=0.5,arrowlength=3.75,arrowwidth=1.5,arrowinset=0.2](179,85)(198,96)
    \Text(178,86)[rt]{\normalsize{\Black{$v_i$}}}
  \end{picture}
 }
\end{equation*}
\begin{equation*}
 +\quad
 \parbox{6.3cm}{
  \begin{picture}(183,173) (107,-22)
    \SetWidth{0.5}
    \SetColor{Black}
    \Photon(198,96)(198,29){-6}{3}
    \Line[arrow,arrowpos=0.5,arrowlength=3.75,arrowwidth=1.5,arrowinset=0.2](227,113)(256,130)
    \Line(198,96)(227,113)
    \Line[arrow,arrowpos=0.5,arrowlength=3.75,arrowwidth=1.5,arrowinset=0.2](198,29)(256,-5)
    \Line[arrow,arrowpos=0.5,arrowlength=3.75,arrowwidth=1.5,arrowinset=0.2](140,-5)(198,29)
    \Line[double,sep=4](140,130)(198,96)
    \Line[arrow,arrowpos=1,arrowlength=3.75,arrowwidth=1.5,arrowinset=0.2](238,104)(257,115)
    \Line[arrow,arrowpos=1,arrowlength=3.75,arrowwidth=1.5,arrowinset=0.2](153,107)(172,96)
    \Line[arrow,arrowpos=1,arrowlength=3.75,arrowwidth=1.5,arrowinset=0.2](153,18)(172,29)
    \Line[arrow,arrowpos=1,arrowlength=3.75,arrowwidth=1.5,arrowinset=0.2](223,29)(242,18)
    \Vertex(198,96){2.5}
    \Vertex(198,29){2.5}
    \Line(225,107.001)(229,118.999)\Line(221.001,115)(232.999,111)
    \Text(214,101)[lt]{\normalsize{\Black{$\tilde{H}_{u,\,d}^0$}}}
    \Text(138,131)[rb]{\normalsize{\Black{$\psi_{3/2}$}}}
    \Text(257,131)[lb]{\normalsize{\Black{$\nu_i$}}}
    \Text(206,62)[l]{\normalsize{\Black{$Z$}}}
    \Text(138,-6)[rt]{\normalsize{\Black{$N$}}}
    \Text(257,-6)[lt]{\normalsize{\Black{$N$}}}
    \Text(164,99)[rt]{\normalsize{\Black{$p$}}}
    \Text(247,110)[lt]{\normalsize{\Black{$p'$}}}
    \Text(164,25)[rb]{\normalsize{\Black{$k$}}}
    \Text(233,25)[lb]{\normalsize{\Black{$k'$}}}
    \Arc[arrow,arrowpos=0.5,arrowlength=3.75,arrowwidth=1.5,arrowinset=0.2](197.5,185.75)(67.752,-124.628,-55.372)
    \Line[dash,dashsize=4](179,85)(198,96)
    \Text(178,86)[rt]{\normalsize{\Black{$v_{u,\,d}$}}}
  \end{picture}
 }.
\end{equation*}
The scattering amplitude for the first diagram reads
\begin{align}
  i\mathcal{M}_{Z_1} &=\sum_q\bar{u}^t(p')\,P_R\,N_{\nu_i\,\tilde{Z}}^{7*}\,\frac{i}{4\,\MP}\,\gamma^{\mu}\left[ \slashed{p}-\slashed{p}',\,\gamma^{\rho}\right] \psi_{\mu}^{+\,s}(p) \nonumber\\
  &\qquad\quad\times\frac{i\left( g_{\rho\nu}-\frac{(p_\rho-p'_\rho)(p_\nu-p'_\nu)}{m_Z^2}\right) }{(p-p')^2-m_Z^2+im_Z\Gamma_Z}\,\bar{u}^q(k')\,\frac{ig}{\cos\theta_W}\,\gamma^\nu\,C_V\,u^r(k) \\
  &\simeq\sum_q\frac{ig_Z\,\xi_i\,C_V\,U^*_{\tilde{Z}\tilde{Z}}}{m_Z^2\,\MP}\,\bar{u}^t(p')\,P_R\left( p'^\mu\gamma^\rho-g^{\mu\rho}(\slashed{p}'-\slashed{p})\right) \psi_{\mu}^{+\,s}(p)\,\bar{u}^q(k')\,\gamma_\rho\,u^r(k)\,, \nonumber
\end{align}
where we only take into account the vector part of the $Z$ coupling to quarks and consider only cases where the propagator reduces to a contact interaction. In addition, we need to sum over the valence quarks inside the nucleon. The amplitudes for the second and third diagram can be combined and are, in analogy to the case of the three-body decay, given by
\begin{align}
  i\mathcal{M}_{Z_{2+3}} &=-\sum_q\frac{ig_Z\,m_Z\,\xi_i\,C_V}{m_Z^2\,\MP}\left( g^{\mu\rho}+\frac{p'^\mu(p^\rho-p'^\rho)}{m_Z^2}\right) \left( 1+\sin\beta\,U^*_{\tilde{H}_u^0\tilde{Z}}-\cos\beta\,U^*_{\tilde{H}_d^0\tilde{Z}}\right) \nonumber\\
  &\qquad\qquad\times\bar{u}^t(p')\,P_R\,\psi_\mu^{+\,s}(p)\,\bar{u}^q(k')\,\gamma_\rho\,u^r(k)\,.
\end{align}
We can then write down the squared matrix element for this process:
\begin{align}
  \bar{\abs{\mathcal{M}}^2} &=\frac{1}{8}\sum_{s,t,q,r}(i\mathcal{M}_{Z_1}+i\mathcal{M}_{Z_{2+3}})(i\mathcal{M}_{Z_1}+i\mathcal{M}_{Z_{2+3}})^* \nonumber\\
  &=-\frac{g_Z^2\,\xi_i^2\left( \sum_qC_V\right) ^2}{8\,m_Z^4\,\MP^2}\Bigg[ U_{\tilde{Z}\tilde{Z}}^2\Tr\left[ \left( \slashed{k}'+m_q\right) \gamma_\rho\left( \slashed{k}+m_q\right) \gamma_\sigma\right] \\
  &\quad\qquad\times\Tr\left[ \slashed{p}'\,P_R\left( p'^\mu\gamma^\rho-g^{\mu\rho}(\slashed{p}'-\slashed{p})\right) \slashed{p}\,\Phi_{\mu\nu}(p)\left( p'^\nu\gamma^\sigma-g^{\nu\sigma}(\slashed{p}'-\slashed{p})\right) \right] \nonumber\\
  &\qquad-2\,m_{3/2}\,m_Z\,U_{\tilde{Z}\tilde{Z}}\left( 1+s_\beta \RE U_{\tilde{H}_u^0\tilde{Z}}-c_\beta \RE U_{\tilde{H}_d^0\tilde{Z}}\right) \left( g^{\nu\sigma}+\frac{p'^\nu(p^\sigma-p'^\sigma)}{m_Z^2}\right) \nonumber\\
  &\quad\qquad\times\Tr\left[ \left( \slashed{k}'+m_q\right) \gamma_\rho\left( \slashed{k}+m_q\right) \gamma_\sigma\right] \Tr\left[ \slashed{p}'P_R\left( p'^\mu\gamma^\rho-g^{\mu\rho}(\slashed{p}'-\slashed{p})\right) \Phi_{\mu\nu}(p)\right] \nonumber\\
  &\qquad+m_Z^2\abs{1+s_\beta\,U_{\tilde{H}_u^0\tilde{Z}}-c_\beta\,U_{\tilde{H}_d^0\tilde{Z}}}^2\left( g^{\mu\rho}+\frac{p'^\mu(p^\rho-p'^\rho)}{m_Z^2}\right) \left( g^{\nu\sigma}+\frac{p'^\nu(p^\sigma-p'^\sigma)}{m_Z^2}\right) \nonumber\\
  &\quad\qquad\times\Tr\left[ \left( \slashed{k}'+m_q\right) \gamma_\rho\left( \slashed{k}+m_q\right) \gamma_\sigma\right] \Tr\left[ \slashed{p}'P_R\,\slashed{p}\,\Phi_{\mu\nu}(p)\right] \Bigg]\,. \nonumber
\end{align}
A couple of traces of gamma matrices appear in this expression. Neglecting the masses of the individual quarks the trace including the nucleon current turns out to be
\begin{equation}
  \Tr\left[ \left( \slashed{k}'+m_q\right) \gamma_\rho\left( \slashed{k}+m_q\right) \gamma_\sigma\right] \simeq4\left[ k_\rho k'_\sigma+k_\sigma k'_\rho-g_{\rho\sigma}\,(k\cdot k')\right] ,
  \label{nucleontrace}
\end{equation}
while the traces including the gravitino field are given by
\begin{equation}
 \begin{split}
  &\Tr\left[ \slashed{p}'P_R\left( p'^\mu\gamma^\rho-g^{\mu\rho}(\slashed{p}'-\slashed{p})\right) \slashed{p}\,\Phi_{\mu\nu}(p)\left( p'^\nu\gamma^\sigma-g^{\nu\sigma}(\slashed{p}'-\slashed{p})\right) \right] \\
  &\qquad=\frac{4}{3\,m_{3/2}^2}\left[ \left( p\cdot p'\right) \left( g^{\rho\sigma}\left( m_{3/2}^4-m_{3/2}^2\left( p\cdot p'\right) +\left( p\cdot p'\right) ^2\right) \right. \right. \\
  &\qquad\quad\quad-p^\rho\left. \left( m_{3/2}^2\,p^\sigma+p'^\sigma\left( p\cdot p'\right) \right) \Big) +p'^\rho\left( m_{3/2}^4\,p'^\sigma-p^\sigma\left( p\cdot p'\right) ^2\right) \right] \\
  &\qquad\quad+\frac{2\,i}{3\,m_{3/2}^2}\left( m_{3/2}^4+2\,m_{3/2}^2\left( p\cdot p'\right) -2\left( p\cdot p'\right) ^2\right) \varepsilon^{\rho\sigma\delta\lambda}p_\delta\,p'_\lambda\,, \\
  &\Tr\left[ \slashed{p}'P_R\left( p'^\mu\gamma^\rho-g^{\mu\rho}(\slashed{p}'-\slashed{p})\right) \Phi_{\mu\nu}(p)\right] \\
  &\qquad=\frac{2}{3\,m_{3/2}^2}\left[ \left( p\cdot p'\right) \left( \delta_\nu^\rho\left( 2\,m_{3/2}^2-\left( p\cdot p'\right) \right) +p_\nu\left( 2\,p^\rho+p'^\rho\right) \right. \right. \\
  &\left. \qquad\quad\quad+\left. p'_\nu\left( m_{3/2}^2\,p'^\rho+p^\rho\left( p\cdot p'\right) \right) \right) -i\left( m_{3/2}^2+\left( p\cdot p'\right) \right) g_{\nu\sigma}\,\varepsilon^{\sigma\rho\delta\lambda}p_\delta\,p'_\lambda\right] , \\
  &\Tr\left[ \slashed{p}'P_R\,\slashed{p}\,\Phi_{\mu\nu}(p)\right] \\
  &\qquad=\frac{2}{3\,m_{3/2}^2}\left[ 2\left( m_{3/2}^2\,g_{\mu\nu}-p_\mu p_\nu\right) \left( p\cdot p'\right) +i\,m_{3/2}^2\,\varepsilon_{\mu\nu\rho\sigma}p^\rho\,p'^\sigma\right] .
 \end{split}
 \label{gravitinotraces}
\end{equation}
In these expressions we already replaced the squared four-momenta with the squared particle masses:
\begin{equation}
  p^2=m_{3/2}^2\,,\quad p'^2=m_\nu^2\simeq 0\,,\quad k^2=k'^2=m_N^2\,.
\end{equation}
Now we want to use the complete kinematics of the scattering process in order to calculate the scattering cross section. Using equation~(\ref{scatteringscalarproduct}) we find
\begin{equation}
 \begin{split}
  (p\cdot p') &=\frac{1}{2}\left( m_{3/2}^2+m_\nu^2-t\right) \simeq\frac{1}{2}\left( m_{3/2}^2-t\right) ,\qquad(p\cdot k)=m_NE\,, \\
  (p\cdot k') &=\frac{1}{2}\left( m_{3/2}^2+t-m_\nu^2+2\,m_NE\right) \simeq\frac{1}{2}\left( m_{3/2}^2+t+2\,m_NE\right) , \\
  (k\cdot k') &=\frac{1}{2}\left( 2\,m_N^2-t\right) ,\quad\qquad\qquad\qquad\qquad\qquad\!(k\cdot p')=\frac{1}{2}\left( t+2\,m_NE\right) , \\
  (p'\cdot k') &=\frac{1}{2}\left( m_{3/2}^2+2\,m_NE-m_\nu^2\right) \simeq\frac{1}{2}\left( m_{3/2}^2+2\,m_NE\right) .
 \end{split}
\end{equation}
Here we used the Mandelstam variables
\begin{equation}
  s=(p+k)^2=(p'+k')^2=m_{3/2}^2+m_N^2+2\,m_NE\quad\text{and}\quad t=(p-p')^2=(k-k')^2.
\end{equation}
The differential cross section is finally given by:
\begin{align}
  \frac{d\sigma}{dt}
  &=\frac{g_Z^2\,\xi_i^2\left( \sum_qC_V\right) ^2t}{192\,\pi\,m_Z^4\,\MP^2\abs{\vec{v}}^2} \Bigg[\abs{U_{\tilde{Z}\tilde{Z}}}^2\Bigg( 1-\frac{3}{4}\,\frac{m_{3/2}^2}{m_N^2}-5\,\frac{E}{m_N}-\frac{3}{4}\,\frac{t}{m_N^2}-\frac{t}{m_{3/2}^2}-6\,\frac{E^2}{m_{3/2}^2} \nonumber\\
  &\qquad\quad-\frac{t\,E}{m_N\,m_{3/2}^2}+\frac{1}{4}\,\frac{t^2}{m_N^2\,m_{3/2}^2}+\frac{t}{m_{3/2}^2}\left( 2\,\left( \frac{E^2}{m_{3/2}^2}+\frac{t\,E}{m_N\,m_{3/2}^2}\right) +\frac{1}{4}\,\frac{t^2}{m_N^2\,m_{3/2}^2}\right) \!\Bigg) \nonumber\\
  &\qquad-8\,\frac{m_N\,m_Z}{t}\,U_{\tilde{Z}\tilde{Z}}\left( 1+s_\beta \RE U_{\tilde{H}_u^0\tilde{Z}}-c_\beta \RE U_{\tilde{H}_d^0\tilde{Z}}\right) \Bigg( \frac{E^3}{m_{3/2}^3}+\frac{1}{2}\,\frac{E}{m_{3/2}}-\frac{1}{2}\,\frac{t\,E}{m_{3/2}^3} \nonumber\\
  &\qquad\quad+\frac{1}{8}\,\frac{m_{3/2}}{m_N}+\frac{t}{m_N\,m_{3/2}}\left( -\frac{E^2}{m_{3/2}^2}-\frac{1}{8}\,\frac{t}{m_{3/2}^2}-\frac{1}{8}\,\frac{t\,E}{m_N\,m_{3/2}^2}-\frac{5}{8}\,\frac{E}{m_N}\right) \nonumber\\
  &\qquad\quad+\frac{1}{16}\,\frac{t}{m_N^2}\left( \frac{t^2}{m_N\,m_{3/2}^3}-3\,\frac{m_{3/2}}{m_N}\right) -\frac{1}{2}\,\frac{E^2}{m_N\,m_{3/2}} \\
  &\qquad\quad+\frac{1}{4}\,\frac{t\,E}{m_N\,m_Z^2}\left( \frac{t\,E}{m_{3/2}^3}-\frac{E}{m_{3/2}}+\frac{3}{4}\,\frac{t^2}{m_N\,m_{3/2}^3}-\frac{1}{4}\,\frac{m_{3/2}}{m_N}-\frac{1}{2}\,\frac{t}{m_N\,m_{3/2}}\right) \!\Bigg) \nonumber\\
  &\qquad+\frac{m_Z^2}{t}\abs{1+s_\beta\,U_{\tilde{H}_u^0\tilde{Z}}-c_\beta\,U_{\tilde{H}_d^0\tilde{Z}}}^2\Bigg(1+2\,\frac{E^2}{m_{3/2}^2}-\frac{t}{m_{3/2}^2}-2\,\frac{t\,E^2}{m_{3/2}^4} \nonumber\\
  &\qquad\quad+\frac{E}{m_N}-\frac{1}{2}\,\frac{t}{m_N^2}+\frac{1}{2}\,\frac{t^2}{m_N^2\,m_{3/2}^2}-\frac{t^2E}{m_N\,m_{3/2}^4} \nonumber\\
  &\qquad\quad+\frac{t}{m_Z^2}\Bigg( 4\,\frac{E^2}{m_{3/2}^2}-\frac{3}{2}-\frac{E}{m_N}+\frac{3}{2}\,\frac{t}{m_{3/2}^2}+4\,\frac{t\,E^2}{m_{3/2}^4}+\frac{1}{2}\,\frac{t^2}{m_{3/2}^4}-4\,\frac{m_N\,t\,E}{m_{3/2}^4} \nonumber\\
  &\qquad\qquad+\frac{t^2\,E}{m_N\,m_{3/2}^4}-8\,\frac{m_N^2\,E^2}{m_{3/2}^4}+4\,\frac{m_N\,E}{t}+\frac{1}{2}\,\frac{m_{3/2}^2}{t}+8\,\frac{m_N^2\,E^2}{m_{3/2}^2\,t}\Bigg) \nonumber\\
  &\qquad\quad+\frac{t^2}{m_Z^4}\Bigg( \frac{3}{4}-2\,\frac{m_N\,E}{t}-\frac{1}{4}\,\frac{m_{3/2}^2}{t}-\frac{3}{4}\,\frac{t}{m_{3/2}^2}-4\,\frac{m_N^2\,E^2}{t\,m_{3/2}^2}+\frac{1}{4}\,\frac{t^2}{m_{3/2}^4} \nonumber\\
  &\qquad\qquad+2\,\frac{m_N\,t\,E}{m_{3/2}^4}+4\,\frac{m_N^2\,E^2}{m_{3/2}^4}\Bigg) \Bigg]\,. \nonumber
\end{align}

\newpage

\section[Inelastic Gravitino--Nucleon Scattering via Photon Exchange]{Inelastic Gravitino--Nucleon Scattering via Photon Exchange}

At tree level there is only one diagram contributing to the inelastic gravitino--nucleon scattering via the exchange of a photon:
\begin{equation*}
 \parbox{6.3cm}{
  \begin{picture}(183,173) (107,-22)
    \SetWidth{0.5}
    \SetColor{Black}
    \Photon(198,96)(198,29){-6}{3}
    \Line[arrow,arrowpos=0.5,arrowlength=3.75,arrowwidth=1.5,arrowinset=0.2](227,113)(256,130)
    \Line[arrow,arrowpos=0.5,arrowlength=3.75,arrowwidth=1.5,arrowinset=0.2](198,29)(256,-5)
    \Line[arrow,arrowpos=0.5,arrowlength=3.75,arrowwidth=1.5,arrowinset=0.2](140,-5)(198,29)
    \Line[double,sep=4](140,130)(198,96)
    \Line[arrow,arrowpos=1,arrowlength=3.75,arrowwidth=1.5,arrowinset=0.2](238,104)(257,115)
    \Line[arrow,arrowpos=1,arrowlength=3.75,arrowwidth=1.5,arrowinset=0.2](153,107)(172,96)
    \Line[arrow,arrowpos=1,arrowlength=3.75,arrowwidth=1.5,arrowinset=0.2](153,18)(172,29)
    \Line[arrow,arrowpos=1,arrowlength=3.75,arrowwidth=1.5,arrowinset=0.2](223,29)(242,18)
    \Vertex(198,96){2.5}
    \Vertex(198,29){2.5}
    \Text(138,131)[rb]{\normalsize{\Black{$\psi_{3/2}$}}}
    \Text(257,131)[lb]{\normalsize{\Black{$\nu_i$}}}
    \Line(198,96)(227,113)
    \Photon(198,96)(227,113){-6}{2}
    \Line(225,107.001)(229,118.999)\Line(221.001,115)(232.999,111)
    \Text(216,101)[lt]{\normalsize{\Black{$\tilde{\gamma}$}}}
    \Text(206,62)[l]{\normalsize{\Black{$\gamma$}}}
    \Text(138,-6)[rt]{\normalsize{\Black{$N$}}}
    \Text(257,-6)[lt]{\normalsize{\Black{$N$}}}
    \Text(164,99)[rt]{\normalsize{\Black{$p$}}}
    \Text(247,110)[lt]{\normalsize{\Black{$p'$}}}
    \Text(164,25)[rb]{\normalsize{\Black{$k$}}}
    \Text(233,25)[lb]{\normalsize{\Black{$k'$}}}
    \Arc[arrow,arrowpos=0.5,arrowlength=3.75,arrowwidth=1.5,arrowinset=0.2](197.5,185.75)(67.752,-124.628,-55.372)
  \end{picture}
 }.
\end{equation*}
The scattering amplitude for this diagram is similar to the respective three-body decay amplitude and reads
\begin{align}
  i\mathcal{M}_\gamma &=\sum_q\bar{u}^t(p')\,P_R\,N_{\nu_i\,\tilde{\gamma}}^{7*}\,\frac{i}{4\,\MP}\,\gamma^{\mu}\left[ \slashed{p}-\slashed{p}',\,\gamma^{\rho}\right] \psi_{\mu}^{+\,s}(p)\,\frac{ig_{\rho\nu}}{(p-p')^2}\,\bar{u}^q(k')\,i\,Q\,e\,\gamma^\nu u^r(k) \nonumber\\
  &\simeq-\sum_q\frac{i\,Q\,e\,\xi_i\,U^*_{\tilde{\gamma}\tilde{Z}}}{\MP\,(p-p')^2}\,\bar{u}^t(p')P_R\left( p'^\mu\gamma^\rho-g^{\mu\rho}(\slashed{p}'-\slashed{p})\right) \psi_{\mu}^{+\,s}(p)\,\bar{u}^q(k')\,\gamma_\rho\,u^r(k)\,,
\end{align}
where we need to sum over the valence quarks in the nucleon as in the case of scattering via $Z$ exchange. We find then the following squared matrix element:
\begin{equation}
 \begin{split}
  \bar{\abs{\mathcal{M}}^2} &=\frac{1}{8}\sum_{s,t,q,r}i\mathcal{M}_\gamma\,(i\mathcal{M}_\gamma)^* \\
  &=-\frac{\xi_i^2\left( \sum_qQ\right) ^2e^2\abs{U_{\tilde{\gamma}\tilde{Z}}}^2}{8\,\MP^2\,(p-p')^4}\,\Tr\left[ \left( \slashed{k}'+m_q\right) \gamma_\rho\left( \slashed{k}+m_q\right) \gamma_\sigma\right] \\
  &\qquad\qquad\times\Tr\left[ \slashed{p}'P_R\left( p'^\mu\gamma^\rho-g^{\mu\rho}(\slashed{p}'-\slashed{p})\right) \slashed{p}\,\Phi_{\mu\nu}(p)\left( p'^\nu\gamma^\sigma-g^{\nu\sigma}(\slashed{p}'-\slashed{p})\right) \right] .
 \end{split}
\end{equation}
The traces appearing in this matrix element have already been presented in equations~(\ref{nucleontrace}) and~(\ref{gravitinotraces}), and the kinematics of this scattering process is the same as for the case of $Z$ boson and Higgs exchange. This finally leads to the differential cross section
\begin{align}
  \frac{d\sigma}{dt}
  &=\frac{\xi_i^2\left( \sum_qQ\right) ^2e^2\abs{U_{\tilde{\gamma}\tilde{Z}}}^2}{192\,\pi\,t\,\MP^2\abs{\vec{v}}^2} \Bigg( 1-\frac{3}{4}\,\frac{m_{3/2}^2}{m_N^2}-5\,\frac{E}{m_N}-\frac{3}{4}\,\frac{t}{m_N^2}-\frac{t}{m_{3/2}^2}-6\,\frac{E^2}{m_{3/2}^2} \\
  &\qquad\quad-\frac{t\,E}{m_N\,m_{3/2}^2}+\frac{1}{4}\,\frac{t^2}{m_N^2\,m_{3/2}^2}+\frac{t}{m_{3/2}^2}\left( 2\,\left( \frac{E^2}{m_{3/2}^2}+\frac{t\,E}{m_N\,m_{3/2}^2}\right) +\frac{1}{4}\,\frac{t^2}{m_N^2\,m_{3/2}^2}\right) \!\Bigg)\,. \nonumber
\end{align}
}

\phantomsection

\end{document}